\documentclass[a4paper,11pt]{article}
\pdfoutput=1 % if your are submitting a pdflatex (i.e. if you have
\usepackage{jheppub} % for details on the use of the package, please
\usepackage{bm} 
\usepackage{multirow}

\usepackage[T1]{fontenc} % if needed
\usepackage[pdftex]{color}

\title{\boldmath 
$P$-wave quarkonium wavefunctions at the origin 
\\
in the $\overline{\rm MS}$ scheme}
\preprint{TUM-EFT 145/21}

\author{Hee~Sok~Chung}
\affiliation{Physik-Department, Technische Universit\"at M\"unchen,
James-Franck-Str. 1, 85748 Garching, Germany}
\affiliation{Excellence Cluster ORIGINS,
Boltzmannstrasse 2, D-85748 Garching, Germany}

\emailAdd{heesok.chung@tum.de}

\abstract{
We compute $P$-wave quarkonium wavefunctions at the origin in the
$\overline{\rm MS}$ scheme based on nonrelativistic effective field theories. 
We include nonperturbative effects from the long-distance behaviors of the
potential, while the short-distance behaviors are determined from perturbative
QCD. We obtain $\overline{\rm MS}$-renormalized $P$-wave quarkonium
wavefunctions at the origin that have the correct scale dependences that are 
expected
from factorization formalisms, so that the dependences on the scheme and scale
cancel in physical quantities. 
This greatly reduces the theoretical uncertainties associated with scheme and
scale dependences in predictions of decay and production rates. 
Based on the calculation of the $P$-wave wavefunctions at the origin in this
work, we make
first-principles predictions of electromagnetic decay rates and 
exclusive electromagnetic production rates of $P$-wave charmonia and 
bottomonia, and compare them with measurements. 
}

\begin{document} 
\maketitle
\flushbottom

%==============================================================================
\section{Introduction}
\label{sec:intro}
%==============================================================================

Production and decay of $P$-wave heavy quarkonia have played important roles 
in studies of nonrelativistic effective field theories of QCD. 
An early triumph of rigorous QCD analyses of heavy quarkonia based on the 
nonrelativistic QCD (NRQCD) effective field theory is the resolution of 
the problem of infrared divergences in $P$-wave heavy quarkonium production 
and decay rates~\cite{Bodwin:1992ye, Bodwin:1992qr, Bodwin:1992eb, 
Bodwin:1994jh}. The infrared divergences in perturbative QCD calculations are
removed through renormalization of 
NRQCD long-distance matrix elements (LDMEs), which are well-defined
universal quantities that encode the nonperturbative nature of heavy quarkonium
states. 
This way, nonrelativistic effective field theory methods eliminate 
the need for ad hoc remedies for cutting off singularities, 
and make possible computations of production and 
decay rates of $P$-wave heavy quarkonia based on first principles. 
Ongoing experimental activities 
in lepton and hadron colliders 
call for continued theoretical efforts for making 
accurate QCD-based predictions of $P$-wave quarkonium production and
decay~\cite{Chung:2008km, Brambilla:2010cs, 
Brambilla:2014jmp, Brambilla:2020ojz, Brambilla:2021abf}. 

NRQCD describes hard processes like decays of quarkonia into light particles
in terms of factorization formulae, which separate the
perturbative short-distance contributions from the
nonperturbative physics that the LDMEs describe~\cite{Bodwin:1994jh}. 
The LDMEs have known scalings in $v$, the typical heavy-quark velocity inside
the quarkonium, and the factorization formulae are organized in powers of $v$.
First-principles calculations of decay and production rates 
can be made by computing the short-distance coefficients and LDMEs in QCD. 
The short-distance coefficients can be determined perturbatively by matching
the factorization formulae with perturbative QCD amplitudes. Infrared 
divergences in perturbative QCD amplitudes are absorbed into the LDMEs, and
leave their trace in the short-distance coefficients as dependences on the
scheme and scale at which the LDMEs are renormalized. 
The scheme and scale dependences cancel between the LDMEs and short-distance
coefficients in the factorization formulae, as long as the LDMEs are
renormalized in the same scheme. 
The scale at which the LDMEs are renormalized is
often called the NRQCD factorization scale. 
In order to make accurate predictions of heavy quarkonium production and 
decay rates, it is crucial that we obtain LDMEs that have the 
correct dependences on the factorization scale, especially when the
short-distance coefficients have strong scheme and scale dependences. 
Loop-level calculations of short-distance coefficients show that
strong dependences on the factorization scale 
in the $\overline{\rm MS}$ scheme begin to appear 
from next-to-next-to-leading order in the strong coupling $\alpha_s$
in electromagnetic decay rates and exclusive electromagnetic production cross
sections of heavy quarkonia~\cite{Czarnecki:1997vz,
Beneke:1997jm, Czarnecki:2001zc, Kniehl:2006qw, Sang:2015uxg, Sang:2020fql}. 
Cancellation of the factorization scale dependence in the 
factorization formulae requires calculations of the LDMEs 
in the same scheme that has the correct dependence
on the $\overline{\rm MS}$ scale. 
This has so far not been possible for $P$-wave heavy quarkonia, 
and most phenomenological studies have relied on model calculations of LDMEs,
which are not accurate enough to reproduce their scheme and scale dependences 
correctly. 

For $S$-wave heavy quarkonium states, 
a recent progress in ref.~\cite{Chung:2020zqc} 
has made possible first-principles calculations of LDMEs 
in the $\overline{\rm MS}$ scheme that 
correctly reproduce the factorization scale dependences at two-loop level. 
The factorization scale dependences in the LDMEs arise from inclusion of the 
ultraviolet (UV) divergent effects that come from interactions in NRQCD 
that are suppressed by powers of $1/m$, where $m$ is the heavy quark pole mass. 
This lead to accurate predictions of decay rates and decay constants of 
$S$-wave heavy quarkonia through exact cancellations of factorization scale 
dependences in the factorization formulas at two-loop accuracy, 
which has not been possible outside of perturbative QCD~\cite{Hoang:1997ui}. 
An unexpected consequence of this calculation is that large cancellations 
occur between the scheme-dependent finite parts of the
two-loop short-distance coefficients and the finite parts in the LDMEs that
remain after subtraction of the UV divergences, which improves the convergence
of the corrections and enhances the reliability of the theoretical predictions
that are based on fixed-order calculations of short-distance coefficients. 
As short-distance coefficients for electromagnetic decay and exclusive
electromagnetic production rates of $P$-wave heavy quarkonia have recently
become available to two-loop accuracy~\cite{Sang:2015uxg,Sang:2020fql}, it is
highly desirable to extend the calculation in ref.~\cite{Chung:2020zqc} to
$P$-wave states. 

In this paper, we compute NRQCD LDMEs of $P$-wave heavy quarkonia in the
$\overline{\rm MS}$ scheme by extending the calculation in 
ref.~\cite{Chung:2020zqc} to $P$-wave states. 
We focus on the LDMEs that appear at leading order in $v$, 
because short-distance coefficients associated with LDMEs beyond 
leading order in $v$ are available only
at tree level~\cite{Brambilla:2017kgw}. 
We compute the LDMEs by using the potential NRQCD (pNRQCD) effective field
theory framework, which provides expressions of NRQCD LDMEs in terms of
quarkonium wavefunctions and their derivatives at the 
origin~\cite{Pineda:1997bj, Brambilla:1999xf, Brambilla:2001xy,
Brambilla:2002nu, Brambilla:2004jw}. The quarkonium wavefunctions 
are bound-state solutions of a Schr\"odinger equation, whose potential is
obtained by matching NRQCD and pNRQCD order by order in expansion in powers of
$1/m$. 
We work in the strong coupling regime, where $mv \gtrsim \Lambda_{\rm QCD} \gg
mv^2$, which is adequate for describing $P$-wave heavy
quarkonia~\cite{Brambilla:2002nu, Brambilla:2020xod}. 
Here, $mv$ and $mv^2$ correspond to the typical sizes of 
the momentum and binding energy of the
heavy quark $Q$ and antiquark $\bar Q$ inside the quarkonium at rest,
respectively. 
In this case, the nonperturbative long-distance behavior of the potential 
is important in calculation of quarkonium wavefunctions. 

In pNRQCD calculations of NRQCD LDMEs, 
the UV divergences are reproduced by singularities in
the wavefunctions or their derivatives at the origin. 
The singularities in the wavefunctions are generated by the corrections to 
the potential beyond leading power in $1/m$, which can diverge faster than the
potential at leading power in $1/m$ in the limit where the distance between the
$Q$ and $\bar Q$ vanishes. Hence, we reproduce the UV
divergences in the NRQCD LDMEs by including the corrections to the 
wavefunctions from the potentials at higher orders in $1/m$. 
The singularities are sensitive to the divergent short-distance behaviors of
the potential, so that in the strong coupling regime, it is necessary to keep
both the nonperturbative long-distance behavior and the perturbative
short-distance behavior of the potential. 
In this case, a difficulty arises from the fact that the nonperturbative 
long-distance behavior of the potential is known only in position space, while
renormalization of the UV divergence in the $\overline{\rm MS}$ scheme is
done in momentum space. 
This difficulty can be overcome by using the method adopted in
refs.~\cite{Hoang:1997ui, Kiyo:2010jm, Chung:2020zqc}, 
where the singularities in the wavefunctions are
first regularized in position space, which can then be converted to dimensional
regularization (DR) by
computing the scheme conversion. The scheme conversion depends only on the
divergent behavior of the wavefunctions at the origin, and can be computed in
perturbative QCD. Then, the dimensionally regulated wavefunctions at the origin
can be renormalized in the $\overline{\rm MS}$ scheme. 
The result can be combined with the $\overline{\rm MS}$ calculations of the
short-distance coefficients to make predictions of decay and production rates
of $P$-wave heavy quarkonia. 

This paper is organized as follows. In section~\ref{sec:LDMEs}, we define the 
NRQCD LDMEs that appear in $P$-wave production and decay rates in forms that
are suitable for calculations in DR. 
The pNRQCD expressions for the NRQCD LDMEs are given in 
section~\ref{sec:pnrqcd}. 
We outline the position-space calculation of the wavefunctions in
section~\ref{sec:wavefunctions}, and compute the scheme conversion from
position-space regularization to the $\overline{\rm MS}$ scheme 
in section~\ref{sec:conversion}.
Based on the calculations of the $\overline{\rm MS}$ wavefunctions, we compute
decay rates, production cross sections, and decay constants of $P$-wave
charmonia and bottomonia in section~\ref{sec:results}. We conclude in 
section~\ref{sec:summary}.

%==============================================================================
\section{\boldmath NRQCD long-distance matrix elements and their renormalization} 
\label{sec:LDMEs}
%==============================================================================

Electromagnetic decay rates and exclusive electromagnetic production rates
of $P$-wave heavy quarkonia at leading orders in $v$ 
involve the following dimension-5 
LDMEs~\cite{Bodwin:1994jh}:
%-------------
\begin{subequations}
\label{eq:ldmes_def_4d}
\begin{eqnarray}
%-------------
\langle {\cal Q} | {\cal O} (^3P_0) | {\cal Q} \rangle &=& 
\frac{1}{3} 
\left| 
\langle 0 | \chi^\dag \left(- \frac{i}{2} \overleftrightarrow{\bm{D}} \right) 
\cdot \bm{\sigma} \psi | {\cal Q} \rangle \right|^2, 
\\
\langle {\cal Q} | {\cal O} (^3P_1) | {\cal Q} \rangle &=& 
\frac{1}{2} 
\left| 
\langle 0 | \chi^\dag \left(- \frac{i}{2} \overleftrightarrow{\bm{D}} \right) 
\times \bm{\sigma} \psi | {\cal Q} \rangle
\right|^2,
\\
\langle {\cal Q} | {\cal O} (^3P_2) | {\cal Q} \rangle &=&
\left|
\langle 0 | \chi^\dag \left(- \frac{i}{2} \overleftrightarrow{D}^{(i} \right)
\sigma^{j)} \psi | {\cal Q} \rangle
\right|^2,
\\
\langle {\cal Q} | {\cal O} (^1P_1) | {\cal Q} \rangle &=&
\left|
\langle 0 | \chi^\dag \left(- \frac{i}{2} \overleftrightarrow{\bm{D}} \right)
 \psi | {\cal Q} \rangle \right|^2.
%-------------
\end{eqnarray}
\end{subequations}
%-------------
Here, $|{\cal Q} \rangle$ is a nonrelativistically normalized 
heavy quarkonium state at rest, 
$|0\rangle$ is the QCD vacuum, 
$\bm{D} = \bm{\nabla} -i g_s \bm{A}$ is the covariant derivative 
with $\chi^\dag \overleftrightarrow{\bm{D}} \psi 
= \chi^\dag \bm{D} \psi - (\bm{D} \chi)^\dag \psi$,  
$\bm{A}$ is the gluon field, 
$\bm{\sigma}$ is a Pauli matrix, 
and $\psi$ and $\chi$ are Pauli spinor fields
that annihilate and create the heavy quark and antiquark, respectively. 
$T^{(ij)}$ is the symmetric traceless part of a tensor $T^{ij}$, which we
define in $d-1$ spatial dimensions by 
$T^{(ij)} \equiv \frac{1}{2} ( T^{ij} + T^{ji} ) -\frac{1}{d-1} \delta^{ij}
T^{kk}$. 
The spectroscopic notation $^{2 S +1} P_J$ denote the spin ($S$), orbital
($P$), and total angular momentum ($J$) 
of the $Q \bar Q$ in the leading $Q \bar Q$ Fock state of the quarkonium state
$|\cal Q \rangle$. Hence, 
the LDMEs $\langle {\cal Q} | {\cal O} (^3P_J) | {\cal Q}
\rangle$, where $J=0$, 1, or 2, appear in processes involving heavy quarkonia
with positive parity and charge conjugation with total spin $J$, such as 
$\chi_{cJ}$ and $\chi_{bJ}$. 
The $\langle {\cal Q} | {\cal O} (^1P_1) | {\cal Q}
\rangle$ appear in processes involving states with positive parity and 
negative charge conjugation with total spin $1$, such as the $h_c$ and $h_b$. 

The LDMEs in eqs.~(\ref{eq:ldmes_def_4d}) depend on the scheme and scale
$\Lambda$ at which they are renormalized. 
The anomalous dimensions of the NRQCD LDMEs at leading order in $v$ 
are given by~\cite{Hoang:2006ty}
%-------------
\begin{subequations}
\label{eq:RG_leading}
\begin{eqnarray}
%-------------
\frac{d}{d \log \Lambda} \log 
\langle 0 | \chi^\dag \left(- \frac{i}{2} \overleftrightarrow{\bm{D}} \right)
\cdot \bm{\sigma} \psi | {\cal Q} \rangle
&=& 
\alpha_s^2 \gamma_{^3P_0}^{(2)} + O(\alpha_s^3),
\\
\frac{d}{d \log \Lambda} \log 
\langle 0 | \chi^\dag \left(- \frac{i}{2} \overleftrightarrow{\bm{D}} \right)
\times \bm{\sigma} \psi | {\cal Q} \rangle
&=& 
\alpha_s^2 \gamma_{^3P_1}^{(2)} + O(\alpha_s^3),
\\
\frac{d}{d \log \Lambda} \log 
\langle 0 | \chi^\dag \left(- \frac{i}{2} \overleftrightarrow{D}^{(i} \right)
{\sigma}^{j)} \psi | {\cal Q} \rangle
&=& 
\alpha_s^2 \gamma_{^3P_2}^{(2)} + O(\alpha_s^3),
\\
\frac{d}{d \log \Lambda} \log 
\langle 0 | \chi^\dag \left(- \frac{i}{2} \overleftrightarrow{\bm{D}}\right)
\psi | {\cal Q} \rangle
&=& 
\alpha_s^2 \gamma_{^1P_1}^{(2)} + O(\alpha_s^3), 
%-------------
\end{eqnarray}
\end{subequations}
%-------------
where\footnote{Our expressions of the anomalous dimensions have opposite signs
compared to refs.~\cite{Hoang:2006ty,Sang:2015uxg,Sang:2020fql}, 
because we define them from the scale dependences of the LDMEs, rather than the
short-distance coefficients.}
%-------------
\begin{subequations}
\label{eq:NRQCDRG}
\begin{eqnarray}
%-------------
\gamma_{^3P_0}^{(2)} &=&
\frac{2}{3} C_F^2 + \frac{1}{6} C_F C_A, 
\\
\gamma_{^3P_1}^{(2)} &=&
\frac{5}{12} C_F^2 + \frac{1}{6} C_F C_A, 
\\
\gamma_{^3P_2}^{(2)} &=&
\frac{13}{60} C_F^2 + \frac{1}{6} C_F C_A , 
\\
\gamma_{^1P_1}^{(2)} &=&
\frac{1}{3} C_F^2 + \frac{1}{6} C_F C_A. 
%-------------
\end{eqnarray}
\end{subequations}
%-------------
Here, $C_F = (N_c^2-1)/(2 N_c)$, $C_A = N_c$, and $N_c = 3$ is the number of
colors. 
We display the evolution equations at the amplitude level, because matching
calculations for electromagnetic decay rates and exclusive electromagnetic
production rates can be done by comparing $Q \bar Q$ amplitudes in QCD and
NRQCD. 
We note that, similarly to the $S$-wave case, the anomalous dimensions of the
$P$-wave LDMEs begin at two-loop level (order $\alpha_s^2$).
We neglect the contributions to the right-hand sides of
eqs.~(\ref{eq:RG_leading}) that come from LDMEs of
higher orders in $v$, because their contributions to decay and production rates
have so far been computed only at tree level~\cite{Brambilla:2017kgw}. 
The nonvanishing of the anomalous dimensions at order $\alpha_s^2$ imply that,
in carrying out the perturbative matching at two-loop level,
the NRQCD LDMEs contain UV divergences that must be renormalized. 
In order to renormalize the LDMEs in the $\overline{\rm MS}$ scheme, we use
DR in $d=4-2 \epsilon$ spacetime dimensions to regulate the UV divergences. 
In perturbative matching calculations, renormalization of the LDMEs in the 
$\overline{\rm MS}$ scheme at two-loop level is carried out in the 
following form :
%-------------
\begin{equation}
\label{eq:ldme_renormalization}%
%-------------
\langle Q \bar Q | {\cal O}(^{2 S+1}P_J) | Q \bar Q \rangle 
^{\overline{\rm MS}}
= 
\left[ 1 - \frac{\alpha_s^2}{2 \epsilon_{\rm UV}} \gamma_{^{2S+1}P_J}^{(2)} 
+ O(\alpha_s^3) 
\right] 
\langle Q \bar Q | {\cal O}(^{2S+1}P_J) | Q \bar Q \rangle
_{\textrm{bare}} ^{\rm DR}. 
%-------------
\end{equation}
%-------------
Here the LDMEs are written in terms of perturbative $Q \bar Q$ states, 
and computed in perturbation theory as series in $\alpha_s$. 
We use the subscript UV to emphasize the ultraviolet origin of the 
$1/\epsilon$ pole. 
The terms in the square brackets on the
right-hand side correspond to the multiplicative renormalization factor 
to two-loop accuracy, and the bare LDME is computed in DR in $d=4-2 \epsilon$ 
spacetime dimensions. In computing the perturbative LDMEs, we associate 
a factor of $( \Lambda^2 \frac{e^{\gamma_{\rm E}}}{4 \pi} )^\epsilon$ for 
every loop integral, so that the LDME on the left-hand side of
eq.~(\ref{eq:ldme_renormalization}) is renormalized in the $\overline{\rm MS}$
scheme at scale $\Lambda$. 
In case of the nonperturbative LDMEs that are computed on quarkonium states,
the $\overline{\rm MS}$-renormalized LDMEs can also be defined by using 
eq.~(\ref{eq:ldme_renormalization}) with the
perturbative $Q \bar Q$ states replaced by quarkonium states ${\cal Q}$, 
provided that the UV divergences in the (nonperturbative) bare LDMEs on
quarkonium states are regularized in DR and expanded in powers of $\alpha_s$. 

If we expand the right-hand side of eq.~(\ref{eq:ldme_renormalization}) in
powers of $\alpha_s$ to two-loop accuracy, we obtain 
%-------------
\begin{eqnarray}
\label{eq:ldme_renormalization2}%
%-------------
\langle Q \bar Q | {\cal O}(^{2 S+1}P_J) | Q \bar Q \rangle
^{\overline{\rm MS}}
&=&
\langle Q \bar Q | {\cal O}(^{2S+1}P_J) | Q \bar Q \rangle
_{\textrm{bare, two-loop level}} ^{\rm DR}
\nonumber \\ && 
- \frac{\alpha_s^2}{2 \epsilon_{\rm UV}} \gamma_{^{2S+1}P_J}^{(2)}
\langle Q \bar Q | {\cal O}(^{2S+1}P_J) | Q \bar Q \rangle
_{\textrm{bare, tree level}} ^{\rm DR},
%-------------
\end{eqnarray}
%-------------
where the subtraction term in the second line removes the UV pole in the
two-loop level bare LDME. The subtraction term can have finite parts of order
$\epsilon^0$, which comes from the order-$\epsilon$ contribution to the
tree-level bare LDME. Hence, both the tree- and two-loop level bare 
LDMEs must be computed in $d$ spacetime dimensions in order to obtain the
correct finite parts. 
In both the spin triplet and spin singlet cases, the LDMEs defined in 
eqs.~(\ref{eq:ldmes_def_4d}) 
do not generalize to $d$ spacetime dimensions in a straightforward way; 
in the $S=1$ case, the operators ${\cal O}(^{3}P_J)$
are defined by using the irreducible representations of a rank-2 tensor under
rotation in 3 spatial dimensions, which do not generalize to arbitrary spatial
dimensions. 
In the $S=0$ case, matching calculations in DR 
using the standard threshold expansion method in 
refs.~\cite{Braaten:1996jt, Braaten:1996rp} lead to the totally antisymmetric 
product of three Pauli matrices given by $\{ \sigma^i,[\sigma^j,\sigma^k] \}$
that appear between the $\chi^\dag$ and $\psi$ fields.
While in 3 spatial dimensions this matrix 
is proportional to the $2 \times 2$ identity matrix, this is not generally 
true in DR. 
A similar combination of Pauli matrices is obtained in the 
covariant spin projector method~\cite{Kuhn:1979bb, Guberina:1980dc}, if one
uses $\gamma_5$ in the t'Hooft-Veltman scheme. 
In principle, any scheme that defines the LDMEs in $d$ spacetime dimensions 
is valid as long as the LDMEs reduce to the 3-dimensional expressions in 
eqs.~(\ref{eq:ldmes_def_4d}) in the limit $d \to 4$, but the scheme dependence
will cancel in decay and production rates only when the short-distance
coefficients are computed in the same scheme. 

We now define the scheme that we use in this paper to compute the LDMEs in 
$d$ spacetime dimensions, which is the same scheme that is used in existing
calculations of short-distance coefficients at two-loop level. 

In the $S=0$ case, the following definition is consistent with loop-level
calculations of short-distance coefficients that employs the 
t'Hooft-Veltman scheme for $\gamma_5$:
%-------------
\begin{equation}
%-------------
\langle {\cal Q} | {\cal O} (^1P_1) | {\cal Q} \rangle ^{\rm DR} =
\left| 
\langle 0 | \chi^\dag \left(- \frac{i}{2} \overleftrightarrow{\bm{D}} \right)
\sigma_5 \psi | {\cal Q} \rangle \right|^2, 
%-------------
\end{equation}
%-------------
where $\sigma_5$ is a $2 \times 2$ matrix that reduces to the identity matrix
in the limit $d \to 4$, and commutes with $\sigma^i$ with $3$-dimensional
indices, while it anticommutes with $\sigma^i$ with $d-4$-dimensional indices.
This definition is also consistent with the threshold expansion method in
refs.~\cite{Braaten:1996jt, Braaten:1996rp}, as long as we assume 
$\{ \sigma^i,[\sigma^j,\sigma^k] \}$ have same commutation and anticommutation
properties as $\sigma_5$. 
This form of $\sigma_5$ has been used in calculations of short-distance
coefficients for $S$-wave spin-singlet quarkonia in
refs.~\cite{Czarnecki:2001zc, Feng:2015uha}. 
We note that an explicit form of $\sigma_5$ is unnecessary in dimensionally 
regulated matching calculations and also in calculations in this paper. 

In the $S=1$ case, we use the fact that any rank-2 tensor in arbitrary spatial
dimensions can be written as the sum of its trace, antisymmetric, and symmetric
traceless parts; that is, we use the identity 
$T^{ij} = T^{ij}_S + T^{ij}_V + T^{ij}_T$, where 
%-------------
\begin{subequations}
\begin{eqnarray}
%-------------
T^{ij}_S &=& \frac{\delta^{ij}}{d-1} T^{kk}, \\
T^{ij}_V &=& T^{[ij]} \equiv \frac{1}{2} \left( T^{ij} - T^{ji} \right), \\
T^{ij}_T &=& T^{(ij)} \equiv \frac{1}{2} \left( T^{ij} + T^{ji} \right) 
- \frac{\delta^{ij}}{d-1} T^{kk}.
%-------------
\end{eqnarray}
\end{subequations}
%-------------
In 3 spatial dimensions, $T^{ij}_S$, $T^{ij}_V$, and $T^{ij}_T$ are just the
irreducible representations of $T^{ij}$ under rotations with spin 0, 1, and 2,
respectively. 
We note that the tensors $T^{ij}_S$, $T^{ij}_V$, and $T^{ij}_T$ are orthogonal,
in the sense that 
%-------------
\begin{equation}
\label{eq:tensor_orthogonality}
%-------------
T^{ij}_S T^{ij}_V = T^{ij}_V T^{ij}_T = T^{ij}_T T^{ij}_S = 0,
%-------------
\end{equation}
%-------------
which is valid in $d-1$ spatial dimensions.
These relations imply that 
$T^{ij}_S T^{ij} = (T^{ij}_S)^2$, 
$T^{ij}_V T^{ij} = (T^{ij}_V)^2$, and
$T^{ij}_T T^{ij} = (T^{ij}_T)^2$. 
We define the $S=1$ LDMEs by applying this decomposition to the rank-2 tensor 
$\overleftrightarrow{D}^i \sigma^j$ in $d-1$ spatial dimensions.
We begin with the expression
%-------------
\begin{equation}
%-------------
\sum_{J=0,1,2} 
{\cal O} (^3P_J) 
= 
\psi ^\dag 
\left( - \frac{i}{2} \overleftrightarrow{D}^i \right)
\sigma^j \chi | 0 \rangle 
\langle 0 | \chi^\dag \left( - \frac{i}{2} \overleftrightarrow{D}^i \right)
\sigma^j \psi 
,
%-------------
\end{equation}
%-------------
which is valid in 3 spatial dimensions, and generalize it to $d-1$ spatial
dimensions. 
We decompose the tensor $\overleftrightarrow{D}^i \sigma^j$ 
into the trace, antisymmetric, and symmetric traceless parts, 
and define each contribution as the LDME of total spin 0, 1, and 2,
respectively. We obtain, for $J=0$, 
%-------------
\begin{eqnarray}
%-------------
\langle {\cal Q} | {\cal O} (^3P_0) | {\cal Q} \rangle^{\rm DR}
&=& 
\langle {\cal Q} | \psi ^\dag
\left( - \frac{i}{2} \overleftrightarrow{D}^i \right)
\sigma^j \chi | 0 \rangle
\times 
\frac{\delta^{ij}}{d-1} 
\langle 0 | \chi^\dag \left( - \frac{i}{2} \overleftrightarrow{\bm{D}} \right)
\cdot 
\bm{\sigma} \psi | {\cal Q} \rangle
\nonumber \\
&=& \frac{1}{d-1} \left| \langle 0 | \chi^\dag \left( - \frac{i}{2} \overleftrightarrow{\bm{D}} \right)
\cdot
\bm{\sigma} \psi | {\cal Q} \rangle\right|^2. 
%-------------
\end{eqnarray}
%-------------
For $J=1$, we have 
%-------------
\begin{eqnarray}
%-------------
\langle {\cal Q} | {\cal O} (^3P_1) | {\cal Q} \rangle^{\rm DR}
&=&
\langle {\cal Q} | \psi ^\dag
\left( - \frac{i}{2} \overleftrightarrow{D}^i \right)
\sigma^j \chi | 0 \rangle
\langle 0 | \chi^\dag \left( - \frac{i}{2} \overleftrightarrow{D}^{[i} \right)
\sigma^{j]} \psi | {\cal Q} \rangle
\nonumber \\
&=& 
\left| 
\langle 0 | \chi^\dag \left( - \frac{i}{2} \overleftrightarrow{D}^{[i} \right)
\sigma^{j]} \psi | {\cal Q} \rangle
\right|^2
.
%-------------
\end{eqnarray}
%-------------
It can be shown easily that this expression reduces to the $3$-dimensional one
by using the identity $\delta_{ik} \delta_{j\ell} - \delta_{i\ell} \delta_{jk} 
= \epsilon_{n ij} \epsilon_{n k \ell}$, 
where $ \epsilon_{ijk} $ is the totally antisymmetric tensor with
$\epsilon_{123} = 1$ in 3 spatial dimensions. 
Finally, for $J=2$, we obtain 
%-------------
\begin{eqnarray}
%-------------
\langle {\cal Q} | {\cal O} (^3P_2) | {\cal Q} \rangle^{\rm DR}
&=&
\langle {\cal Q} | \psi ^\dag
\left( - \frac{i}{2} \overleftrightarrow{D}^i \right)
\sigma^j \chi | 0 \rangle
\langle 0 | \chi^\dag \left( - \frac{i}{2} \overleftrightarrow{D}^{(i} \right)
\sigma^{j)} \psi | {\cal Q} \rangle
\nonumber \\
&=&
\left| 
\langle 0 | \chi^\dag \left( - \frac{i}{2} \overleftrightarrow{D}^{(i} \right)
\sigma^{j)} \psi | {\cal Q} \rangle \right|^2.
%-------------
\end{eqnarray}
%-------------
We note that these definitions of $^3P_J$ LDMEs are consistent with the 
calculations of short-distance coefficients in DR 
in refs.~\cite{Petrelli:1997ge, Sang:2015uxg, Sang:2020fql}.

We have now established the definitions of NRQCD LDMEs 
that are suitable for calculations in DR. These
definitions are consistent with the $\overline{\rm MS}$ calculations of
the short-distance coefficients for production and decay of $P$-wave heavy
quarkonia. 
In this paper, we compute the $\overline{\rm MS}$-renormalized LDMEs using 
these definitions, so that we can obtain theoretical predictions for production 
and decay rates of $P$-wave heavy quarkonia that are independent of the scheme 
and scale in which the NRQCD LDMEs are renormalized.

%==============================================================================$
\section{\boldmath $P$-wave long-distance matrix elements in pNRQCD}
\label{sec:pnrqcd} 
%==============================================================================

We now compute the NRQCD LDMEs in eqs.~(\ref{eq:ldmes_def_4d}) in 
pNRQCD~\cite{Pineda:1997bj, Brambilla:1999xf, Brambilla:2001xy, 
Brambilla:2002nu, Brambilla:2004jw}. 
In strongly coupled pNRQCD, the LDMEs are given
by~\cite{Brambilla:2002nu, Brambilla:2020xod} 
%-------------
\begin{eqnarray}
%-------------
\langle {\cal Q} | {\cal O}(^{2S+1} P_J) | {\cal Q} \rangle 
&=& 
\int d^3 r \int d^3 r' \int d^3 R \, \Psi_{\cal Q}^* (\bm{r}) 
\nonumber \\ 
&& \times \left[ - V_{{\cal O}} 
(\bm{x}_1, \bm{x}_2; \bm{\nabla}_1, \bm{\nabla}_2) 
\delta^{(3)} (\bm{x}_1-\bm{x}_1') \delta^{(3)} (\bm{x}_2-\bm{x}_2') 
\right] \Psi_{\cal Q} (\bm{r}'), 
%-------------
\end{eqnarray}
%-------------
where the contact term $V_{{\cal O}}$ is the matching coefficient 
obtained by matching
nonperturbatively to NRQCD in an expansion in powers of $1/m$. 
Here, $\bm{r} = \bm{x}_1-\bm{x}_2$ and $\bm{r}'=\bm{x}_1'-\bm{x}_2'$ are the
relative coordinates between the quark and antiquark. 
The quarkonium wavefunction $\Psi_{\cal Q} (\bm{r})$, which we take to be 
unit normalized, satisfies the Schr\"odinger equation 
%-------------
\begin{equation}
%-------------
\bigg[ - \frac{\bm{\nabla}}{m} + V(\bm{r}, \bm{\nabla}) \bigg] 
\Psi_{\cal Q} (\bm{r}) = E_{\cal Q} \Psi_{\cal Q} (\bm{r}),
%-------------
\end{equation}
%-------------
where $\bm{\nabla} = \bm{\nabla}_{\bm{r}}$. 
The potential $V(\bm{r}, \bm{\nabla})$ is obtained by matching pNRQCD to
NRQCD order by order in expansion in powers of $1/m$, 
and $E_{\cal Q}$ is the binding energy of the $\cal Q$ state. 
For the LDMEs in eq.~(\ref{eq:ldmes_def_4d}), the $V_{{\cal O}}$ are given
by~~\cite{Brambilla:2002nu, Brambilla:2020xod} 
%-------------
\begin{equation}
\label{eq:contact_3d}
%-------------
V_{{\cal O}(^{2 S+1} P_J)}
= N_c T_{SJ}^{ij} \nabla_{\bm{r}}^i 
\left[ 1 + \frac{2}{3} \frac{i {\cal E}_2}{m} + O(1/m^2) \right] 
\delta^{(3)} (\bm{r}) 
\nabla_{\bm{r}}^j ,
%-------------
\end{equation}
%-------------
where $T_{SJ}^{ij}$ are the spin projections in 3 spatial dimensions that are
given by 
%-------------
\begin{subequations}
\begin{eqnarray}
%-------------
T_{10}^{ij} &=& \frac{1}{3} \sigma^i \otimes \sigma^j, \\
T_{11}^{ij} &=& 
\frac{1}{2} \epsilon_{kim} \epsilon_{kjn} 
\sigma^m \otimes \sigma^n, \\
T_{12}^{ij} &=& 
\left( \frac{\delta_{im} \sigma^n + \sigma_{in} \sigma^m}{2} 
- \frac{\delta_{mn}}{3} \sigma^i \right) \otimes 
\left( \frac{\delta_{jm} \sigma^n + \sigma_{jn} \sigma^m}{2} 
- \frac{\delta_{mn}}{3} \sigma^j \right), \\
T_{01}^{ij} &=& \delta^{ij} 1 \otimes 1, 
%-------------
\end{eqnarray}
\end{subequations}
%-------------
and $i {\cal E}_2$ is a gluonic correlator defined by~\cite{Brambilla:2002nu,
Brambilla:2020xod} 
%-------------
\begin{equation}
%-------------
i {\cal E}_2 = \frac{i T_F}{N_c} 
\int_0^\infty dt \, t^2 \langle 0 | g_s \bm{E}^{i,a}(t,\bm{0})
\Phi_{ab} (t,0) g_s \bm{E}^{i,b} (0,\bm{0})| 0 \rangle, 
%-------------
\end{equation}
%-------------
where $T_F = 1/2$, 
$\bm{E}^{i,a} T^a = G^{i0}$ is the chromoelectric field, 
$G^{\mu \nu}$ is the gluon field-strength tensor, 
and $\Phi_{ab}(t,0)$ is an adjoint Wilson line in the temporal
direction connecting the points $(0,\bm{0})$ and $(t,\bm{0})$. 
The expressions for $T^{ij}_{SJ}$ follow from the 3-dimensional definitions 
of the LDMEs in eqs.~(\ref{eq:ldmes_def_4d}). The spin indices 
on the Pauli matrices in $T^{ij}_{SJ}$ on the left and right of
the $\otimes$ symbol contract with the implicit 
$Q$ and $\bar Q$ spin indices on the wavefunctions $\Psi_{\cal Q}^*(\bm{r})$
and $\Psi_{\cal Q} (\bm{r})$, respectively. 
Since the gluonic correlator $i {\cal E}_2$ scales like $\Lambda_{\rm QCD}$, 
the LDMEs computed from 
eq.~(\ref{eq:contact_3d}) are accurate to relative order $\Lambda_{\rm QCD}/m$, 
and the uncalculated order $1/m^2$ terms in the contact term give corrections
to the LDMEs of at most order $v^2$. 
Hence, eq.~(\ref{eq:contact_3d}) implies that 
%-------------
\begin{equation}
\label{eq:ldme_pnrqcd_3d}
%-------------
\langle {\cal Q} | {\cal O}(^{2S+1} P_J) | {\cal Q} \rangle 
= 
N_c \left[ 
\left(\bm{\nabla}_{\bm{r}}^i \Psi_{\cal Q}^* (\bm{r}) \right)
T_{SJ}^{ij} 
\left(\bm{\nabla}_{\bm{r}}^j \Psi_{\cal Q}(\bm{r}) \right)
\right]_{\bm{r}=\bm{0}}
\times
\left[ 1+ \frac{2}{3} \frac{i {\cal E}_2}{m} + O(v^2) \right]. 
%-------------
\end{equation}
%-------------

The spin projections in eq.~(\ref{eq:ldme_pnrqcd_3d}) can be simplified 
when the wavefunctions 
$\Psi_{\cal Q}(\bm{r})$ have definite $^{2 S +1}P_J$ quantum numbers, 
so that the wavefunction at the origin 
$\bm{\nabla} \Psi_{\cal Q}(\bm{0}) \equiv
\bm{\nabla}_{\bm{r}} \Psi_{\cal Q}(\bm{r}=\bm{0})$ 
is annihilated by all but one of the $T_{SJ}^{ij}$. 
In this case, from the completeness of spin projections we have 
$(\bm{\nabla}^i \Psi_{\cal Q}^*(\bm{0}) ) T_{SJ}^{ij}
(\bm{\nabla}^j \Psi_{\cal Q}(\bm{0}) )
=2 | \bm{\nabla} \Psi_{\cal Q} (\bm{0})|^2$. 
Here, the factor 2 comes from the trace over the spin indices of 
$\sum_{S,J} T_{SJ}^{ij}$. 
It is useful to compute the average over the polarization of the ${\cal Q}$ 
state, because of rotational symmetry, the LDME for each polarization is the
same as the average of the LDMEs over polarizations of the ${\cal Q}$ state. 
We write the average of $| \bm{\nabla} \Psi_{\cal Q} (\bm{0})|^2$ over
polarizations as $\overline{| \bm{\nabla} \Psi_{\cal Q} (\bm{0})|^2}$. 
In this case, since there is no preferred direction, we obtain 
%-------------
\begin{equation}
%-------------
\overline{| \bm{\nabla} \Psi_{\cal Q} (\bm{0})|^2}
= 
\overline{| \hat{\bm{r}} \cdot \bm{\nabla} \Psi_{\cal Q} (\bm{0})|^2}, 
%-------------
\end{equation}
%-------------
where $\hat{\bm{r}} = \bm{r}/|\bm{r}|$. 

For the limit $\bm{r} \to \bm{0}$ to be meaningful, 
$\bm{\nabla}_{\bm{r}} \Psi_{\cal Q}(\bm{r})$ must be regular at
$\bm{r}=\bm{0}$.
That is, the limit $\bm{r} \to \bm{0}$ cannot be taken if 
$\bm{\nabla}_{\bm{r}} \Psi_{\cal Q} (\bm{r})$ is divergent at $\bm{r}=\bm{0}$. 
Since ${\cal E}_2$ does not have logarithmic UV divergences, the anomalous
dimension of the LDME $\langle {\cal Q} | {\cal O}(^{2S+1} P_J) | {\cal Q}
\rangle$ must come from $\bm{\nabla} \Psi_{\cal Q} (\bm{0})$\footnote{
This is consistent with the calculation of the anomalous dimensions in
ref.~\cite{Hoang:2006ty}, where the UV divergences in the NRQCD LDMEs 
are computed in terms of potential exchanges between $Q$ and $\bar Q$.}.
That is, the derivative of the wavefunction at the origin contains 
UV divergences, so that the limit $|\bm{r}| \to 0$ can only be taken after the
divergences are subtracted through renormalization. 

Since $\bm{\nabla} \Psi_{\cal Q} (\bm{0})$ is UV divergent, 
we need $d$-dimensional expressions of the LDMEs in terms of wavefunctions
in order to compute the correct $\overline {\rm MS}$-renormalized LDMEs. 
Given our $d$-dimensional definitions of LDMEs, it is straightforward to find
the $d$-dimensional versions of $V_{\cal O}$:
%-------------
\begin{equation}
\label{eq:contact_DR} 
%-------------
V_{{\cal O}(^{2 S+1} P_J)}^{\rm DR} 
= N_c {\bar T}_{SJ}^{ij} \nabla_{\bm{r}}^i
\left[ 1 + \frac{2}{3} \frac{i {\cal E}_2}{m} + O(1/m^2) \right] 
\delta^{(3)} (\bm{r}) \nabla_{\bm{r}}^j ,
%-------------
\end{equation}
%-------------
where ${\bar T}_{SJ}^{ij}$ are the spin projections defined in 
$d-1$ spatial dimensions, given by
%-------------
\begin{subequations}
\begin{eqnarray}
%-------------
{\bar T}_{10}^{ij} &=& \frac{1}{d-1} \sigma^i \otimes \sigma^j, \\
{\bar T}_{11}^{ij} &=&
\left( \frac{\delta_{im} \sigma^n - \delta_{in} \sigma^m}{2} \right)
\otimes
\left( \frac{\delta_{jm} \sigma^n - \delta_{jn} \sigma^m}{2} \right), \\
{\bar T}_{12}^{ij} &=&
\left( \frac{\delta_{im} \sigma^n + \delta_{in} \sigma^m}{2}
- \frac{\delta_{mn}}{d-1} \sigma^i \right) \otimes
\left( \frac{\delta_{jm} \sigma^n + \delta_{jn} \sigma^m}{2}
- \frac{\delta_{mn}}{d-1} \sigma^j \right), \\
{\bar T}_{01}^{ij} &=& \delta^{ij} \sigma_5 \otimes \sigma_5.
%-------------
\end{eqnarray}
\end{subequations}
%-------------
It is easy to see that ${\bar T}^{ij}_{SJ}$ reduce to $T^{ij}_{SJ}$ in 
3 spatial dimensions. 
Strictly speaking, since the denominator factor 3 in the coefficient of 
$i {\cal E}_2/m$ in eq.~(\ref{eq:contact_3d}) comes from the number of
spatial dimensions, the term $\frac{2}{3} \frac{i {\cal E}_2}{m}$
in eq.~(\ref{eq:contact_DR}) should instead read 
$\frac{2}{d-1} \frac{i {\cal E}_2}{m}$. However, since there are no poles in 
$\epsilon$ associated with ${\cal E}_2$, we can safely set $d=4$ in this term. 
The dimensionally regulated NRQCD LDMEs are then given by 
%-------------
\begin{equation}
\label{eq:ldme_pnrqcd_dr}% 
%-------------
\langle {\cal Q} | {\cal O}(^{2S+1} P_J) | {\cal Q} \rangle^{\rm DR} 
=
N_c \left[
\left(\bm{\nabla}^i \Psi_{\cal Q}^* (\bm{0}) \right)
{\bar T}_{SJ}^{ij}
\left(\bm{\nabla}^j \Psi_{\cal Q}(\bm{0}) \right)
\right]
\times
\left[ 1+ \frac{2}{3} \frac{i {\cal E}_2}{m} + O(v^2) \right].
%-------------
\end{equation}
%-------------
With these expressions at hand, we can write the $\overline{\rm
MS}$-renormalization of the wavefunction at the origin at two-loop level 
as follows:
%-------------
\begin{subequations}
\label{eq:wf_renormalization}
\begin{eqnarray}
%-------------
\bigg[ \frac{\delta^{ij}\bm{\sigma} \cdot \bm{\nabla} 
}{d-1} 
\Psi_{{\cal Q}} (\bm{0})
\bigg]^{\overline{\rm MS}}
&=&
\left[ 1 - \frac{\alpha_s^2}{4 \epsilon_{\rm UV}} \gamma_{^3P_0}^{(2)} 
+O(\alpha_s^3) \right] 
\bigg[ \frac{\delta^{ij}\bm{\sigma} \cdot \bm{\nabla} 
}{d-1} 
\Psi_{{\cal Q}} (\bm{0}) \bigg]^{\rm DR}, 
\quad
\\
\big[ \sigma^{[i} \nabla^{j]} \Psi_{{\cal Q}} (\bm{0})
\big]^{\overline{\rm MS}}
&=&
\left[ 1- \frac{\alpha_s^2}{4 \epsilon_{\rm UV}} \gamma_{^3P_1}^{(2)}
+O(\alpha_s^3) \right]
\big[ \sigma^{[i} \nabla^{j]} \Psi_{{\cal Q}} (\bm{0})
\big]^{\rm DR}, 
\\
\big[ \sigma^{(i} \nabla^{j)} \Psi_{{\cal Q}} (\bm{0})
\big]^{\overline{\rm MS}}
&=&
\left[ 1- \frac{\alpha_s^2}{4 \epsilon_{\rm UV}} \gamma_{^3P_2}^{(2)}
+O(\alpha_s^3) \right]
\big[ \sigma^{(i} \nabla^{j)} \Psi_{{\cal Q}} (\bm{0})
\big]^{\rm DR}, 
\\
\big[ \sigma_5 \bm{\nabla} \Psi_{{\cal Q}} (\bm{0})
\big]^{\overline{\rm MS}}
&=&
\left[ 1- \frac{\alpha_s^2}{4 \epsilon_{\rm UV}} \gamma_{^1P_1}^{(2)}
+O(\alpha_s^3) \right]
\big[ \sigma_5 \bm{\nabla} \Psi_{{\cal Q}} (\bm{0})
\big]^{\rm DR}, 
%-------------
\end{eqnarray}
\end{subequations}
%-------------
where the spin indices on $\bm{\sigma}$ and $\sigma_5$ contract with the
implicit $Q$ and $\bar Q$ spin indices on the wavefunctions.
These expressions provide definitions of the spin projections onto states
with definite $^{2 S+1}P_J$ quantum numbers in $d-1$ spatial dimensions, 
and are consistent with the $\overline{\rm MS}$ renormalization 
of the NRQCD LDMEs. That is, when we compute the NRQCD LDMEs from the 
$\overline{\rm MS}$-renormalized wavefunctions at the origin in
eq.~(\ref{eq:wf_renormalization}), 
we automatically obtain the correct $\overline{\rm MS}$-renormalized LDMEs. 
Once the wavefunctions are renormalized, so that the UV divergences regulated
by nonzero $d-4$ are removed, the spin projections can now be computed in 
3 spatial dimensions. 
The $\overline{\rm MS}$-renormalized LDMEs are then given by 
%-------------
\begin{equation}
\label{eq:ldme_pnrqcd_msbar}%
%-------------
\langle {\cal Q} | {\cal O}(^{2S+1} P_J) | {\cal Q} \rangle^{\overline{\rm MS}}
= 
2 N_c 
\overline{ 
\left| \left[ \hat{\bm{r}} \cdot \bm{\nabla}
\Psi_{\cal Q}(\bm{0})
\right]^{\overline{\rm MS}} \right|^2 }
\times
\left[ 1+ \frac{2}{3} \frac{i {\cal E}_2}{m} + O(v^2) \right],
%-------------
\end{equation}
%-------------
which is valid when the wavefunction $\Psi_{\cal Q}(\bm{r})$ has 
definite $^{2 S+1}P_J$ quantum numbers. 

In order to compute the $\overline{\rm MS}$-renormalized wavefunctions at the
origin, we need to obtain the $\bm{\nabla} \Psi(\bm{0})$ in DR, where the UV
divergences are given by poles in $\epsilon$ and organized in powers of
$\alpha_s$. In general, this is difficult because dimensionally
regulated calculations are done in perturbation theory in momentum space, while
nonperturbative calculations of the wavefunctions are best done in
position space, as nonperturbative potentials are usually given as 
functions of $\bm{r}$. In position-space calculations of
$\bm{\nabla} \Psi(\bm{0})$, the UV divergences can be regulated by introducing
a position-space cutoff. The relation between the position-space regularized 
wavefunction at the origin and the dimensionally regulated one can be written as
%-------------
\begin{equation}
%-------------
\big[ \bm{\nabla}^i \Psi(\bm{0}) \big]^{\rm DR}
= \bar{Z}^{ij} 
\big[ \bm{\nabla}^j \Psi(\bm{0}) \big]_{(r_0)}, 
%-------------
\end{equation}
%-------------
where the subscript $(r_0)$ on $\bm{\nabla} \Psi(\bm{0})$ denotes that the UV
divergences are regulated in position space, and $\bar{Z}^{ij}$ 
is the scheme conversion coefficient between position-space regularization and
DR. The definition of the position-space regularization that we use in this
paper will be given in the next section. The scheme conversion coefficient
$Z^{ij}$
between the position-space regularization and $\overline{\rm MS}$ can then be
found from $\bar{Z}^{ij}$ by subtracting the UV poles according
to eq.~(\ref{eq:wf_renormalization}), which leads to the relation 
%-------------
\begin{equation}
%-------------
\big[ \bm{\hat{r}} \cdot \bm{\nabla} \Psi(\bm{0}) \big]^{\overline{\rm MS}}
= 
\bm{\hat{r}}^i 
Z^{ij}
\big[ \bm{\nabla}^j \Psi(\bm{0}) \big]_{(r_0)}.
%-------------
\end{equation}
%-------------
We discuss the calculation of the wavefunctions at the origin in position-space
regularization and the scheme conversion coefficient in the following sections.

%==============================================================================$
\section{\boldmath $P$-wave quarkonium wavefunctions in position space}
\label{sec:wavefunctions} 
%==============================================================================

In this section, we compute the $P$-wave quarkonium wavefunctions by solving
the Schr\"odinger equation in position space. 
The potential in position space can be written generically as 
%---------------
\begin{eqnarray}
\label{eq:potential_generic}
%---------------
V(\bm{r},\bm{\nabla}) &=&  
V^{(0)} (r) + \frac{1}{m} V^{(1)} (r) 
+ \frac{1}{m^2} \bigg[ V_r^{(2)} (r) + \frac{1}{2} \{ V_{p^2}^{(2)} (r) ,
-\bm{\nabla}^2 \} 
+ V_{L^2} (r) \bm{L}^2  
\nonumber \\ && 
+ V_{S^2}^{(2)} (r) \bm{S}^2 + V_{S_{12}}^{(2)} (r) S_{12} 
+ V_{so}^{(2)} (r) \bm{L} \cdot \bm{S} \bigg] 
- \frac{\bm{\nabla}^4}{4 m^3} + \cdots, 
%---------------
\end{eqnarray}
%---------------
where $\bm{L} = -i \bm{r} \times \bm{\nabla}$, 
$\bm{S}$ is the $Q \bar Q$ spin, 
and $S_{12} = \sigma^i \otimes \sigma^i -3 \hat{\bm{r}} \cdot \bm{\sigma}
\otimes \hat{\bm{r}} \cdot \bm{\sigma}$, 
where the Pauli matrices on the left and
right of $\otimes$ apply to the antiquark and quark spin indices of the
wavefunction\footnote{We note that the form of $S_{12}$ that we use differs from
ref.~\cite{Pineda:2000sz}, because we use the particle-antiparticle basis 
for $Q$ and $\bar Q$ spin indices instead of the particle-particle basis used
in ref.~\cite{Pineda:2000sz}.}. 
The contribution at leading power in $1/m$ is the static potential, which has a
nonperturbative expression in terms of a vacuum expectation value of a 
Wilson loop~\cite{Wilson:1974sk, Susskind:1976pi, Brown:1979ya,
Brambilla:1999xf}. At short distances ($r \ll 1/\Lambda_{\rm QCD}$), 
$V^{(0)} (r)$ can be expressed as series in $\alpha_s$, which reads 
$V^{(0)} (r) = - \alpha_s C_F/r$
at leading order in $\alpha_s$~\cite{Pineda:2003jv, Bazavov:2014soa}.
The explicit expressions for the order $1/m$ and $1/m^2$ 
potentials at short distances are shown in
appendix~\ref{appendix:potentials}. The potentials of orders $1/m$
and higher generally depend on the matching scheme, while the static potential 
$V^{(0)} (r)$ is scheme independent. 
In order to obtain results that are consistent with the 
short-distance coefficients appearing in 
NRQCD factorization formulae, the wavefunctions must be computed by using
potentials from the on-shell matching scheme, because matching calculations in
NRQCD are done by comparing on-shell $Q \bar Q$ amplitudes with QCD
counterparts. On the other hand, the nonperturbative long-distance behaviors of
the potentials are available in the Wilson loop matching scheme, which provides
nonperturbative expressions of the potentials that allow lattice QCD
determinations~\cite{Brambilla:2000gk, Pineda:2000sz}. 
The wavefunctions in the on-shell matching scheme can 
be obtained from the ones in the Wilson loop matching scheme 
by unitary transformations; explicit calculations of the unitary transformation
will be done in sec.~\ref{sec:unitary_transformations}. 

Our goal is to compute $P$-wave wavefunctions including the effect of the
potential to leading nontrivial order in $1/m$. 
Because unitary transformations can
reshuffle the $1/m$ and $1/m^2$ terms in the potential, it is necessary to
include also the $1/m^2$ terms in order to determine the effect of the 
$1/m$ potential
unambiguously. The last term in eq.~(\ref{eq:potential_generic}), 
which comes from the relativistic correction to
the kinetic energy, should be regarded as a order $1/m$ correction, because by
using the Schr\"odinger equation, a power of $\bm{\nabla}^2$ can be traded with
a power of $m (E_{\cal Q} - V^{(0)} (r) )$. 

Since the $V^{(1)}(r)/m$ term in the potential comes from the spin-independent
dimension-5 terms in the NRQCD Lagrangian, in the standard NRQCD power
counting,
the effects of the $1/m$ potential to wavefunctions are suppressed by $v^2$. 
Computation of the corrections from the $1/m$ and $1/m^2$ potentials
correspond to computing the order $v^2$ corrections to the
wavefunctions.
We compute the effect from the higher order potentials by using 
the quantum-mechanical perturbation theory (QMPT) to first order, 
where corrections to the wavefunctions are computed by using 
the Rayleigh-Schr\"odinger perturbation theory\footnote{
It has been suggested in refs.~\cite{Brambilla:2000gk, Pineda:2000sz, 
Brambilla:2004jw} that in 
a more conservative power counting, $V^{(1)}(r)/m$ can be the same order as the
static potential, and so, its effect must be included at leading order.
However, it is possible that this power counting overestimates the effect of
the $V^{(1)}(r)/m$ term to the wavefunctions, because wavefunctions depend only
on the shape of the potential. In order to address this issue, in
appendix~\ref{appendix:1mallorder} we compute
numerically the bound-state wavefunctions from the potential $V^{(0)} (r) +
V^{(1)}(r)/m$ without expanding the $1/m$ potential, and compare with the
result from first order Rayleigh-Schr\"odinger perturbation theory. 
The calculation in appendix~\ref{appendix:1mallorder} shows that the bulk of
the effects of the $V^{(1)}(r)/m$ term to the wavefunctions 
are well reproduced at first order in the Rayleigh-Schr\"odinger 
perturbation theory, and hence, the expansion of the $V^{(1)}(r)/m$ term 
in powers of $1/m$ is well justified.
}.
We define the leading-order potential as the static potential, minus the loop
corrections at short distances beyond leading order in $\alpha_s$. That is, 
$V_{\rm LO} (r) = V^{(0)} (r) - \delta V_C (r)$, where $\delta V_C (r)$
corresponds to the corrections to the static potential at orders $\alpha_s^2$
and beyond computed in perturbative QCD. 
Then, $\delta V(\bm{r}, \bm{\nabla}) \equiv 
V(\bm{r},\bm{\nabla}) - V_{\rm LO} (r)$ is given by 
$\delta V_C (r)$ plus the potentials of higher orders in $1/m$. 

The leading-order Hamiltonian reads 
%---------------
\begin{equation}
%---------------
h_{\rm LO} (r,\bm{\nabla}) = - \frac{\bm{\nabla}^2}{m} + V_{\rm LO} (r).
%---------------
\end{equation}
%---------------
The LO wavefunctions $\Psi_n^{\rm LO}$ are normalized bound-state
solutions of the LO Schr\"odinger equation
%---------------
\begin{equation}
\label{eq:LO_schroedinger}
%---------------
h_{\rm LO} (r,\bm{\nabla}) \Psi_n^{\rm LO} (\bm{r}) 
= E_n^{\rm LO} \Psi_n^{\rm LO} (\bm{r}),
%---------------
\end{equation}
%---------------
with corresponding binding energies $E_n^{\rm LO}$. 
Before we move onto the calculation
of the corrections, we first discuss some properties of the LO wavefunctions. 
Since the LO potential $V_{\rm LO} (r)$ diverges like $1/r$ at $r \to 0$, 
the wavefunctions are regular at $r=0$. 
From the series solution method for solving Schr\"odinger equations 
in nonrelativistic quantum mechanics, we know that in this case, the $P$-wave
LO wavefunctions vanish linearly as $r \to 0$, while the $D$- and $F$-wave
LO wavefunctions vanish quadratically and cubically, respectively. 
That is, the first derivative of the $P$-wave LO wavefunction is finite at the
origin, while $\bm{\nabla} \Psi_n^{\rm LO} (\bm{0})$ 
vanishes for states with higher orbital angular momentum. 
While $S$-wave wavefunctions are finite at the origin, its first derivative 
vanishes at the origin due to reflection symmetry. 
Hence, at leading order in the QMPT, only $P$-wave wavefunctions contribute
to the LDMEs in eqs.~(\ref{eq:ldme_pnrqcd_3d}).  
We also note that since spin and orbital angular momenta are conserved 
separately by $h_{\rm LO}$, we can write the LO wavefunctions as 
linear combinations of spherical harmonics times a function of $|\bm{r}| = r$:
%---------------
\begin{equation}
%---------------
R_{\rm LO} (r) Y_1^{\lambda_L} (\hat{\bm{r}}) \Sigma_S, 
%---------------
\end{equation}
%---------------
where $R_{\rm LO} (r)$ is the radial wavefunction, 
and $Y_1^{\lambda_L} (\hat{\bm{r}})$ is the spherical harmonics of 
orbital angular
momentum $1$ that encodes the dependence on the angles of $\bm{r}$, with
$\lambda_L=-1$, 0, or 1. 
Because $V_{\rm LO}(r)$ is real, we can always choose the overall phase of the 
wavefunction $\Psi_n^{\rm LO} (\bm{r})$ so that $R_{\rm LO} (r)$ is real. 
The $Q \bar Q$ spin indices of the wavefunction 
are carried by a $2 \times 2$ matrix $\Sigma_S$,
which we normalize as ${\rm tr} (\Sigma^\dag_S \Sigma_S) = 1$. 
In this normalization, the radial wavefunction is normalized as 
$\int_0^\infty dr \, r^2 |R_{\rm LO} (r)|^2 = 1$. 
In 3 spatial dimensions, $\Sigma_{S=0}$ is proportional to the identity matrix 
$1_{2 \times 2}$, while the $\Sigma_{S=1}$ are linear combinations of Pauli
matrices. 
Wavefunctions with specific angular momentum quantum numbers 
$^{2 S+1}P_J$ can be constructed from linear combinations of 
$R_{\rm LO} (r) Y_1^{\lambda_L} (\hat{\bm{r}}) \Sigma_S$ by using 
Clebsch-Gordan coefficients. By using the orthonormality of Clebsch-Gordan
coefficients, we find by direct computation 
%---------------
\begin{equation}
\label{eq:wforg_LO}
%---------------
\left| \bm{\nabla} \Psi_n^{\rm LO} (\bm{0}) \right|^2
= 
\frac{3}{4 \pi} 
\left| R'_{\rm LO} (0) \right|^2 
%---------------
\end{equation}
%---------------
where $R'_{\rm LO}(0)$ is the first derivative of $R_{\rm LO}(r)$ at $r=0$. 

For analytical calculations, it is convenient to use a different basis for 
the $\hat{\bm{r}}$ dependence based on irreducible Cartesian tensors. 
In this basis, the wavefunctions with quantum numbers 
$^1P_1$, $^3P_0$, $^3P_1$, and $^3P_2$
are given by the tensors $1_{2 \times 2} \hat{\bm{r}}^i$, 
$\frac{\delta^{ij}}{3} 
\hat{\bm{r}} \cdot \bm{\sigma}$, 
$\hat{\bm{r}}^{[i} \sigma^{j]}$, and 
$\hat{\bm{r}}^{(i} \sigma^{j)}$,
respectively, multiplied by the radial wavefunction $R_{\rm LO}(r)$, 
up to a normalization. Here $1_{2 \times 2}$ is an identity in the 
$Q \bar Q$ spin. 
In each case, spatial indices of the tensors
represent polarizations of the $^{2 S +1}P_J$ state.
In particular, the tensors for the $^3P_J$ states are just the irreducible
representations of $\hat{\bm{r}}^i \sigma^j$ for spin $J$, 
so that for all spin triplet cases, we can consider wavefunctions that 
are proportional to $\hat{\bm{r}}^i \sigma^j$ and project
onto the spin-$J$ components whenever necessary. 
The average over polarizations of the square of a wavefunction is given by
summing over repeated indices of these tensors and taking the trace over the $Q
\bar Q$ spin. 
In this case, we obtain 
$\overline{\left| \bm{\nabla} \Psi_n^{\rm LO} (\bm{0}) \right|^2} 
= \overline{\left| \hat{\bm{r}} \cdot \bm{\nabla} \Psi_n^{\rm LO} (\bm{0})
\right|^2}
= \frac{3}{4 \pi} \left| R'_{\rm LO} (0) \right|^2$. 
In the calculations in this paper, the Cartesian basis will be used 
in analytical calculations of the scheme conversion,  while the 
spherical harmonics is more appropriate in organizing the numerical
calculations of the Schr\"odinger equation.

While $\bm{\nabla} \Psi_n^{\rm LO} (\bm{r})$ is regular at $\bm{r}=\bm{0}$, 
the corrections from $\delta V(\bm{r},\bm{\nabla})$ can produce divergences,
because the $1/m$ and $1/m^2$ potentials contain terms that diverge faster 
than $1/r$ at $r=0$. For example, the $1/m$ potential diverges like $1/r^2$ at
$r=0$, and so, the correction from $V^{(1)}(r)/m$ will make 
$\bm{\nabla} \Psi (\bm{0})$ diverge logarithmically. Similarly, the
spin-dependent $1/m^2$ potentials diverge like $1/r^3$, which can make the
wavefunctions diverge linearly at $r=0$. 

Now we discuss the calculation of the correction to the wavefunctions in the 
QMPT. To first order in the Rayleigh-Schr\"odinger perturbation theory, 
the wavefunction $\Psi_n(\bm{r})$ is given in terms
of the LO wavefunction $\Psi_n^{\rm LO} (\bm{r})$ by 
%---------------
\begin{equation}
%---------------
\Psi_n(\bm{r}') = 
\Psi_n^{\rm LO}(\bm{r}') + \delta \Psi_n(\bm{r}')
= 
\Psi_n^{\rm LO}(\bm{r}') - \int d^3 r \, \hat{G}_n (\bm{r}',\bm{r})
\delta V(\bm{r},\bm{\nabla}) \Psi_n^{\rm LO} (\bm{r}),
%---------------
\end{equation}
%---------------
where $ \hat{G}_n (\bm{r}',\bm{r})$ is the reduced Green's function for the
eigenstate $n$, defined by 
%---------------
\begin{equation}
\label{eq:redgreen}
%---------------
\hat{G}_n (\bm{r}',\bm{r}) = \sum_{k \neq n} 
\frac{\Psi_k^{\rm LO} (\bm{r}') \Psi_k^{\rm LO}{}^* (\bm{r})}
{E_k^{\rm LO} - E_n^{\rm LO}},
%---------------
\end{equation}
%---------------
where the sum runs over all eigenstates of the LO Schr\"odinger equation except
for the state $n$. 
The reduced Green's function can be defined in terms of the Green's function
$G(\bm{r}',\bm{r};E)$, defined for arbitrary complex $E$ by 
%---------------
\begin{equation}
%---------------
G(\bm{r}',\bm{r};E) = \sum_{k}
\frac{\Psi_k^{\rm LO} (\bm{r}') \Psi_k^{\rm LO}{}^* (\bm{r})}
{E_k^{\rm LO} - E},
%---------------
\end{equation}
%---------------
where the sum is over all eigenstates of $h_{\rm LO}$. From this we have 
%---------------
\begin{equation}
\label{eq:reducedgreen_relation1}
%---------------
\hat{G}_n (\bm{r}',\bm{r}) = \lim_{E \to E_n^{\rm LO}} 
\left[ G(\bm{r}',\bm{r};E) 
- \frac{\Psi_n^{\rm LO} (\bm{r}') \Psi_n^{\rm LO}{}^* (\bm{r})}
{E_n^{\rm LO} - E}\right], 
%---------------
\end{equation}
%---------------
and 
%---------------
\begin{equation}
\label{eq:reducedgreen_relation2}
%---------------
\hat{G}_n (\bm{r}',\bm{r}) = \lim_{\eta \to 0} \frac{1}{2} 
\left[ G(\bm{r}',\bm{r};E_n^{\rm LO}+\eta) + 
G(\bm{r}',\bm{r};E_n^{\rm LO}-\eta) \right].
%---------------
\end{equation}
%---------------
The correction to $\bm{\nabla} \Psi(\bm{0})$ at first order 
is given by  $-\bm{\nabla}_{\bm{r}'} \int d^3 r\, \hat{G}_n (\bm{r}',\bm{r})
\delta V(\bm{r},\bm{\nabla}) \Psi_n^{\rm LO} (\bm{r})$ at $\bm{r}'=\bm{0}$. 
As argued previously, this integral contains divergences at
$\bm{r}'=\bm{0}$ when $\delta V(\bm{r},\bm{\nabla})$
contains terms that diverge faster than $1/r$ at $r = 0$. 
The contribution in $\delta V(\bm{r},\bm{\nabla})$ that diverges like 
$1/r^2$ at $r=0$ such as the $1/m$ potential produces a logarithmic divergence
that is proportional to $\log r$ at first order in QMPT. Similarly,
contributions that diverge like $1/r^3$ such as the spin-dependent terms in 
$\delta V(\bm{r},\bm{\nabla})$ produce linear UV divergences. 
In the case of $P$-wave states, the delta functions $\delta^{(3)} (\bm{r})$ in 
 $\delta V(\bm{r},\bm{\nabla})$ do not contribute to 
$\delta \Psi_n(\bm{r}')$ to first order in the QMPT, 
because the $P$-wave wavefunctions $\Psi_n^{\rm LO} (\bm{r})$ vanish at
$\bm{r}=\bm{0}$. 

The $\delta V(\bm{r},\bm{\nabla})$ contains differentiation that act on the
wavefunctions $\Psi_n^{\rm LO} (\bm{r})$. 
They come from the $\bm{\nabla}^2$ in the velocity-dependent potential and the 
relativistic corrections to the kinetic energy, as well as the spin-dependent
terms in the potential, which also depend on the angles of $\bm{r}$ and the $Q
\bar Q$ spin. 
The $\bm{\nabla}^2$ can be reduced into functions of $|\bm{r}|=r$ by
using the Schr\"odinger equation, similarly to what has been done in
ref.~\cite{Chung:2020zqc} for
the $S$-wave case. In the case of the spin-dependent contributions, the
terms that depend on $\bm{r}$, $\bm{\nabla}$, and the $Q \bar Q$ spin 
can also be reduced into functions of $r$ by diagonalizing their 
matrix elements on states with definite angular momentum quantum numbers. 
The spin-dependent contributions in 
$\delta V(\bm{r},\bm{\nabla})$ that apply to $P$-wave wavefunctions come from 
$\bm{L} \cdot \bm{S}$ and $S_{12}$. As is known from nonrelativistic quantum
mechanics, the operator 
$\bm{L} \cdot \bm{S}$ takes values $0$, $-2$, $-1$, and $1$ when applied to
wavefunctions with angular momentum quantum numbers 
$^1P_1$, $^3P_0$, $^3P_1$, and $^3P_2$, respectively. 
The operator $S_{12}$ is more involved, because it can change the orbital
angular momentum by two units. 
Explicit calculations of the action of $S_{12}$ on $P$-wave wavefunctions are
done in appendix~\ref{appendix:s12}. 
The operator $S_{12}$ takes values 
$0$, $-4$, and $2$ when applied to wavefunctions with angular momentum quantum
numbers $^1P_1$, $^3P_0$, and $^3P_1$, respectively, 
while when we apply $S_{12}$ to a $^3P_2$ wavefunction, 
we obtain $-\frac{2}{5}$ times the $^3P_2$ wavefunction plus a $F$-wave
contribution. The exact form of the $F$-wave contribution in the Cartesian
basis is shown in eq.~(\ref{eq:Fwavetensor}).
Because the tensor structure of the $F$-wave
contribution is orthogonal to the tensor structures of $P$-wave wavefunctions, 
the $F$-wave contribution vanishes in 
$\overline{|\bm{\nabla} \Psi_n (\bm{0})|^2}$ to first order in the QMPT. 
Therefore, at the current level of accuracy, we can neglect the $F$-wave
contribution, and consider only $P$-wave states when computing the
LDMEs\footnote{If there are degeneracies in the LO Hamiltonian, corrections in 
the Rayleigh-Schr\"odinger perturbation theory must be computed in the basis
that diagonalizes the higher order potentials. For example, in perturbative QCD
where $V_{\rm LO}(r) = -\alpha_s C_F/r$, 
radially excited $P$-wave states can have same
LO binding energies as $F$-wave states. However, the degeneracies between
$P$-wave and $F$-wave states disappear
when we include the long-distance nonperturbative contributions to the 
static potential. }. 
In this case, the operators 
$\bm{L} \cdot \bm{S}$ and $S_{12}$ take definite values when applied to
wavefunctions of definite $^{2 S +1}P_J$ quantum numbers, so that the tensor
structures of the $P$-wave wavefunctions in the Cartesian basis 
remain unchanged. Hence, the corrections from $\delta V(\bm{r},\bm{\nabla})$
affect only the radial wavefunctions, which acquire dependence on $S$ and $J$
from first order in the QMPT. The same is true in terms of spherical harmonics. 

While in position space in $3$ spatial dimensions it is possible to reduce 
the action of $\delta V(\bm{r},\bm{\nabla})$ to $P$-wave wavefunctions of
definite $^{2 S+1} P_J$ quantum numbers into functions of $r$, the reduction
of spin-dependent potentials do not easily generalize to $d-1$ 
spatial dimensions. Later, we will need to obtain definite expressions in DR
that match with the position-space expressions in order to identify the UV
divergences in $d$ dimensions, so that we can carry out renormalization in the
${\overline{\rm MS}}$ scheme. Hence, we keep the spin-dependent operators in 
$\delta V(\bm{r},\bm{\nabla})$ as is, while we can still reduce the
$\bm{\nabla}^2$ in the velocity-dependent potential and the relativistic
corrections to the kinetic energy. We obtain 
%---------------
\begin{eqnarray}
\label{eq:corr_reduced}
%---------------
\delta \Psi_n(\bm{r}') &=& 
- 
\int d^3 r \, \hat{G}_n (\bm{r}', \bm{r}) 
\delta {\cal V}(\bm{r},\bm{\nabla}) \Psi_n^{\rm LO} (\bm{r}) 
\nonumber \\ && 
- 
\int d^3 r \, \hat{G}_n (\bm{r}', \bm{r}) 
\left\{ \delta V_C(r) + \frac{E_n^{\rm LO}}{m} \left[ V_{p^2}^{(2)} (r) 
+ \frac{1}{2} V_{\rm LO} (r) \right] \right\} \Psi_n^{\rm LO} (\bm{r}) 
\nonumber \\ && 
+ \frac{1}{2 m} \Psi_n^{\rm LO} (\bm{r}')
\int d^3r \, \left[ V_{p^2}^{(2)}(r)+ \frac{1}{2} V_{\rm LO} (r) \right] 
\left| \Psi_n^{\rm LO} (\bm{r})  \right|^2
\nonumber \\ && 
- \frac{1}{2 m} V_{p^2}^{(2)} (r')  \Psi_n^{\rm LO} (\bm{r}')
- \frac{1}{4 m} V_{\rm LO} (r')  \Psi_n^{\rm LO} (\bm{r}'),
%---------------
\end{eqnarray}
%---------------
where $\delta V_C(r) = V^{(0)} (r)-V_{\rm LO} (r)$, and 
%---------------
\begin{equation}
%---------------
\delta {\cal V} (\bm{r},\bm{\nabla})
= \frac{V^{(1)}(r)}{m} 
- \frac{V_{p^2}^{(2)}(r) V_{\rm LO} (r)}{m} 
- \frac{(V_{\rm LO} (r))^2}{4 m}
+ \frac{V^{(2)}(\bm{r},\bm{\nabla})}{m^2} . 
%---------------
\end{equation}
%---------------
In eq.~(\ref{eq:corr_reduced}), the UV-divergent contributions to 
$\bm{\nabla} \delta \Psi_n(\bm{0})$ are contained in the first integral. 
The second integral gives a finite contribution to $\bm{\nabla} \delta
\Psi_n(\bm{0})$, because the terms in the curly brackets diverge like $1/r$ at
$r=0$. The third integral is also finite, because the terms in the square
brackets diverge like $1/r$ at $r=0$, while $|\Psi_n^{\rm LO} (\bm{r})|^2$
vanishes quadratically at $r=0$. 
The terms in the last line requires some investigation, because their
contributions to $\bm{\nabla} \delta \Psi_n(\bm{0})$ involve 
$V_{p^2}^{(2)} (r)$ and $V_{\rm LO} (r)$ and their derivatives at $r=0$. 
In the case of the LO potential, since the static potential is perturbative at
$r=0$, so is $V_{\rm LO} (r)$, and hence, both 
$V_{\rm LO} (0)$ and $\bm{\nabla} V_{\rm LO} (0)$ vanish in DR because they 
are scaleless~\cite{Brown:1979ya, Brambilla:2002nu, Pineda:2003jv}. 
On the other hand, while $\bm{\nabla} V_{p^2}^{(2)} (r)$ is
perturbative at $r=0$, $V_{p^2}^{(2)} (0)$ has a nonperturbative expression
in Wilson-loop matching given by~\cite{Pineda:2000sz} 
%---------------
\begin{equation}
%---------------
V_{p^2}^{(2)} (0) |^{\rm WL} =
2 i \hat{\bm{r}}^i \hat{\bm{r}}^j \frac{T_F}{N_c}
\int_0^\infty dt \, t^2 \langle 0 | g_s \bm{E}^{i,a}(t,\bm{0})
\Phi_{ab} (t,0) g_s \bm{E}^{j,b} (0,\bm{0})| 0 \rangle
= \frac{2 i}{3} {\cal E}_2,
%---------------
\end{equation}
%---------------
where the second equality follows from refs.~\cite{Brambilla:2002nu, 
Brambilla:2020xod}. 
Interestingly, the correction from 
$V_{p^2}^{(2)} (0)$ to the $P$-wave wavefunction at the origin cancels the 
order-$1/m$ correction to $V_{\cal O}$, so that the NRQCD LDMEs are independent
of ${\cal E}_2$, up to corrections of order $v^2$. 

Now we compute $\bm{\nabla} \delta \Psi_n(\bm{0})$ from
eq.~(\ref{eq:corr_reduced}). We first regulate the divergent integral in
position space, by setting $|\bm{r}'|= r_0$ small but nonzero. 
This defines finite-$r$ regularization, which is the position-space
regularization that we use in this work. This is a straightforward
generalization of the position-space regularization used in
refs.~\cite{Hoang:1997ui, Kiyo:2010jm, Chung:2020zqc} to $P$-wave states. 
We define the finite-$r$ regularized wavefunction at the origin 
by the following expression 
%---------------
\begin{eqnarray}
\label{eq:corr_finiter}
%---------------
\bm{\nabla} \delta \Psi_n(\bm{0}) |_{(r_0)} &=&
-
\int d^3 r \, 
\bm{\nabla}_{\bm{r}'} 
\hat{G}_n (\bm{r}', \bm{r})
\delta {\cal V}(\bm{r},\bm{\nabla}) \Psi_n^{\rm LO} (\bm{r})
\Big|_{|\bm{r}'| = r_0} 
\nonumber \\ &&
- 
\int d^3 r \, 
\bm{\nabla}_{\bm{r}'} 
\hat{G}_n (\bm{r}', \bm{r})
\left\{ \delta V_C(r) + \frac{E_n^{\rm LO}}{m} \left[ V_{p^2}^{(2)} (r)
+ \frac{1}{2} V_{\rm LO} (r) \right] \right\} \Psi_n^{\rm LO} (\bm{r})
\Big|_{|\bm{r}'| = 0} 
\nonumber \\ &&
+ \frac{
\left[ \bm{\nabla} \Psi_n^{\rm LO} (\bm{0}) \right] 
}{2 m} 
\left\{ - V_{p^2}^{(2)} (0)
+ \int d^3r \, \left[ V_{p^2}^{(2)}(r)+ \frac{1}{2} V_{\rm LO} (r) \right]
\left| \Psi_n^{\rm LO} (\bm{r})  \right|^2
\right\} , 
%---------------
\end{eqnarray}
%---------------
where we use the subscript $(r_0)$ to denote that the UV divergences are
regulated by using finite-$r$ regularization. 
We keep the divergences and finite contributions in the limit $r_0 \to 0$,
while we neglect any contributions that vanish as $r_0 \to 0$, such as positive
powers of $r_0$. 
The UV divergences that are regulated by nonzero $r_0$ are isolated in the
first integral. As a result, the difference in $\bm{\nabla} \Psi_n(\bm{0})$
between finite-$r$ regularization and DR at first order in the QMPT is given by 
%---------------
\begin{eqnarray}
\label{eq:deltaZ_def2}
%---------------
\delta \bar{Z}^{ij} 
\left[ \bm{\nabla}^j \Psi_n(\bm{0}) \right]_{(r_0)} 
&=& 
\bm{\nabla}^i \Psi_n(\bm{0}) \big|_{(r_0)} 
- 
\bm{\nabla}^i \Psi_n(\bm{0}) \big|^{\rm DR} 
\nonumber \\ 
&=&
\int d^3 r \, 
\bm{\nabla}_{\bm{r}'}^i 
\hat{G}_n (\bm{r}', \bm{r})
\delta {\cal V}(\bm{r},\bm{\nabla}) \Psi_n^{\rm LO} (\bm{r})
\Big|_{|\bm{r}'| = 0}^{\rm DR} 
\nonumber \\ &&
- 
\int d^3 r \, 
\bm{\nabla}_{\bm{r}'}^i 
\hat{G}_n (\bm{r}', \bm{r})
\delta {\cal V}(\bm{r},\bm{\nabla}) \Psi_n^{\rm LO} (\bm{r})
\Big|_{|\bm{r}'| = r_0}, 
%---------------
\end{eqnarray}
%---------------
where $\delta \bar{Z}^{ij}$ is defined through the relation
$\bar{Z}^{ij} = \delta^{ij} - \delta \bar{Z}^{ij}$, 
so that $\delta \bar{Z}^{ij}$ corresponds to the subtraction term that appears
from first order in the QMPT. 
Because the right-hand side of eq.~(\ref{eq:deltaZ_def2}) 
is linear in $\delta {\cal V}$, 
eq.~(\ref{eq:deltaZ_def2}) is equal to 
$\delta \bar{Z}^{ij} \left[ \bm{\nabla}^j \Psi_n^{\rm LO} (\bm{0}) \right]$ 
to first order in the QMPT. 
Hence, the wavefunction at the origin in DR is 
%---------------
\begin{equation}
\label{eq:wf_DR_final}
%---------------
\big[ \bm{\nabla}^i \Psi_n(\bm{0}) \big]_{\rm DR} 
=
\big[ \bm{\nabla}^i \Psi_n(\bm{0}) \big]_{(r_0)}
- \delta \bar{Z}^{ij}
\big[ \bm{\nabla}^j \Psi_n^{\rm LO} (\bm{0}) \big].
%---------------
\end{equation}
%---------------
We compute $\delta \bar{Z}^{ij}$ in the next section by examining the UV
diveregences in the integrands of eq.~(\ref{eq:deltaZ_def2}). 
We also define $\delta Z^{ij}$ through the relation 
$Z^{ij} = \delta^{ij} - \delta Z^{ij}$, so that $\delta Z^{ij}$ is obtained 
from $\delta \bar{Z}^{ij}$ by subtracting the poles in
$\epsilon$. The $\overline{\rm MS}$-renormalized wavefunctions at the
origin are then given by 
%---------------
\begin{equation}
\label{eq:wf_MSren_final}
%---------------
\big[ \hat{\bm{r}} \cdot \bm{\nabla} \Psi_n(\bm{0}) \big]^{\overline{\rm MS}} 
= 
\hat{\bm{r}} \cdot \big[ \bm{\nabla} \Psi_n(\bm{0}) \big]_{(r_0)} 
- \hat{\bm{r}}^i \delta Z^{ij} 
\big[ \bm{\nabla}^j \Psi_n^{\rm LO} (\bm{0}) \big]. 
%---------------
\end{equation}
%---------------
As was discussed in the previous section, in order to compute the NRQCD LDMEs
from eq.~(\ref{eq:wf_MSren_final}), the wavefunctions must carry definite 
$^{2 S +1} P_J$ quantum numbers. While the spin projections on the finite-$r$
regularized wavefunction at the origin 
$\big[ \bm{\nabla} \Psi_n(\bm{0}) \big]_{(r_0)}$ can be done in $3$ spatial
dimensions, 
the spin projections must be done for the subtraction term 
$\delta \bar{Z}^{ij} \big[ \bm{\nabla}^j \Psi_n^{\rm LO} (\bm{0}) \big]$ 
in DR before renormalization in $\overline{\rm MS}$, 
because from order $\epsilon$ 
the spin-dependent potentials can induce transitions between
states with different $^{2 S +1} P_J$ quantum numbers, 
which can affect the finite parts in $\delta Z^{ij}$. 
This can be done by diagonalizing $\delta \bar{Z}^{ij}$ to order $\epsilon^0$ 
accuracy according to the spin projections of the wavefunctions in DR, 
by using the $d$-dimensional definitions in eq.~(\ref{eq:wf_renormalization}).
Explicit calculations of $\delta Z^{ij}$ will be done in the next section. 

We have argued that the corrections to the wavefunctions at the origin from
the $1/m$ potential scales like $v^2$. We have found that the correction 
from the velocity-dependent potential at order $1/m^2$ involves a correction 
that scales like $\Lambda_{\rm QCD}/m$. 
Since corrections of similar form may arise at second order in the QMPT, we
assume that the wavefunctions at the origin that we compute in this section are
accurate up to corrections of relative orders $v^3$ and $\Lambda_{\rm
QCD}^2/m^2$.

%==============================================================================
\section{\boldmath Scheme conversion}
\label{sec:conversion}
%==============================================================================

In this section we compute the scheme conversion coefficient 
$\delta Z_{ij} = Z_{ij} - \delta_{ij}$, which is obtained by 
subtracting the $1/\epsilon$ poles in $\delta \bar{Z}_{ij}$ according to the
$\overline{\rm MS}$ prescription. 
Equation~(\ref{eq:deltaZ_def2}) implies that $\delta \bar{Z}_{ij}$ is
determined by the UV-divergent behavior of the integral 
%---------------
\begin{equation}
%---------------
- \int d^3 r \,
\bm{\nabla}_{\bm{r}'}
\hat{G}_n (\bm{r}', \bm{r})
\delta {\cal V}(\bm{r},\bm{\nabla}) \Psi_n^{\rm LO} (\bm{r})
%---------------
\end{equation}
%---------------
at $\bm{r}'=\bm{0}$. If we split this integral into a UV-divergent part plus a
UV-finite part, the UV-finite part vanishes in 
$\delta \bar{Z}^{ij}$, 
because the UV-finite part does not depend on the UV regulator. 
That is, for the purpose of computing $\delta \bar{Z}_{ij}$, it suffices to
identify the UV-divergent part of the integral, whose integrand is same in both
DR and finite-$r$ regularization when the regularizations are lifted 
($\epsilon \to 0$ and $r_0 \to 0$). 
Since $\hat{G}_n(\bm{r}', \bm{r})$ can be obtained from 
$G(\bm{r}',\bm{r};E)$ by using the relation in
eq.~(\ref{eq:reducedgreen_relation2}), it suffices to compute the UV divergence
of the integral 
%---------------
\begin{equation}
%---------------
- \int d^3 r \,
\bm{\nabla}_{\bm{r}'}
G(\bm{r}', \bm{r};E)
\delta {\cal V}(\bm{r},\bm{\nabla}) \Psi_n^{\rm LO} (\bm{r})
%---------------
\end{equation}
%---------------
at $\bm{r}'=\bm{0}$. We first identify the divergence of this integral in
position space, and then derive the equivalent expression in momentum space in
$d$ spacetime dimensions.

%------------------------------------------------------------------------------
\subsection{\boldmath
Position-space divergences in the wavefunctions at the origin}
%------------------------------------------------------------------------------

We investigate the UV divergence in the finite-$r$ regularized integral 
%---------------
\begin{equation}
\label{eq:basic_uv_int}
%---------------
- \int d^3 r \, 
\bm{\nabla}_{\bm{r}'}
G (\bm{r}', \bm{r}; E)
\delta {\cal V}(\bm{r},\bm{\nabla}) \Psi_n^{\rm LO} (\bm{r})
\Big|_{|\bm{r}'| = r_0}
%---------------
\end{equation}
%---------------
at small $|\bm{r}'| = r_0$. 
To find a simplified expression of the UV divergence in this integral, 
we examine the small-$r$ behavior of the integrand. 
We note that the contribution from 
$V^{(1)}(r)$ is logarithmically divergent, because 
$V^{(1)}(r)$ diverges like $1/r^2$ at $r=0$.
The same applies to contribution from 
$V_{p^2}^{(2)}(r) V_{\rm LO} (r)$ and $( V_{\rm LO} (r) )^2$ terms in 
$\delta {\cal V}(\bm{r},\bm{\nabla})$. 
On the other hand, the contribution from $V^{(2)} (r,\bm{\nabla})$ is 
linearly divergent, because 
$V^{(2)} (r,\bm{\nabla})$ contains contributions
that diverge like $1/r^3$ at $r=0$. 
If we expand $\Psi_n^{\rm LO} (\bm{r})$ in powers of $r$ in the integrand, 
the UV-divergent contributions are contained in the order $r$ and $r^2$ terms 
of the expansion\footnote{In the calculation of $S$-wave wavefunctions in
ref.~\cite{Chung:2020zqc}, 
only the leading-power contribution was kept, because 
power divergent contributions to $S$-wave wavefunctions at the origin
only arise from delta functions in the potential, so that any positive powers
of $r$ can be neglected.}. That is, 
%---------------
\begin{eqnarray}
\label{eq:uvdiv_expansion1}
%---------------
&& \hspace{-5ex} 
- \int d^3 r \, 
\bm{\nabla}_{\bm{r}'}
G(\bm{r}', \bm{r}; E)
\delta {\cal V}(\bm{r}, \bm{\nabla}) \Psi_n^{\rm LO} (\bm{r})
\nonumber\\ &=& 
- \int d^3 r \, 
\bm{\nabla}_{\bm{r}'}
G(\bm{r}', \bm{r}; E)
\delta {\cal V}(\bm{r}, \bm{\nabla}) 
\left\{ r \left[ \frac{\partial}{\partial r} \Psi_n^{\rm LO} (\bm{r}) 
\right]_{r=0} 
+ \frac{1}{2} r^2 
\left[ \frac{\partial^2}{\partial r^2} \Psi_n^{\rm LO} (\bm{r}) \right]_{r=0}
\right\} 
\nonumber \\ && + \textrm{UV-finite contributions}, 
%---------------
\end{eqnarray}
%---------------
where the UV-finite contributions are finite in the limit 
$|\bm{r}'| \to 0$. 
We now express the terms in the curly brackets in terms of 
$\bm{\nabla} \Psi_n^{\rm LO} (\bm{0})$. The first term in the curly brackets is
given by 
%---------------
\begin{equation}
%---------------
r \left[ \frac{\partial}{\partial r} \Psi_n^{\rm LO} (r) \right]_{r=0}
= \bm{r} \cdot \left[ \bm{\nabla} \Psi_n^{\rm LO} (\bm{0}) \right], 
%---------------
\end{equation}
%---------------
where $r = |\bm{r}|$. 
For the second term, we use the relation 
%---------------
\begin{equation}
%---------------
\frac{\partial^2}{\partial r^2} = \bm{\nabla}^2 - \frac{2}{r}
\frac{\partial}{\partial r} + \frac{\bm{L}^2}{r^2}, 
%---------------
\end{equation}
%---------------
which gives 
%---------------
\begin{equation}
%---------------
\frac{\partial^2}{\partial r^2} \Psi_n^{\rm LO} (\bm{r}) 
= m [V_{\rm LO} (r)- E] \Psi_n^{\rm LO} (\bm{r}) 
- 2 \frac{\partial}{\partial r} \left( \frac{1}{r} \Psi_n^{\rm LO} (\bm{r})
\right) , 
%---------------
\end{equation}
%---------------
where we used $\bm{L}^2 \Psi_n^{\rm LO} (\bm{r}) = 2 \Psi_n^{\rm LO} (\bm{r})$
for $P$-wave states.
Since $\Psi_n^{\rm LO} (\bm{r})$ vanishes linearly as $r\to 0$, 
we can write the limit $r \to 0$ as 
%---------------
\begin{equation}
%---------------
\lim_{r \to 0} \frac{\partial^2}{\partial r^2} \Psi_n^{\rm LO} (\bm{r})
= m [r V_{\rm LO} (r)]_{r=0} 
\left[ \frac{\partial}{\partial r} \Psi_n^{\rm LO} (\bm{r}) \right]_{r=0} 
- 2 \left[ \frac{\partial}{\partial r} \left( \frac{1}{r} 
\Psi_n^{\rm LO} (\bm{r}) \right) \right]_{r=0}.
%---------------
\end{equation}
%---------------
On the other hand, by using the fact that $\Psi_n^{\rm LO} (\bm{r})$ is
regular and vanishes linearly at $r=0$, 
we obtain from Taylor series expansion, 
%---------------
\begin{equation}
%---------------
\lim_{r \to 0} \frac{\partial^2}{\partial r^2} \Psi_n^{\rm LO} (\bm{r})
= 2 \left[ \frac{\partial}{\partial r} \left( \frac{1}{r}
\Psi_n^{\rm LO} (\bm{r}) \right) \right]_{r=0},
%---------------
\end{equation}
%---------------
which implies 
%---------------
\begin{equation}
\label{eq:wfseconderivative}
%---------------
\lim_{r \to 0} \frac{\partial^2}{\partial r^2} \Psi_n^{\rm LO} (\bm{r})
= \frac{1}{2} m [r V_{\rm LO} (r)]_{r=0}
\left[ \frac{\partial}{\partial r} \Psi_n^{\rm LO} (\bm{r}) \right]_{r=0}. 
%---------------
\end{equation}
%---------------
We note that $[r V_{\rm LO} (r)]_{r=0}$ is nothing but the Coulomb strength
of the LO potential at short distances, 
which is given by $-\alpha_s C_F$. 
Then, the second term in the curly brackets in 
eq.~(\ref{eq:uvdiv_expansion1}) can be written as 
%---------------
\begin{equation}
%---------------
\frac{1}{2} r^2
\bigg[ \frac{\partial^2}{\partial r^2} \Psi_n^{\rm LO} (\bm{r}) \bigg]_{r=0} 
= - \frac{\alpha_s C_F }{4} m r 
\bm{r} \cdot \left[ \bm{\nabla} 
\Psi_n^{\rm LO} (\bm{0}) \right], 
%---------------
\end{equation}
%---------------
so that 
%---------------
\begin{eqnarray}
\label{eq:uvdiv_expansion2}
%---------------
&& \hspace{-5ex}
- \int d^3 r \, 
\bm{\nabla}_{\bm{r}'}
G(\bm{r}', \bm{r}; E)
\delta {\cal V}(\bm{r}, \bm{\nabla}) \Psi_n^{\rm LO} (\bm{r})
\nonumber\\ &=&
- \int d^3 r \, 
\bm{\nabla}_{\bm{r}'}
G(\bm{r}', \bm{r}; E)
\delta {\cal V}(\bm{r}, \bm{\nabla})
\left( 
1 - \frac{\alpha_s C_F}{4} m r 
\right)
\bm{r} \cdot \left[ \bm{\nabla}
\Psi_n^{\rm LO} (\bm{0}) \right]
\nonumber \\ && + \textrm{UV-finite contributions}.
%---------------
\end{eqnarray}
%---------------
In order to be able to compare with dimensionally regulated expressions of the
UV divergence, it is useful to find
the momentum-space expression of the UV-divergent part. 
In 3 spatial dimensions, 
the momentum-space Green's function $\tilde G(\bm{p}', \bm{p}; E)$ 
is related to the position-space counterpart by 
%---------------
\begin{equation}
%---------------
G(\bm{r}', \bm{r}; E) = 
\int_{\bm{p}} \int_{\bm{p}'} 
e^{i \bm{p}' \cdot \bm{r}'} e^{-i \bm{p} \cdot \bm{r}} 
\tilde G(\bm{p}', \bm{p}; E) \Big|_{d=4}, 
%---------------
\end{equation}
%---------------
where 
%---------------
\begin{equation}
%---------------
\int_{\bm{p}} \equiv \int \frac{d ^{d-1} \bm{p}}{(2 \pi)^{d-1}}. 
%---------------
\end{equation}
%---------------
This lets us write the leading term as 
%---------------
\begin{eqnarray}
%---------------
&& \hspace{-5ex}
-\int d^3 r \, 
\bm{\nabla}_{\bm{r}'}^i 
G(\bm{r}', \bm{r}; E)
\delta {\cal V}(\bm{r}, \bm{\nabla})
\bm{r}^j 
\nonumber \\ &=& 
-\int_{\bm{p}'} \int_{\bm{p}} e^{i \bm{p}' \cdot \bm{r}'} 
\bm{p}'^i \tilde G(\bm{p}', \bm{p}; E) 
[ \bm{\nabla}_{\bm{q}}^j \delta \tilde{{\cal V}} (\bm{p}, \bm{q}) 
] \Big|_{\bm{q}=\bm{0}, \, d=4}.
%---------------
\end{eqnarray}
%---------------
Here $\delta \tilde{{\cal V}} (\bm{p}, \bm{q})$ is related to the
position-space counterpart in 3 spatial dimensions by 
%---------------
\begin{equation}
%---------------
\delta \tilde{{\cal V}} (\bm{p}, \bm{q}) \Big|_{d=4} = \int d^3r \,
e^{i \bm{p} \cdot \bm{r}} \delta {\cal V} (\bm{r}, \bm{\nabla}) 
e^{-i \bm{q} \cdot \bm{r}} . 
%---------------
\end{equation}
%---------------
The $d$-dimensional expression for $\delta \tilde{{\cal V}} (\bm{p}, \bm{q})$
will be given later. 
For the subleading term, we need 
the Fourier transform of $-\frac{1}{4} \alpha_s C_F m r \bm{r}$. 
We first compute the Fourier transform of $-\frac{1}{4} \alpha_s C_F m r$ as 
%---------------
\begin{equation}
%---------------
- \frac{\alpha_s C_F m}{4} \int d^3 r \, e^{i \bm{k} \cdot \bm{r}} r 
= 
\frac{\alpha_s C_F m}{4} \bm{\nabla}_{\bm{k}}^2 
\int d^3 r \frac{e^{i \bm{k} \cdot \bm{r}} }{r}
= 
\frac{2 \pi \alpha_s C_F m}{\bm{k}^4},
%---------------
\end{equation}
%---------------
which gives us 
%---------------
\begin{equation}
%---------------
- \frac{\alpha_s C_F m}{4} \int d^3 r e^{i \bm{k} \cdot \bm{r}} r 
\bm{r} 
= i \bm{\nabla}_{\bm{q}} \frac{2 \pi \alpha_s C_Fm }{(\bm{k}-\bm{q})^4}
\Big|_{\bm{q}=\bm{0}}
= i \frac{1}{-\bm{k}^2/m} 
\bm{\nabla}_{\bm{q}} \frac{-4 \pi \alpha_s C_F}{(\bm{k}-\bm{q})^2} 
\Big|_{\bm{q}=\bm{0}}.
%---------------
\end{equation}
%---------------
From this we obtain 
%---------------
\begin{eqnarray}
\label{eq:uvdiv_expansion2}
%---------------
&& \hspace{-5ex}
- \int d^3 r \, 
\bm{\nabla}_{\bm{r}'}^i
G(\bm{r}', \bm{r}; E)
\delta {\cal V}(\bm{r}, \bm{\nabla}) \Psi_n^{\rm LO} (\bm{r})
\nonumber\\ &=&
[\bm{\nabla}^j \Psi_n^{\rm LO} (\bm{0})] \bigg[ 
-\int_{\bm{p}'} \int_{\bm{p}} e^{i \bm{p}' \cdot \bm{r}'}
\bm{p}'^i \tilde G(\bm{p}', \bm{p}; E)
[ \bm{\nabla}_{\bm{q}}^j 
\delta \tilde{{\cal V}} (\bm{p}, \bm{q}) ]_{\bm{q}=\bm{0}} 
\nonumber \\ &&  
- \int_{\bm{p}'} \int_{\bm{p}}
e^{i \bm{p}' \cdot \bm{r}'} \bm{p}'^i \tilde G(\bm{p}', \bm{p}; E)
\int_{\bm{k}}
\delta \tilde{{\cal V}} (\bm{p}, \bm{k})
\frac{1}{-\bm{k}^2/m} 
\left[ \bm{\nabla}_{\bm{q}}^j 
\frac{-4 \pi \alpha_s C_F}{(\bm{k}-\bm{q})^2} 
\right]_{\bm{q}=\bm{0}}
\bigg]
_{d=4} 
\nonumber \\ && + \textrm{UV-finite contributions}.
%---------------
\end{eqnarray}
%---------------
Since the last integral in the brackets is at most logarithmically
divergent, we are free to make modifications to this integrand as long as we
keep its large loop momentum behavior unchanged. If we make the replacements 
$1/(-\bm{k}^2/m) \to 1/(E-\bm{k}^2/m)$ and 
$\frac{-4 \pi \alpha_s C_F}{(\bm{k}-\bm{q})^2} \to 
\tilde V_{\rm LO} (\bm{k}-\bm{q})$, we obtain 
%---------------
\begin{eqnarray}
\label{eq:uvdiv_expansion3}
%---------------
&& \hspace{-5ex}
- \bm{\nabla}_{\bm{r}'}^i
\int d^3 r \, G(\bm{r}', \bm{r}; E)
\delta {\cal V}(\bm{r}, \bm{\nabla}) \Psi_n^{\rm LO} (\bm{r})
\nonumber\\ &=&
[\bm{\nabla}^j \Psi_n^{\rm LO} (\bm{0})] \bigg[ -
\int_{\bm{p}'} \int_{\bm{p}} e^{i \bm{p}' \cdot \bm{r}'}
\bm{p}'^i \tilde G(\bm{p}', \bm{p}; E)
[ \bm{\nabla}_{\bm{q}}^j
\delta \tilde{{\cal V}} (\bm{p}, \bm{q}) ]_{\bm{q}=\bm{0}}
\nonumber \\ &&
- \int_{\bm{p}'} \int_{\bm{p}}
e^{i \bm{p}' \cdot \bm{r}'} \bm{p}'^i \tilde G(\bm{p}', \bm{p}; E)
\int_{\bm{k}}
\frac{
\delta \tilde{{\cal V}} (\bm{p}, \bm{k})
[ \bm{\nabla}_{\bm{q}}^j \tilde{V}_{\rm LO} (\bm{k}-\bm{q}) ]_{\bm{q}=\bm{0}}
}{E-\bm{k}^2/m} 
\bigg] 
_{d=4} 
\nonumber \\ && + \textrm{UV-finite contributions}.
%---------------
\end{eqnarray}
%---------------
Finally, we use the following form of the momentum-space Green's function 
%---------------
\begin{eqnarray}
\label{eq:greenfunction_mom}
%---------------
\tilde G(\bm{p}',\bm{p};E) &= &
- \frac{(2 \pi)^{d-1} \delta^{(d-1)} (\bm{p}-\bm{p}')}{E-\bm{p}^2/m} 
- \frac{1}{E-\bm{p}'^2/m} \tilde V_{\rm LO} (\bm{p}'-\bm{p}) 
\frac{1}{E-\bm{p}^2/m} 
\nonumber \\ 
&&  
- \frac{1}{E-\bm{p}'^2/m} T(\bm{p}',\bm{p},E) 
\frac{1}{E-\bm{p}^2/m},
%---------------
\end{eqnarray}
%---------------
where\footnote{Here we correct a typo in ref.~\cite{Chung:2020zqc} where 
$E-(\bm{p}'+\bm{k}_i)^2/(2 m)$ was used instead of 
$E-(\bm{p}'+\bm{k}_i)^2/m$ in the denominators of $T (\bm{p}',\bm{p},E)$.}
%---------------
\begin{equation}
%---------------
T (\bm{p}',\bm{p},E) =
\sum_{n=1}^\infty
\int_{\bm{k}_1}
\int_{\bm{k}_2}
\cdots
\int_{\bm{k}_n}
\tilde V_{\rm LO} (\bm{k}_1)
\prod_{i=1}^n \frac{
\tilde V_{\rm LO} ( \bm{k}_{i+1}-\bm{k}_i)
}{\left[E - \frac{(\bm{p}'+\bm{k}_i)^2}{m} \right]}, 
%---------------
\end{equation}
%---------------
with $\bm{k}_{n+1} = \bm{p}-\bm{p}'$. 
Equation~(\ref{eq:greenfunction_mom}) is the solution of the momentum-space
Lippmann-Schwinger equation
%---------------
\begin{equation}
%---------------
\left( \frac{\bm{p}'^2}{m}- E \right) \tilde G (\bm{p}',\bm{p};E)
+ \int_{\bm{k}} \tilde V_{\rm LO}(\bm{k}) \tilde G (\bm{p}'-\bm{k},\bm{p};E)
= (2 \pi)^{d-1} \delta^{(d-1)} (\bm{p}-\bm{p}').
%---------------
\end{equation}
%---------------
For later convenience, we write eq.~(\ref{eq:greenfunction_mom}) in a form 
that is valid in $d-1$ spatial dimensions. 
Equation~(\ref{eq:greenfunction_mom}) is organized in a way that the large
$\bm{p}$ and $\bm{p}'$ behavior becomes less divergent as the number of
the LO potential increases. Hence, in the first momentum integral in 
eq.~(\ref{eq:uvdiv_expansion3}), which is at most linearly divergent, 
it suffices to keep only the first two terms
of eq.~(\ref{eq:greenfunction_mom}), while in the second momentum integral in 
eq.~(\ref{eq:uvdiv_expansion3}), which is at most logarithmically divergent, 
we keep only the first term of the iterative solution for the Green's function. 
From this we find our final form of the UV-divergent part of the finite-$r$
regularized integral: 
%---------------
\begin{eqnarray}
\label{eq:uvdiv_expansion4}
%---------------
&& \hspace{-5ex}
- \bm{\nabla}_{\bm{r}'}^i
\int d^3 r \, G(\bm{r}', \bm{r}; E)
\delta {\cal V}(\bm{r}, \bm{\nabla}) \Psi_n^{\rm LO} (\bm{r})
\Big|_{|\bm{r}'|=r_0} 
\nonumber\\ &=&
\bigg[ 
\int_{\bm{p}} e^{i \bm{p} \cdot \bm{r}'}
\bm{p}^i \frac{
\big[ \bm{\nabla}_{\bm{q}}^j \delta \tilde{{\cal V}} (\bm{p}, \bm{q}) \big]
_{\bm{q}=\bm{0}} 
}{E-\bm{p}^2/m}
+ \int_{\bm{p}'} \int_{\bm{p}} e^{i \bm{p}' \cdot \bm{r}'}
\bm{p}'^i \frac{
\tilde{V}_{\rm LO} (\bm{p}'-\bm{p}) 
\big[ \bm{\nabla}_{\bm{q}}^j \delta \tilde{{\cal V}} (\bm{p}, \bm{q}) \big]
_{\bm{q}=\bm{0}}
}{(E-\bm{p}'^2/m) (E-\bm{p}^2/m)} 
\nonumber \\ && \hspace{5ex} 
+ \int_{\bm{p}'}
\int_{\bm{p}}
e^{i \bm{p}' \cdot \bm{r}'} \bm{p}'^i 
\frac{
\delta \tilde{{\cal V}} (\bm{p}', \bm{p})
\big[ \bm{\nabla}_{\bm{q}}^j \tilde{V}_{\rm LO} (\bm{p}-\bm{q}) \big]
_{\bm{q}=\bm{0}}
}{(E-\bm{p}'^2/m) (E-\bm{p}^2/m)} 
\bigg]_{|\bm{r}'|=r_0}
[\bm{\nabla}^j \Psi_n^{\rm LO} (\bm{0})] 
\nonumber \\ && + \textrm{UV-finite contributions}.
%---------------
\end{eqnarray}
%---------------
The UV divergences that are regulated by nonzero values of $|\bm{r}'| = r_0$ 
are now contained in the momentum integrals. 
Since we are only interested in the divergent contributions and finite terms in
the limit $r_0 \to 0$, we neglect any contributions that vanish as $r_0 \to 0$
such as positive powers of $r_0$ in the UV-divergent integral in the square
brackets. 

We note that the UV-finite contributions can contain infrared divergences; 
in fact, the logarithmically divergent contribution in
eq.~(\ref{eq:uvdiv_expansion4}) does contain IR divergences, because we
obtained the momentum-space expression by computing the Fourier transformation 
of a position-space expression that does not vanish as $r \to \infty$. 
Since the IR divergence in the UV-divergent part cancels with the UV-finite
part, it also cancels in the scheme conversion coefficient between the
finite-$r$ regularized integral and the dimensionally regulated one, 
as long as the IR divergences are regulated in the same way in both DR and
finite-$r$ regularization.

%------------------------------------------------------------------------------
\subsection{Divergences in the wavefunctions at the origin in DR}
%------------------------------------------------------------------------------

We now find the UV-divergent contribution of the dimensionally regulated
integral. The correct dimensionally regulated expression must reproduce the
loop corrections to the NRQCD LDMEs in the perturbative expansion in powers of
$\alpha_s$, because the renormalization of the LDMEs are carried out according
to eq.~(\ref{eq:ldme_renormalization}). 

%%%%%%%%%%%%%%%%%%%%%%%%%%%%%%%%%%%%%%%%%%%%%%%%%%%%%%%%%%%%%%%%%%%%%%%%%%%%%%%
\begin{figure}[tbp]
\centering
\includegraphics[width=.9\textwidth]{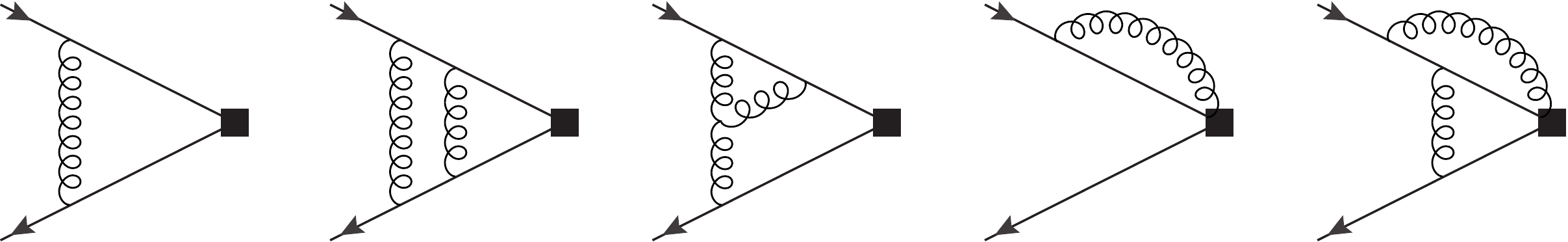}
\caption{\label{fig:pwavediags}
Some 
representative Feynman diagrams for one- and two-loop corrections to the NRQCD
LDME $\langle 0 | \chi^\dag (-\frac{i}{2} 
\protect \overleftrightarrow{\bm{D}} ) \psi | Q \bar Q \rangle$. 
Solid lines are heavy quarks and antiquarks, curly lines
are gluons, and filled squares represent the operator $\chi^\dag (-\frac{i}{2}
\protect \overleftrightarrow{\bm{D}} ) \psi$.
}
\end{figure}
%%%%%%%%%%%%%%%%%%%%%%%%%%%%%%%%%%%%%%%%%%%%%%%%%%%%%%%%%%%%%%%%%%%%%%%%%%%%%%%

We first examine the diagrams that contribute to the perturbative
calculation of the NRQCD LDMEs at two-loop accuracy. 
We consider the perturbative matrix element 
$\langle 0 | \chi^\dag (-\frac{i}{2} \overleftrightarrow{\bm{D}} ) \psi | Q
\bar Q \rangle$, where we keep the spin indices on $\chi$ and $\psi$ 
unconstrained, while the color indices are contracted, 
and $|Q \bar Q \rangle$ is in a color-singlet state. 
The $Q$ and $\bar Q$ are on shell, and have nonrelativistic
spatial momenta $\bm{q}$ and $-\bm{q}$ in the rest frame of the $Q \bar Q$. 
Some representative one- and two-loop diagrams are shown in
fig.~\ref{fig:pwavediags}. 
A general way to compute these
diagrams in DR is to first integrate over the temporal components of the loop 
momenta, and then expand the resulting integrand in powers of the spatial
components of momenta divided by the heavy quark mass, before
integrating over the spatial components of the loop momenta\footnote{An
alternative way to compute these diagrams in DR is to use the method of
regions~\cite{Beneke:1997zp}, 
where the loop integrals are partitioned by regions of loop momenta,
and expanded according to the power counting appropriate for each momentum
region before integrating over the temporal and spatial components of loop
momenta.}. The integration over the temporal components of loop momenta is
carried out by using contour integration, where one picks up contributions from
the poles coming from propagators of virtual lines. 
This enables us to separate the on-shell quark or antiquark contributions 
from on-shell gluon contributions. 
Note that, since in NRQCD the quark and 
antiquark propagator denominators are linear in the temporal components of the
momenta they carry, contribution from a on-shell quark line cannot be defined
unless there is an antiquark line that carries a momentum that contains the 
same loop momentum as the quark line. 
That is, if we have a virtual quark line whose
propagator is $i/[k_0 - \bm{k}^2/(2 m)+i \varepsilon]$, 
when integrating over $k^0$
we can neglect the contribution from its pole by closing the
contour on the upper half plane. However, when there is also a virtual 
antiquark line that carries the momentum $k+\ell$, where $\ell$ is the relative
momentum between the virtual quark and antiquark lines, the propagator of the
antiquark line is given by 
$i/[-(k_0+\ell_0) - (\bm{k}+\bm{\ell})^2/(2 m) +i\varepsilon]$, and so, 
the contour integration over $k^0$ must enclose the pole from either the 
quark propagator or the antiquark propagator. Therefore, we define the
contribution from on-shell quark or antiquark lines only when there are 
quark and antiquark lines that contain the same loop momentum. 

In general, after integration over the temporal components of the loop momenta, 
we can define two subdiagrams of a generic loop diagram 
by the on-shell $Q \bar Q$ pair that is most adjacent to 
the NRQCD operator $\chi^\dag (-\frac{i}{2} \overleftrightarrow{\bm{D}} )
\psi$ which, when removed, separates the diagram into two disconnected
diagrams. 
Let us label the subdiagram that contains the NRQCD operator as $S_O$, and
the subdiagram that contains the initial-state on-shell $Q \bar Q$ as $S_V$.
We note that $S_V$ depends on the relative spatial momenta of the initial and
final $Q \bar Q$, while $S_O$ depends on the relative spatial momenta of the
on-shell $Q$ and $\bar Q$ that connects $S_O$ and $S_V$; the temporal
components of these momenta are constrained by the on-shell condition. 
We include the on-shell $Q\bar Q$ lines between $S_O$ and $S_V$ 
in the subdiagram $S_V$. 
In this case, the sum of all Feynman diagrams is given by 
$\int_{\bm{p}} S_O(\bm{p}) S_V(\bm{p},\bm{q})$. 
We note that the subdiagram $S_V$ 
is given by scattering of on-shell $Q$ and $\bar Q$ via gluon exchanges. 
This is nothing but the matching condition for the
momentum-space potential in the on-shell matching scheme; 
that is, the potential in the on-shell matching scheme is the
irreducible contribution in $S_V$ that is obtained by amputating the
external legs and removing on-shell heavy quark and antiquark
lines~\cite{Pineda:2011dg}. 
The subdiagram $S_V$ is then given to all orders in perturbation theory by 
%---------------
\begin{eqnarray}
\label{eq:potential_OSmatching}
%---------------
S_V (\bm{p},\bm{q}) &=&
(2 \pi)^{d-1} \delta^{(d-1)} (\bm{p}-\bm{q})
+
\frac{1}{E_{\bm{q}}-\bm{p}^2/m} \bigg[ 
\tilde V (\bm{p},\bm{q})
+ \int_{\bm{k}_1}
\tilde V(\bm{p}, \bm{k}_1)
\frac{
\tilde V(\bm{k}_1, \bm{q})
}{E_{\bm{q}}-\bm{k}_1^2/m}
\nonumber \\ &&
+
\int_{\bm{k}_1} \int_{\bm{k}_2}
\tilde V(\bm{p}, \bm{k}_1)
\frac{
\tilde V(\bm{k}_1, \bm{k}_2)
}{E_{\bm{q}}-\bm{k}_1^2/m}
\frac{
\tilde V(\bm{k}_2, \bm{q})
}{E_{\bm{q}}-\bm{k}_2^2/m}
+\cdots\bigg],
%---------------
\end{eqnarray}
%---------------
where $E_{\bm{q}} = \bm{q}^2/m-i \varepsilon$. Here, $\tilde V(\bm{p},\bm{q})$ 
is the $d$-dimensional momentum-space potential in the on-shell 
matching scheme. Explicit expressions for the potential 
$\tilde V(\bm{p},\bm{q})$ in $d-1$ spatial dimensions 
is shown in eq.~(\ref{eq:pot_pert_DR}). 
The first term in eq.~(\ref{eq:potential_OSmatching}) 
is the tree-level contribution that corresponds to the case
where none of the virtual quark or antiquark lines are on shell. 
At tree level, the subdiagram
$S_O$ is just the tree-level NRQCD LDME.  The contribution to the NRQCD LDME
from $S_O$ at tree level and $S_V$ to all orders in $\alpha_s$ is given by 
%---------------
\begin{eqnarray}
\label{eq:ldme_wavefunction}
%---------------
\int_{\bm{p}} S_O^{\rm tree} (\bm{p}) S_V(\bm{p},\bm{q}) &=& 
\bm{q} + 
\int_{\bm{p}} \frac{\bm{p} }{E_{\bm{q}} - \bm{p}^2/m} 
\bigg[ 
\tilde V (\bm{p},\bm{q}) 
+ \int_{\bm{k}_1}
\frac{
\tilde V(\bm{p}, \bm{k}_1)
\tilde V(\bm{k}_1, \bm{q}) 
}{E_{\bm{q}}-\bm{k}_1^2/m}
\nonumber \\ && \hspace{5ex} 
+ \int_{\bm{k}_1} \int_{\bm{k}_2}
\frac{ \tilde V(\bm{p}, \bm{k}_1) \tilde V(\bm{k}_1, \bm{k}_2)
\tilde V(\bm{k}_2, \bm{q})
}{(E_{\bm{q}}-\bm{k}_1^2/m) (E_{\bm{q}}-\bm{k}_2^2/m) }
+ \cdots \bigg].
%---------------
\end{eqnarray}
%---------------
Here, we suppress the spin, color, and kinematical factors in $S_O$, 
so that the tree-level LDME is just $\bm{q}$. 
We note that 
$S_O^{\rm tree} (\bm{p})$ has $Q \bar Q$ color indices that are
proportional to the $SU(3)$ identity matrix, and the $|Q \bar Q\rangle$ state
is also in a color-singlet state. Hence, we only need to consider the 
color-singlet projection of the potential. 

We can identify the contributions to eq.~(\ref{eq:ldme_wavefunction}) from the
Feynman diagrams in fig.~\ref{fig:pwavediags} as follows. 
The first diagram in fig.~\ref{fig:pwavediags} contributes to a single
insertion of the tree-level potential when the virtual $Q \bar Q$ lines are on
shell. The second diagram in fig.~\ref{fig:pwavediags} contributes to two
insertions of the tree-level potential when all of the $Q \bar Q$ lines are on
shell; if the $Q \bar Q$ lines between the gluon vertices are off shell, 
the diagram contributes to one insertion of the one-loop Abelian term
of the potential. 
The third diagram contributes to one insertion of the one-loop non-Abelian
term of the potential proportional to $C_A$ when the virtual $Q \bar Q$ lines
adjacent to the NRQCD operator $\chi^\dag (- \frac{i}{2}
\overleftrightarrow{\bm{D} } ) \psi$ are on shell (the virtual quark line
between the two gluon vertices does not have a corresponding antiquark line, 
so that we can always neglect the contribution from the pole of its
propagator). The remaining diagrams, as well as contributions from the first
three diagrams in which the virtual $Q \bar Q$ lines adjacent to the NRQCD
operator are off shell, involves loop corrections to $S_O$ and are not included
in eq.~(\ref{eq:ldme_wavefunction}). 
Analyses of Feynman diagrams that are not shown in
fig.~\ref{fig:pwavediags} can be carried out in the same way. 

From scattering theory in nonrelativistic quantum mechanics, it is known that 
eq.~(\ref{eq:ldme_wavefunction}) is the contribution from the wavefunction.
We can rewrite eq.~(\ref{eq:ldme_wavefunction}) as 
%---------------
\begin{eqnarray}
\label{eq:ldme_wavefunction_wf}
%---------------
\int_{\bm{p}} S_O^{\rm tree} (\bm{p}) S_V(\bm{p},\bm{q}) =
\int_{\bm{p}} 
\bm{p} \lim_{E \to \bm{q}^2/m} \left[ \tilde G_{\rm full}(\bm{p},\bm{q};E) 
(\bm{q}^2/m-E) \right],  
%---------------
\end{eqnarray}
%---------------
where $\tilde G_{\rm full}(\bm{p},\bm{q};E)$ is the momentum-space Green's
function that satisfies 
%---------------
\begin{equation}
%---------------
\left( \frac{\bm{p}'^2}{m}- E \right) \tilde G_{\rm full} (\bm{p}',\bm{p};E)
+\int_{\bm{k}} \tilde V (\bm{p}',\bm{k}) \tilde G_{\rm full} (\bm{k},\bm{p};E)
= (2 \pi)^{d-1} \delta^{(d-1)} (\bm{p}-\bm{p}').
%---------------
\end{equation}
%---------------
The quantity in the square brackets in eq.~(\ref{eq:ldme_wavefunction_wf})
satisfies the momentum-space Schr\"odinger equation with respect to $\bm{p}$, 
with energy eigenvalue $E = \bm{q}^2/m$. Hence, 
eq.~(\ref{eq:ldme_wavefunction_wf}) represents $i \bm{\nabla} \Psi
(\bm{0})$ in momentum space in $d-1$ spatial dimensions. 

There are still contributions that come from loop corrections to the subdiagram
$S_O$ beyond tree level. For example, the last two Feynman diagrams in
fig.~\ref{fig:pwavediags} correspond to loop corrections to $S_O$. 
The first three Feynman diagrams in fig.~\ref{fig:pwavediags} can also 
contribute to $S_O$, when the quark lines are off shell. 
In general, loop corrections to $S_O$ come from regions of loop momenta
where the virtual quark lines are off shell.
In the pNRQCD expression for the NRQCD
LDMEs, the contribution from this momentum region corresponds to the 
corrections to the contact term $V_{\cal O}$, namely the term $\frac{2}{3}
\frac{i {\cal E}_2}{m}$ and corrections of higher orders in $1/m$ that appear
in eq.~(\ref{eq:contact_DR}). 
Since we are only interested in 
the contributions that correspond to the wavefunctions, we do not need to 
consider $S_O$ beyond tree level.

Having established that eq.~(\ref{eq:ldme_wavefunction}) corresponds to the
NRQCD LDME that comes from the wavefunction, we now use 
$\tilde V(\bm{p},\bm{q}) = \tilde V_{\rm LO} (\bm{p}-\bm{q}) + 
\tilde{\delta V} (\bm{p},\bm{q})$ and expand in powers of 
$\tilde{\delta V} (\bm{p},\bm{q})$. 
This expansion 
corresponds to the 
Rayleigh-Schr\"odinger perturbation theory in momentum space, once the effect
of $\tilde{\delta V} (\bm{p},\bm{q})$ to the energy eigenvalues are also taken
into account. 
We can reduce the $\bm{p}^2$ and $\bm{q}^2$
terms in $\tilde{\delta V} (\bm{p},\bm{q})$ using the Schr\"odinger equation
and the Lippmann-Schwinger equation in momentum space to obtain 
$\tilde{\delta {\cal V}}(\bm{p},\bm{q})$, which is the momentum-space version
of $\delta {\cal V} (\bm{r},\bm{\nabla})$ in $d-1$ spatial dimensions. 
We display the explicit expression for $\tilde{\delta {\cal V}}(\bm{p},\bm{q})$
in eq.~(\ref{eq:potvar_pert_DR}).
Since we are only interested in the divergent contributions, we replace 
$\tilde{\delta V} (\bm{p},\bm{q})$ by $\tilde{\delta {\cal V}}(\bm{p},\bm{q})$ 
in the integrand. 
Then, the divergences in the loop corrections in
eq.~(\ref{eq:ldme_wavefunction}) at first order in the QMPT 
is given by 
%---------------
\begin{equation}
\label{eq:ldme_twoloop1}
%---------------
- \int_{\bm{p}} \bm{p} 
\lim_{E \to \bm{q}^2/m} 
\left[ \int_{\bm{k}_1} \int_{\bm{k}_2}
\tilde G(\bm{p},\bm{k}_1;E) 
\tilde {\delta {\cal V}} (\bm{k}_1,\bm{k}_2)
\tilde G(\bm{k}_2,\bm{q};E) (\bm{q}^2/m-E) 
\right].
%---------------
\end{equation}
%---------------
In perturbative calculations of the LDMEs, we are interested in the 
UV divergences
in this integral that is proportional to the tree-level LDME $\bm{q}$. 
By using the fact that this integral is at most linearly UV divergent, 
we can keep only the first two terms in the iterative solution 
in eq.~(\ref{eq:greenfunction_mom}), which gives 
%---------------
\begin{eqnarray}
\label{eq:ldme_twoloop2}
%---------------
\int_{\bm{p}} S_O^{\rm tree} (\bm{p}) S_V(\bm{p},\bm{q}) \Big|_{\rm UV} 
&=& 
\int_{\bm{p}} \frac{\bm{p} \, 
\tilde {\delta {\cal V}} (\bm{p},\bm{q})
}{E_{\bm{q}} -\bm{p}^2/m} 
+\int_{\bm{p}'} \int_{\bm{p}} 
\frac{\bm{p}'
\tilde V_{\rm LO} (\bm{p}'-\bm{p}) 
\tilde{\delta {\cal V}} (\bm{p},\bm{q}) 
}{(E_{\bm{q}} -\bm{p}'^2/m) (E_{\bm{q}}-\bm{p}^2/m)} 
\nonumber \\ && 
+\int_{\bm{p}'} \int_{\bm{p}} \frac{\bm{p}'
\tilde {\delta {\cal V}} (\bm{p}',\bm{p}) 
\tilde V_{\rm LO} (\bm{p}-\bm{q}) 
}{(E_{\bm{q}} -\bm{p}'^2/m) (E_{\bm{q}}-\bm{p}^2/m)}. 
%---------------
\end{eqnarray}
%---------------
Since in perturbation theory both $\tilde V_{\rm LO}$ and 
$\tilde {\delta {\cal V}}$ are suppressed by at least $\alpha_s$, 
this also corresponds to the expansion of eq.~(\ref{eq:ldme_twoloop1}) to
two-loop accuracy. 
Equation~(\ref{eq:ldme_twoloop2}) will generally depend on $\bm{q}$, while
we are only interested in the piece that is linear in $\bm{q}$, which is the
contribution that is proportional to the tree-level LDME. 
We note that, due to rotational symmetry, eq.~(\ref{eq:ldme_twoloop2}) vanishes
when we set $\bm{q}=\bm{0}$ in $\tilde {\delta {\cal V}} (\bm{p},\bm{q})$ and 
$\tilde V_{\rm LO} (\bm{p}-\bm{q})$. Hence, the nonvanishing contribution 
that is linear in $\bm{q}$ comes from the expansion of 
$\tilde {\delta {\cal V}} (\bm{p},\bm{q})$ and
$\tilde V_{\rm LO} (\bm{p}-\bm{q})$ to linear order in $\bm{q}$, which is given
by 
%---------------
\begin{eqnarray}
\label{eq:ldme_twoloop3}
%---------------
\int_{\bm{p}} S_O^{\rm tree} (\bm{p}) S_V(\bm{p},\bm{q}) \Big|_{\rm UV} 
&=& 
\bigg[ 
\int_{\bm{p}} \frac{\bm{p}
\big[ \bm{\nabla}_{\bm{q}}^j \tilde {\delta {\cal V}} (\bm{p},\bm{q})
\big]_{\bm{q}=\bm{0}} 
}{E -\bm{p}^2/m}
+\int_{\bm{p}'} \int_{\bm{p}} \frac{\bm{p}'
\tilde V_{\rm LO} (\bm{p}'-\bm{p}) 
\big[ \bm{\nabla}_{\bm{q}}^j \tilde{\delta {\cal V}} (\bm{p},\bm{q})
\big]_{\bm{q}=\bm{0}} 
}{(E -\bm{p}'^2/m) (E-\bm{p}^2/m)}
\nonumber \\ && \hspace{5ex}
+\int_{\bm{p}'} \int_{\bm{p}} \frac{\bm{p}'
\tilde {\delta {\cal V}} (\bm{p}',\bm{p})
\big[ \bm{\nabla}_{\bm{q}}^j \tilde V_{\rm LO} (\bm{p}-\bm{q}) 
\big]_{\bm{q}=\bm{0}}
}{(E -\bm{p}'^2/m) (E-\bm{p}^2/m)}
\bigg] 
\bm{q}^j.
%---------------
\end{eqnarray}
%---------------
This expression can also be regarded as the leading asymptotic behavior of the
integrand of eq.~(\ref{eq:ldme_twoloop2}) for large loop momenta. 
Because $\tilde{ \delta {\cal V}} (\bm{p},\bm{q})$ is 
given as expansion in powers of $1/m$, each coefficient of this expansion 
is a function of $\bm{p}$ and $\bm{q}$ and has a definite mass dimension. 
The same applies to $\tilde V_{\rm LO} (\bm{p}-\bm{q})$, 
which is a function of $|\bm{p}-\bm{q}|$ and also has a definite mass
dimension. Hence, the large $\bm{p}$ behavior of the integrand is given by 
the expansions of $\tilde V_{\rm LO} (\bm{p}'-\bm{p})$ and 
$\tilde{ \delta {\cal V}} (\bm{p},\bm{q})$ in
powers of $\bm{q}$ at leading nonvanishing order. 
At this point, $E$ in eq.~(\ref{eq:ldme_twoloop3}) can be regarded as an
arbitrary complex number, and we can take the limit $E \to E_n^{\rm LO}$ when
necessary. 

The quantity in the square brackets in 
eq.~(\ref{eq:ldme_twoloop3}) is our expression for the UV-divergent part
of the correction to the wavefunction in DR. 
Since we obtained
eq.~(\ref{eq:ldme_twoloop3}) from loop corrections to the NRQCD LDMEs in $d$
spacetime dimensions, our result is valid in dimensionally regulated matching
calculations in NRQCD. We also note that the loop
integrands in eqs.~(\ref{eq:ldme_twoloop3}) and 
(\ref{eq:uvdiv_expansion4}) are same once the UV regulators are lifted 
($\epsilon \to 0$ and $r_0 \to 0$). Hence, eq.~(\ref{eq:ldme_twoloop3})
is the DR counterpart of eq.~(\ref{eq:uvdiv_expansion4}). 
We can now compute the 
$\delta \bar{Z}^{ij}$ from the difference
in the divergent integrals in eqs.~(\ref{eq:uvdiv_expansion4}) and
(\ref{eq:ldme_twoloop3}) between the two different regularizations. 
We define 
%---------------
\begin{equation}
\label{eq:schemeconversion_calc1}
%---------------
\delta Z_E^{ij} = {\cal J}_{(r_0)}^{ij} - {\cal J}_{\rm DR}^{ij},
%---------------
\end{equation}
%---------------
where 
%---------------
\begin{subequations}
\label{eq:schemeconversion_calc2}
\begin{eqnarray}
%---------------
{\cal J}_{(r_0)}^{ij} &=&
\bigg[ \int_{\bm{p}} \frac{ 
e^{i \bm{p} \cdot \bm{r}}
\bm{p}^i \big[ \bm{\nabla}_{\bm{q}}^j
\tilde {\delta {\cal V}} (\bm{p},\bm{q}) \big]_{\bm{q}=\bm{0}}
}{E -\bm{p}^2/m}
+\int_{\bm{p}'} \int_{\bm{p}}
\frac{ e^{i \bm{p}' \cdot \bm{r}} \bm{p}'^i
\tilde V_{\rm LO} (\bm{p}'-\bm{p})
 \big[ \bm{\nabla}_{\bm{q}}^j \tilde{\delta {\cal V}}
(\bm{p},\bm{q}) \big]_{\bm{q}=\bm{0}}
}{( E -\bm{p}'^2/m) (E -\bm{p}^2/m)}
\nonumber \\ && \hspace{2ex}
+\int_{\bm{p}'} \int_{\bm{p}} \frac{
e^{i \bm{p}' \cdot \bm{r}}
\bm{p}'^i \tilde {\delta {\cal V}} (\bm{p}',\bm{p})
\big[ \bm{\nabla}_{\bm{q}}^j
\tilde V_{\rm LO} (\bm{p}-\bm{q}) \big]_{\bm{q}=\bm{0}}
}{(E -\bm{p}'^2/m) (E-\bm{p}^2/m)} \bigg]_{|\bm{r}| = r_0} , 
\\
{\cal J}_{\rm DR}^{ij} &=&
\int_{\bm{p}} \frac{ \bm{p}^i \big[ \bm{\nabla}_{\bm{q}}^j
\tilde {\delta {\cal V}} (\bm{p},\bm{q}) \big]_{\bm{q}=\bm{0}}
}{E -\bm{p}^2/m}
+\int_{\bm{p}'} \int_{\bm{p}}
\frac{\bm{p}'^i
\tilde V_{\rm LO} (\bm{p}'-\bm{p})
 \big[ \bm{\nabla}_{\bm{q}}^j \tilde{\delta {\cal V}}
(\bm{p},\bm{q}) \big]_{\bm{q}=\bm{0}}
}{( E -\bm{p}'^2/m) (E -\bm{p}^2/m)}
\nonumber \\ && \hspace{2ex}
+\int_{\bm{p}'} \int_{\bm{p}} \frac{
\bm{p}'^i \tilde {\delta {\cal V}} (\bm{p}',\bm{p})
\big[ \bm{\nabla}_{\bm{q}}^j
\tilde V_{\rm LO} (\bm{p}-\bm{q}) \big]_{\bm{q}=\bm{0}}
}{(E -\bm{p}'^2/m) (E-\bm{p}^2/m)}. 
%---------------
\end{eqnarray}
\end{subequations}
%---------------
Here, the momentum integrals in 
${\cal J}_{(r_0)}^{ij}$ are computed in finite-$r$
regularization with $|\bm{r}| = r_0$, 
and the integrals in ${\cal J}_{\rm DR}^{ij}$ are computed in $d-1$ spatial
dimensions. 
Note that we have now dropped the prime in $\bm{r}'$. 
We refer to the first terms in our definitions of
${\cal J}_{\rm DR}^{ij}$ and ${\cal J}_{(r_0)}^{ij}$ as the leading divergent
contributions, and the remaining terms as the subleading divergent
contributions, based on the analysis in the previous section.
In case the integrals contain IR divergences, 
both ${\cal J}_{\rm DR}^{ij}$ and ${\cal J}_{(r_0)}^{ij}$ can be computed 
in $d-1$ spatial dimensions, so that the IR divergences in both integrals 
are regulated dimensionally, and cancel in $\delta Z_E^{ij}$. 
Once we obtain $\delta Z_E^{ij}$, 
$\delta \bar{Z}^{ij}$ is given by $\delta Z_E^{ij}$ at values
of $E$ that coincide with the LO energy eigenvalues. 
Then, $\delta Z^{ij} = Z^{ij} - \delta^{ij}$ in the $\overline{\rm MS}$ scheme 
is found by subtracting the UV poles in $\delta \bar{Z}^{ij}$.

%------------------------------------------------------------------------------
\subsection{Calculation of the scheme conversion coefficient}
%------------------------------------------------------------------------------

In this section we compute the scheme conversion coefficient 
$Z^{ij}$ by computing 
$\delta Z_E^{ij}$ from the divergent integrals 
${\cal J}_{\rm DR}^{ij}$ and ${\cal J}_{(r_0)}^{ij}$, 
which are regulated in DR and finite-$r$ regularization, respectively. 
We note that since the NRQCD LDMEs are given in terms of 
$ [\hat{\bm{r}} \cdot\bm{\nabla} \Psi (\bm{0})]^{\overline{\rm MS}}$, 
we only need to obtain $\hat{\bm{r}}^i \delta Z^{ij}$. 
This, in turn, implies that we may contract $\hat{\bm{r}}^i$ with 
$\delta Z_E^{ij}$ before renormalization is carried out, as long as doing so 
does not modify the finite parts. This is especially useful in calculations of 
the finite-$r$ regularized integrals, because in this case, 
contracting $\hat{\bm{r}}^i$
in the integrand reduces the rank of the $\bm{r}$-dependent tensor integrals.

We first investigate the contribution to $\delta Z_E^{ij}$ from the
spin-independent terms in $\tilde{\delta {\cal V}}$. 
The leading divergent contribution to the dimensionally regulated integral 
${\cal J}_{\rm DR}^{ij}$ from the $1/m$ potential is given by 
%---------------
\begin{equation}
\label{eq:jcal_dr_1mpot}
%---------------
\frac{\pi^2 \alpha_s^2 C_F}{m} 
\left( \frac{C_F}{2} (1-2 \epsilon) - C_A (1-\epsilon) \right) c_\epsilon
\int_{\bm{p}} \frac{\bm{p}^i}{E-\bm{p}^2/m} 
\left[ \bm{\nabla}_{\bm{q}}^j 
\frac{\Lambda^{2 \epsilon}}{|\bm{p}-\bm{q}|^{1+2 \epsilon}} 
\right]_{\bm{q}=\bm{0}}, 
%---------------
\end{equation}
%---------------
where $c_\epsilon$ is defined in eq.~(\ref{eq:cep}). 
The integral over $\bm{p}$ is computed as 
%---------------
\begin{eqnarray}
%---------------
&& \hspace{-10ex} 
\int_{\bm{p}} \frac{\bm{p}^i}{E-\bm{p}^2/m}
\left[ \bm{\nabla}_{\bm{q}}^j
\frac{\Lambda^{2 \epsilon}}{|\bm{p}-\bm{q}|^{1+2 \epsilon}}
\right]_{\bm{q}=\bm{0}}
=
\int_{\bm{p}} \frac{(1+2 \epsilon) \hat{\bm{p}}^i \hat{\bm{p}}^j
}{E-\bm{p}^2/m}
\frac{\Lambda^{2 \epsilon}}{|\bm{p}|^{1+2 \epsilon}}
\nonumber \\
&=&
- \frac{\delta^{ij}}{d-1} \frac{m (1+2 \epsilon) }{8 \pi^2} 
\left[ \frac{1}{\epsilon_{\rm UV}} +2 + 
2 \log \left( - \frac{\Lambda^2}{2 m E} \right) 
+ O(\epsilon) \right]. 
%---------------
\end{eqnarray}
%---------------
Since this integral is logarithmically divergent, the contributions to
the subleading terms in ${\cal J}_{\rm DR}^{ij}$ from the $1/m$ potential 
are finite,
so that we do not need to consider them. 
The leading divergent contribution to the finite-$r$ regularized integral 
${\cal J}_{(r_0)}^{ij}$ is given by 
%---------------
\begin{equation}
\label{eq:jcal_fr_1mpot}
%---------------
\frac{\pi^2 \alpha_s^2 C_F}{m}
\left( \frac{C_F}{2} - C_A \right) 
\int_{\bm{p}} \frac{e^{i \bm{p} \cdot \bm{r}} \bm{p}^i}{E-\bm{p}^2/m}
\left[ \bm{\nabla}_{\bm{q}}^j
\frac{\Lambda^{2 \epsilon}}{|\bm{p}-\bm{q}|^{1+2 \epsilon}}
\right]_{\bm{q}=\bm{0}}.
%---------------
\end{equation}
%---------------
The integral over $\bm{p}$, when contracted with $\hat{\bm{r}}^i$, is given by 
%---------------
\begin{eqnarray}
%---------------
&& \hspace{-10ex}
\int_{\bm{p}} \frac{e^{i \bm{p} \cdot \bm{r}} \bm{p} \cdot \hat{\bm{r}} 
}{E-\bm{p}^2/m}
\left[ \bm{\nabla}_{\bm{q}}^j
\frac{1}{|\bm{p}-\bm{q}|}
\right]_{\bm{q}=\bm{0}}
=
\int_{\bm{p}} \frac{
e^{i \bm{p} \cdot \bm{r}}
\hat{\bm{p}} \cdot \hat{\bm{r}} \hat{\bm{p}}^j
}{E-\bm{p}^2/m}
\frac{1}{|\bm{p}|}
\nonumber \\
&=&
- m \hat{\bm{r}}^j 
\frac{1}{24 \pi^2} \left[ \frac{4}{3} -4 \gamma_{\rm E} -2 \log (-m E r_0^2)
\right]
+O(r_0),
%---------------
\end{eqnarray}
%---------------
where we set $|\bm{r}|=r_0$ in the last equality. 
From this we obtain the contribution from the $1/m$ potential to 
$\hat{\bm{r}}^i \delta Z_E^{ij}$ by subtracting
eq.~(\ref{eq:jcal_dr_1mpot}) from eq.~(\ref{eq:jcal_fr_1mpot}), which gives 
%---------------
\begin{equation}
\label{eq:ze_1mpot} 
%---------------
\hat{\bm{r}}^i \delta Z_E^{ij} \big|_{V^{(1)}} 
= 
\frac{\alpha_s C_F}{6} \hat{\bm{r}}^j 
\left[ \frac{C_F}{6} - \frac{7 C_A}{12} 
+ 
\left( \frac{C_F}{2}- C_A \right) 
\left( \frac{1}{4 \epsilon_{\rm UV}} + 
\log (\Lambda r_0 e^{\gamma_{\rm E}}) 
\right) 
\right]. 
%---------------
\end{equation}
%---------------
We see that the dependence on $E$ has cancelled between the dimensionally
regulated integral and the finite-$r$ regularized integral. 
Hence, eq.~(\ref{eq:ze_1mpot}) is valid for all $P$-wave eigenstates; that is,
the contribution to $\delta \bar{Z}^{ij}$ from the $1/m$ potential is given
by eq.~(\ref{eq:ze_1mpot}) for all $P$-wave states. 
The scheme conversion coefficient $\hat{\bm{r}}^i \delta Z^{ij}$ 
is then given by subtracting the $1/\epsilon$ pole:
%---------------
\begin{equation}
\label{eq:zmsbar_1mpot}
%---------------
\hat{\bm{r}}^i \delta Z^{ij} \big|_{V^{(1)}}
=
\frac{\alpha_s^2 C_F}{6} \hat{\bm{r}}^j
\left[ \frac{C_F}{6} - \frac{7 C_A}{12}
+
\left( \frac{C_F}{2}- C_A \right)
\log (\Lambda r_0 e^{\gamma_{\rm E}})
\right].
%---------------
\end{equation}
%---------------

The contributions from the velocity-dependent potential and the relativistic
correction to the kinetic energy are computed in the same way, as they appear 
in $\tilde{\delta {\cal V}}(\bm{p},\bm{q})$ in the same form as the $1/m$
potential, and differ only in the $\epsilon$-dependent coefficients. 
We obtain 
%---------------
\begin{equation}
\label{eq:zmsbar_velpot}
%---------------
\hat{\bm{r}}^i \delta Z^{ij} \big|_{V^{(2)}_{p^2}}
=
-\frac{\alpha_s^2 C_F^2}{3} \hat{\bm{r}}^j
\left[ \frac{5}{6} 
+
\log (\Lambda r_0 e^{\gamma_{\rm E}})
\right],
%---------------
\end{equation}
%---------------
and 
%---------------
\begin{equation}
\label{eq:zmsbar_relkin}
%---------------
\hat{\bm{r}}^i \delta Z^{ij} \big|_{- \frac{\bm{\nabla}^4}{4 m^3} }
=
-\frac{\alpha_s^2 C_F^2}{12} \hat{\bm{r}}^j
\left[ \frac{5}{6}
+
\log (\Lambda r_0 e^{\gamma_{\rm E}})
\right].
%---------------
\end{equation}
%---------------

We note that the $\bm{p}$ and $\bm{p}'$-independent term in the 
momentum-space potential does not contribute to $\delta Z_E$. 
In position space, this term corresponds to the
delta function potential, which vanishes on $P$-wave LO wavefunctions. 

Now we move over to the spin-dependent contributions, which come from the
hyperfine and the spin-orbit potentials. 
The leading divergent contribution to ${\cal J}_{\rm DR}^{ij}$ from the
hyperfine potential is given by 
%---------------
\begin{eqnarray}
\label{eq:HF_powerUV}
%---------------
&& \hspace{-5ex} 
\frac{\pi \alpha_s C_F}{4 m^2} 
[\sigma_{i_0}, \sigma_{i_1}] \otimes [\sigma_{i_0}, \sigma_{i_2}] 
\int_{\bm{p}} \frac{\bm{p}^i}{E-\bm{p}^2/m}
\left[ \bm{\nabla}_{\bm{q}}^j
\frac{(\bm{p}-\bm{q})^{i_1} (\bm{p}-\bm{q})^{i_2}}{(\bm{p}-\bm{q})^2} 
\right]_{\bm{q}=\bm{0}}
\nonumber \\ && = 
\frac{\pi \alpha_s C_F}{4 m^2}
[\sigma_{i_0}, \sigma_{i_1}] \otimes [\sigma_{i_0}, \sigma_{i_2}]
\left[ 2 I_3^{i i_1 i_2 j} - 
\delta^{i_1 j} I_1^{i i_2}
- \delta^{i_2 j} I_1^{i i_1} \right]_{\rm DR}, 
%---------------
\end{eqnarray}
%---------------
and the leading divergent contribution from the spin-orbit potential to
${\cal J}_{\rm DR}^{ij}$ is given by
%---------------
\begin{eqnarray}
\label{eq:SO_powerUV}
%---------------
&& \hspace{-5ex}
- \frac{3 \pi \alpha_s C_F}{2 m^2}
\left( [\sigma_{i_1}, \sigma_{i_2}] \otimes 1 -
1 \otimes [\sigma_{i_1}, \sigma_{i_2}] \right)
\int_{\bm{p}} \frac{\bm{p}^i}{E-\bm{p}^2/m}
\left[ \bm{\nabla}_{\bm{q}}^j
\frac{(\bm{p}-\bm{q})^{i_1} \bm{q}^{i_2}}{(\bm{p}-\bm{q})^2}
\right]_{\bm{q}=\bm{0}}
\nonumber \\ && =
- \frac{3 \pi \alpha_s C_F}{2 m^2}
\left( [\sigma_{i_1}, \sigma_{i_2}] \otimes 1 -
1 \otimes [\sigma_{i_1}, \sigma_{i_2}] \right)
\delta^{i i_2} I_1^{i i_1} |_{\rm DR},
%---------------
\end{eqnarray}
%---------------
where the tensor integrals $I_3$ and $I_1$ are defined in DR by 
%---------------
\begin{subequations}
\begin{eqnarray}
%---------------
I_3^{ijkl} |_{\rm DR} &=&
\int_{\bm{p}} \frac{\bm{p}^i \bm{p}^j \bm{p}^k \bm{p}^l}{E-\bm{p}^2/m}
\frac{1}{\bm{p}^4},
\\
I_1^{ij} |_{\rm DR} &=& 
\delta^{kl} I_3^{ijkl} |_{\rm DR}.
%---------------
\end{eqnarray}
\end{subequations}
%---------------
The corresponding contributions to ${\cal J}_{(r_0)}^{ij}$ are given by 
eqs.~(\ref{eq:HF_powerUV}) and (\ref{eq:SO_powerUV}), with the tensor integrals
computed in finite-$r$ regularization:
%---------------
\begin{subequations}
\begin{eqnarray}
%---------------
I_3^{ijkl} |_{(r_0)} &=&
\int_{\bm{p}} \frac{e^{i \bm{p} \cdot \bm{r}} 
\bm{p}^i \bm{p}^j \bm{p}^k \bm{p}^l}{E-\bm{p}^2/m}
\frac{1}{\bm{p}^4} 
\bigg|_{|\bm{r}|=r_0}, 
\\
I_1^{ij} |_{(r_0)} &=&
\delta^{kl} I_3^{ijkl}  |_{(r_0)}.
%---------------
\end{eqnarray}
\end{subequations}
%---------------
The dimensionally regulated integrals are evaluated as 
%---------------
\begin{subequations}
\begin{eqnarray}
%---------------
I_3^{ijkl} |_{\rm DR} &=&
\frac{1}{d^2-1} ( \delta^{ij} \delta^{kl} + \delta^{ik} \delta^{jl} +
\delta^{il} \delta^{jk} ) \int_{\bm{p}} \frac{1}{E-\bm{p}^2/m} 
\nonumber \\
&=& 
( \delta^{ij} \delta^{kl} + \delta^{ik} \delta^{jl} +
\delta^{il} \delta^{jk} ) \frac{m^2}{60 \pi^2}  \sqrt{-\frac{E}{m}} +
O(\epsilon), 
\\
I_1^{ij} |_{\rm DR} &=&
\delta^{ij} \frac{m^2}{12 \pi} \sqrt{-\frac{E}{m}} + O(\epsilon),
%---------------
\end{eqnarray}
\end{subequations}
%---------------
We note that both tensor integrals are power UV divergent. Since in DR, power
UV divergences are subtracted automatically, they do not appear in the final
results, especially after expansion in powers of $\epsilon$. We note that
neither tensor integrals contain logarithmic divergences. 
Now we compute the finite-$r$ regularized integrals. 
As we have done in the case of the spin-independent contributions, we will
compute the finite-$r$ regularized integrals contracted with $\hat{\bm{r}}^i$
in the integrand,
which reduces the rank of the tensor integral that we need to consider. 
They are computed as 
%---------------
\begin{subequations}
\begin{eqnarray}
%---------------
\hat{\bm{r}}^i I_3^{ijkl} |_{(r_0)} &=&
\left( - i \frac{\partial}{\partial r} \right)
\int_{\bm{p}} \frac{e^{i \bm{p} \cdot \bm{r}}}{E-\bm{p}^2/m}
\frac{\bm{p}^j \bm{p}^k \bm{p}^l}{\bm{p}^4}
\bigg|_{|\bm{r}|=r_0} 
\nonumber \\
&=& 
-i m (\delta^{jk} \hat{\bm{r}}^l + \delta^{kl} \hat{\bm{r}}^j + \delta^{lm}
\hat{\bm{r}}^k ) 
\left( - i \frac{\partial}{\partial r_0} \right) 
\left[ \frac{1}{32 \pi} - \frac{\sqrt{-m E} r_0}{60 \pi} + O(r_0^2) \right]
\nonumber \\
&& - i \hat{\bm{r}}^j \hat{\bm{r}}^k \hat{\bm{r}}^l 
\left( - i \frac{\partial}{\partial r_0} \right)
\left[ \frac{1}{32 \pi} + O(r_0^2) \right]
\nonumber \\
&=&
\frac{m^2}{60 \pi} (\delta^{jk} \hat{\bm{r}}^l + \delta^{kl} \hat{\bm{r}}^j 
+ \delta^{lm} \hat{\bm{r}}^k )
\sqrt{- \frac{E}{m} } + O(r_0),
\\
\hat{\bm{r}}^i I_1^{ij} |_{(r_0)} &=&
m^2 \hat{\bm{r}}^j \frac{1}{12 \pi} \sqrt{- \frac{E}{m}} + O(r_0).
%---------------
\end{eqnarray}
\end{subequations}
%---------------
Here, we used $- i \frac{\partial}{\partial r} e^{i \bm{p} \cdot \bm{r}}
= \bm{p} \cdot \hat{\bm{r}} e^{i \bm{p} \cdot \bm{r}}$. 
From the quantities in the square brackets, we see that power UV divergent
corrections to the $P$-wave wavefunction can produce nonzero values of 
$\Psi_n (\bm{0})$, which vanish in $\bm{\nabla} \Psi_n(\bm{0})$. 
The difference in the tensor integrals between 
DR and finite-$r$ regularization vanish:
%---------------
\begin{subequations}
\begin{eqnarray}
%---------------
\hat{\bm{r}}^i I_3^{ijkl} |_{\rm DR} 
- \hat{\bm{r}}^i I_3^{ijkl} |_{(r_0)} 
&=&
0 + O(\epsilon,r_0),
\\
\hat{\bm{r}}^i I_1^{ij} |_{\rm DR} 
- \hat{\bm{r}}^i I_1^{ij} |_{(r_0)}
&=&
0 + O(\epsilon,r_0). 
%---------------
\end{eqnarray}
\end{subequations}
%---------------
Hence, the leading divergent contributions from the spin-dependent 
potentials cancel between $\hat{\bm{r}}^i {\cal J}_{\rm DR}^{ij}$ 
and $\hat{\bm{r}}^i {\cal J}_{(r_0)}^{ij}$. 

Because the leading divergent contributions from the spin-dependent potentials 
are power divergent, the subleading divergent contributions may contain 
logarithmic divergences, and must be included in the calculation of the scheme
conversion coefficient. 
The subleading divergent contributions to ${\cal J}_{\rm DR}^{ij}$ from the
hyperfine and the spin-orbit potentials are given by 
%---------------
\begin{eqnarray}
\label{eq:hyperfine_correction_log}
%---------------
&&
\hspace{-3ex}
\frac{\pi \alpha_s C_F}{4 m^2}
[\sigma_{i_0}, \sigma_{i_1}] \otimes [\sigma_{i_0}, \sigma_{i_2}]
\nonumber \\ && \times 
\bigg\{ 
\int_{\bm{p}'} 
\int_{\bm{p}} \frac{\bm{p}'^i \tilde{V}_{\rm LO} (\bm{p}'-\bm{p})}
{(E-\bm{p}'^2/m) (E-\bm{p}^2/m) }
\left[ \bm{\nabla}_{\bm{q}}^j
\frac{(\bm{p}-\bm{q})^{i_1} (\bm{p}-\bm{q})^{i_2}}{(\bm{p}-\bm{q})^2}
\right]_{\bm{q}=\bm{0}}
\nonumber \\ 
&& \hspace{5ex} + 
\int_{\bm{p}'} 
\int_{\bm{p}} \frac{\bm{p}'^i 
(\bm{p}'-\bm{p})^{i_1} (\bm{p}'-\bm{p})^{i_2} 
}
{(E-\bm{p}'^2/m) (\bm{p}'-\bm{p})^2 (E-\bm{p}^2/m) }
\left[ \bm{\nabla}_{\bm{q}}^j
\tilde V_{\rm LO}(\bm{p}-\bm{q}) 
\right]_{\bm{q}=\bm{0}}
\bigg\} 
\nonumber \\
&&=
-\frac{\pi^2 \alpha_s^2 C_F^2 }{m^2}
[\sigma_{i_0}, \sigma_{i_1}] \otimes [\sigma_{i_0}, \sigma_{i_2}]
\left[ 2 J_{3a}^{i i_1 i_2 j} + 2 J_{3b}^{i i_1 i_2 j} 
- \delta^{i_1 j} J_1^{i i_2} - \delta^{i_2 j} J_1^{i i_1} \right]_{\rm DR}, 
%---------------
\end{eqnarray}
%---------------
and 
%---------------
\begin{eqnarray}
\label{eq:spinorbit_correction_log}
%---------------
&& \hspace{-3ex} 
- \frac{3 \pi \alpha_s C_F}{2 m^2}
\left( [\sigma_{i_1}, \sigma_{i_2}] \otimes 1 -
1 \otimes [\sigma_{i_1}, \sigma_{i_2}] \right)
\nonumber \\ && \times 
\bigg\{
\int_{\bm{p}'} 
\int_{\bm{p}} \frac{\bm{p}'^i \tilde{V}_{\rm LO} (\bm{p}'-\bm{p})}
{(E-\bm{p}'^2/m) (E-\bm{p}^2/m) }
\left[ \bm{\nabla}_{\bm{q}}^j
\frac{(\bm{p}-\bm{q})^{i_1} \bm{q}^{i_2}}{(\bm{p}-\bm{q})^2}
\right]_{\bm{q}=\bm{0}}
\nonumber \\
&& \hspace{5ex}  +
\int_{\bm{p}'} 
\int_{\bm{p}} \frac{\bm{p}'^i
(\bm{p}'-\bm{p})^{i_1} \bm{p}^{i_2} }
{(E-\bm{p}'^2/m) (\bm{p}'-\bm{p})^2 (E-\bm{p}^2/m) }
\left[ \bm{\nabla}_{\bm{q}}^j
\tilde V_{\rm LO}(\bm{p}-\bm{q})
\right]_{\bm{q}=\bm{0}}
\bigg\}
\nonumber \\ 
&&=
\frac{6 \pi^2 \alpha_s^2 C_F^2}{m^2}
\left( [\sigma_{i_1}, \sigma_{i_2}] \otimes 1 -
1 \otimes [\sigma_{i_1}, \sigma_{i_2}] \right)
\left[ 2 J_{3c}^{i i_1 i_2 j} + \delta^{i_2 j} J_1^{i i_1} \right]_{\rm DR},
%---------------
\end{eqnarray}
%---------------
respectively, where $J_{3a}$, $J_{3b}$, $J_{3c}$, and $J_1$ are 
logarithmically UV divergent tensor integrals defined in DR by 
%---------------
\begin{subequations}
\label{eq:tenint_DR}
\begin{eqnarray}
%---------------
J_{3a}^{ijkl} |_{\rm DR} &=&
\int_{\bm{p}'}
\int_{\bm{p}}
\frac{\bm{p}'^i}{(E-\bm{p}'^2/m) (\bm{p}'-\bm{p})^2 (E-\bm{p}^2/m)}
\frac{\bm{p}^j \bm{p}^k \bm{p}^l}{\bm{p}^4},
\\
J_{3b}^{ijkl} |_{\rm DR} &=&
\int_{\bm{p}'}
\int_{\bm{p}}
\frac{\bm{p}'^i
(\bm{p}'-\bm{p})^j 
(\bm{p}'-\bm{p})^k 
}{(E-\bm{p}'^2/m) (\bm{p}'-\bm{p})^2 (E-\bm{p}^2/m)}
\frac{\bm{p}^l}{\bm{p}^4}, 
\\
J_{3c}^{ijkl} |_{\rm DR} &=&
\int_{\bm{p}'}
\int_{\bm{p}}
\frac{\bm{p}'^i
(\bm{p}'-\bm{p})^j
\bm{p}^k
}{(E-\bm{p}'^2/m) (\bm{p}'-\bm{p})^2 (E-\bm{p}^2/m)}
\frac{\bm{p}^l}{\bm{p}^4},
\\
J_{1}^{ij} |_{\rm DR} &=& \delta^{kl} J_{3a}^{ijkl} |_{\rm DR}. 
%---------------
\end{eqnarray}
\end{subequations}
%---------------
The subleading divergent contributions to ${\cal J}_{(r_0)}^{ij}$
are also given by eqs.~(\ref{eq:hyperfine_correction_log}) and
(\ref{eq:spinorbit_correction_log}), with the UV-divergent tensor integrals 
regulated in finite-$r$ regularization: 
%---------------
\begin{subequations}
\label{eq:tenint_r}
\begin{eqnarray}
%---------------
J_{3a}^{ijkl} |_{(r_0)} &=&
\int_{\bm{p}'}
\int_{\bm{p}}
\frac{e^{i \bm{p}'\cdot \bm{r}} 
\bm{p}'^i}{(E-\bm{p}'^2/m) (\bm{p}'-\bm{p})^2 (E-\bm{p}^2/m)}
\frac{\bm{p}^j \bm{p}^k \bm{p}^l}{\bm{p}^4} \Big|_{|\bm{r}|=r_0},
\\
J_{3b}^{ijkl} |_{(r_0)} &=&
\int_{\bm{p}'}
\int_{\bm{p}}
\frac{e^{i \bm{p}'\cdot \bm{r}} \bm{p}'^i (\bm{p}'-\bm{p})^j (\bm{p}'-\bm{p})^k
}{(E-\bm{p}'^2/m) (\bm{p}'-\bm{p})^2 (E-\bm{p}^2/m)}
\frac{\bm{p}^l}{\bm{p}^4} \Big|_{|\bm{r}|=r_0},
\\
J_{3c}^{ijkl} |_{(r_0)} &=&
\int_{\bm{p}'}
\int_{\bm{p}}
\frac{e^{i \bm{p}'\cdot \bm{r}} \bm{p}'^i (\bm{p}'-\bm{p})^j \bm{p}^k
}{(E-\bm{p}'^2/m) (\bm{p}'-\bm{p})^2 (E-\bm{p}^2/m)}
\frac{\bm{p}^l}{\bm{p}^4} \Big|_{|\bm{r}|=r_0},
\\
J_{1}^{ij} |_{(r_0)} &=& \delta^{kl} J_{3a}^{ijkl} |_{(r_0)}. 
%---------------
\end{eqnarray}
\end{subequations}
%---------------
As the leading divergent contributions from the spin-dependent potentials 
cancel in $\delta Z_E^{ij}$, 
the spin-dependent contributions 
are determined by the logarithmically UV divergent 
tensor integrals in eqs.~(\ref{eq:tenint_DR}) and (\ref{eq:tenint_r}). 
We compute the differences in the tensor integrals between finite-$r$
regularization and DR in appendix~\ref{appendix:tensorint}. 
We write the results in terms of the spin and angular momentum basis given 
in eqs.~(\ref{eq:wf_renormalization}). This requires computation of the
$d-1$-dimensional Pauli matrix combinations of the form $a \otimes b$ that
appear in the spin-dependent potentials in momentum space. 
We do this by applying the projections onto spin triplet and spin singlet 
to the $a \otimes b$ as is done in eqs.~(\ref{eq:wf_renormalization}). 
That is, we contract the $Q$ and $\bar Q$ indices of 
$a \otimes b$ with $\bm{\sigma}$ for spin triplet and $\sigma_5$ for spin
singlet, which give $a \bm{\sigma} b$ and $a \sigma_5 b$, respectively. 
The reduction of Pauli matrices in $d-1$ spatial dimensions is done by
repeated application of the identity 
$\{ \sigma^i, \sigma^j \} = 2 \delta^{ij}$, keeping in
mind that $\sigma_5$ commutes with Pauli matrices with 3-dimensional indices, 
while it anticommutes with ones carrying $d-4$-dimensional indices. 
The Pauli matrix algebra in $d-1$ spatial dimensions can be done easily by
using the \textsc{Mathematica} package \textsc{FeynCalc} with
the \textsc{FeynOnium} addon~\cite{Mertig:1990an, Shtabovenko:2016sxi, 
Shtabovenko:2020gxv, Brambilla:2020fla}. 
In the spin triplet case, we apply the
decomposition of the rank-2 tensor carrying indices $i$ and $j$ 
from $\sigma^i \hat{\bm{r}}^k \delta Z_E^{kj}$ into the trace,
antisymmetric, and symmetric traceless parts, following
eqs.~(\ref{eq:wf_renormalization}). 
We have, for spin triplet, 
%---------------
\begin{subequations}
\label{eq:ze_triplet}
\begin{eqnarray}
%---------------
\sigma^i \hat{\bm{r}}^k \delta Z_E^{kj} \big|_{V_{HF}^{(2)}} 
&=& 
\alpha_s^2 C_F^2 
\frac{\delta^{ij}}{d-1} \bm{\sigma} \cdot \hat{\bm{r}}
\left( - \frac{1}{48 \epsilon_{\rm UV}} 
- \frac{1}{12} \log (\Lambda r_0 e^{\gamma_{\rm E}}) 
- \frac{13}{144}  \right) 
\nonumber \\ && 
+ \alpha_s^2 C_F^2 \sigma^{[i} \hat{\bm{r}}^{j]}
\left(\frac{1}{96 \epsilon_{\rm UV}} 
+ \frac{1}{24} \log (\Lambda r_0 e^{\gamma_{\rm E}})
+ \frac{1}{18}  \right) 
\nonumber \\ && 
+ \alpha_s^2 C_F^2 \sigma^{(i} \hat{\bm{r}}^{j)}
\left( - \frac{1}{480 \epsilon_{\rm UV}} 
- \frac{1}{120} \log (\Lambda r_0 e^{\gamma_{\rm E}})
-\frac{7}{900} \right)
\nonumber \\ && + 
\frac{\alpha_s^2 C_F^2 }{200} f^{ij} (\hat{\bm{r}},\bm{\sigma}) 
+ O(\epsilon,r_0), 
%---------------
\end{eqnarray}
%---------------
\begin{eqnarray}
%---------------
\sigma^i \hat{\bm{r}}^k \delta Z_E^{kj} \big|_{V_{SO}^{(2)}} 
&=&
\alpha_s^2 C_F^2
\frac{\delta^{ij}}{d-1} \bm{\sigma} \cdot \hat{\bm{r}}
\left( 
- \frac{1}{16 \epsilon_{\rm UV}} 
- \frac{1}{4} \log (\Lambda r_0 e^{\gamma_{\rm E}}) 
- \frac{19}{48} 
\right)
\nonumber \\ &&
+ \alpha_s^2 C_F^2 \sigma^{[i} \hat{\bm{r}}^{j]}
\left( - \frac{1}{32 \epsilon_{\rm UV}} 
- \frac{1}{8} \log (\Lambda r_0 e^{\gamma_{\rm E}})
- \frac{11}{48} 
\right)
\nonumber \\ &&
+ \alpha_s^2 C_F^2 \sigma^{(i} \hat{\bm{r}}^{j)}
\left( 
\frac{1}{32 \epsilon_{\rm UV}} 
+ \frac{1}{8} \log (\Lambda r_0 e^{\gamma_{\rm E}})
+ \frac{11}{48} 
\right)
+ O(\epsilon,r_0), 
%---------------
\end{eqnarray}
\end{subequations}
%---------------
where $f^{ij} (\hat{\bm{r}}, \bm{\sigma})$ is the $F$-wave contribution defined
in $3$ spatial dimensions in eq.~(\ref{eq:Fwavetensor}). 
As we have argued in section~\ref{sec:wavefunctions}, this
contribution vanishes in the LDMEs to first order in QMPT, and so, we can
neglect the $F$-wave contribution from the scheme conversion coefficient. 
For the spin singlet case, we obtain 
%---------------
\begin{subequations}
\label{eq:ze_singlet}
\begin{eqnarray}
%---------------
\sigma_5 \hat{\bm{r}}^i \delta Z_E^{ij} \big|_{V_{HF}^{(2)}} 
&=& 
- \frac{\alpha_s^2 C_F^2}{40} \sigma_5 \hat{\bm{r}}^j 
+O(\epsilon,r_0), 
\\
\sigma_5 \hat{\bm{r}}^i \delta Z_E^{ij} \big|_{V_{SO}^{(2)}} 
&=&
0 + O(\epsilon,r_0).
%---------------
\end{eqnarray}
\end{subequations}
%---------------
To obtain these results, we need to compute the following Pauli 
matrix combinations containing $\sigma_5$ in $d-1$ spatial dimensions
%---------------
\begin{subequations}
\begin{eqnarray}
%---------------
&&
[\sigma^{i_0}, \sigma^{i}] \sigma_5 [\sigma^{i_0}, \sigma^{j}] 
+
[\sigma^{i_0}, \sigma^{j}] \sigma_5 [\sigma^{i_0}, \sigma^{i}] 
= 
-16 (1+\epsilon) \sigma_5 \delta^{ij} 
,
\\ && 
[\sigma^{i_0}, \sigma^{i_1}] \sigma_5 [\sigma^{i_0}, \sigma^{i_1}] 
= -8 (3+7 \epsilon+2 \epsilon^2) \sigma_5, 
\\ &&
[\sigma^{i}, \sigma^{j}] \sigma_5 
- \sigma_5 [\sigma^{i}, \sigma^{j}] 
= 0, 
\\ && 
[\sigma^{i_0}, \sigma^{i_1}] \sigma_5 [\sigma^{i_0}, \sigma^{i_2}]
\hat{\bm{r}}^{i_1} \hat{\bm{r}}^{i_2}
= - 8 \sigma_5 + O(\epsilon),
%---------------
\end{eqnarray}
\end{subequations}
%---------------
where $i$ and $j$ are 3-dimensional indices. 
We note that the second relation cannot be obtained from the first one, because 
dummy indices must be summed over in $d-1$ spatial
dimensions. We compute the last relation in 3 spatial dimensions, because there
are no poles in $\epsilon$ associated with this term. 
We note that if we were to use a scheme for spin singlet where $\sigma_5$
commutes with all Pauli matrices, similarly to what is done in 
na\"ive dimensional regularization, the spin-dependent contributions to 
$\delta Z_E^{ij}$ for the $^1P_1$ state vanish to order $\epsilon^0$. 

We note that our results for the spin-dependent contributions in
$\delta Z_E^{ij}$ in eqs.~(\ref{eq:ze_triplet}) and (\ref{eq:ze_singlet})
are independent of $E$, and therefore, they are the spin-dependent
contributions to $\delta \bar{Z}^{ij}$ for all $P$-wave states. 
The decomposition into irreducible tensors in eqs.~(\ref{eq:ze_triplet})
diagonalizes $\delta \bar{Z}^{ij}$ in terms of wavefunctions of definite 
$^{2 S+1}P_J$ quantum numbers in DR, because they are exactly in the form of 
the spin projections in DR defined in eqs.~(\ref{eq:wf_renormalization}). 
That is, the coefficients of the tensors 
$\delta^{ij} \bm{\sigma} \cdot \hat{\bm{r}}/(d-1)$, 
$\sigma^{[i} \hat{\bm{r}}^{j]}$, 
$\sigma^{(i} \hat{\bm{r}}^{j)}$, 
and $\sigma_5 \hat{\bm{r}}^j$ 
apply individually to wavefunctions 
with quantum numbers $^3P_0$, $^3P_1$, $^3P_2$, and $^1P_1$, respectively, 
in the subtraction term $\hat{\bm{r}}^i \delta \bar{Z}^{ij} 
\left[ \bm{\nabla}^j \Psi_n^{\rm LO} (\bm{0}) \right]$. 
Hence, renormalization in the $\overline{\rm MS}$ scheme is carried out simply 
by subtracting the $1/\epsilon$ poles in the coefficients of these tensors. 
We combine the spin-dependent and spin-independent 
contributions to obtain $\hat{\bm{r}}^i \delta Z^{ij}$, which is given by 
%---------------
\begin{subequations}
\label{eq:zms_final1}
\begin{eqnarray}
%---------------
\sigma^i \hat{\bm{r}}^k \delta Z^{kj} 
&=& \frac{\delta^{ij}}{d-1} \bm{\sigma} \cdot \hat{\bm{r}} \delta Z_{^3P_0}
+ \sigma^{[i} \hat{\bm{r}}^{j]} \delta Z_{^3P_1}
+ \sigma^{(i} \hat{\bm{r}}^{j)} \delta Z_{^3P_2}, 
\\
\sigma_5 \hat{\bm{r}}^k \delta Z^{kj}
&=& \sigma_5 \hat{\bm{r}}^{j}
\delta Z_{^1P_1}, 
%---------------
\end{eqnarray}
\end{subequations}
%---------------
where 
%---------------
\begin{subequations}
\label{eq:zms_final2}
\begin{eqnarray}
%---------------
\delta Z_{^3P_0} 
&=&
- \alpha_s^2 
\left[
\left( \frac{2 C_F^2}{3} + \frac{C_F C_A}{6} \right) 
\log (\Lambda r_0 e^{\gamma_{\rm E}})
+ \frac{7 C_F C_A}{72} + \frac{29 C_F^2}{36} 
\right] ,
\\ 
\delta Z_{^3P_1} &=& 
- \alpha_s^2 
\left[ 
\left( \frac{5 C_F^2}{12} + \frac{C_F C_A}{6} \right)
\log (\Lambda r_0 e^{\gamma_{\rm E}})
+ \frac{7 C_F C_A}{72} + \frac{71 C_F^2}{144}
\right],
\\ 
\delta Z_{^3P_2} &=& 
- \alpha_s^2 
\left[ 
\left( \frac{13 C_F^2}{60} + \frac{C_F C_A}{6} \right)
\log (\Lambda r_0 e^{\gamma_{\rm E}})
+ \frac{7 C_F C_A}{72} + \frac{353 C_F^2}{3600} 
\right],
\\
\delta Z_{^1P_1} &=& 
- \alpha_s^2 
\left[
\left( \frac{C_F^2}{3} + \frac{C_F C_A}{6} \right)
\log (\Lambda r_0 e^{\gamma_{\rm E}})
+ \frac{7 C_F C_A}{72} + \frac{31 C_F^2}{90}
\right].
%---------------
\end{eqnarray}
\end{subequations}
%---------------
Equations~(\ref{eq:zms_final1}) and (\ref{eq:zms_final2}) are our final
results for the scheme conversion coefficient. The 
$\delta Z_{^{2 S+1} P_J}$ are the diagonal elements of $\delta Z^{ij}$ in terms
of the $^{2 S+1} P_J$ quantum numbers, so that 
$\hat{\bm{r}}^i \delta Z^{ij} [\bm{\nabla}^j 
\Psi_{n(^{2 S+1}P_J)}^{\rm LO} (\bm{0}) ] = \delta Z_{^{2 S+1}P_J}
\hat{\bm{r}} \cdot [\bm{\nabla}
\Psi_{n(^{2 S+1}P_J)}^{\rm LO} (\bm{0}) ]$. 
This result allows us to compute the $\overline{\rm MS}$-renormalized 
wavefunctions at the origin by using eq.~(\ref{eq:wf_MSren_final}).

Similarly to the $S$-wave calculation in ref.~\cite{Chung:2020zqc}, our
calculation of the scheme-conversion coefficient is unaffected by the
nonperturbative long-distance behaviors of the potential, because $\delta Z$
only depends on the short-distance behaviors that determine the UV divergences
of the integrals ${\cal J}^{ij}_{\rm DR}$ and ${\cal J}^{ij}_{(r_0)}$. 

Equation~(\ref{eq:wf_MSren_final}) implies that the $\Lambda$ dependence 
of the $\overline{\rm MS}$-renormalized $P$-wave wavefunction is determined by 
$\delta Z$. From this we obtain the following evolution equations 
%---------------
\begin{subequations}
\begin{eqnarray}
\label{eq:zRG}
%---------------
\frac{d \log \left[ \hat{\bm{r}} \cdot \bm{\nabla} \Psi_{n(^3P_0)}(\bm{0})
\right]^{\overline{\rm MS}} }{d \log \Lambda}
&=& 
- \frac{d \delta Z_{^3P_0} 
}{d \log \Lambda}
= \alpha_s^2 \left( \frac{2}{3} C_F^2 + \frac{1}{6} C_F C_A \right),
\\
\frac{d \log \left[ \hat{\bm{r}} \cdot \bm{\nabla} \Psi_{n(^3P_1)}(\bm{0})
\right]^{\overline{\rm MS}} }{d \log \Lambda}
&=&
- \frac{d \delta Z_{^3P_1}
}{d \log \Lambda}
= \alpha_s^2 \left( \frac{5}{12} C_F^2 + \frac{1}{6} C_F C_A \right),
\\
\frac{d \log \left[ \hat{\bm{r}} \cdot \bm{\nabla} \Psi_{n(^3P_2)}(\bm{0})
\right]^{\overline{\rm MS}} }{d \log \Lambda}
&=&
- \frac{d \delta Z_{^3P_2}
}{d \log \Lambda}
= \alpha_s^2 \left( \frac{13}{60} C_F^2 + \frac{1}{6} C_F C_A \right),
\\
\frac{d \log \left[ \hat{\bm{r}} \cdot \bm{\nabla} \Psi_{n(^1P_1)}(\bm{0})
\right]^{\overline{\rm MS}} }{d \log \Lambda}
&=&
- \frac{d \delta Z_{^1P_1}
}{d \log \Lambda}
= \alpha_s^2 \left( \frac{1}{3} C_F^2 + \frac{1}{6} C_F C_A \right),
%---------------
\end{eqnarray}
\end{subequations}
%---------------
which reproduce the anomalous dimensions in eqs.~(\ref{eq:NRQCDRG}).

%------------------------------------------------------------------------------
\subsection{Unitary transformation}
\label{sec:unitary_transformations}
%------------------------------------------------------------------------------

As previously described, our calculation of the scheme-conversion coefficient 
$\delta Z$ is valid when we compute the finite-$r$ regularized wavefunctions at
the origin by using the potentials in the on-shell matching scheme. 
However, long-distance nonperturbative behaviors of the potentials are given in
position space in the Wilson loop matching scheme. The short-distance behaviors
of the potentials differ in the two schemes, as shown in
appendix~\ref{appendix:potentials}. 
The two matching schemes are related by a unitary transformation. 
If $\Psi_n^{\rm OS}(\bm{r})$ is a solution of the Schr\"odinger equation 
with the potentials from on-shell matching, the wavefunction 
$\Psi_n^{\rm WL}(\bm{r})$ that satisfies the Schr\"odinger equation with
the potentials from Wilson loop matching is given by 
%---------------
\begin{equation}
\label{eq:unitary_transform}
%---------------
\Psi_n^{\rm WL} (\bm{r}) = U^{-1}(r) \Psi_n^{\rm OS} (\bm{r}), 
%---------------
\end{equation}
%---------------
where the unitary transformation $U(r)$ is given at short distances
by~\cite{Brambilla:2000gk, Peset:2015vvi, Chung:2020zqc}
%---------------
\begin{equation}
%---------------
U(r) = 1 + \frac{\alpha_s C_F}{4 m} 
\left( \frac{1}{r} + \frac{\partial}{\partial r} \right) + O(1/m^2). 
%---------------
\end{equation}
%---------------
Here we have expanded $U(r)$ in powers of $1/m$, 
consistently with our calculation of the wavefunctions in the QMPT. 
Since for $P$-wave states, $\bm{\nabla} \Psi_n^{\rm WL} (\bm{0}) |_{(r_0)}$ and 
$\bm{\nabla} \Psi_n^{\rm OS} (\bm{0}) |_{(r_0)}$ have same 
logarithmic divergences at $r_0 \to 0$, 
and power divergences disappear in $\bm{\nabla} \Psi_n(\bm{0})$, 
we expect the difference between $\bm{\nabla} \Psi_n^{\rm WL} (\bm{0})$ 
and $\bm{\nabla} \Psi_n^{\rm OS} (\bm{0})$ to be suppressed by at least $1/m$.
We compute the leading nonvanishing contribution to 
$\bm{\nabla} \Psi_n^{\rm OS} (\bm{0})
- \bm{\nabla} \Psi_n^{\rm WL} (\bm{0})$ by 
%---------------
\begin{eqnarray}
%---------------
\hat{\bm{r}} \cdot \bm{\nabla} \Psi_n^{\rm OS} (\bm{0}) 
\big|_{(r_0)} 
&=& 
\hat{\bm{r}} \cdot \bm{\nabla} U(r) \Psi_n^{\rm WL} (\bm{0}) 
\big|_{(r_0)} 
\nonumber \\ &=& 
\hat{\bm{r}} \cdot\bm{\nabla}\delta \Psi_n^{\rm WL} (\bm{0})
\big|_{(r_0)} 
+ \hat{\bm{r}} \cdot\bm{\nabla} U(r) \Psi_n^{\rm LO} (\bm{0})
+O(1/m^2) 
\nonumber \\ 
&=& \hat{\bm{r}} \cdot\bm{\nabla} \Psi_n^{\rm WL} (\bm{0}) 
\big|_{(r_0)} 
- \frac{3 \alpha_s C_F}{8 m} \frac{\partial^2}{\partial r^2} 
\Psi_n^{\rm LO} (\bm{0})
+O(1/m^2) , 
%---------------
\end{eqnarray}
%---------------
where in the second and third lines we used 
$\Psi_n^{\rm WL} (\bm{r})
= \Psi_n^{\rm LO} (\bm{r}) + \delta \Psi_n^{\rm WL} (\bm{r})$. 
If we use eq.~(\ref{eq:wfseconderivative}) to compute the second derivative 
of $\Psi_n^{\rm LO} (\bm{r})$ at $r=0$, we obtain 
%---------------
\begin{equation}
%---------------
\hat{\bm{r}} \cdot \bm{\nabla} \Psi_n^{\rm OS} (\bm{0}) \big|_{(r_0)} 
=
\hat{\bm{r}} \cdot \bm{\nabla} \Psi_n^{\rm WL} (\bm{0}) \big|_{(r_0)}
- \frac{3 \alpha_s^2 C_F^2}{16} 
\hat{\bm{r}} \cdot \bm{\nabla} \Psi_n^{\rm LO} (\bm{0}),
%---------------
\end{equation}
%---------------
which lets us compute $\hat{\bm{r}} \cdot \bm{\nabla} \Psi_n (\bm{0})$ in the
on-shell matching scheme from the result in the Wilson-loop matching scheme.
Similarly to the $S$-wave case, since the difference in the wavefunctions at 
the origin between the two schemes depend only on the behaviors of the
potentials at short distances, we can write the following approximate 
relation for $P$-wave states:
%---------------
\begin{equation}
\label{eq:approx_relation}
%---------------
- \hat{\bm{r}} \cdot \bm{\nabla}_{\bm{r}'} 
\int d^3r \, \hat{G}_n(\bm{r}',\bm{r}) 
\left( \frac{\alpha_s C_F \bm{L}^2}{2 m^2 r^3}
- \frac{\alpha_s^2 C_F^2}{4 m r^2} \right) 
\Psi_n^{\rm LO} (\bm{r}) \Big|_{|\bm{r}'| = 0} 
= 
\frac{3 \alpha_s^2 C_F^2}{16} \hat{\bm{r}} \cdot \bm{\nabla} \Psi_n^{\rm LO}
(\bm{0}), 
%---------------
\end{equation}
%---------------
which is accurate up to corrections from second order in the QMPT. 
The terms in the parenthesis come from the difference in the 
potential in on-shell matching and Wilson-loop matching schemes at 
short distances. 
We have neglected the delta function term, because it does not contribute 
to $P$-wave states in position space. 
If this approximate relation holds, we may compute $P$-wave wavefunctions in
position space in the on-shell matching scheme by using the following 
prescription for the potential
%---------------
\begin{equation}
\label{eq:1mpot_osprescription}
%---------------
\delta V (\bm{r},\bm{\nabla}) |^{\rm OS}
= \delta V(\bm{r},\bm{\nabla}) |^{\rm WL}
+ \frac{\alpha_s^2 C_F^2}{4 m r^2} 
- \frac{\alpha_s C_F}{2 m^2 r^3} \bm{L}^2, 
%---------------
\end{equation}
%---------------
so that while the potential in the on-shell scheme at short distance is given
by the expressions in eqs.~(\ref{eq:potentials_pert}), long-distance
nonperturbative behavior of the potential is given by Wilson loop matching.

%==============================================================================
\section{\boldmath Numerical results}
\label{sec:results}
%==============================================================================

We now compute electromagnetic decay rates, exclusive production cross
sections, and decay constants of $P$-wave quarkonia, 
based on NRQCD factorization formulae
with short-distance coefficients at two-loop accuracy and our 
calculation of the wavefunctions at the origin. 
We consider electromagnetic decay rates of $\chi_{Q0}$ and $\chi_{Q2}$ 
into $\gamma \gamma$, where $Q = c$ or $b$. 
We also consider exclusive production cross sections 
$\sigma (e^+ e^- \to \chi_{cJ} + \gamma)$ at $\sqrt{s}=10.58$~GeV, 
which was first computed in ref.~\cite{Chung:2008km}. 
Finally, we compute scalar and axialvector decay constants $f_{\chi_{Q0}}$ and 
$f_{\chi_{Q1}}$, which we define in QCD by 
%---------------
\begin{subequations}
\label{decayconst_defs}
\begin{eqnarray}
%---------------
\langle 0 | \bar Q Q | \chi_{Q0} \rangle &=& - m_{\chi_{Q0}} f_{\chi_{Q0}}, 
\\
\langle 0 | \bar Q \bm{\gamma} \gamma_5 Q | \chi_{Q1} \rangle
&=& i f_{\chi_{Q1}} m_{\chi_{Q1}} \bm{\epsilon}_{\chi_{Q1}},
%---------------
\end{eqnarray}
\end{subequations}
%---------------
where $Q$ is the quark field in QCD, and 
$\bm{\epsilon}_{\chi_{Q1}}$ is the polarization vector of $\chi_{Q1}$.
In the QCD definitions of the decay constants, the quarkonium states are
relativistically normalized, while the states in the NRQCD LDMEs are
normalized nonrelativistically. Although these decay constants cannot be
measured directly in experiments, the axialvector decay constant appears in
hard exclusive production rates of $\chi_{Q1}$~\cite{Brodsky:1989pv,
Chernyak:1983ej, Jia:2008ep, Wang:2013ywc}, and the scalar decay constant could 
be measured in lattice QCD. 
The NRQCD factorization formulae for these quantities at leading order in $v$,
and corresponding short-distance coefficients at two-loop accuracies are
summarized in appendix~\ref{appendix:sdcs}.

%------------------------------------------------------------------------------
\subsection{Numerical inputs}
%------------------------------------------------------------------------------

We first describe the numerical inputs that we use for obtaining our results. 
We work consistently with the $S$-wave calculation in
ref.~\cite{Chung:2020zqc}, except that we
extend the calculation of the wavefunctions and the reduced Green's functions
to nonzero orbital angular momentum.

%------------------------------------------------------------------------------
\subsubsection{Heavy quark mass and strong coupling}
%------------------------------------------------------------------------------

We compute $\alpha_s$ in the $\overline{\rm MS}$ scheme by using 
{\sf RunDec}~\cite{Chetyrkin:2000yt, Herren:2017osy} at 4-loop accuracy 
at a fixed QCD renormalization scale $\mu_R$, which facilitates exact order by
order cancellation of the logarithm of the NRQCD factorization scale between
the wavefunctions at the origin and the short-distance coefficients. 
We set the active number of flavors $n_f$ to be $n_f = 3$ for calculations 
involving charmonia, and $n_f =4$ for bottomonia, counting only the light quark
flavors. 
The value of $\mu_R$ that we use are $\mu_R = 2.5^{+1.5}_{-1.0}$~GeV for
charmonium and $\mu_R = 5^{+3}_{-3}$~GeV for bottomonium; these ranges are
obtained in ref.~\cite{Peset:2018ria} from theoretical descriptions of 
masses of lowest-lying quarkonium states that have mild dependences on the
scale, and have also been used for calculations of $S$-wave quarkonium
wavefunctions in ref.~\cite{Chung:2020zqc}. 
As will be discussed in the next section, we also compute $\alpha_s$ at
different scales when considering resummation of logarithms that appear in loop
corrections to the potentials. 

The mass $m$ that appears in the pNRQCD expressions of the LDMEs, as well as
the Schr\"odinger equation, is the heavy quark pole mass. As is well known, the
pole mass contains a renormalon ambiguity, which makes it impossible to assign 
a precise numerical value to $m$. 
In order to circumvent this issue, we use the modified
renormalon subtracted ($\rm RS'$) mass~\cite{Pineda:2001zq}, 
which is related to the pole mass $m$ by 
%---------------
\begin{equation}
\label{eq:mrsprime}
%---------------
m = m_{\rm RS'} (\nu_f) + \delta m_{\rm RS'}(\nu_f), 
%---------------
\end{equation}
%---------------
where $\delta m_{\rm RS'}(\nu_f)$ is the renormalon subtraction term that has a
perturbative expansion in powers of $\alpha_s$, and 
$\nu_f$ is the scale associated with the subtraction. 
The subtraction term $\delta m_{\rm RS'}$ contains the same renormalon
ambiguity as the pole mass at leading power in $1/m$, so that 
unlike the pole mass, 
$m_{\rm RS'} (\nu_f)$ has a well-defined value. 
We use eq.~(\ref{eq:mrsprime}) to replace $m$ in the Schr\"odinger equation 
by $m_{\rm RS'}+ \delta m_{\rm RS'}$ 
and expand in powers of $\delta m_{\rm RS'}$, and then compute 
corrections from $\delta m_{\rm RS'}$ by using the
Rayleigh-Schr\"odinger perturbation theory. 
The expression for $\delta m_{\rm RS'}$ can be found in
refs.~\cite{Pineda:2001zq, Peset:2018ria}. 
The scale $\nu_f$ can be different from $\mu_R$, but a value of $\nu_f$ that is
too different from $\mu_R$ can produce large logarithms of $\nu_f/\mu_R$ in the
perturbative expansion of the renormalon subtraction term; 
on the other hand, a large $\nu_f$ would result in a large value of 
$\delta m_{\rm RS'}$, which could negatively impact the convergence of the
perturbation series. 
We choose 
$\nu_f = 2$~GeV, and expand $\delta m_{\rm RS'}$ in powers of 
$\alpha_s(\mu_R)$. 
At $\nu_f = 2$~GeV, the $\rm RS'$ quark masses are given by 
$m_{c, {\rm RS'}} = 1316(41)$~MeV for charm, and 
$m_{b, {\rm RS'}} = 4743(41)$~MeV for bottom~\cite{Peset:2018ria}. 
Because $\delta m_{\rm RS'}$ begins at order $\alpha_s^2$, 
we only need to consider the correction to the wavefunction that comes from the
kinetic energy term in the Schr\"odinger equation. 

Equation~(\ref{eq:mrsprime}) can also be used in the NRQCD factorization
formulae, where inverse powers of the heavy quark pole mass appear. 
In this case, replacing $m$ by $m_{\rm RS'} + \delta m_{\rm RS'}$ 
and expanding in powers of $\delta m_{\rm RS'}$ produces corrections to the
short-distance coefficients at relative order $\alpha_s^2$. 

%------------------------------------------------------------------------------
\subsubsection{Potentials at long distances}
%------------------------------------------------------------------------------

For the static potential and the $1/m$ potential, we adopt the approach in
ref.~\cite{Chung:2020zqc} to combine the perturbative expressions that are 
valid at short distances and lattice determinations at long distances. 
The potentials can be written as 
%---------------
\begin{subequations}
\label{eq:potentials_longdistancematch}
\begin{eqnarray}
%---------------
V^{(0)} (r) &=& V^{(0)} (r) |_{\rm pert} + V^{(0)} (r) |_{\rm long}, 
\\
\label{eq:1mpotential_WL_longdistancematch}
V^{(1)} (r) |^{\rm WL} &=& V^{(1)} (r) |_{\rm pert}^{\rm WL} 
+ V^{(1)} (r) |_{\rm long}^{\rm WL}, 
%---------------
\end{eqnarray}
\end{subequations}
%---------------
where the superscript WL denotes that the $1/m$ potential is computed in the
Wilson loop matching scheme. The $V^{(0)} (r) |_{\rm long}$
and $V^{(1)} (r) |_{\rm long}^{\rm WL}$ are smooth functions of $r$ that vanish
at short distances, which are chosen so that the expressions 
in eqs.~(\ref{eq:potentials_longdistancematch}) reproduce the lattice QCD
determinations in refs.~\cite{Bali:2000vr, Koma:2012bc} at long distances. 
We use the explicit expressions for 
the potentials in ref.~\cite{Chung:2020zqc}, 
where perturbative corrections are included up to order $\alpha_s^3$ (relative
order $\alpha_s^2$) in 
$V^{(0)} (r) |_{\rm pert}$~\cite{Fischler:1977yf, Schroder:1998vy}, 
and $V^{(1)} (r) |_{\rm pert}^{\rm WL}$ is computed
at leading order in $\alpha_s$ (order $\alpha_s^2$)~\cite{Brambilla:2000gk}. 
In computing the $V^{(0)} (r) |_{\rm pert}$ and 
$V^{(1)} (r) |_{\rm pert}^{\rm WL}$, we choose the 
renormalization scale as $\mu_r = (r^{-2} + \mu_R^2)^{1/2}$, so that $\mu_r
\approx 1/r$ at short distances; this resums the logarithms in $r$ 
that appear in the loop corrections to the static potential. This
resummation makes the
effects of higher order perturbative corrections to $V^{(0)} (r) |_{\rm pert}$
beyond what we include our calculations 
numerically insignificant~\cite{Chung:2020zqc}.

We use the expressions in eqs.~(\ref{eq:potentials_longdistancematch})
to define the LO potential and the $1/m$ potential that we use for computing 
the wavefunctions in the on-shell matching scheme. 
We define the LO potential by 
%---------------
\begin{equation}
\label{eq:LOpotential_explicit}
%---------------
V_{\rm LO} (r) = - \frac{\alpha_s(\mu_R) C_F}{r} 
+ V^{(0)} (r) |_{\rm long},
%---------------
\end{equation}
%---------------
where in the first term, we compute $\alpha_s$ at a fixed renormalization scale
$\mu_R$. In this case, the Coulombic correction term $\delta V_C(r)$ 
is given by 
%---------------
\begin{equation}
\label{eq:Coulombcorr_explicit}
%---------------
\delta V_C (r) = V^{(0)} (r) - V_{\rm LO}(r) = 
V^{(0)} (r) |_{\rm pert} + \frac{\alpha_s(\mu_R) C_F}{r}. 
%---------------
\end{equation}
%---------------
Here, the last term combined with the leading-order (order-$\alpha_s$) term in 
the perturbative expression for the static potential 
$V^{(0)} (r) |_{\rm pert}$ gives 
$[\alpha_s (\mu_R)-\alpha_s(\mu_r)] C_F/r$, 
which cancels the $\mu_R$ dependence of the LO potential. 
Hence, corrections from $\delta V_C (r)$ cancel the $\mu_R$ dependences of the
LO wavefunctions. 
Since this term and the order-$\alpha_s^2$ term in $V^{(0)} (r) |_{\rm pert}$ 
correspond to the change of the Coulomb strength of the LO
potential of relative order $\alpha_s$, we consider the correction from 
$\delta V_C(r)$ to the wavefunction as a correction of relative order
$\alpha_s$. 

Following the prescription in eq.~(\ref{eq:1mpot_osprescription}), 
we write the $1/m$ potential in the on-shell scheme by 
%---------------
\begin{equation}
\label{eq:1mpot_explicit}
%---------------
V^{(1)} (r) |^{\rm OS} = \frac{\alpha_s^2(\mu_R) C_F (\tfrac{1}{2}
C_F-C_A)}{2 r^2} 
+ V^{(1)} (r) |_{\rm long}^{\rm WL},
%---------------
\end{equation}
%---------------
where in the first term, we compute $\alpha_s$ at a fixed scale $\mu_R$. 
This allows us to compute the wavefunctions in the on-shell matching scheme
directly without computing the unitary transformation of the wavefunction in
the Wilson loop matching scheme, provided that the
approximate relation in eq.~(\ref{eq:approx_relation}) is satisfied.

%------------------------------------------------------------------------------
\subsubsection{Reduced Green's function}
\label{sec:redgreen_numerical}
%------------------------------------------------------------------------------

We now describe the method that we use to compute the 
reduced Green's function $\hat{G}_n(\bm{r}',\bm{r})$. 
We use two different methods for numerical computation depending on the region
of $|\bm{r}|$ and $|\bm{r}'|$. In the first method, which is valid for small
$|\bm{r}|$ or $|\bm{r}'|$, we obtain the reduced Green's
function from the relation in eq.~(\ref{eq:reducedgreen_relation2}) by
computing numerically the Green's function $G(\bm{r},\bm{r}';E)$ for different
values of $E$. 
Since we only need $P$-wave contributions, 
we first decompose the Green's function into specific 
orbital angular momentum contributions as 
%---------------
\begin{equation}
%---------------
G(\bm{r}',\bm{r};E) = \sum_{L=0}^\infty \sum_{M=-L}^{+L} 
G^{L} (r',r;E) Y_L^{M} (\hat{\bm{r}}') Y_L^{M*} (\hat{\bm{r}}), 
%---------------
\end{equation}
%---------------
where $r= |\bm{r}|$ and $r'=|\bm{r}'|$. 
For each $L$, $G^{L} (r',r;E)$ satisfies the differential equation
%---------------
\begin{equation}
\label{eq:lippmann_L}
%---------------
\left[ - \frac{1}{m} \left( \frac{\partial^2}{\partial r^2}
+ \frac{2}{r} \frac{\partial}{\partial r} \right) + V_{\rm LO} (r) 
+ \frac{L (L+1)}{m r^2} -E \right] 
G^L (r',r;E) = \frac{1}{r^2} \delta(r-r').
%---------------
\end{equation}
%---------------
To obtain solutions of eq.~(\ref{eq:lippmann_L}), 
we extend the method used in refs.~\cite{Strassler:1990nw, Kiyo:2010jm,
Chung:2020zqc} for $S$-wave states to arbitrary orbital angular momentum. 
The solution of eq.~(\ref{eq:lippmann_L}) can be written in the form 
%---------------
\begin{equation}
\label{eq:greenf_numerical}
%---------------
G^L (r',r;E) = m \frac{u_>(r_>)}{r_>} \frac{ u_<(r_<)}{r_<}, 
%---------------
\end{equation}
%---------------
where $r_> = \max (r,r')$, $r_< = \min(r,r')$, 
and $u_>$, $u_<$ are two linearly independent solutions of 
the homogeneous equation 
%---------------
\begin{equation}
\label{eq:1dschrodinger}
%---------------
\left[ \frac{d^2}{dr^2} + m (E-V_{\rm LO} (r) )
- \frac{L (L+1)}{r^2} \right] u(r)  = 0,
%---------------
\end{equation}
%---------------
with the boundary conditions 
%---------------
\begin{equation}
%---------------
W (u_>,u_<) (r) = 
u_>(r) u_<'(r) - u_>'(r) u_<(r) = 1, 
%---------------
\end{equation}
%---------------
for any $r$. We note that the Wro\'nskian $W (u_>,u_<) (r)$ of the two solutions 
is independent of $r$, since eq.~(\ref{eq:1dschrodinger}) does not contain 
$d u(r)/dr$. 
The conditions that uniquely determine $u_>(r)$ and $u_<(r)$ read 
%---------------
\begin{subequations}
\label{eq:uboundaryconditions}
\begin{eqnarray}
%---------------
&& \lim_{r \to 0} r^{-L} u_<(r) = 0, \quad 
\lim_{r \to 0} \frac{d}{dr} \left( r^{-L} u_<(r) \right) =1, \\
&& \lim_{r \to \infty} u_>(r) = 0, \quad 
\lim_{r\to 0} r^L u_>(r) = \frac{1}{2 L+1}.
%---------------
\end{eqnarray}
\end{subequations}
%---------------
The conditions on the regular solution $u_<(r)$ 
are obtained from the fact that a solution of 
eq.~(\ref{eq:1dschrodinger}) that is regular at $r=0$ 
must vanish like $r^{L+1}$ as $r\to 0$. 
The condition that $u_>(r)$ is square integrable unambiguously fixes $u_>(r)$, 
because in general the second solution is a linear combination of $u_>(r)$ and
$u_<(r)$. 
We obtain $u_<(r)$ and $u_>(r)$ by solving eq.~(\ref{eq:1dschrodinger})
numerically in \textsc{Mathematica}, from which we determine 
$G^L (r',r;E)$ for a given $E$. 
Then, by using the relation in eq.~(\ref{eq:reducedgreen_relation2}), 
we obtain the reduced Green's function for small $r_<$. 
Since when $E = E_n^{\rm LO}$, $u_<(r)$ is a 
bound-state solution of eq.~(\ref{eq:1dschrodinger}), 
the square-integrable solution $u_>(r)$ does not exist if $E$ coincides with
the eigenenergies of the Schr\"odinger equation. 
Hence, when using eq.~(\ref{eq:reducedgreen_relation2}),
we cannot take $E$ to be too close to $E_n^{\rm LO}$. In our numerical
calculations, we use $\eta = 10^{-2}$~GeV. 

Since in the first method, $u_<(r)$ is determined by initial conditions 
at $r=0$, and $E$ cannot take values that are too close to the eigenenergies of
the Schr\"odinger equation, the calculation of the Green's function 
becomes unstable when $r$ and $r'$ are both large. 
Hence, for large $r$ and $r'$, we use the formal expression in 
eq.~(\ref{eq:redgreen}) to 
compute the reduced Green's function by truncating the series by 
including a limited number of lowest eigenstates. 
In our numerical calculations, we include 9 lowest $P$-wave eigenstates in the
truncated series. 
Similarly to the $S$-wave case, we expect this method to
become unreliable for small $r$ or $r'$, because there the series may 
not converge well. 
Hence, we combine the reduced Green's functions computed from the two different
methods by using 
%---------------
\begin{equation}
\label{eq:green_allr}
%---------------
\hat{G}_n (\bm{r}',\bm{r}) = 
b(r_<) \times \hat{G}_n(\bm{r}',\bm{r})|_{\rm short} 
+ [1-b(r_<)] \times \hat{G}_n(\bm{r}',\bm{r})|_{\rm long}, 
%---------------
\end{equation}
%---------------
where $\hat{G}_n(\bm{r}',\bm{r})|_{\rm short}$ is computed from
eqs.~(\ref{eq:greenf_numerical}) and (\ref{eq:reducedgreen_relation2}),
while $\hat{G}_n(\bm{r}',\bm{r})|_{\rm long}$ is obtained by truncating the
series in eq.~(\ref{eq:redgreen}). Here, $b(r)$ is a smooth function with 
$b(0) = 1$ and $b(\infty) = 0$, so that eq.~(\ref{eq:green_allr}) is 
reliable for all $r$ and $r'$. The explicit form of $b(r)$ that we use is 
%---------------
\begin{equation}
%---------------
b(r) = \frac{1}{\pi} \left[ \tan^{-1} (4 m (r_b -r)) - \tan^{-1} (4 m r_b)
\right] +1, 
%---------------
\end{equation}
%---------------
with $r_b = 1$~GeV$^{-1}$. 
We note that although the above form of $b(r)$ has a nonzero $b(\infty)$, 
it is still adequate for our numerical calculations, because in practice 
we work in a finite range of $r$, and $b(r)$ is much smaller 
than 1 at large $r$ for both charm and bottom masses. 
The validity of the reduced Green's function that
we obtain can be tested numerically by checking the relations
%---------------
\begin{subequations}
\label{eq:green_check}
\begin{eqnarray}
%---------------
\left( E_k^{\rm LO} - E_n^{\rm LO} \right) 
\int d^3r \, \hat{G}_n (\bm{r}',\bm{r}) \Psi_k^{\rm LO} (\bm{r}) 
 &=& \Psi_k^{\rm LO} (\bm{r}') , 
\\
\int d^3r \, \hat{G}_n (\bm{r}',\bm{r}) \Psi_n^{\rm LO} (\bm{r}) 
 &=& 0, 
%---------------
\end{eqnarray}
\end{subequations}
%---------------
for $k \neq n$. 
We note that $b(r)$ has negligible effects in calculations of the corrections
$\delta \bm{\nabla} \Psi_n(\bm{0})|_{(r_0)}$, 
because the position-space integrals in
eq.~(\ref{eq:corr_finiter}) are dominated by contributions at small $r$. 
On the other hand, 
the second term in eq.~(\ref{eq:green_allr}) becomes necessary for computing 
$\delta \Psi_n(\bm{r})$ at large $r$, or for reproducing the relations in
eq.~(\ref{eq:green_check}) numerically at large $r'$. 

We compute the eigenenergies and wavefunctions of the LO Schr\"odinger equation 
by finding values of $E$ that makes $u_<(r)$ square integrable; 
in this case, $E$ is a LO eigenenergy and the corresponding radial wavefunction
is given by $u_<(r)/r$ times a normalization coefficient. 
An alternative way for finding $\Psi_n^{\rm LO} (\bm{r})$ and $E_n^{\rm LO}$ 
is to use the Crank-Nicolson method~\cite{crank_nicolson_1947}. 
The advantage of this method is that it does
not depend on an initial condition, so that this method is free of accumulating 
numerical errors for large $r$. 
We use the Crank-Nicolson method to test the validity of the
numerical results for $E_n^{\rm LO}$ and $\Psi_n^{\rm LO} (\bm{r})$ that we 
obtain. We use the modified Crank-Nicolson method developed in
ref.~\cite{Kang:2006jd}, which 
has the advantage that it does not require the Gram-Schmidt process when
computing eigenenergies and wavefunctions for excited states. 

We also use the Crank-Nicolson method to test the convergence of the
corrections to the wavefunctions that we compute. By using the Crank-Nicolson
method, we can compute corrections to
the wavefunctions to all orders in the Rayleigh-Schr\"odinger perturbation
theory by finding bound-state solutions of the Schr\"odinger equation
including the correction terms to the potential, as long as the correction
terms are smooth functions of $r$; this is possible because the Crank-Nicolson
method remains reliable even when the potential diverges faster than $1/r$. 
This result can be compared with the wavefunction computed to first order in
the Rayleigh-Schr\"odinger perturbation theory to estimate the size of higher
order corrections. We present the numerical test of convergence of the 
corrections from the $1/m$ potential in Rayleigh-Schr\"odinger perturbation
theory in appendix~\ref{appendix:1mallorder}.
On the other hand, the all-orders calculation using the Crank-Nicolson method
does not allow an order-by-order calculation that is necessary in 
establishing exact two-loop level cancellations of the divergent small 
$r$ behavior between the finite-$r$ regularized wavefunctions at the origin 
and the scheme conversion coefficient $\delta Z$. 
Therefore, the all-orders calculation will only be used to complement the 
order-by-order calculation in the Rayleigh-Schr\"odinger perturbation theory.

%------------------------------------------------------------------------------
\subsection[Numerical results for $P$-wave charmonia]
{\boldmath  Numerical results for $P$-wave charmonia}
%------------------------------------------------------------------------------

We now present our numerical results for $P$-wave charmonia. 
We list the central values of %the radial wavefunctions at the origin 
$|R'_{\rm LO}(0)|$ for the two lowest $P$-wave states in
table~\ref{tab:charmcorrections}. 
For brevity, we refer to $|R'_{\rm LO}(0)|$ as the wavefunction at the origin
in discussions of $P$-wave quarkonia. 
The LO binding energies for the $1P$ and $2P$ states are 
$E^{\rm LO}_{1P} = 0.568$~GeV and 
$E^{\rm LO}_{2P} = 1.03$~GeV, respectively.
We note that the LO binding energy for the $1P$ state 
is consistent with the $\chi_{cJ}$ and $h_c$ masses, when compared 
with the LO binding energies of the $S$-wave states 
$E^{\rm LO}_{1S} = 0.233$~GeV and 
$E^{\rm LO}_{2S} = 0.769$~GeV in ref.~\cite{Chung:2020zqc}.
The $1P$ charmonium mass $m_{1P} = 3.47$~GeV computed from 
$m_{1P} = 2 m_{\rm RS'} + 2 \delta m_{\rm RS'} + E^{\rm LO}_{1P}$ 
is in good agreement with the PDG values of $\chi_{cJ}$ and $h_c$ masses 
within $0.1$~GeV~\cite{Zyla:2020zbs}. 
Because we find good agreement in the $1P$ charmonium masses with
measurement, we use the measured charmonium masses from
ref.~\cite{Zyla:2020zbs} when computing decay rates and decay constants,
because the experimental values have uncertainties that are negligible compared
to the theoretical uncertainties.  
We identify the $1P$ states with angular momentum quantum numbers $^3P_J$
and $^1P_1$ as $\chi_{cJ}$ and $h_c$, respectively. 
While the binding energy for the $2P$ state is also consistent with the mass 
of the $X(3872)$, we do not identify the $2P$ states with the $X(3872)$ or
other states of
similar masses, because those states have masses that are heavier than the 
open flavor threshold, and are unlikely to be pure quarkonium
states\footnote{This is also supported by the fact that 
the radial size of the LO wavefunction for the $2P$ state 
is not small compared to the distance $r_c \approx 5.8$~GeV$^{-1}$ 
at which the static potential is expected to suffer from color
screening~\cite{Bali:2000vr}. 
For example, 
in the case of the $2P$ wavefunction, the contributions to the 
normalization $\int d^3r \, |\Psi^{\rm LO}(\bm{r})|^2 =1$ 
from the regions $r<r_c$ and $r>r_c$ are comparable in size, 
while the normalization of the $1P$ wavefunction 
is dominated by $r < r_c$. 
}. 

Unsurprisingly, our results for $|R'_{\rm LO}(0)|$ for charmonium states are
much larger than what we obtain in perturbative QCD, where the long-distance
nonperturbative effects in the potentials are neglected. For the $1P$ state,
a perturbative QCD calculation at same values of $\alpha_s$ and $m$ gives a 
value of $|R'_{\rm LO}(0)|$ that is only a few percent of our result in
table~\ref{tab:charmcorrections}. The discrepancy is even stronger for the $2P$
state, because in perturbative QCD, $|R'_{\rm LO}(0)|$ decreases with
increasing radial excitation, which is opposite to what we find when we include
the long-distance nonperturbative effects in the static potential. 

We list the corrections to the wavefunctions at the origin relative to the 
leading-order wavefunction at the origin in table~\ref{tab:charmcorrections}. 
The corrections are
classified as the non-Coulombic correction $\delta_\Psi^{\rm NC}$ coming from
the $1/m$ and $1/m^2$ potentials, 
the Coulombic correction $\delta_\Psi^{\rm C}$ that comes from 
$\delta V_C(r)$, and the correction $\delta_\Psi^{\rm RS'}$ from the 
$\rm RS'$ subtraction term. Explicit expressions for 
$\delta_\Psi^{\rm NC}$, $\delta_\Psi^{\rm C}$, and $\delta_\Psi^{\rm RS'}$
are given by 
%---------------
\begin{subequations}
\label{eq:delpsi_expressions}
\begin{eqnarray}
%---------------
\delta_\Psi^{\rm NC} &=& -\delta Z_{^{2 S+1}P_J}- 
\frac{1}{\hat{\bm{r}} \cdot \bm{\nabla} \Psi_n^{\rm LO} (\bm{0})} 
\nonumber \\ && \times 
\bigg[ 
\hat{\bm{r}}' \cdot 
\bm{\nabla}_{\bm{r}'}
\int d^3 r \,
\hat{G}_n (\bm{r}', \bm{r})
\delta {\cal V}(\bm{r},\bm{\nabla}) \Psi_n^{\rm LO} (\bm{r})
\Big|_{|\bm{r}'| = r_0}
\nonumber \\ && \hspace{4ex} 
+
\frac{E_n^{\rm LO}}{m} 
\hat{\bm{r}}' \cdot 
\bm{\nabla}_{\bm{r}'}
\int d^3 r \,
\hat{G}_n (\bm{r}', \bm{r})
\left( V_{p^2}^{(2)} (r)
+ \frac{1}{2} V_{\rm LO} (r) \right) \Psi_n^{\rm LO} (\bm{r})
\Big|_{|\bm{r}'| = 0}
\bigg] 
\nonumber \\ &&
+ \frac{1}{2 m}
\left[ - V_{p^2}^{(2)} (0)
+ \int d^3r \, \left( V_{p^2}^{(2)}(r)+ \frac{1}{2} V_{\rm LO} (r) \right)
\left| \Psi_n^{\rm LO} (\bm{r})  \right|^2
\right] ,
\end{eqnarray}
\begin{equation}
\delta_\Psi^{\rm C} =
- \frac{1}{\hat{\bm{r}} \cdot \bm{\nabla} \Psi_n^{\rm LO} (\bm{0})}
\hat{\bm{r}}' \cdot
\bm{\nabla}_{\bm{r}'}
\int d^3 r \,
\hat{G}_n (\bm{r}', \bm{r})
\delta V_C(r) \Psi_n^{\rm LO} (\bm{r})
\Big|_{|\bm{r}'| = 0}, 
\end{equation}
\begin{equation}
\delta_\Psi^{\rm RS'} =
- \frac{1}{\hat{\bm{r}} \cdot \bm{\nabla} \Psi_n^{\rm LO} (\bm{0})}
\frac{\delta m_{\rm RS'}}{m_{\rm RS'}} 
\hat{\bm{r}}' \cdot \bm{\nabla}_{\bm{r}'}
\int d^3 r \, \hat{G}_n (\bm{r}', \bm{r})
V_{\rm LO} (r) \Psi_n^{\rm LO} (\bm{r})
\Big|_{|\bm{r}'| = 0},
%---------------
\end{equation}
\end{subequations}
%---------------
so that the $\overline{\rm MS}$-renormalized wavefunctions at the origin are
given by 
%---------------
\begin{equation}
%---------------
\hat{\bm{r}} \cdot \bm{\nabla} \Psi_n(\bm{0}) |_{\overline{\rm MS}} 
= 
\hat{\bm{r}} \cdot \bm{\nabla} \Psi_n^{\rm LO} (\bm{0}) 
\times \left( 1+ \delta_\Psi^{\rm NC} + \delta_\Psi^{\rm C} 
+ \delta_\Psi^{\rm RS'} 
+ O(\Lambda_{\rm QCD}^2/m^2, v^3)
\right). 
%---------------
\end{equation}
%---------------
We compute $\delta_\Psi^{\rm NC}$ at the $\overline{\rm MS}$ scale 
$\Lambda = m$. We neglect the term $- \frac{1}{2 m} V_{p^2}^{(2)} (0)$ in 
$\delta_\Psi^{\rm NC}$, because it cancels the $i {\cal E}_2/m$ term in the 
pNRQCD expressions for the LDMEs, and does not contribute to decay or
production rates. 

%%%%%%%%%%%%%%%%%%%%%%%%%%%%%%%%%%%%%%%%%%%%%%%%%%%%%%%%%%%%%%%%%%%%%%%%%%%%%%%
\begin{table}[tbp]
\centering
\begin{tabular}{|c|c|c|c|c|c|c|c|}
\hline
State & $|R'_{\rm LO} (0)|$~(GeV$^{5/2}$) & 
$\delta_\Psi^{\rm NC}|_{^3P_0}$ & $\delta_\Psi^{\rm NC}|_{^3P_1}$ &
$\delta_\Psi^{\rm NC}|_{^3P_2}$ & 
$\delta_\Psi^{\rm NC}|_{^1P_1}$ 
& $\delta_\Psi^{\rm C}$ & $\delta_\Psi^{\rm RS'}$ \\
\hline
$1P$ & 0.184 & 0.453 & 0.493 & 0.505 & 0.500 & 0.266 & 0.103 
\\
\hline
$2P$ & 0.243 & 0.513 & 0.547 & 0.569 & 0.558 & 0.201 & 0.102
\\
\hline
\end{tabular}
\caption{\label{tab:charmcorrections}
LO wavefunctions at the origin $|R'(0)|$ 
and relative corrections to the wavefunctions at the origin
in the $\overline{\rm MS}$ scheme at scale $\Lambda=m$ for $1P$ and $2P$
charmonium states.
$\delta_\Psi^{\rm NC}$ is the correction from the $1/m$ and $1/m^2$ potentials,
$\delta_\Psi^{\rm C}$ is the Coulombic correction, 
and $\delta_\Psi^{\rm RS'}$ is the correction from the $\rm RS'$
subtraction term.
The $\delta_\Psi^{\rm NC}$, $\delta_\Psi^{\rm C}$, 
and $\delta_\Psi^{\rm RS'}$ are dimensionless.
}
\end{table}
%%%%%%%%%%%%%%%%%%%%%%%%%%%%%%%%%%%%%%%%%%%%%%%%%%%%%%%%%%%%%%%%%%%%%%%%%%%%%%%

%%%%%%%%%%%%%%%%%%%%%%%%%%%%%%%%%%%%%%%%%%%%%%%%%%%%%%%%%%%%%%%%%%%%%%%%%%%%%%%
\begin{figure}[tbp]
\centering
\includegraphics[width=.49\textwidth]{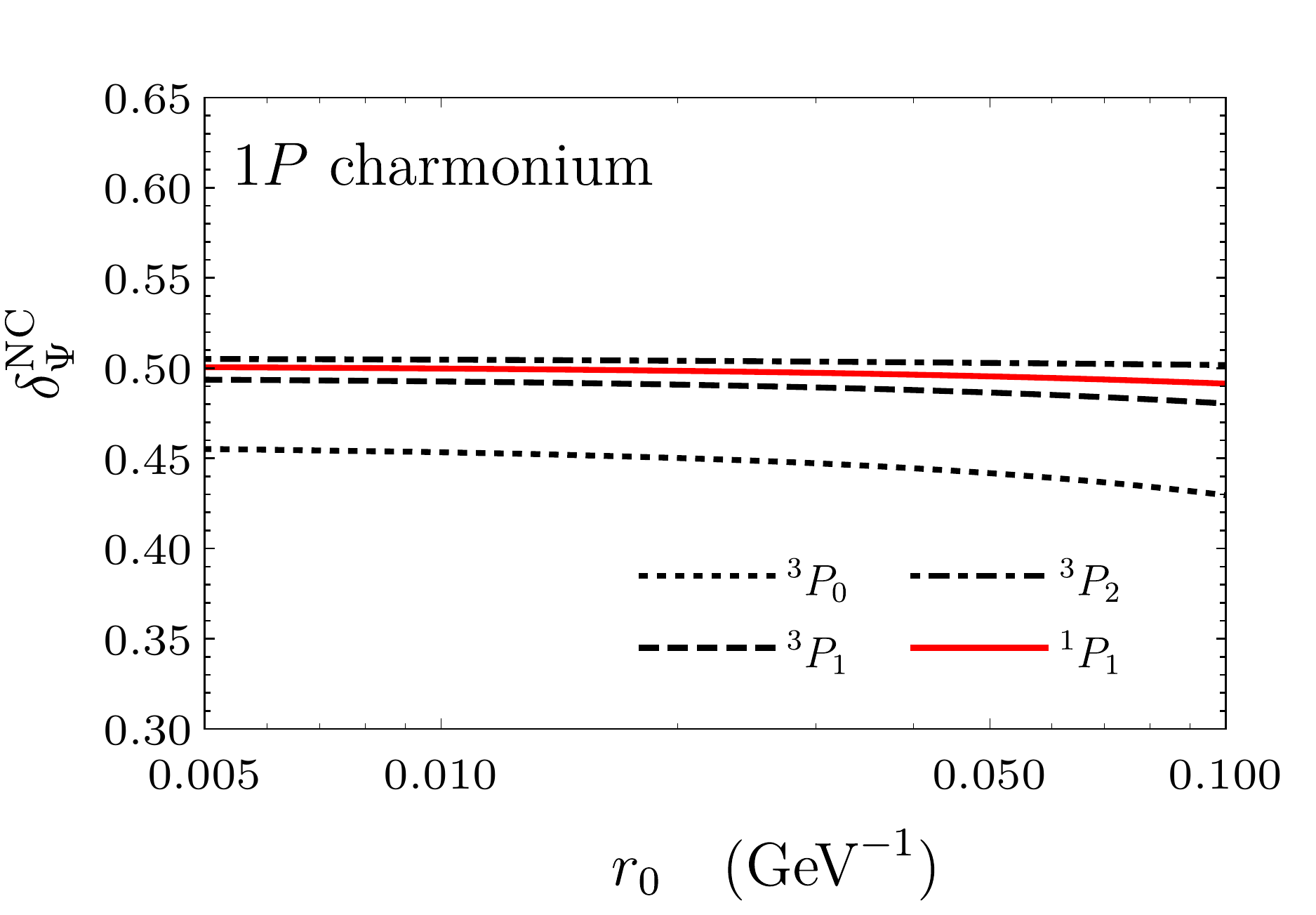}
\includegraphics[width=.49\textwidth]{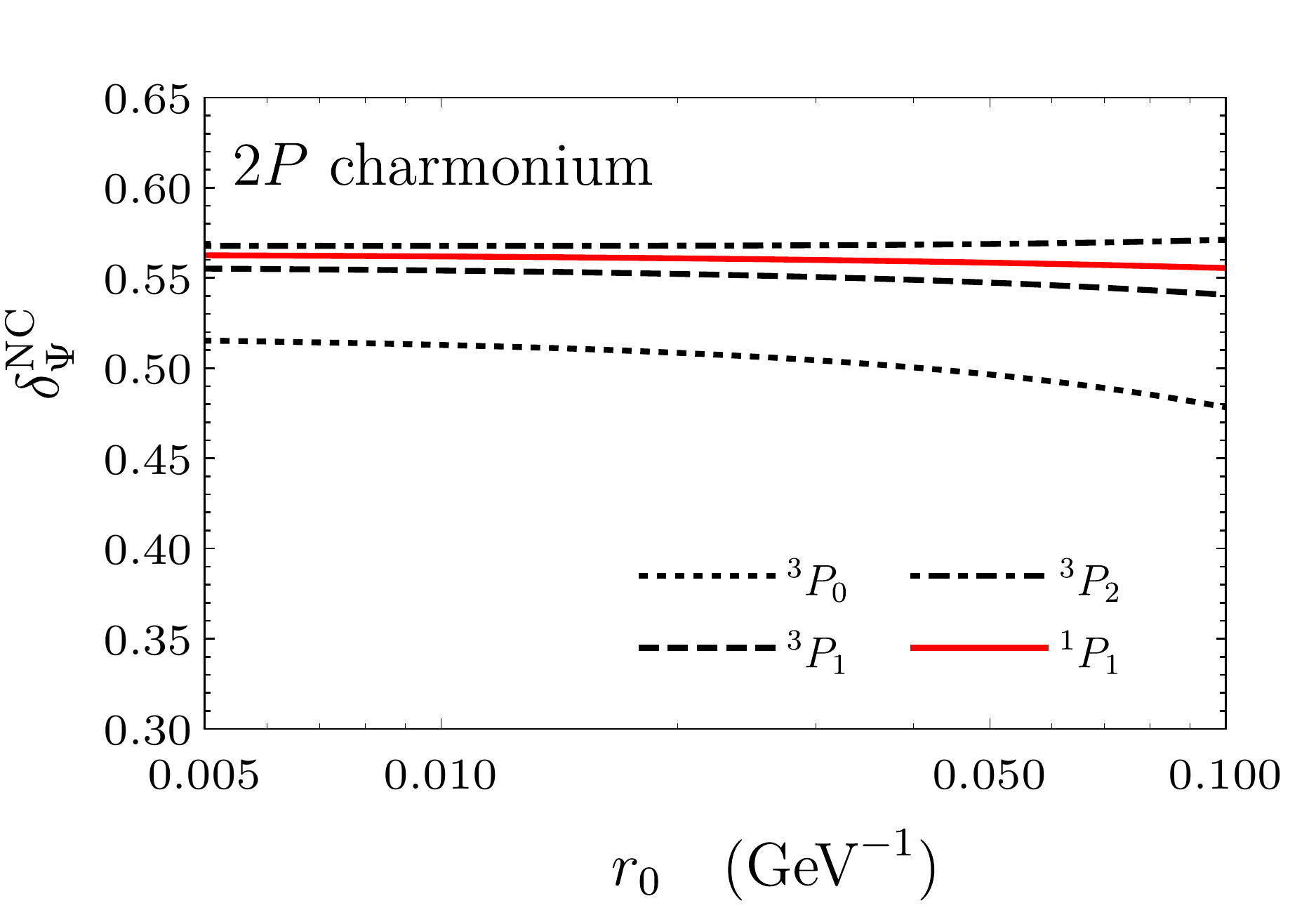}
\caption{\label{fig:charmdelNC}
Non-Coulombic corrections $\delta_\Psi^{\rm NC}$ at finite $r_0$ for the
charmonium $1P$ (left panel) and $2P$ (right panel) states, 
for angular momentum quantum numbers
$^3P_0$ (dotted lines), $^3P_1$ (dashed lines), $^3P_2$ (dot-dashed lines),
and $^1P_1$ (red solid lines).
The $r_0$ dependences are mild for the range $r_0 < 0.02$~GeV$^{-1}$ that we
consider.
}
\end{figure}
%%%%%%%%%%%%%%%%%%%%%%%%%%%%%%%%%%%%%%%%%%%%%%%%%%%%%%%%%%%%%%%%%%%%%%%%%%%%%%%

In computing $\delta_\Psi^{\rm NC}$, the finite-$r$ regulator $r_0$ must be
taken to be as small as possible to suppress terms of positive powers of $r_0$, 
as long as the numerical calculation is stable. 
Because in the $P$-wave case, there are no power divergences in the 
finite-$r$ regularized wavefunction at the origin, we can take $r_0$ to be 
much smaller than the $S$-wave calculation in ref.~\cite{Chung:2020zqc}, 
as long as the approximate
relation in eq.~(\ref{eq:approx_relation}) is well reproduced. 
We find that eq.~(\ref{eq:approx_relation}) remains valid for 
both $1P$ and $2P$ states with errors of about 1.5\% even at 
very small values of $r_0$ of about $0.001$~GeV$^{-1}$. 
The numerical results for $\delta_\Psi^{\rm NC}$ are almost insensitive to 
$r_0$ for values of $r_0$ less than about $0.02$~GeV$^{-1}$. 
We show the $r_0$ dependence of non-Coulombic corrections 
for $1P$ and $2P$ states in fig.~\ref{fig:charmdelNC}. 
We fix $r_0 =0.01$~GeV$^{-1}$ for computing the central values of $\delta_\Psi^{\rm NC}$ 
and neglect the uncertainty in the numerical calculation of 
finite-$r$ regularized wavefunctions at the origin compared to other
uncertainties. 

We note that the non-Coulombic corrections $\delta_\Psi^{\rm NC}$ are positive
and sizable for all $^{2 S+1} P_J$ states. 
This is in contrast with the negative order-$\alpha_s^2$ corrections to the
short-distance coefficients, as can be seen from the
NRQCD factorization formulae appendix~\ref{appendix:sdcs}. 
Hence, similarly to the $S$-wave case in ref.~\cite{Chung:2020zqc}, we expect
sizable cancellations in the $\Lambda$-independent finite parts between 
the order-$\alpha_s^2$ short-distance coefficients and $\delta_\Psi^{\rm NC}$, 
and reductions in the size of the corrections to the decay and production rates
beyond tree level. 

The differences in $\delta_\Psi^{\rm NC}$ between different
angular momentum states are small
at the $\overline{\rm MS}$ scale $\Lambda = m$. 
The largest deviation occurs for the
$^3P_0$ state, for which the non-Coulombic correction is smaller by about 5\%
compared to other angular momentum states. 
This shows that heavy quark spin symmetry is conserved approximately 
between $P$-wave quarkonium wavefunctions at the origin. 
The contributions from the long-distance part of the $1/m$ potential are tiny, 
and amount to less than 1\% for both $1P$ and $2P$ states. 
It is worth noting that, as can be seen from the results in
appendix~\ref{appendix:pertQCD}, 
the results for $\delta_\Psi^{\rm NC}$ obtained from perturbative QCD
calculations are comparable in order of magnitude to the results in  
table~\ref{tab:charmcorrections} for a similar value of $\alpha_s$
[$\alpha_s(\mu_R = 2.5{\rm ~GeV}) \approx 0.27$], even though the 
specific values, especially the dependence on angular momentum quantum numbers, 
are quite different. 
This, and the fact that $\delta_\Psi^{\rm NC}$ contain finite parts of order
$\alpha_s^2$ that remain after subtracting the $1/\epsilon_{\rm UV}$ poles, 
suggest that the bulk of $\delta_\Psi^{\rm NC}$ may come
from corrections to the wavefunctions at short distances. 

In ref.~\cite{Sang:2015uxg}, the authors argued that 
$|R'(0)|$ is larger for the $^3P_2$ state compared to the $^3P_0$ state, 
based on the fact that the spin-dependent potential is attractive 
for the $^3P_2$ state, while the sign is opposite for the $^3P_0$ state. 
This is opposite to what we find in the calculation of $\delta_\Psi^{\rm NC}$.
However, the analysis in ref.~\cite{Sang:2015uxg} does not take into 
account the fact that the corrections from the spin-dependent potential are 
UV divergent, and so, conclusions on the sizes or even the signs of 
the corrections can only be obtained after the divergences are 
subtracted through renormalization, as we have done 
in this work in the $\overline{\rm MS}$ scheme.

Since $\delta_\Psi^{\rm C}$ is of relative order $\alpha_s$, 
the Coulombic correction at second order in the Rayleigh-Schr\"odinger
perturbation theory is of relative order $\alpha_s^2$, and so, 
it may be important to include this correction. 
Explicit numerical calculations of the second order correction 
show that this correction is negligibly small. We also confirm this from 
the calculation of Coulombic corrections to all orders by using the 
modified Crank-Nicolson method~\cite{Kang:2006jd}, 
which also shows that the Coulombic corrections converge rapidly. 
Hence, we neglect the Coulombic corrections beyond first order in the QMPT. 
The corrections from the $\rm RS'$ subtraction term are small for both $1P$ and
$2P$ states. 

Before we compute decay and production rates of $P$-wave charmonia at two-loop
level, let us compare our results at leading order in $1/m$ 
with potential model calculations that are often used in phenomenological
studies of $P$-wave charmonium production and decay\footnote{
While potential-model calculations may attempt to capture the effect of
long-distance behavior of higher order potentials, this requires an arbitrary 
separation between short and long-distance contributions which is model
dependent and can become uncontrollable. Moreover, it is likely that the
long-distance behaviors of higher order potentials have insignificant effects
to $|R'(0)|$ compared to uncertainties and model dependences of potential-model
results, because the slope of the potential at long distances do not change 
appreciably by inclusion of the higher order potentials. 
}. 
Our first-principles calculation gives the value 
$|R'_{\rm LO}(0)|^2 = 0.034^{+0.03}_{-0.00}$~GeV$^5$ for the $1P$ state. 
Here, the uncertainties come mainly from the variation of $\mu_R$.
If we include the Coulombic correction $\delta_\Psi^{\rm C}$, which would be 
appropriate for calculations at one-loop level, we obtain
$|R'_{\rm LO}(0)|^2 (1+\delta_\Psi^{\rm C})^2 = 0.054^{+0.01}_{-0.00}$~GeV$^5$.
This is close to the values obtained phenomenologically in
refs.~\cite{Chung:2008km, Brambilla:2020ojz} 
from two-photon decay rates of $\chi_{c0}$ and $\chi_{c2}$ at one-loop level,
and also to the value obtained from inclusive hadroproduction rates of
$\chi_{c1}$ and $\chi_{c2}$ in ref.~\cite{Bodwin:2015iua}.
On the other hand, potential-model calculations usually give larger values of 
the wavefunction at the origin; 
the results for $|R'(0)|^2$ for the $1P$ state 
from several widely used potential models in refs.~\cite{Buchmuller:1980su, 
Eichten:1995ch, Bodwin:2007fz, Eichten:2019hbb} 
range from $0.068$~GeV$^5$ to $0.131$~GeV$^5$. 
We note that, however, potential models have charm quark masses whose numerical
values differ wildly from the $\rm RS'$ mass that we use. 
The model-dependent charm quark mass also affects 
NRQCD factorization formulae through their 
dependences on the heavy quark pole mass. 
As can be seen from the factorization formulae 
in appendix~\ref{appendix:sdcs}, 
the heavy quark pole mass appears in electromagnetic decay rates and 
exclusive electromagnetic production rates at leading orders in $v$ 
in the form 
$\langle {\cal Q} | {\cal O} (^{2 S+1}P_J) | {\cal Q} \rangle / m^3$. 
Hence, for the purpose of comparing with phenomenological models, 
it makes more sense to consider the combination $|R'(0)|^2/m^3$. 
Our result including the Coulombic correction gives 
$|R'_{\rm LO}(0)|^2 (1+\delta_\Psi^{\rm C})^2 /m_{\rm RS'}^3 
= 0.024^{+0.01}_{-0.00}$~GeV$^2$, 
while potential-model calculations in refs.~\cite{Buchmuller:1980su,
Eichten:1995ch, Eichten:2019hbb, Bodwin:2007fz} give values of 
$|R'(0)|^2/m^3$ that are between $0.021$~GeV$^2$ and $0.023$~GeV$^2$. 
Considering that these values are 
only valid up to corrections of relative order $v^2$, 
we conclude that at one-loop level, 
the potential-model results agree well with our first-principles calculation 
of the wavefunctions at the origin for $1P$ charmonia, as long as 
the heavy quark mass is chosen appropriately in model calculations of decay
and production rates, similarly to what has been done in the model-dependent
analysis in ref.~\cite{Brambilla:2020xod}. 

In the case of inclusive production processes, large $p_T$ 
cross sections of $P$-wave quarkonia depend on the dimensionless 
combination of the
leading-order LDMEs and the heavy quark mass given by $|R'(0)|^2/m^5$, 
which can be seen from the calculation of single-parton fragmentation
functions into $P$-wave quarkonium~\cite{Braaten:1994kd}. 
Our first-principles calculation gives 
$|R'_{\rm LO}(0)|^2 (1+\delta_\Psi^{\rm C})^2 /m_{\rm RS'}^5 = 0.014$, 
while the potential-model calculations based on refs.~\cite{Buchmuller:1980su,
Eichten:1995ch, Eichten:2019hbb, Bodwin:2007fz} give values of $|R'(0)|^2/m^5$ 
that range
from $0.006$ to $0.011$. Especially, the Buchm\"uller-Tye potential model in
ref.~\cite{Buchmuller:1980su} that
is often adopted in phenomenological studies of $P$-wave charmonium
hadroproduction (see, for example, refs.~\cite{Ma:2010vd, Gong:2012ug})
gives the value $|R'(0)|^2/m^5 = 0.011$. While this model calculation
is smaller than our
first-principles calculation by about 23\%, it is fair to say that they are
consistent, considering the current level of accuracy of 
inclusive charmonium production phenomenology. 

Even though our first-principles calculations of the $P$-wave wavefunctions 
at the origin are consistent with model-dependent calculations at one-loop
level and at leading order in $v$, 
both radiative and relativistic corrections can be sizable 
in processes involving $P$-wave charmonia. 
Our calculations of the corrections to the $P$-wave wavefunctions at the origin
from $1/m$ and $1/m^2$ potentials allow us to include consistently the
radiative corrections to relative order $\alpha_s^2$ in decay and production
rates of $P$-wave quarkonia, where the dependence on the NRQCD factorization
scale cancels exactly at two-loop level between the short-distance coefficients
and the corrections to the wavefunctions at the origin. 
On the other hand, 
because our pNRQCD expressions for the leading-order LDMEs are valid up to
corrections of relative order $v^2$, due to uncalculated corrections of order
$1/m^2$ in the matching coefficients $V_{\cal O}$, it is not possible
to fully incorporate the effects of relativistic corrections 
through relative order $v^2$. 
Hence, in computing the absolute decay and production rates, we work at leading
order in $v$, neglecting the contributions from LDMEs of dimensions 6 and above in
the NRQCD factorization formulae. 
On the other hand, we expect the uncalculated order-$v^2$ corrections 
coming from the matching coefficients $V_{\cal O}$ 
to cancel in ratios of LDMEs for different angular momentum
quantum numbers, and the heavy-quark spin symmetry breaking effects in the 
matching coefficient $V_{\cal O}$ to be suppressed by 
$\Lambda_{\rm QCD}^2/m^2$. Hence, when considering ratios of decay 
rates, we include the order-$v^2$ corrections coming from higher
dimensional LDMEs. 

We begin by considering the two-photon decay rates of $\chi_{c0}$ and
$\chi_{c2}$. 
By using the NRQCD factorization formula in eq.~(\ref{eq:chi0gamgam_nrqcd}) 
and the pNRQCD formula for the leading-order LDME, 
we obtain the following expression for 
$\Gamma(\chi_{c0} \to \gamma \gamma)$:
%---------------
\begin{eqnarray}
\label{eq:chi0gam_pnrqcd}
%---------------
\Gamma(\chi_{c0} \to \gamma \gamma) &=& 
\frac{12 \pi e_c^4 \alpha^2}{m_{\chi_{c0}}}
\frac{3 N_c}{2 \pi} 
\frac{|R'_{\rm LO}(0)|^2}{m_{\rm RS'}^3} 
\left( 1 - \frac{3 \delta m_{\rm RS'}}{m_{\rm RS'}} \right) 
\left( 1 + \delta_\Psi^{\rm C} + \delta_\Psi^{\rm NC}|_{^3P_0} 
+ \delta_\Psi^{\rm RS'} \right)^2 
\nonumber\\ && \times 
\left| 1 + \alpha_s c_{\gamma \gamma(00)}^{(1)} 
+ \alpha_s^2 c_{\gamma \gamma(00)}^{(2)} \right|^2 
+ O(\alpha_s^3, v^2)
\nonumber\\ 
&=& 
\frac{12 \pi e_c^4 \alpha^2}{m_{\chi_{c0}}}
\frac{3 N_c}{2 \pi} 
\frac{|R'_{\rm LO}(0)|^2}{m_{\rm RS'}^3} \bigg[ 
1 + 2 \alpha_s c_{\gamma \gamma(00)}^{(1)} + 2 \delta_\Psi^{\rm C} + 
\left( \alpha_s c_{\gamma \gamma(00)}^{(1)} \right)^2 
\nonumber \\ &&  \hspace{5ex} 
+ \left( \delta_\Psi^{\rm C} \right)^2 
+ 4 \alpha_s c_{\gamma \gamma(00)}^{(1)} \delta_\Psi^{\rm C} 
+ 2 \alpha_s^2 \, {\rm Re} ( c_{\gamma \gamma(00)}^{(2)} ) 
+ 2 \delta_\Psi^{\rm NC}|_{^3P_0} 
\nonumber \\ && \hspace{5ex} 
+ 2 \delta_\Psi^{\rm RS'} - \frac{3 \delta m_{\rm RS'}}{m_{\rm RS'}}
\bigg] 
+ O(\alpha_s^3, v^2). 
%---------------
\end{eqnarray}
%---------------
Here, $\alpha=1/137.0$ is the electromagnetic coupling constant. 
The factor $(1 - {3 \delta m_{\rm RS'}}/{m_{\rm RS'}})$
comes from replacing $1/m^3$ in the NRQCD factorization formula by 
$1/(m_{\rm RS'} + \delta m_{\rm RS'})^3$ and expanding in powers of 
$\delta m_{\rm RS'}$ to relative order $\alpha_s^2$. 
In the last equality, we have expanded the factors in the parentheses to
relative order $\alpha_s^2$; the terms in the square brackets then correspond
to the correction factors coming from radiative corrections to the
short-distance coefficients and the corrections to the wavefunctions at the
origin. 

Similarly, we obtain the following expression for 
$\Gamma(\chi_{c2} \to \gamma \gamma)$:
%---------------
\begin{eqnarray}
\label{eq:chi2gam_pnrqcd}
%---------------
\Gamma(\chi_{c2} \to \gamma \gamma) &=& 
\frac{16 \pi e_c^4 \alpha^2}{5 m_{\chi_{c2}}} 
\frac{3 N_c}{2 \pi} 
\frac{|R'_{\rm LO}(0)|^2}{m_{\rm RS'}^3}
\left( 1 - \frac{3 \delta m_{\rm RS'}}{m_{\rm RS'}} \right)
\left( 1 + \delta_\Psi^{\rm C} + \delta_\Psi^{\rm NC}|_{^3P_2} 
+ \delta_\Psi^{\rm RS'} \right)^2 
\nonumber\ \\ && \times 
\bigg( \left| 1 + \alpha_s c_{\gamma \gamma(22)}^{(1)}
+ \alpha_s^2 c_{\gamma \gamma(22)}^{(2)} \right|^2
+ \left| \alpha_s c_{\gamma \gamma(20)}^{(1)} \right|^2 \bigg) 
+ O(\alpha_s^3, v^2)
\nonumber\\
&=&
\frac{16 \pi e_c^4 \alpha^2}{5 m_{\chi_{c2}}}
\frac{3 N_c}{2 \pi} 
\frac{|R'_{\rm LO}(0)|^2}{m_{\rm RS'}^3} \bigg[
1 + 2 \alpha_s c_{\gamma \gamma(22)}^{(1)} + 2 \delta_\Psi^{\rm C} +
\left( \alpha_s c_{\gamma \gamma(22)}^{(1)} \right)^2
\nonumber \\ && \hspace{5ex} 
+ \left( \alpha_s c_{\gamma \gamma(20)}^{(1)} \right)^2
+ 4 \alpha_s c_{\gamma \gamma(22)}^{(1)} \delta_\Psi^{\rm C}
+ \left( \delta_\Psi^{\rm C} \right)^2 
+ 2 \alpha_s^2 \, {\rm Re} ( c_{\gamma \gamma(22)}^{(2)} ) 
\nonumber \\ && \hspace{5ex} 
+ 2 \delta_\Psi^{\rm NC}|_{^3P_2}
+ 2 \delta_\Psi^{\rm RS'} - \frac{3 \delta m_{\rm RS'}}{m_{\rm RS'}}
\bigg]
+ O(\alpha_s^3, v^2). 
%---------------
\end{eqnarray}
%---------------
Again, we obtain the last equality by expanding the correction factors to
relative order $\alpha_s^2$. 

Based on the expressions in eqs.~(\ref{eq:chi0gam_pnrqcd}) and 
(\ref{eq:chi2gam_pnrqcd}), we obtain the numerical results for the two-photon
decay rates given by 
%---------------
\begin{subequations}
\label{eq:chigamresults}
\begin{eqnarray}
%---------------
\Gamma(\chi_{c0} \to \gamma \gamma) 
&=& 
5.07 ^{+0.46}_{-0.00} \pm 1.52 \textrm{~keV}
= 
5.07 ^{+1.59}_{-1.52} \textrm{~keV},
\\
\Gamma(\chi_{c2} \to \gamma \gamma) 
&=& 
0.95 ^{+0.04}_{-0.01} \pm 0.29 \textrm{~keV}
=
0.95 \pm 0.29 \textrm{~keV},
%---------------
\end{eqnarray}
\end{subequations}
%---------------
where the first uncertainties come from varying $\mu_R$ between $1.5$~GeV and 
4~GeV, which represent the uncertainties from uncalculated corrections of
higher orders in $\alpha_s$, and the second uncertainties come from uncalculated
corrections of order $v^2$, which we take to be $0.3$ times the central values,
based on the typical size of $v^2 \approx 0.3$ for charmonium states. 
In the last equalities, we add the uncertainties in quadrature. 
We note that the logarithms of the NRQCD factorization scale $\Lambda$ cancel
exactly at two-loop level between $\delta_\Psi^{\rm NC}$ and the two-loop
corrections to the short-distance coefficients, so that there is no uncertainty
due to the dependence on $\Lambda$ in our numerical results.

We first discuss the convergence of the corrections that we include in our
calculation. In the case of the decay amplitude for 
$\chi_{c0} \to \gamma \gamma$, the one-loop
correction to the short-distance coefficients coming from $\alpha_s
c_{\gamma \gamma (00)}^{(1)}$ is about 0.008, while the two-loop correction
coming from $\alpha_s^2 c_{\gamma \gamma (00)}^{(2)}$ is about $-0.18$. 
When combined with the corrections to the wavefunctions at the origin and the 
$\rm RS'$ subtraction term, at order $\alpha_s$, the corrections coming from 
$\alpha_s c_{\gamma \gamma (00)}^{(1)}$ and $\delta_\Psi^{\rm C}$ add up to
0.27, and the corrections coming from 
$\alpha_s^2 c_{\gamma \gamma (00)}^{(2)}$, 
$\delta_\Psi^{\rm NC}$, $\delta_\Psi^{\rm RS'}$, and $\delta m_{\rm RS'}$ 
are about 0.21. Although the corrections of order $\alpha_s^2$ and $v^2$ 
are less in size than the order-$\alpha_s$ correction, 
the signs of the corrections are same. The correction factor at the 
squared amplitude level, given by the terms in the square brackets in
eq.~(\ref{eq:chi0gam_pnrqcd}), is about 2.1. 
Hence, in the case of the two-photon decay rate of $\chi_{c0}$, the convergence
of the corrections is not improved by inclusion of the corrections to the
wavefunctions at the origin. 

On the other hand, in the decay amplitudes for 
$\chi_{c2} \to \gamma \gamma$, the one-loop correction from 
$\alpha_s c_{\gamma \gamma(22)}^{(1)}$ is about $-0.23$, while the 
two-loop correction from $\alpha_s^2 c_{\gamma \gamma (22)}^{(2)}$ is about 
$-0.29$. When combined with the corrections to the wavefunctions at the origin
and the $\rm RS'$ subtraction term, the order-$\alpha_s$ corrections coming from
$\alpha_s c_{\gamma \gamma(22)}^{(1)}$ and $\delta_\Psi^{\rm C}$ add up to
0.03, and the corrections from
$\alpha_s^2 c_{\gamma \gamma(22)}^{(2)}$,
$\delta_\Psi^{\rm NC}$, $\delta_\Psi^{\rm RS'}$, and $\delta m_{\rm RS'}$
add to about 0.15. 
The correction factor at the squared amplitude level, given by the terms in the 
square brackets in the last equality of eq.~(\ref{eq:chi2gam_pnrqcd}), 
is about 1.2. 

If we compute the decay rates by expanding the corrections at the amplitude
level instead of working at the squared amplitude level, which can be
done by expanding the square roots of the terms in the square brackets 
in the last equalities in eqs.~(\ref{eq:chi0gam_pnrqcd}) and
(\ref{eq:chi2gam_pnrqcd}), the central value of
$\Gamma(\chi_{c0} \to \gamma \gamma)$ increases to 5.48~keV, 
and the central value of 
$\Gamma(\chi_{c2} \to \gamma \gamma)$ decreases to 0.80~keV. 
These values are within the uncertainties of our numerical results. 

We note that our numerical results have central values that are larger than 
the measured two-photon decay rates from BESIII~\cite{Ablikim:2012xi}, 
which are given by 
$\Gamma(\chi_{c0} \to \gamma \gamma) = 
2.33 \pm 0.20 \pm 0.22$~keV and 
$\Gamma(\chi_{c2} \to \gamma \gamma) = 
0.63 \pm 0.04 \pm 0.06$~keV. 
While the measured two-photon decay rate of the $\chi_{c2}$ is compatible with 
our numerical result within uncertainties, 
our result overestimates the $\chi_{c0}$ two-photon rate by about 1.8 times
the theoretical uncertainty. 

It would be interesting if the relativistic corrections of relative order $v^2$
reduces the discrepancy between the measured decay rates and the theoretical
values. Even though the order-$v^2$ corrections to the leading-order LDMEs are
currently unknown, we expect the spin-dependent corrections 
to be suppressed by $\Lambda_{\rm QCD}^2/m^2$, 
so that the order-$v^2$ corrections to the ratio 
${\cal R}_{\chi_c} = \Gamma(\chi_{c2} \to \gamma \gamma)/
\Gamma(\chi_{c0} \to \gamma \gamma)$ 
comes solely from higher dimensional LDMEs. 
By using the results for the tree-level short-distance coefficients associated
with the higher dimensional LDMEs~\cite{Brambilla:2017kgw}, 
and the pNRQCD calculations of the LDMEs at
leading nonvanishing orders in $1/m$~\cite{Brambilla:2020xod}, 
we obtain the following expression for
${\cal R}_{\chi_c}$ that is valid through order $\alpha_s^2$ and $v^2$:
%---------------
\begin{eqnarray}
\label{eq:chigamgamratio_pnrqcd}
%---------------
{\cal R}_{\chi_c} &=& 
\frac{4}{15} 
\bigg\{ 1+2 \alpha_s \left( c_{\gamma \gamma(22)}^{(1)} 
- c_{\gamma \gamma(00)}^{(1)} \right) 
+ \alpha_s^2 \bigg[ 3 \left( c_{\gamma \gamma(00)}^{(1)} \right)^2
+ \left( c_{\gamma \gamma(20)}^{(1)} \right)^2 
\nonumber \\  && \hspace{5ex} 
+ \left( c_{\gamma \gamma(22)}^{(1)} \right)^2 
-4 c_{\gamma \gamma(00)}^{(1)} c_{\gamma \gamma(22)}^{(1)}
+2 \, {\rm Re} \big( c_{\gamma \gamma(22)}^{(2)} 
- c_{\gamma \gamma(00)}^{(2)} \big)
\bigg] 
\nonumber \\  && \hspace{5ex} 
+2 \left( \delta_\Psi^{\rm NC}|_{^3P_2} - \delta_\Psi^{\rm NC}|_{^3P_0} \right) 
+ \frac{E^{\rm LO}}{3 m_{\rm RS'}} 
\bigg\}
+ O(\alpha_s^3, \Lambda_{\rm QCD}^2/m^2). 
%---------------
\end{eqnarray}
%---------------
The last term in the curly brackets comes from the contributions of 
LDMEs of dimensions 6 and 7 to the NRQCD factorization
formulae~\cite{Brambilla:2017kgw, Brambilla:2020xod}. 
The logarithm of $\Lambda$ in $c_{\gamma \gamma( 22)}^{(2)}
- c_{\gamma \gamma (00)}^{(2)}$ is cancelled exactly by 
$\delta_\Psi^{\rm NC}|_{^3P_2} - \delta_\Psi^{\rm NC}|_{^3P_0}$ 
through order $\alpha_s^2$. 
By using our numerical results for $\delta_\Psi^{\rm NC}$, we obtain 
%---------------
\begin{eqnarray}
\label{eq:chigamgamratio_numerical}
%---------------
{\cal R}_{\chi_c} &=&
0.16 {}^{+0.02}_{-0.04} \pm 0.02
= 0.16^{+0.03}_{-0.04}, 
%---------------
\end{eqnarray}
%---------------
where the first uncertainty comes from varying $\mu_R$ between 1.5~GeV and 
4~GeV, and the second uncertainty comes from uncalculated corrections
of order $\Lambda_{\rm QCD}^2/m^2$, which we estimate to be 
$(500\textrm{~MeV}/m)^2 \approx 0.14$ times the central value. 
In the last equality, we add the uncertainties in quadrature. 

The quantity in the curly brackets in eq.~(\ref{eq:chigamgamratio_pnrqcd}) 
contains the corrections of order $\alpha_s$, $\alpha_s^2$, and $v^2$. The
order-$\alpha_s$ contribution coming from the one-loop corrections to the
short-distance coefficients is about $-0.48$.
The two-loop corrections to the short-distance coefficients amount to about
$-0.16$, 
while the corrections to the wavefunctions at the origin is about $0.10$ 
at $\Lambda = m$. Finally, the order-$v^2$ correction is about 
$0.14$. While the corrections to the wavefunctions and the order-$v^2$
correction from higher dimensional LDMEs are moderate in size, they do help
counter the effect of the large negative corrections from the short-distance
coefficients: if we keep only the loop corrections to the short-distance
coefficients, the correction factor given by the quantity in the curly brackets
in eq.~(\ref{eq:chigamgamratio_pnrqcd}) is about 0.35. Inclusion of the
corrections to the wavefunctions at the origin and the order-$v^2$
corrections from higher dimensional LDMEs increase this correction factor to
about 0.60.
Nevertheless, our numerical result is still smaller than the measured value 
${\cal R}_{\chi_{c}} = 0.27 \pm 0.04$ from BESIII~\cite{Ablikim:2012xi}, 
even though the discrepancy is reduced by inclusion of the corrections 
to the wavefunctions at the origin and the order-$v^2$ corrections 
considered in this work. 

Next, we consider the cross sections 
$\sigma(e^+ e^- \to \chi_{cJ} + \gamma)$ ($J =0$, 1, and 2) at 
$\sqrt{s} = 10.58$~GeV. 
By using the NRQCD factorization formula in appendix~\ref{appendix:sdcs}, 
we obtain the following expression
%---------------
\begin{eqnarray}
\label{eq:crosssection_pnrqcd}
%---------------
\sigma(e^+ e^- \to \chi_{cJ} + \gamma) 
&=& 
\frac{3 N_c}{2 \pi} |R'_{\rm LO}(0)|^2 \, 
{\sigma}_{cJ}^{(0)} (s, m_{\rm RS'}) 
\nonumber \\ && \times 
\bigg[ 1 + 
\alpha_s \hat{\sigma}_{cJ}^{(1)} (r) 
+ \alpha_s^2 \hat{\sigma}_{cJ}^{(2)} (r) 
+ 2 \delta_\Psi^{\rm C} 
+ 2 \alpha_s 
\delta_\Psi^{\rm C} \hat{\sigma}_{cJ}^{(1)} (r)
+ \left(\delta_\Psi^{\rm C} \right)^2
\nonumber \\ && \hspace{5ex} 
+ 2 \delta_\Psi^{\rm NC} |_{^3P_J}
+ 2 \delta_\Psi^{\rm RS'} 
+ 
c_m^J 
\frac{\delta m_{\rm RS'}}{m_{\rm RS'}} 
\bigg] 
+ O(\alpha_s^3, v^2),
%---------------
\end{eqnarray}
%---------------
where $r = 4 m^2/s$ 
and $c_m^J = m \frac{\partial }{\partial m}
\log {\sigma}_{cJ}^{(0)} (s, m) |_{m=m_{\rm RS'}}$.
The last term in the square brackets comes from replacing
the heavy quark pole mass in $\hat{\sigma}_{cJ}^{(0)}$ 
by $m_{\rm RS'} + \delta m_{\rm RS'}$ and expanding in
powers of $\delta m_{\rm RS'}$ to relative order $\alpha_s^2$. 
Note that in the limit $s \to \infty$, $\lim_{m^2/s \to 0} c_m^J = -3$. 
The two-loop short-distance coefficients $\hat{\sigma}_{cJ}^{(2)} (r)$ 
contain logarithms in
$\Lambda$, which cancel exactly with the $\Lambda$ dependence in 
$\delta_\Psi^{\rm NC}$ through order $\alpha_s^2$. 

Our numerical results for the cross sections are 
%---------------
\begin{subequations}
\begin{eqnarray}
%---------------
\sigma(e^+ e^- \to \chi_{c0} + \gamma) &=& 
3.11 ^{+0.47}_{-0.09} \pm 0.93\textrm{~fb}
= 3.11 _{-0.94}^{+1.05}\textrm{~fb}, 
\\
\sigma(e^+ e^- \to \chi_{c1} + \gamma) &=& 
23.3 ^{+3.8}_{-0.4}  \pm 7.0\textrm{~fb}
= 23.3^{+7.9}_{-7.0}\textrm{~fb}, 
\\
\sigma(e^+ e^- \to \chi_{c2} + \gamma) &=& 
4.93 ^{+0.59}_{-0.00}  \pm 1.48\textrm{~fb} 
= 4.93 ^{+1.59}_{-1.48}\textrm{~fb}, 
%---------------
\end{eqnarray}
\end{subequations}
%---------------
where the uncertainties are as in eq.~(\ref{eq:chigamresults}).
Again, since the $\Lambda$ dependences cancel exactly through order
$\alpha_s^2$, there is no uncertainty from dependence on the NRQCD
factorization scale. 
In the case of $\chi_{c0}$, the one-loop correction to the short-distance
coefficients is positive, while the two-loop correction is negative; the
radiative corrections are small in size. On the other hand, for $\chi_{c1}$ and
$\chi_{c2}$, the loop corrections are negative at both order $\alpha_s$ and
$\alpha_s^2$, and are sizable, especially for $\chi_{c2}$. 
The corrections to the wavefunctions at the origin and the correction coming
from the use of the $\rm RS'$ mass are positive, which counteract the negative
corrections from the short-distance coefficients. The correction factor, given
by the quantity in the square brackets in eq.~(\ref{eq:crosssection_pnrqcd}), 
is about 2.5, 1.9, and 1.1 for $\chi_{c0}$, $\chi_{c1}$, and $\chi_{c2}$,
respectively. 
Our result for the $\chi_{c1}+\gamma$ production rate is consistent with the
Belle measurement 
$\sigma(e^+ e^- \to \chi_{c1} + \gamma) = 17.3^{+4.2}_{-3.9} \pm 1.7$~fb~\cite{Jia:2018xsy} within
uncertainties, while our result for the $\chi_{c2} + \gamma$ cross section is 
lower than, but close to the upper limit 
$\sigma(e^+ e^- \to \chi_{c2} + \gamma) < 5.7$~fb from 
Belle~\cite{Jia:2018xsy}. 

Finally, we compute the scalar decay constant $f_{\chi_{c0}}$ and the 
axialvector decay constant $f_{\chi_{c1}}$, 
which are defined in eqs.~(\ref{decayconst_defs}). 
By using the NRQCD factorization formulae in appendix~\ref{appendix:sdcs}, we
obtain 
%---------------
\begin{subequations}
\label{eq:decayconst_pnrqcd}
\begin{eqnarray}
%---------------
\label{eq:decayconst_scalar_pnrqcd}
m f_{\chi_{c0}}^{\rm OS} &=& 
m_{\overline{\rm MS}} (\mu)  f_{\chi_{c0}}^{\overline{\rm MS}} (\mu) 
\nonumber \\ 
&=& 
\sqrt{\frac{6 N_c}{ \pi m_{\chi_{c0}}}} 
|R'_{\rm LO}(0)|
\bigg( 1 + \alpha_s c_s^{(1)} + \alpha_s^2c_s^{(2)} 
+ \delta_\Psi^{\rm C} 
+ \alpha_s c_s^{(1)} \delta_\Psi^{\rm C}
\nonumber \\ &&  \hspace{20ex} 
+ \delta_\Psi^{\rm NC} |_{^3P_0}
+ \delta_\Psi^{\rm RS'} \bigg)
+O(\alpha_s^3,v^2), 
\\
\label{eq:decayconst_axial_pnrqcd}
f_{\chi_{c1}} &=& 
\sqrt{\frac{9 N_c}{ \pi m_{\chi_{c1}}}}
\frac{|R'_{\rm LO}(0)|}{m_{\rm RS'}}
\bigg( 1 + \alpha_s c_a^{(1)} + \alpha_s^2 c_a^{(2)}
+ \delta_\Psi^{\rm C}
+ \alpha_s c_a^{(1)} \delta_\Psi^{\rm C}
\nonumber \\ &&  \hspace{20ex}
+ \delta_\Psi^{\rm NC} |_{^3P_1}
+ \delta_\Psi^{\rm RS'}
- \frac{\delta m_{\rm RS'}}{m_{\rm RS'}}
\bigg)
+O(\alpha_s^3,v^2),
%---------------
\end{eqnarray}
\end{subequations}
%---------------
where $m_{\overline{\rm MS}} (\mu)$ is the charm quark mass 
in the $\overline{\rm MS}$ scheme at scale $\mu$, and 
the superscripts OS and $\overline{\rm MS}$ on $f_{\chi_{c0}}$ denote
the scheme in which the scalar decay constants are renormalized. 
Since the scalar decay constant is renormalized in the same way as the quark
mass term in the QCD Lagrangian, the combination 
$m f_{\chi_{c0}}^{\rm OS} = 
\bar{m}(\mu) f_{\chi_{c0}}^{\overline{\rm MS}} (\mu)$ is scheme and scale
independent. 
The axialvector decay constant is renormalization scheme and scale independent, 
although the short-distance coefficients do depend on the scheme in which
$\gamma_5$ is defined in DR. We work with na\"ive dimensional regularization,
because the two-loop short-distance coefficient $c_a^{(2)}$
is only available in this scheme. The one-loop coefficient 
$c_a^{(1)}$ has also been computed in the t'Hooft-Veltman scheme in
ref.~\cite{Wang:2013ywc}, where the authors find 
$c_a^{(1)}|_{\rm HV} = 0$. 

The QCD matrix elements that define the decay constants can have imaginary
parts coming from diagrams that contain on-shell intermediate states. 
The short-distance coefficients explicitly contain the imaginary parts from
relative order $\alpha_s^2$, which affect the size of the decay constants from
relative order $\alpha_s^4$.
In practice, we are only interested in the size of
the decay constants, because we can always absorb the phase into the quarkonium
state; hence, we neglect the imaginary parts of the short-distance
coefficients at the current level of accuracy. 

Our numerical results for the decay constants are 
%---------------
\begin{subequations}
\label{eq:decayconst_results}
\begin{eqnarray}
%---------------
m f_{\chi_{c0}}^{\rm OS} &=& 
0.439 ^{+0.042}_{-0.006} \pm 0.132 \textrm{~GeV}^2
= 0.439^{+0.138}_{-0.132} \textrm{~GeV}^2, 
\\
f_{\chi_{c1}} &=& 
0.236 ^{+0.011}_{-0.000} \pm 0.071 \textrm{~GeV}
=
0.236\pm 0.071 \textrm{~GeV}, 
%---------------
\end{eqnarray}
\end{subequations}
%---------------
where the uncertainties are as in eq.~(\ref{eq:chigamresults}).
By dividing $m f_{\chi_{c0}}^{\rm OS} = 
m_{\overline{\rm MS}} (\mu)  f_{\chi_{c0}}^{\overline{\rm MS}} (\mu)$ 
by the $\overline{\rm MS}$ charm quark mass
$\overline{m} = m_{\overline{\rm MS}} (\overline{m}) 
= 1.27 \pm 0.02$~GeV~\cite{Zyla:2020zbs}, 
we obtain the $\overline{\rm MS}$-renormalized 
scalar decay constant at scale $\mu = \overline{m}$:
%---------------
\begin{equation}
%---------------
f_{\chi_{c0}}^{\overline{\rm MS}} (\mu = \overline{m}) 
= 0.346^{+0.109}_{-0.104}\textrm{~GeV}. 
%---------------
\end{equation}
%---------------
The correction factor given by the terms in the parenthesis in
eq.~(\ref{eq:decayconst_scalar_pnrqcd}) is about 1.5. The order-$\alpha_s$
correction from $c_s^{(1)}$ and $\delta_\Psi^{\rm C}$ amounts to about 0.2,
while the two-loop correction $c_s^{(2)}$ and the corrections to the
wavefunctions at the origin add up to about 0.3.  
In the case of the axialvector decay constant, the order-$\alpha_s$ correction
from $c_a^{(1)}$ and $\delta_\Psi^{\rm C}$ is about 0.15, and the two-loop
correction from $c_a^{(2)}$, combined with the corrections to the 
wavefunctions at the origin and the $-\delta m_{\rm RS'}/m$ term, 
amount to about 0.17. 
As a result, the correction factor given by the terms in the parenthesis in 
eq.~(\ref{eq:decayconst_axial_pnrqcd}) is about 1.32, which is a bit milder
than that of the scalar decay constant. 
Since the $c_a^{(1)}$ and $c_a^{(2)}$ have been computed by using 
na\"ive dimensional regularization for $\gamma_5$, the numerical results will
be different in the t'Hooft-Veltman scheme. 
For example, $c_a^{(1)} = 0$ in the t'Hooft-Veltman scheme, 
and in this case, the order-$\alpha_s$ correction increases to about 0.27. 

Our numerical results have been computed with the $\rm RS'$ charm quark mass 
$m_{\rm RS'}$ at
the scale $\nu_f=2$~GeV. While this removes the ambiguity in the pole mass, the
$\rm RS'$ mass itself depends on the scale $\nu_f$. For example, setting the
scale $\nu_f = 1$~GeV increases $m_{\rm RS'}$ to 1.496(41)~GeV. 
In decay and production rates, as well as decay constants, 
the change in $m_{\rm RS'}$ is compensated by the correction from 
$\delta m_{\rm RS'}$, which also affects the correction $\delta_\Psi^{\rm RS'}$
to the wavefunctions at the origin. 
We find that if we use $\nu_f = 1$~GeV, the $\chi_{c1}+\gamma$ and 
$\chi_{c2}+\gamma$ production cross sections, the two-photon rates of
$\chi_{c0}$ and $\chi_{c2}$, as well as the ratio ${\cal R}_{\chi_c}$,  
reduces by less than 10\%, while 
the decay constants increase by less than 10\%. While the 
$\chi_{c0}+\gamma$ production rate decreases by more than 15\%, all of these
changes are well within the estimated uncertainties of our results. 
Hence, it is fair to say that, by using the $\rm RS'$ mass, 
the heavy quark mass
dependences in our numerical results are well under control.

%------------------------------------------------------------------------------
\subsection[Numerical results for $P$-wave bottomonia]
{\boldmath Numerical results for $P$-wave bottomonia}
%------------------------------------------------------------------------------

We now present our numerical results for $P$-wave bottomonia. We list the
central values of the radial wavefunctions at the origin $R'(0)$ for the three
lowest $P$-wave states in table~\ref{tab:bottomcorrections}. 
The LO binding energies for the $1P$, $2P$, and $3P$ states are 
$E^{\rm LO}_{1P} = 0.298$~GeV, 
$E^{\rm LO}_{2P} = 0.614$~GeV, and 
$E^{\rm LO}_{3P} = 0.888$~GeV, respectively. 
Compared to the LO binding energies for $2S$ and $3S$ bottomonia in
ref.~\cite{Chung:2020zqc}, given by 
$E^{\rm LO}_{2S} = 0.417$~GeV and 
$E^{\rm LO}_{3S} = 0.723$~GeV, respectively, 
the $P$-wave binding energies are consistent with the measured masses of the 
$\chi_{cJ}(nP)$ states. 
The $nP$ bottomonium masses 
$m_{1P} = 9.94$~GeV, 
$m_{2P} = 10.26$~GeV, and 
$m_{3P} = 10.53$~GeV, 
computed from 
$m_{nP} = 2 m_{\rm RS'} + 2 \delta m_{\rm RS'} + E^{\rm LO}_{nP}$ 
($n=1$, 2, and 3) 
are in good agreement with the PDG values within $0.1$~GeV~\cite{Zyla:2020zbs}. 
In computing decay rates and decay constants, we use the PDG values of the
bottomonium masses in ref.~\cite{Zyla:2020zbs}, because the measured values 
have uncertainties that are negligible compared to the theoretical
uncertainties. 
We identify the bottomonium $nP$ states with angular momentum quantum numbers
$^3P_J$ and $^1P_1$ by $\chi_{bJ}(nP)$ and $h_b(nP)$, respectively.

The LO radial wavefunctions at the origin in table~\ref{tab:bottomcorrections} 
are much larger than what we would get if we neglect the long-distance
nonperturbative behavior of the static potential. Even for the $1P$ state, our
result for $|R_{\rm LO}'(0)|$ is almost 10 times larger than what we obtain in
perturbative QCD calculations of the wavefunctions at same values of $\alpha_s$
and $m$. Hence, even for the lowest-lying $P$-wave bottomonium states, the 
long-distance nonperturbative behavior of the potentials cannot be neglected. 

\begin{figure}[tbp]
\centering
\includegraphics[width=.32\textwidth]{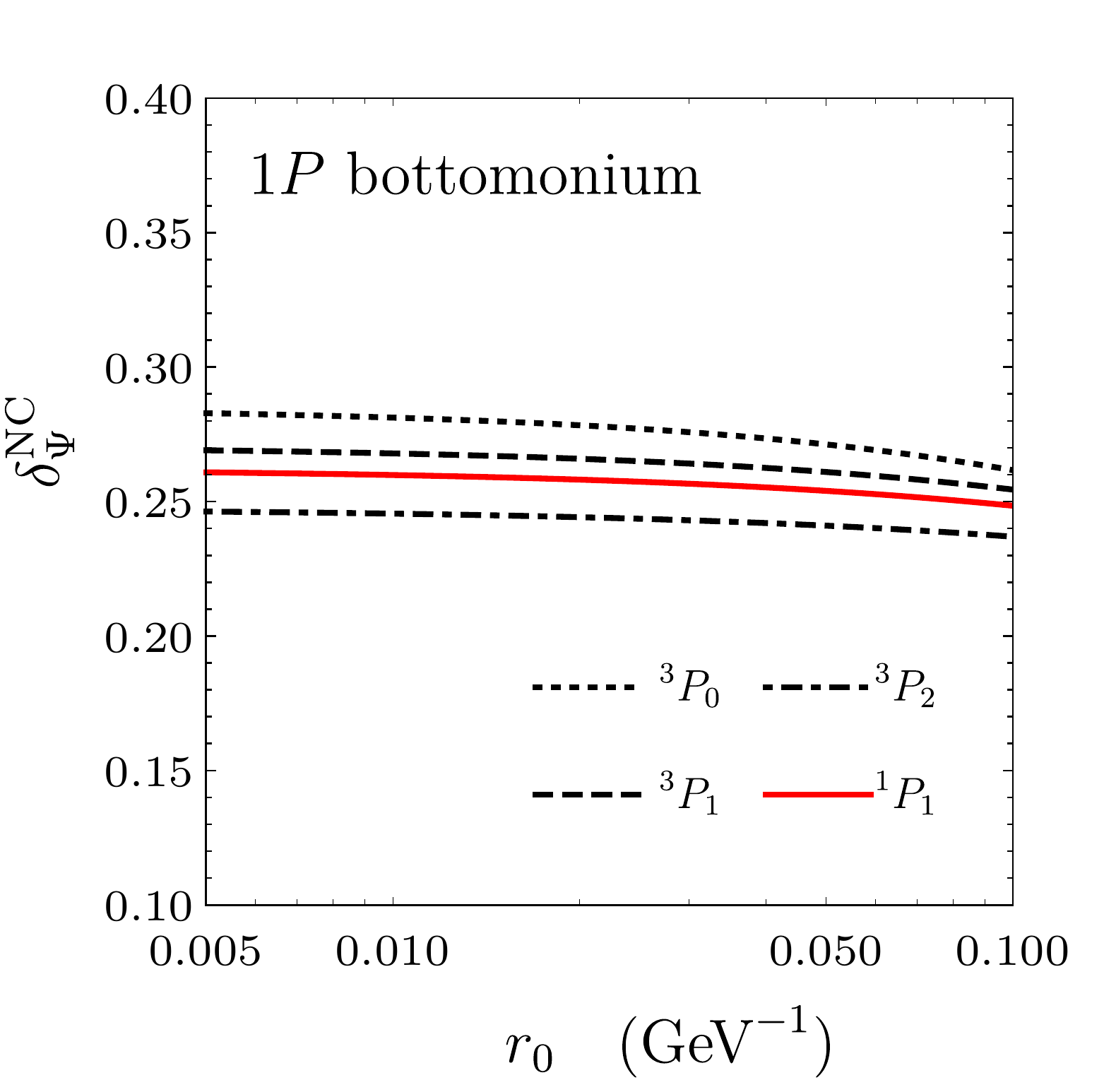}
\includegraphics[width=.32\textwidth]{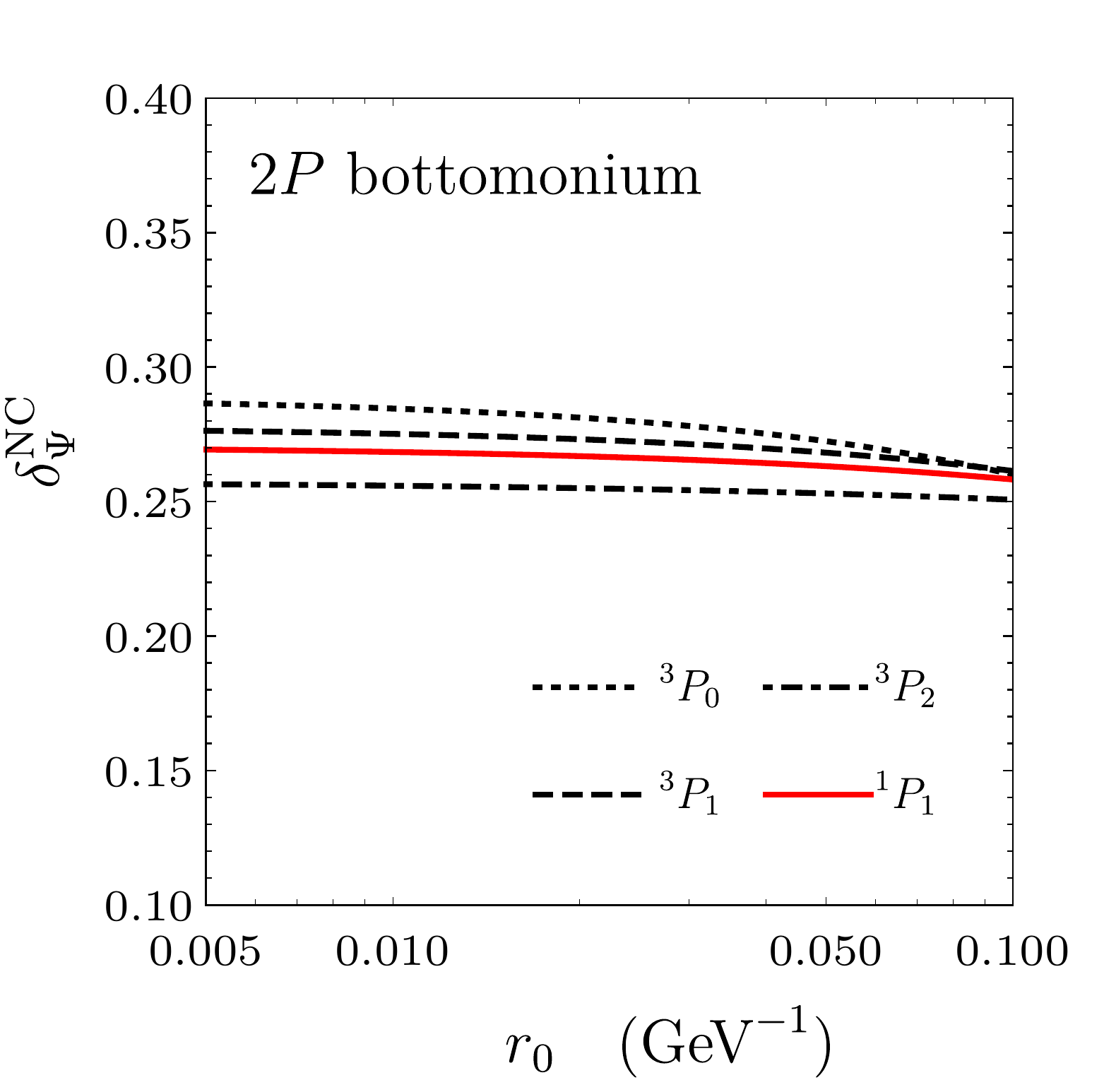}
\includegraphics[width=.32\textwidth]{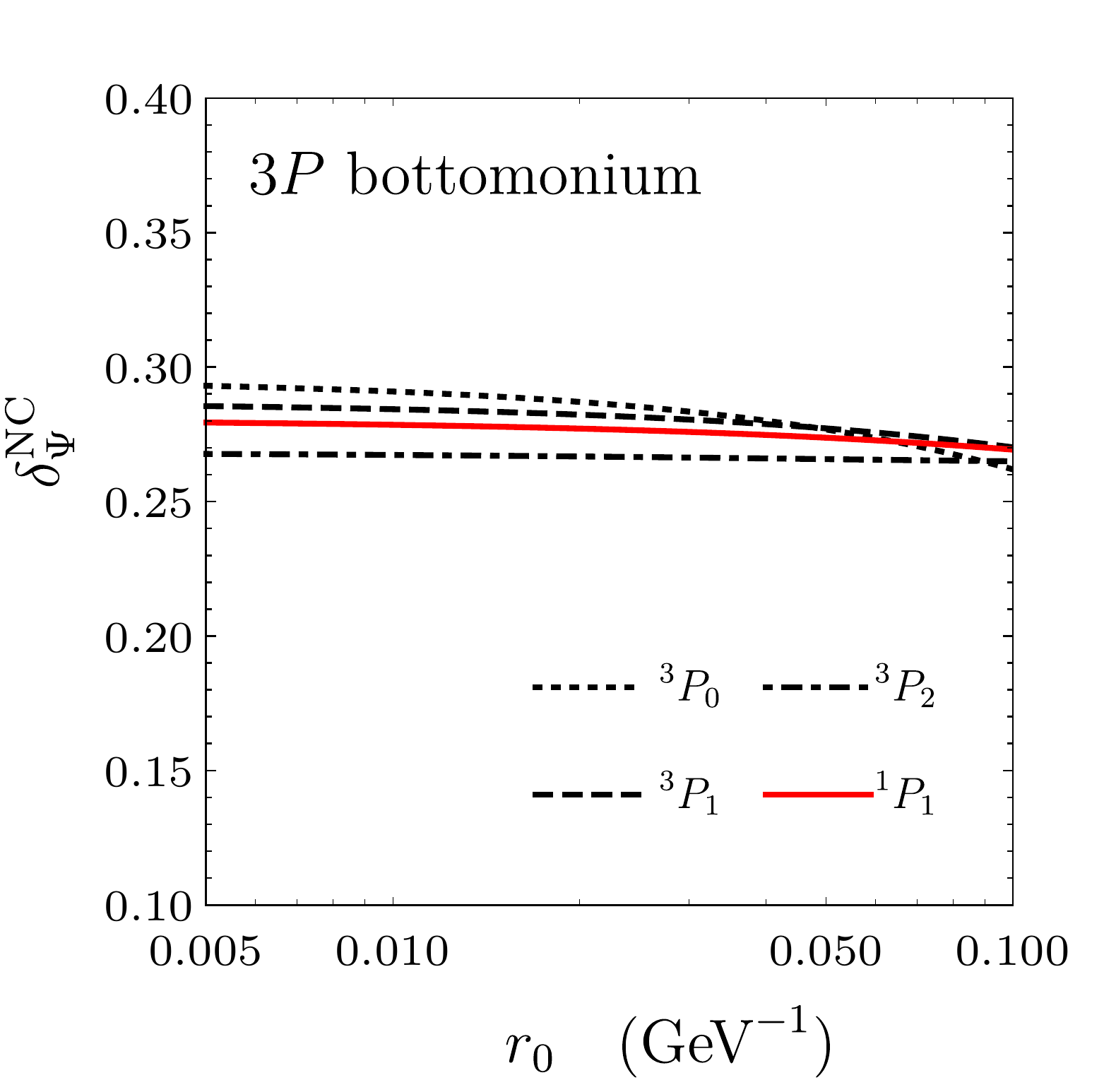}
\caption{\label{fig:bottommdelNC}
Non-Coulombic corrections $\delta_\Psi^{\rm NC}$ at finite $r_0$ for the
bottomonium $1P$ (left panel), $2P$ (middle panel), and 
$3P$ (right panel) states, for angular momentum quantum numbers
$^3P_0$ (dotted lines), $^3P_1$ (dashed lines), $^3P_2$ (dot-dashed lines),
and $^1P_1$ (red solid lines).
The $r_0$ dependences are mild for the range $r_0 < 0.02$~GeV$^{-1}$ that we
consider.
}
\end{figure}

In the case of bottomonia, the approximate 
relation in eq.~(\ref{eq:approx_relation}) is well reproduced for all $1P$,
$2P$, and $3P$ states by much better than 1\%.
The numerical results for $\delta_\Psi^{\rm NC}$ are almost insensitive to
$r_0$ for values of $r_0$ less than about $0.02$~GeV$^{-1}$.
We show the $r_0$ dependence of non-Coulombic corrections
for $1P$, $2P$, and $3P$ states in fig.~\ref{fig:bottommdelNC}.
We fix $r_0 =0.01$~GeV$^{-1}$ for computing the central values of $\delta_\Psi^{\rm NC}$
and neglect the uncertainty in the numerical calculation of
finite-$r$ regularized wavefunctions at the origin. 

We list the corrections to $|R'_{\rm LO} (0)|$ from the $1/m$ and 
$1/m^2$ potentials at $\Lambda = m$ in the $\overline{\rm MS}$ scheme 
($\delta_\Psi^{\rm NC}$), the Coulombic correction 
($\delta_\Psi^{\rm C}$), and the correction from the $\rm RS'$ subtraction term
($\delta_\Psi^{\rm RS'}$) in table~\ref{tab:bottomcorrections}.
The non-Coulombic corrections are smaller compared to the charmonium case, but
are nevertheless significant. Similarly to the charmonium case, the values of 
$\delta_\Psi^{\rm NC}$ that we obtain are similar in order of magnitude to what
we obtain from perturbative QCD calculations in
appendix~\ref{appendix:pertQCD} for similar values of $\alpha_s$ 
[$\alpha_s (\mu_R = 5{\rm ~GeV}) \approx 0.22$]. 
The Coulombic corrections are also sizable,
especially for the $1P$ state. The corrections from the $\rm RS'$ subtraction
term are small. All of the corrections are positive. 

Since the Coulombic corrections are large, especially for the $1P$ state, 
it is worth investigating its 
convergence at higher orders in the Rayleigh-Schr\"odinger perturbation theory.
Compared to the all-orders calculation of the Coulombic corrections using the
modified Crank-Nicolson method, our result for $\delta_\Psi^{\rm C}$ for the 
$1P$ state differs from the all-orders result only by $3\%$. 
For $2P$ and $3P$ states, the agreement is even better at about $1\%$. 
We conclude that the Coulombic corrections converge rapidly, 
and at the current level of accuracy, it is sufficient to consider only the 
$\delta_\Psi^{\rm C}$ from first order in the Rayleigh-Schr\"odinger
perturbation theory.

%%%%%%%%%%%%%%%%%%%%%%%%%%%%%%%%%%%%%%%%%%%%%%%%%%%%%%%%%%%%%%%%%%%%%%%%%%%%%%%
\begin{table}[tbp]
\centering
\begin{tabular}{|c|c|c|c|c|c|c|c|}
\hline
State & $|R'_{\rm LO} (0)|$~(GeV$^{5/2}$) & 
$\delta_\Psi^{\rm NC}|_{^3P_0}$ & $\delta_\Psi^{\rm NC}|_{^3P_1}$ &
$\delta_\Psi^{\rm NC}|_{^3P_2}$ & 
$\delta_\Psi^{\rm NC}|_{^1P_1}$ 
& $\delta_\Psi^{\rm C}$ & $\delta_\Psi^{\rm RS'}$ \\
\hline
$1P$ & 0.698 & 0.281 & 0.268 & 0.246 & 0.260 & 0.369 & 0.019
\\
\hline
$2P$ & 0.880 & 0.285 & 0.275 & 0.256 & 0.268 & 0.280 & 0.018
\\
\hline
$3P$ &  1.00 & 0.291 & 0.284 & 0.267 & 0.279 & 0.228 & 0.017
\\
\hline
\end{tabular}
\caption{\label{tab:bottomcorrections}
LO wavefunctions at the origin $|R'(0)|$ 
and relative corrections to the wavefunctions at the origin
in the $\overline{\rm MS}$ scheme at scale $\Lambda=m$ for $1P$, $2P$, and $3P$
bottomonium states.
$\delta_\Psi^{\rm NC}$ is the correction from the $1/m$ and $1/m^2$ potentials,
$\delta_\Psi^{\rm C}$ is the Coulombic correction, 
and $\delta_\Psi^{\rm RS'}$ is the correction from the $\rm RS'$
subtraction term.
The $\delta_\Psi^{\rm NC}$, $\delta_\Psi^{\rm C}$, 
and $\delta_\Psi^{\rm RS'}$ are dimensionless.
}
\end{table}
%%%%%%%%%%%%%%%%%%%%%%%%%%%%%%%%%%%%%%%%%%%%%%%%%%%%%%%%%%%%%%%%%%%%%%%%%%%%%%%

%%%%%%%%%%%%%%%%%%%%%%%%%%%%%%%%%%%%%%%%%%%%%%%%%%%%%%%%%%%%%%%%%%%%%%%%%%%%%%%
%%%%%%%%%%%%%%%%%%%%%%%%%%%%%%%%%%%%%%%%%%%%%%%%%%%%%%%%%%%%%%%%%%%%%%%%%%%%%%%

Also in the bottomonium case it is worth comparing with potential models that
are widely used for bottomonium phenomenology. At one-loop level, we have  
$|R_{\rm LO}'(0)|^2 (1+\delta_\Psi^{\rm C})^2 = 0.91$~GeV$^5$, 
1.27~GeV$^5$, and 1.51~GeV$^5$ for the $1P$, $2P$, and $3P$ states,
respectively. Potential model results from refs.~\cite{Buchmuller:1980su,
Eichten:1995ch, Chung:2010vz, Eichten:2019hbb} have values of 
$|R_{1P}'(0)|^2$ that range from $0.93$~GeV$^5$ to $2.07$~GeV$^5$, 
$|R_{2P}'(0)|^2$ range from $1.15$~GeV$^5$ to $2.44$~GeV$^5$, and 
$|R_{3P}'(0)|^2$ range from $1.30$~GeV$^5$ to $2.70$~GeV$^5$. 
The model dependence of the values of $|R'(0)|^2$ is as strong as the
charmonium case, with the largest values being more than twice the smallest 
values. Similarly to the charmonium case, the potential models have different
bottom quark masses, and so, it makes more sense to compare the combination 
$|R_{nP}'(0)|^2/m^3$ that appears in decay and exclusive electromagnetic
production rates. Our first-principles calculations at one-loop level give 
$|R_{\rm LO}'(0)|^2 (1+\delta_\Psi^{\rm C})^2/m_{\rm RS'}^3 = 
0.86\times 10^{-2}$~GeV$^2$, $1.19\times 10^{-2}$~GeV$^2$,
and $1.41\times 10^{-2}$~GeV$^2$ for the $1P$, $2P$, and $3P$ states,
respectively.
Potential-model calculations give values of $|R_{nP}'(0)|^2/m^3$ that are 
10--70\% larger than our results. 
In the case of inclusive production, our first-principles calculations at
one-loop level give
$|R_{\rm LO}'(0)|^2 (1+\delta_\Psi^{\rm C})^2/m_{\rm RS'}^5 = 
3.81 \times 10^{-4}$, $5.28 \times 10^{-4}$, 
and $6.29 \times 10^{-4}$ for the $1P$, $2P$, and $3P$ states,
respectively,
while potential-model calculations range about 85--145\% of our results, except
for the model in ref.~\cite{Chung:2010vz}, 
which gives results that are more than twice our
first-principles calculation. 
Considering the current level of precision of inclusive quarkonium
production phenomenology, the strong dependence of these values on 
potential models may not be catastrophic; nevertheless, 
it is difficult to obtain precise predictions from potential-models values 
of the wavefunctions at the origin. 

We first compute the two-photon decay rates of $\chi_{b0} (nP)$ and $\chi_{b2}
(nP)$. By using the formulas in eqs.~(\ref{eq:chi0gam_pnrqcd}) and
(\ref{eq:chi2gam_pnrqcd}), we obtain 
%---------------
\begin{subequations}
\label{eq:chibottomgamresults}
\begin{eqnarray}
%---------------
\Gamma(\chi_{b0}(1P) \to \gamma \gamma)
&=& 37.4 ^{+9.5}_{-1.5} \pm 3.7 \textrm{~eV}
= 37.4 ^{+10.2}_{-4.0} \textrm{~eV},
\\
\Gamma(\chi_{b2}(1P) \to \gamma \gamma)
&=& 7.5 ^{+0.5}_{-0.1} \pm 0.8 \textrm{~eV}
= 7.5 ^{+0.9}_{-0.8} \textrm{~eV},
%---------------
\end{eqnarray}
%---------------
\begin{eqnarray}
%---------------
\Gamma(\chi_{b0}(2P) \to \gamma \gamma)
&=& 51.3 ^{+10.7}_{-2.0} \pm 5.1 \textrm{~eV}
= 51.3^{+11.9}_{-5.5} \textrm{~eV},
\\
\Gamma(\chi_{b2}(2P) \to \gamma \gamma)
&=& 10.0 ^{+0.4}_{-0.11} \pm 1.0 \textrm{~eV}
= 10.0 ^{+1.1}_{-1.0} \textrm{~eV},
%---------------
\end{eqnarray}
%---------------
\begin{eqnarray}
%---------------
\Gamma(\chi_{b0}(3P) \to \gamma \gamma)
&=& 60.8 ^{+11.1}_{-2.4} \pm 6.1 \textrm{~eV}
= 60.8 ^{+12.6}_{-6.5} \textrm{~eV},
\\
\Gamma(\chi_{b2}(3P) \to \gamma \gamma)
&=& 11.7 ^{+0.2}_{-0.1} \pm 1.2 \textrm{~eV}
= 11.7 \pm 1.2 \textrm{~eV},
%---------------
\end{eqnarray}
\end{subequations}
%---------------
where the first uncertainties come form varying $\mu_R$ between 2~GeV and 
8~GeV, and the second uncertainties come from uncalculated corrections of order
$v^2$, which we estimate to be 10\% of the central values; 
this is based on the typical size of $v^2 \approx 0.1$ for bottomonium states. 
We add the uncertainties in quadrature. 
We note that these results are more precise than the model-dependent
calculations in ref.~\cite{Brambilla:2020xod}, 
because the results in ref.~\cite{Brambilla:2020xod} have large uncertainties
from dependence on potential models. 
In the case of bottomonia, we see better convergence of the corrections
compared to the charmonium case. In the $\chi_{b0} \to \gamma \gamma$ decay
rates, the order-$\alpha_s$ correction coming from the one-loop short-distance
coefficient and the Coulombic correction $\delta_\Psi^{\rm C}$ is about 80\% of
the leading-order value, 
and the higher order corrections from the two-loop short-distance coefficient,
non-Coulombic corrections to the wavefunctions at the origin, and the
corrections from the $\rm RS'$ subtraction term add up to about 0.5 times the
leading-order value. 
In the case of the $\chi_{b2} \to \gamma \gamma$ decay rates, the sizes of the
corrections are smaller; the corrections at one loop and higher orders are
both about 0.4 times the leading order values, respectively. 
Because the corrections have same signs at one-loop level and at higher orders,
they are numerically significant.  

We then compute the ratio ${\cal R}_{\chi_b(nP)} = \Gamma(\chi_{b2}(nP)\to
\gamma \gamma)/\Gamma(\chi_{b0}(nP) \to \gamma \gamma)$ 
by using eq.~(\ref{eq:chigamgamratio_pnrqcd}). We obtain 
%---------------
\begin{subequations}
\label{eq:chibottomgamratioresults}
\begin{eqnarray}
%---------------
{\cal R}_{\chi_b(1P)} &=& 0.126 ^{+0.013}_{-0.050} \pm 0.001 \pm 0.004
= 0.126 ^{+0.014}_{-0.051}
, \\
{\cal R}_{\chi_b(2P)} &=& 0.136 ^{+0.012}_{-0.044} \pm 0.002 \pm 0.004
= 0.136^{+0.013}_{-0.044}, \\
{\cal R}_{\chi_b(3P)} &=& 0.144 ^{+0.012}_{-0.039} \pm 0.002 \pm 0.004
= 0.144^{+0.013}_{-0.040}. 
%---------------
\end{eqnarray}
\end{subequations}
%---------------
Here, the first uncertainties come from variation of $\mu_R$ between 2~GeV and
8~GeV, and the second and third uncertainties come from uncalculated
corrections of order $\Lambda_{\rm QCD}^2/m^2$ and $v^3$, 
which we estimate to be $(500\textrm{~MeV}/m)^2$ and $0.05$ times the central
values, respectively. 
We include uncertainties of order $v^3$, because in the case of bottomonia, 
$v^3$ can be larger than $\Lambda_{\rm QCD}^2/m^2$. 
We add the uncertainties in quadrature. 
The correction at order $\alpha_s$ coming from the short-distance coefficients 
at one loop is about $-0.4$ times the leading-order value, while the higher
order corrections are about $-0.2$ times the leading-order value. 
The effect of the order-$v^2$ corrections is small, amounting to less than 10\%
of the leading-order value. 
We see that 
the uncertainties are dominated by the dependence on $\mu_R$, which is a sign
of poor convergence of the perturbative expansion. The fact that the
convergence of the perturbative corrections is poor can 
also be seen from the fact that the values of ${\cal R}_{\chi_b}$ computed from 
the results in eq.~(\ref{eq:chibottomgamresults}), which are about 
0.17 -- 0.20, disagree with the results in
eq.~(\ref{eq:chibottomgamratioresults}). That is, the numerical results for 
${\cal R}_{\chi_b}$ depend strongly on the way the corrections are organized. 
Hence, it is possible that the uncertainties in
eq.~(\ref{eq:chibottomgamratioresults}) computed from varying $\mu_R$
underestimates the effect of loop corrections of higher orders in $\alpha_s$.  

Next, we compute the scalar decay constants of $\chi_{b0}(nP)$ by using
eq.~(\ref{eq:decayconst_scalar_pnrqcd}). 
We obtain 
%---------------
\begin{subequations}
\label{eq:decayconst0_bottom_results}
\begin{eqnarray}
%---------------
m f_{\chi_{b0}(1P)}^{\rm OS} &=&
0.96^{+0.12}_{-0.02} \pm 0.10 \textrm{~GeV}^2
= 0.96^{+0.15}_{-0.10} \textrm{~GeV}^2,
\\
m f_{\chi_{b0}(2P)}^{\rm OS} &=&
1.12^{+0.11}_{-0.02} \pm 0.11 \textrm{~GeV}^2
= 1.12^{+0.15}_{-0.11} \textrm{~GeV}^2,
\\
m f_{\chi_{b0}(3P)}^{\rm OS} &=&
1.21 ^{+0.09}_{-0.02} \pm 0.12 \textrm{~GeV}^2
= 1.21^{+0.15}_{-0.12} \textrm{~GeV}^2.
%---------------
\end{eqnarray}
\end{subequations}
%---------------
Similarly, we obtain the axialvector decay constant for $\chi_{b1}(nP)$ by
using eq.~(\ref{eq:decayconst_axial_pnrqcd}). We obtain 
%---------------
\begin{subequations}
\label{eq:decayconst1_bottom_results}
\begin{eqnarray}
%---------------
f_{\chi_{b1}(1P)} &=&
0.156 ^{+0.014}_{-0.003} \pm 0.016 \textrm{~GeV}
= 0.156 ^{+0.021}_{-0.016} \textrm{~GeV},
\\
f_{\chi_{b1}(2P)} &=&
0.181 ^{+0.014}_{-0.003} \pm 0.018 \textrm{~GeV}
= 0.181 ^{+0.023}_{-0.018} \textrm{~GeV},
\\
f_{\chi_{b1}(3P)} &=&
0.198 ^{+0.013}_{-0.003} \pm 0.020 \textrm{~GeV}
= 0.198 ^{+0.024}_{-0.020} \textrm{~GeV}.
%---------------
\end{eqnarray}
\end{subequations}
%---------------
In both scalar and axialvector decay constants, the corrections at one loop are
about 30\% of the leading-order values, and the higher order corrections are
about 10--15\% of the leading-order values. 
Even though the corrections are milder compared to decay rates, the central
values of our results are still larger than tree-level results by
40--50\%. 
By using the bottom quark mass in the $\overline{\rm MS}$ scheme 
$\overline{m} = m_{\overline{\rm MS}} (\overline{m}) =
4.18^{+0.03}_{-0.02}$~GeV~\cite{Zyla:2020zbs}, 
we obtain the $\overline{\rm MS}$-renormalized
scalar decay constant at scale $\mu = \overline{m}$:
%---------------
\begin{subequations}
\label{eq:decayconst_bottom_results_msbar}
\begin{eqnarray}
%---------------
f_{\chi_{b0}(1P)}^{\overline{\rm MS}} (\mu = \overline{m})
= 0.229^{+0.036}_{-0.023}\textrm{~GeV},
\\
f_{\chi_{b0}(2P)}^{\overline{\rm MS}} (\mu = \overline{m})
= 0.267 ^{+0.037}_{-0.027}\textrm{~GeV},
\\
f_{\chi_{b0}(3P)}^{\overline{\rm MS}} (\mu = \overline{m})
= 0.290^{+0.037}_{-0.029}\textrm{~GeV}.
%---------------
\end{eqnarray}
\end{subequations}
%---------------

Similarly to the charmonium case, our numerical results have been computed by
using the $\rm RS'$ bottom quark mass at $\nu_f = 2$~GeV. Even though the
bottom quark is heavier than the charm quark, so that we expect weaker
dependence on the bottom quark mass in our calculation of decay rates and decay
constants, it is still worth investigating the $\nu_f$ dependence. We find that
our numerical results change by less than 3\% if we use $\nu_f = 1$~GeV instead
of $\nu_f = 2$~GeV. Hence, the heavy quark mass dependences in our results 
are under good control.

%------------------------------------------------------------------------------
\subsection{Numerical results from Pad\'e approximants}
%------------------------------------------------------------------------------

The numerical results for $P$-wave charmonia and bottomonia in the previous
sections show bad convergence of the loop corrections from short-distance
coefficients and corrections to the wavefunctions at the origin. 
This is in stark contrast with the previous work on $S$-wave quarkonia in
ref.~\cite{Chung:2020zqc},
where the poor convergence of the perturbative expansion of 
short-distance coefficients was significantly improved by inclusion of the
corrections to the $S$-wave wavefunctions at the origin. 
It is known that in many cases, the Pad\'e approximant of a function often
gives a better approximation than a truncated Taylor series. 
Pad\'e approximants have been successfully applied to studies of QCD
perturbation series, where the convergence is often only 
asymptotic~\cite{Samuel:1992xd, Samuel:1992qg, Ellis:1994qf, Ellis:1995jv,
Samuel:1995jc}. It would therefore be interesting to see how the use of 
Pad\'e approximants affects $P$-wave quarkonium decay and production rates. 
Common uses of Pad\'e approximants are to predict the size of unknown
corrections of higher orders, and also to estimate the size of a (possibly
divergent) perturbation series. We will primarily be interested in the latter;
however, the former can serve as a test of the applicability of Pad\'e
approximants by comparing known higher order corrections beyond 
next-to-next-to-leading order with what Pad\'e approximants predict without
knowledge of the correct result.

In general, it is not clear whether Pad\'e approximants are applicable to
short-distance coefficients of NRQCD factorization formulae, which 
depend explicitly on the NRQCD factorization scheme and scale. 
If Pad\'e approximants are used just for short-distance coefficients,
the exact order by order cancellations of NRQCD factorization scheme and 
scale dependences that occur in series expansions in $\alpha_s$ can break down,
which is unfavorable given the strong dependence on the factorization scale of
short-distance coefficients and LDMEs. 
Before we tackle the problem of $P$-wave decay and production rates, 
let us put Pad\'e
approximants to the test: we consider the vector decay constant $f_V$ of 
a $S$-wave vector quarkonium $V$, 
for which the short-distance coefficient is known to three-loop
accuracy~\cite{Barbieri:1975ki,Celmaster:1978yz,Czarnecki:1997vz,Beneke:1997jm,
Marquard:2014pea}:
%---------------
\begin{equation}
%---------------
f_V = \sqrt{\frac{2}{m_V}} c_v \langle 0 | \chi^\dag \bm{\epsilon} \cdot
\bm{\sigma} \psi | V \rangle + O(v^2), 
%---------------
\end{equation}
%---------------
where $\bm{\epsilon}$ is the polarization vector of the quarkonium $V$, and 
$c_v = 1 + c_v^{(1)} \alpha_s + c_v^{(2)} \alpha_s^2 + c_v^{(3)}
\alpha_s^3 + O(\alpha_s^4)$. Explicit expressions for 
$c_v^{(1)}$, $c_v^{(2)}$, and $c_v^{(3)}$ in the $\overline{\rm MS}$ scheme 
can be found in
refs.~\cite{Barbieri:1975ki,Celmaster:1978yz,Czarnecki:1997vz,Beneke:1997jm,
Marquard:2014pea}. 
The two-loop coefficient $c_v^{(2)}$ contains $\log \Lambda$, and 
the three-loop coefficient $c_v^{(3)}$ contains $\log^2 \Lambda$ and $\log
\Lambda$ terms, where $\Lambda$ is the scale at which the LDME 
$\langle 0 | \chi^\dag \bm{\epsilon} \cdot
\bm{\sigma} \psi | V \rangle$ is renormalized. 
The big question is whether Pad\'e approximants can correctly
predict $c_v^{(3)}$ only from knowledge of $c_v^{(1)}$ and $c_v^{(2)}$. 
From the expression for $c_v$ to two-loop accuracy, we obtain the Pad\'e
approximant of order $[1/1]$ given by 
%---------------
\begin{equation}
\label{eq:pade_wrong}
%---------------
\left[ 1 + c_v^{(1)} \alpha_s + c_v^{(2)} \alpha_s^2 \right]_{[1/1]}
= \frac{1+ (c_v^{(1)} -  c_v^{(2)} /c_v^{(1)} )\alpha_s}
{1- (c_v^{(2)} /c_v^{(1)}) \alpha_s}.
%---------------
\end{equation}
%---------------
Here, the order $[M/N]$ of a Pad\'e approximant denotes the orders 
$M$ and $N$ of the polynomials 
in $\alpha_s$ in the numerator and the denominator, respectively. 
The denominator implies a finite radius of convergence of the series expansion,
given by $| (c_v^{(2)} /c_v^{(1)}) \alpha_s| < 1$.
Expanding the right-hand side of eq.~(\ref{eq:pade_wrong}) to order
$\alpha_s^3$ gives the prediction from the Pad\'e approximant 
$c_v^{(3)} |_{\textrm{Pad\'e}} = (c_v^{(2)})^2/c_v^{(1)}$. 
By comparing this with the actual calculation of $c_v^{(3)}$ from
ref.~\cite{Marquard:2014pea}, we find that the prediction is completely 
wrong, giving $c_v^{(3)} |_{\textrm{Pad\'e}} = -24.0$ 
while the correct result is $c_v^{(3)} = -52.2$, at scales $\Lambda =
\mu_R = m$ and $n_f = 4$. 
The discrepancy increases for smaller values of $\Lambda$, as 
$c_v^{(3)} |_{\textrm{Pad\'e}}$ predicts the wrong sign for the 
$\log^2 \Lambda$ term; that is, the Pad\'e approximant 
incorrectly predicts the $\Lambda$ dependence of the
three-loop coefficient. Clearly, applying Pad\'e approximants directly to
the scheme and scale dependent short-distance coefficients will not do. 

There may yet be a chance that Pad\'e approximants could work if applied to 
correction factors that are scheme and scale independent. 
The calculation of the $S$-wave wavefunctions at the origin in 
ref.~\cite{Chung:2020zqc} makes construction of such correction factors
possible. 
In ref.~\cite{Chung:2020zqc}, the LDME 
$\langle 0 | \chi^\dag \bm{\epsilon} \cdot
\bm{\sigma} \psi | V \rangle$ was computed in the $\overline{\rm MS}$ scheme,
in the form 
%---------------
\begin{equation}
%---------------
\langle 0 | \chi^\dag \bm{\epsilon} \cdot
\bm{\sigma} \psi | V \rangle = \sqrt{2 N_c} |\Psi_V (0)| 
\left[1+O(v^2, \Lambda_{\rm QCD}^2/m^2) \right],
%---------------
\end{equation}
%---------------
where $\Psi_V (r)$ is the wavefunction for the quarkonium $V$, and 
%---------------
\begin{equation}
%---------------
\Psi_V (0) = \Psi_V^{\rm LO} (0) \left[ 1+ \delta_\Psi^{\rm C} 
+ \delta_\Psi^{\rm NC} + \delta_\Psi^{\rm RS'} 
+O(v^3, \Lambda_{\rm QCD}^2/m^2) \right], 
%---------------
\end{equation}
%---------------
where $\Psi_V^{\rm LO} (r)$ is a $S$-wave solution to the LO Schr\"odinger
equation [eq.~(\ref{eq:LO_schroedinger})], and 
$\delta_\Psi^{\rm C}$, $\delta_\Psi^{\rm NC}$, and $\delta_\Psi^{\rm RS'}$ 
are corrections at first order in the Rayleigh-Schr\"odinger perturbation
theory from the loop corrections to the static potential, the potentials of
higher orders in $1/m$, and the $\rm RS'$ subtraction term, respectively. 
The non-Coulombic correction $\delta_\Psi^{\rm NC}$ contains a 
UV divergence, which is renormalized in
the $\overline{\rm MS}$ scheme at scale $\Lambda$. 
While the Coulombic correction $\delta_\Psi^{\rm C}$ and the correction from
the $\rm RS'$ subtraction term $\delta_\Psi^{\rm RS'}$ cancel the 
$\mu_R$ and $\nu_f$ dependences of $\Psi_V^{\rm LO} (0)$, 
$\delta_\Psi^{\rm NC}$ cancels the $\Lambda$ dependence in $c_v^{(2)}$ to
two-loop accuracy; hence, if we write 
%---------------
\begin{eqnarray}
%---------------
f_V &=& \sqrt{\frac{2}{m_V}} 
\left[ |\Psi_V^{\rm LO} (0)| \left( 1+ \delta_\Psi^{\rm C} + \delta_\Psi^{\rm
RS'} \right)
\right] 
\nonumber \\ && \times 
\left[ 1 + c_v^{(1)} \alpha_s + c_v^{(2)} \alpha_s^2
+ \delta_\Psi^{\rm NC} +O(\alpha_s^3) \right] 
+ O(v^2) , 
%---------------
\end{eqnarray}
%---------------
every factor in the square brackets is separately invariant under variations of
$\mu_R$, $\Lambda$, and $\nu_f$. We neglect the three-loop coefficient
$c_v^{(3)}$ here, because the calculation of the non-Coulombic correction to
matching accuracy requires calculation of the second order corrections in the
Rayleigh-Schr\"odinger perturbation theory, which has only been done in 
perturbative QCD. 
In order to compare with the available order-$\alpha_s^3$ correction computed
in perturbative QCD in ref.~\cite{Beneke:2014qea}, 
we compute the Pad\'e approximant of the last correction factor as 
%---------------
\begin{equation}
%---------------
\left[ 1 + c_v^{(1)} \alpha_s + c_v^{(2)} \alpha_s^2
+ \delta_\Psi^{\rm NC} +O(\alpha_s^3) \right]_{[1/1]}
= \frac{1+ \left[ c_v^{(1)} - ( c_v^{(2)}+ \alpha_s^{-2} \delta_\Psi^{\rm NC})
/c_v^{(1)} \right] \alpha_s} 
{1- \left[(c_v^{(2)} + \alpha_s^{-2} \delta_\Psi^{\rm NC}) /c_v^{(1)}
\right] \alpha_s},
%---------------
\end{equation}
%---------------
from which we obtain the prediction $\alpha_s^3 ( c_v^{(2)}+ \alpha_s^{-2}
\delta_\Psi^{\rm NC})^2 / c_v^{(1)}$ for the scheme and scale invariant 
order-$\alpha_s^3$ term. Here, we used the fact that in
perturbative QCD, $\delta_\Psi^{\rm NC}$ is of order $\alpha_s^2$. 
This result can be compared with the perturbative QCD
calculation of the leptonic decay rate of $1S$ quarkonium in
ref.~\cite{Beneke:2014qea}. 
By computing $\delta_\Psi^{\rm NC}$ in perturbative QCD\footnote{We note that 
the analytical results for $\delta_\Psi^{\rm NC}$ in perturbative QCD 
can be obtained at order $\alpha_s^2$ indirectly from the calculations in
refs.~\cite{Czarnecki:1997vz,Beneke:1997jm, 
Hoang:1998xf,Melnikov:1998ug,Penin:1998kx,Hoang:1999zc,
Melnikov:1998pr,Yakovlev:1998ke,Beneke:1999qg,Nagano:1999nw,Penin:1998mx,
Penin:2004ay}. 
The numerical calculation in ref.~\cite{Chung:2020zqc} reproduces the
analytical results (see appendix~D of ref.~\cite{Chung:2020zqc}). 
}
at $\alpha_s(\mu_R = 3.5{\rm ~GeV}) = 0.2411$, $m = 4.911$~GeV, 
and $n_f = 4$, as were taken in ref.~\cite{Beneke:2014qea}, we obtain
$\delta_\Psi^{\rm NC} = 0.496$ for the $1S$ state at $\Lambda = m$, 
which leads to $\alpha_s^3 ( c_v^{(2)}+ \alpha_s^{-2}
\delta_\Psi^{\rm NC})^2 / c_v^{(1)} \approx -24.6 \alpha_s^3$. 
The correct result for the order-$\alpha_s^3$ 
correction term can be obtained from the calculation of the
leptonic decay rate of the $1S$ state in ref.~\cite{Beneke:2014qea}, 
by subtracting the Coulombic corrections to three-loop
accuracy~\cite{Beneke:2005hg}, and also subtracting the correction from 
the binding energy~\cite{Keung:1982jb,Luke:1997ys,Beneke:2005hg,Bodwin:2008vp}, 
taking the square root, and expanding in powers of $\alpha_s$, which gives 
$-21.5 \alpha_s^3$. The prediction from the Pad\'e approximant is in fair
agreement with the correct result. 

Following the example of the $S$-wave quarkonium vector decay constant,
we construct correction factors to $P$-wave quarkonium
decay and production rates that are independent on the renormalization,
factorization, and mass renormalon subtraction scales by using the corrections
to the wavefunctions computed in the previous section. 
We write the decay rate 
$\Gamma(\chi_{c0}\to \gamma \gamma)$ as 
%---------------
\begin{eqnarray}
\label{eq:chi0gam_pnrqcd_SI}
%---------------
\Gamma(\chi_{c0} \to \gamma \gamma) &=&
\frac{12 \pi e_c^4 \alpha^2}{m_{\chi_{c0}}}
\frac{3 N_c}{2 \pi}
\left[ |R'_{\rm LO} (0)| 
(1+ \delta_\Psi^{\rm C} + \delta_\Psi^{\rm RS'}) \right]^2 
\left[ \frac{1-3 \delta m_{\rm RS'}/m_{\rm RS'}}{m_{\rm RS'}^3} \right] 
\nonumber \\ &&
\times 
\bigg[
1 + 2 \alpha_s c_{\gamma \gamma(00)}^{(1)} + 
\left( \alpha_s c_{\gamma \gamma(00)}^{(1)} \right)^2
+ 2 \alpha_s^2 \, {\rm Re} ( c_{\gamma \gamma(00)}^{(2)} )
+ 2 \delta_\Psi^{\rm NC}|_{^3P_0}
\bigg]
\nonumber \\ && 
+ O(\alpha_s^3, v^2),
%---------------
\end{eqnarray}
%---------------
where we rearranged the corrections to the wavefunctions at the origin and the 
loop corrections to the short-distance coefficients so that each factor in the
square brackets is separately invariant under variations of $\mu_R$, 
$\Lambda$, and also $\nu_f$, up to corrections of order $\alpha_s^3$. 
In the first factor in the square brackets, 
the $\mu_R$ and $\nu_f$ dependence of 
$|R'_{\rm LO} (0)|$ is cancelled by $\delta_\Psi^{\rm C} + 
\delta_\Psi^{\rm RS'}$. 
In the second factor, the dependence of $m_{\rm RS'}$ on $\nu_f$ is cancelled
by the correction term $3 \delta m_{\rm RS'}/m_{\rm RS'}$. 
Finally, in the third factor, the $\mu_R$ dependence in 
$\alpha_s c_{\gamma \gamma(00)}^{(1)}$ is cancelled explicitly by the 
$\log \mu_R$ term in $c_{\gamma \gamma(00)}^{(2)}$, and the $\Lambda$
dependence cancels exactly between $\alpha_s^2 c_{\gamma \gamma(00)}^{(2)}$
and $\delta_\Psi^{\rm NC}|_{^3P_0}$ at two-loop level. 
Since the corrections in the first two factors are mild and converge rapidly,
we examine the Pad\'e approximant for the dimensionless correction factor 
that includes the loop corrections to the short-distance coefficients. 
If we consider the $\delta_\Psi^{\rm NC}|_{^3P_0}$ term to be of same order as
the order-$\alpha_s^2$ terms, which is necessary in establishing the exact
two-loop level cancellation of the $\Lambda$ dependence, we obtain the Pad\'e
approximant of order $[1/1]$ given by 
%---------------
\begin{eqnarray}
\label{eq:chi0gam_pade}
%---------------
&& \hspace{-5ex} 
\bigg[
1 + 2 \alpha_s c_{\gamma \gamma(00)}^{(1)} +
\left( \alpha_s c_{\gamma \gamma(00)}^{(1)} \right)^2
+ 2 \alpha_s^2 \, {\rm Re} ( c_{\gamma \gamma(00)}^{(2)} )
+ 2 \delta_\Psi^{\rm NC}|_{^3P_0}
\bigg]_{[1/1]}
\nonumber \\ 
&=& 
\frac{1+ (A-B/A) \alpha_s}{1- (B/A) \alpha_s},
%---------------
\end{eqnarray}
%---------------
where $A = 2 c_{\gamma \gamma(00)}^{(1)}$ and 
$B = ( c_{\gamma \gamma(00)}^{(1)} )^2 
+ 2 \, {\rm Re} ( c_{\gamma \gamma(00)}^{(2)} ) 
+ 2 \alpha_s^{-2} \delta_\Psi^{\rm NC}|_{^3P_0}$. 
Even though the Pad\'e approximant in eq.~(\ref{eq:chi0gam_pade}) has the same
series expansion as the third correction factor in
eq.~(\ref{eq:chi0gam_pnrqcd_SI}) to order $\alpha_s^2$, their numerical values
are very different; while the last correction factor in 
eq.~(\ref{eq:chi0gam_pnrqcd_SI}) is about 1.51, the numerical value of the 
Pad\'e approximant is about 1.00. 
If we replace the last correction factor in eq.~(\ref{eq:chi0gam_pnrqcd_SI}) by
its Pad\'e approximant, we obtain the following numerical result:
%---------------
\begin{equation}
\label{eq:chi0gampaderesult}
%---------------
\Gamma(\chi_{c0} \to \gamma \gamma) |_{\textrm{Pad\'e}} =
3.08 {}^{+0.32}_{-0.67} \pm 0.92 \textrm{~keV} =
3.08 {}^{+0.98}_{-1.14} \textrm{~keV}, 
%---------------
\end{equation}
%---------------
where the first uncertainty comes from varying $\mu_R$ between 1.5~GeV and
4~GeV, and the second uncertainty comes from uncalculated corrections of order
$v^2$. We add the uncertainties in quadrature. 
This result is in much better agreement with the BESIII measurement of the
decay rate $\Gamma(\chi_{c0} \to \gamma \gamma) = 2.33 \pm 0.20 \pm 0.22$~keV
from ref.~\cite{Ablikim:2012xi}
than the fixed-order calculation in the previous section. 

We can also perform a similar analysis for the decay rate $\Gamma(\chi_{c2} \to
\gamma \gamma)$, which can be written as 
%---------------
\begin{eqnarray}
\label{eq:chi2gam_pnrqcd_SI}
%---------------
\Gamma(\chi_{c2} \to \gamma \gamma) &=&
\frac{16 \pi e_c^4 \alpha^2}{5 m_{\chi_{c2}}}
\frac{3 N_c}{2 \pi}
\left[ |R'_{\rm LO} (0)| 
(1+ \delta_\Psi^{\rm C} + \delta_\Psi^{\rm RS'}) \right]^2 
\left[ \frac{1-3 \delta m_{\rm RS'}/m_{\rm RS'}}{m_{\rm RS'}^3} \right] 
\nonumber \\ && \times 
\bigg[
1 + 2 \alpha_s c_{\gamma \gamma(22)}^{(1)} + 
\left( \alpha_s c_{\gamma \gamma(22)}^{(1)} \right)^2
+ \left( \alpha_s c_{\gamma \gamma(20)}^{(1)} \right)^2
+ 2 \alpha_s^2 \, {\rm Re} ( c_{\gamma \gamma(22)}^{(2)} )
\nonumber \\ && \hspace{5ex} 
+ 2 \delta_\Psi^{\rm NC}|_{^3P_2}
\bigg]
+ O(\alpha_s^3, v^2),
%---------------
\end{eqnarray}
%---------------
where each factor in the square brackets is again invariant under variations of
$\mu_R$, $\Lambda$, and $\nu_f$ up to corrections of order $\alpha_s^3$. 
By replacing the last correction factor in eq.~(\ref{eq:chi2gam_pnrqcd_SI}) 
by its Pad\'e approximant, we obtain the numerical result 
%---------------
\begin{equation}
\label{eq:chi2gampaderesult}
%---------------
\Gamma(\chi_{c2} \to \gamma \gamma) |_{\textrm{Pad\'e}} =
0.61 {}^{+0.10}_{-0.17} \pm 0.18 \textrm{~keV} =
0.61 {}^{+0.21}_{-0.25} \textrm{~keV}, 
%---------------
\end{equation}
%---------------
where the uncertainties are as in eq.~(\ref{eq:chi0gampaderesult}). 
This result is again in better agreement than the fixed-order calculation 
with the BESIII measurement of the decay
rate $\Gamma(\chi_{c2} \to \gamma \gamma) = 0.63 \pm 0.04 \pm
0.06$~keV~\cite{Ablikim:2012xi}. 

By dividing eq.~(\ref{eq:chi2gampaderesult}) by eq.~(\ref{eq:chi0gam_pade}),
and multiplying a factor of $1+E^{\rm LO}/(3 m)$ coming from the order-$v^2$
correction~\cite{Brambilla:2017kgw, Brambilla:2020ojz}, we obtain 
${\cal R}_{\chi_c} 
= 0.24 \pm 0.04$, where the uncertainties come from variations of $\mu_R$ and
uncalculated corrections of order $\Lambda_{\rm QCD}^2/m^2$. 
This is in good agreement with the BESIII result 
${\cal R}_{\chi_c} = 0.27 \pm 0.04$~\cite{Ablikim:2012xi}. 

We repeat the analysis for the the $\chi_{cJ}+\gamma$ production rates in 
$e^+ e^-$ collisions at $\sqrt{s}=10.58$~GeV, for which we obtain 
%---------------
\begin{subequations}
\begin{eqnarray}
%---------------
\sigma(e^+ e^- \to \chi_{c0} + \gamma) |_{\textrm{Pad\'e}} &=&
1.29 ^{+0.25}_{-0.55} \pm 0.39\textrm{~fb}
= 1.29 ^{+0.46}_{-0.67}\textrm{~fb},
\\
\sigma(e^+ e^- \to \chi_{c1} + \gamma) |_{\textrm{Pad\'e}} &=&
14.3 ^{+1.7}_{-3.2} \pm 4.3 \textrm{~fb}
=14.3 ^{+4.6}_{-5.4} \textrm{~fb}, 
\\
\sigma(e^+ e^- \to \chi_{c2} + \gamma) |_{\textrm{Pad\'e}} &=&
2.80 ^{+1.02}_{-1.97} \pm 0.84 \textrm{~fb}
= 2.80 ^{+1.33}_{-2.14} \textrm{~fb}, 
%---------------
\end{eqnarray}
\end{subequations}
%---------------
where the uncertainties are as in eq.~(\ref{eq:chi0gampaderesult}).
While effects of the use of Pad\'e approximants are milder than the two-photon
decay rates, the central value for $\sigma(e^+ e^- \to \chi_{c1} + \gamma)$
obtained from the Pad\'e approximant is in better agreement with the Belle
measurement $\sigma(e^+ e^- \to \chi_{c1} + \gamma) = 17.3^{+4.2}_{-3.9} \pm
1.7$~fb~\cite{Jia:2018xsy}, and the numerical result for the $\chi_{c2} + \gamma$ production rate
is smaller than the upper limit $\sigma(e^+ e^- \to \chi_{c2} + \gamma) <
5.7$~fb from Belle~\cite{Jia:2018xsy}. 

The use of Pad\'e approximants give the following results for the decay
constants 
%---------------
\begin{subequations}
\label{eq:decayconst_results_pade}
\begin{eqnarray}
%---------------
m f_{\chi_{c0}}^{\rm OS}|_{\textrm{Pad\'e}} &=&
0.392^{+0.120}_{-0.118} \textrm{~GeV}^2,
\\
f_{\chi_{c1}} |_{\textrm{Pad\'e}} &=&
0.206\pm 0.062 \textrm{~GeV},
%---------------
\end{eqnarray}
\end{subequations}
%---------------
where the uncertainties are as in eq.~(\ref{eq:chi0gampaderesult}).
By dividing $m f_{\chi_{c0}}^{\rm OS}$ by $\overline{m} = 1.27\pm0.02$~GeV, 
we obtain 
$f_{\chi_{c0}}^{\overline{\rm MS}} (\mu = \overline{m}) |_{\textrm{Pad\'e}}
= 0.308^{+0.95}_{-0.93}$~GeV.
In the case of decay constants, the effects of the use of Pad\'e approximants 
are mild, and the numerical results are in agreement with the fixed-order
calculations within uncertainties. 

It is remarkable that by using Pad\'e approximants, we obtain values of 
decay and production rates of $P$-wave charmonia that can deviate 
considerably from the fixed-order calculations at two-loop level. 
That is, we find sizable differences between 
the correction factors computed by using Pad\'e approximants 
and values obtained from truncated series expansions in powers of $\alpha_s$. 
In these cases, the Pad\'e approximants have very small radii of
convergence when expanded as series in $\alpha_s$. 
On the other hand, in the calculations of 
$S$-wave heavy quarkonium decay rates and decay constants in 
ref.~\cite{Chung:2020zqc}, 
by combining the loop corrections to the short-distance coefficients with 
the corrections to the wavefunctions at the origin, we find that the NRQCD
factorization scheme and scale invariant corrections become rapidly convergent. 
As a result, Pad\'e approximants give values of $S$-wave heavy quarkonium decay
rates and decay constants that differ by at most about 5\% compared to the 
numerical results in ref.~\cite{Chung:2020zqc}. 
However, even in these cases, the Pad\'e approximants predict series in
$\alpha_s$ that do not converge for usual values of $\alpha_s$, 
although the situation is much less severe than the $P$-wave case. 

We note that, without the corrections to the wavefunctions at the origin
computed in this work, the Pad\'e approximants of the scheme and scale
dependent short-distance coefficients give numerical results that are very 
different from what we have obtained above, and sometimes lead to unphysical,
negative values of the rates. This is consistent with the poor behavior of the
Pad\'e approximant of the short-distance coefficient $c_v$ of the $S$-wave
quarkonium vector decay constant. 

In the case of $P$-wave bottomonia, the effects of the use of Pad\'e
approximants are small for decay rates and
decay constants, and we obtain numerical results that are consistent with 
the results in the previous section within uncertainties. 

While it is interesting that by using Pad\'e approximants, we obtain
two-photon decay rates of $\chi_{c0}$ and $\chi_{c2}$ and 
$\chi_{c1}+\gamma$ production cross sections that are in good agreement 
with the BESIII and Belle measurements, implications of these
agreements are limited because perturbative corrections are available only
up to next-to-next-to leading order accuracy in $\alpha_s$, and so, we can
only obtain Pad\'e approximants of the lowest possible order. 
For example, the value of ${\cal R}_{\chi_c}$ obtained by using the 
Pad\'e approximant of the expression in eq.~(\ref{eq:chigamgamratio_pnrqcd})
agrees with what we obtain by 
dividing eq.~(\ref{eq:chi2gampaderesult}) by eq.~(\ref{eq:chi0gam_pade}),
only when we include corrections of higher orders in $\alpha_s$ that 
the Pad\'e approximants predict.

%=============================================================================
\section{\boldmath Summary and discussion}
\label{sec:summary}
%==============================================================================

In this work, we obtained $P$-wave heavy quarkonium wavefunctions at the origin
in the $\overline{\rm MS}$ scheme based on the pNRQCD effective 
field theory formalism. The results allow computation of 
$\overline{\rm MS}$-renormalized NRQCD LDMEs for
$P$-wave charmonia and bottomonia, which are necessary in making 
scheme-independent predictions of decay and production rates at two-loop level. 
The definitions of the NRQCD LDMEs and wavefunctions in
$d$ spacetime dimensions, which are necessary for carrying out renormalization
in the $\overline{\rm MS}$ scheme, are given in sections~\ref{sec:LDMEs} and
\ref{sec:pnrqcd}. 
The wavefunctions are computed from a potential that is determined by lattice
QCD at long distances, while its short-distance behavior is given by
perturbative QCD. We include corrections to the wavefunctions at
subleading orders in $1/m$, which produce singularities at the origin. 
By generalizing the calculation of $S$-wave quarkonium
wavefunctions in ref.~\cite{Chung:2020zqc}, we compute the wavefunctions at the
origin in position space by regulating the singularity by using finite-$r$
regularization. The position-space expressions for the corrections to the
wavefunctions are given in section~\ref{sec:wavefunctions}. 
The finite-$r$ regularized wavefunctions at the origin are then converted to
the $\overline{\rm MS}$ scheme by computing the scheme conversion
in section~\ref{sec:conversion}. 
We use the results to make first-principles based, model-independent
predictions of electromagnetic decay rates and exclusive electromagnetic
production cross sections of $P$-wave charmonia and bottomonia in
section~\ref{sec:results}.

Because the $P$-wave wavefunctions at the origin that we obtain have the correct
dependences on the scale and scheme at which they are renormalized, we obtain
predictions of decay and production rates that are independent of the NRQCD
factorization scheme and scale, which has not been possible so far. 
Our first-principles based calculation also makes possible a proper 
treatment of the heavy quark pole mass, whose ambiguity can be removed by the
use of renormalon subtracted masses. We also resum the logarithms that appear
in the perturbative QCD corrections to the static potential, which
significantly improves the convergence, as shown in refs.~\cite{Kiyo:2010jm,
Chung:2020zqc}. These allow making much more refined predictions of physical
quantities compared to existing model-dependent methods. Unlike potential-model
based phenomenological studies in refs.~\cite{Sang:2015uxg, Brambilla:2020xod,
Sang:2020fql}, our first-principles based predictions of decay and production
rates are robust under variations of input parameters such as 
renormalization and  factorization scales, and heavy quark masses. 

The corrections to the wavefunctions at the origin computed
in this work allows us to investigate the corrections to the decay and
production rates of $P$-wave heavy quarkonia in a scheme-independent manner. 
The calculation of the $S$-wave quarkonium wavefunctions at the origin in
ref.~\cite{Chung:2020zqc} showed that the poor convergence of the 
scale dependent short-distance coefficients in the $\overline{\rm MS}$ scheme
is significantly improved by including the corrections to the wavefunctions at
the origin, which remove the scheme and scale dependences. 
In the $P$-wave case, however, we find that the convergence of the corrections
are still poor even after including the corrections to the $P$-wave
wavefunctions at the origin, especially for decay and production rates
of charmonia. In order to quantitatively assess the convergence of the
corrections, we examine the Pad\'e approximants of the scheme and
scale-invariant corrections. 
While in the case of $S$-wave heavy quarkonia, the Pad\'e
approximants of the corrections to the decay rates and decay constants in 
ref.~\cite{Chung:2020zqc} have values that are consistent with the truncated
series, we find significant differences between Pad\'e approximants and truncated
series of the scale and scheme independent corrections to $P$-wave charmonium 
decay and production rates. This suggests that the behaviors of 
perturbative corrections to decay and production rates differ between
$S$-wave and $P$-wave heavy quarkonia. 

Interestingly, we find that the Pad\'e approximants lead to values of
two-photon decay rates of $\chi_{c0}$ and $\chi_{c2}$ and exclusive
electromagnetic production rates of $\chi_{c1}$ that agree well with
measurements from BESIII~\cite{Ablikim:2012xi} and Belle~\cite{Jia:2018xsy}, 
while the truncated series give values that are in tension with
experimental values. 
We also note that inclusion of the corrections to the $P$-wave wavefunctions 
at the origin is important in computing the Pad\'e approximants. 
We find, from the available three-loop calculations of the vector decay
constant of $S$-wave heavy quarkonia, that Pad\'e approximants can give poor
descriptions of scheme and scale dependent short-distance coefficients. 
In the $P$-wave case, using Pad\'e approximants for the two-loop level 
short-distance coefficients in the $\overline{\rm MS}$ scheme can 
sometimes lead to unphysical, negative values of the rates, 
and generally give values that are very different from what we obtain from 
the scheme-independent calculations in this work. 
However, we note that this analysis is limited to Pad\'e approximants of the 
lowest possible orders, because the perturbative corrections to the 
short-distance coefficients are available only up to two-loop
accuracies. 

We note that except for the ratio of decay rates, the accuracies
of the phenomenological results involving $P$-wave heavy quarkonia 
in this work are limited to leading order in $v$. This is because the
current level of accuracy of the pNRQCD expressions of the lowest
dimensional NRQCD LDMEs 
is limited to first order in the expansion in powers of $1/m$,
even though the higher dimensional LDMEs and the corresponding short-distance
coefficients are already available~\cite{Brambilla:2017kgw, Brambilla:2020xod}. 
It is possible that the order-$v^2$ corrections are more important in $P$-wave
quarkonium decay and production rates than the $S$-wave case. 
In order to fully incorporate the effect of the order-$v^2$ corrections, the
order-$v^2$ corrections to the lowest dimensional LDMEs 
must be included, which requires calculation of the pNRQCD matching
coefficients $V_{\cal O}$ to second order in the expansion in powers of
$1/m$~\cite{Brambilla:2002nu, Brambilla:2020xod}.
We expect, however, the spin-dependent corrections in the matching coefficients
to be suppressed by $\Lambda_{\rm QCD}^2/m^2$, and the leading spin-dependent
corrections to the LDMEs to come from the corrections to the wavefunctions at
the origin computed in this work. 

The results presented in this work are relevant not only to precision studies
of $P$-wave charmonium and bottomonium phenomenology, but also to calculations
at tree and one-loop level. By comparing the first-principles calculations of
the $P$-wave quarkonium wavefunctions at the origin at one-loop level with
results from potential models, we have found that potential models can provide
reasonable descriptions of the lowest dimensional LDMEs of $P$-wave heavy
quarkonia; potential models can give central values of the LDMEs 
that are consistent with first-principles calculations, 
as long as the heavy quark mass is chosen consistently with the potential model
employed. Nevertheless, due to large uncertainties coming from model
dependences and the fact that the model calculations are incapable of
reproducing the correct scheme and scale dependences of the NRQCD LDMEs, 
accurate QCD-based determinations of the LDMEs 
are essential for making improved predictions of decay and
production rates, as we have done in this work.

%==============================================================================
\acknowledgments
%==============================================================================

The author expresses his gratitude to Nora Brambilla and Antonio Vairo 
for fruitful
discussions and their encouragement in completing this work. 
The author acknowledges the contribution of Saray Arteaga at the initial stages
of this work. 
This work is supported by Deutsche Forschungsgemeinschaft 
(DFG, German Research Foundation) cluster of excellence ``ORIGINS'' under
Germany's Excellence Strategy - EXC-2094 - 390783311.

%==============================================================================
\appendix
%==============================================================================

%==============================================================================
\section{Short-distance coefficients}
\label{appendix:sdcs}
%==============================================================================

In this appendix, we list the NRQCD factorization formulae and short-distance
coefficients for the decay rates, production cross sections, and decay
constants that we consider in section~\ref{sec:results}. 
The NRQCD factorization formula for the two-photon decay rate of $\chi_{Q0}$
($Q=c$ or $b$) is~\cite{Bodwin:1994jh} 
%---------------
\begin{equation}
\label{eq:chi0gamgam_nrqcd}
%---------------
\Gamma(\chi_{Q 0} \to \gamma \gamma) = 
\frac{12 \pi e_Q^4 \alpha^2}{m_{\chi_{Q0}}} 
| c_{\gamma \gamma (00)} |^2 
\frac{\langle \chi_{Q0} | {\cal O}(^3P_0) | \chi_{Q0} \rangle}{m^3} + O(v^2), 
%---------------
\end{equation}
%---------------
where $m_{\chi_{Q0}}$ is the mass of the $\chi_{Q 0}$, 
$e_Q$ is the fractional charge of the quark $Q$, and $\alpha$ is the QED
coupling constant. A factor of $m_{\chi_{Q0}}$ in the denominator 
comes from the two-body phase space. The short-distance
coefficient $c_{\gamma \gamma (00)}$ is given to two-loop accuracy 
in the $\overline{\rm MS}$ scheme by~\cite{Barbieri:1980yp,
Barbieri:1981xz, Bodwin:1994jh, Sang:2015uxg}
%---------------
\begin{equation}
%---------------
c_{\gamma \gamma (00)} 
= 1+ 
\alpha_s (\mu_R) c_{\gamma \gamma (00)} ^{(1)} 
+
\alpha_s^2(\mu_R) c_{\gamma \gamma (00)} ^{(2)} 
+ O(\alpha_s^3), 
%---------------
\end{equation}
%---------------
and 
%---------------
\begin{subequations}
\begin{eqnarray}
%---------------
c_{\gamma \gamma (00)} ^{(1)}
&=& 
\frac{C_F}{\pi} \left( \frac{\pi^2}{8} - \frac{7}{6} \right), 
\\
c_{\gamma \gamma (00)} ^{(2)}
&=& 
\frac{C_F}{\pi^2} \frac{\beta_0}{4} \left( \frac{\pi^2}{8} - \frac{7}{6}
\right) \log \frac{\mu_R^2}{m^2} 
- \gamma_{^3P_0}^{(2)} \log \frac{\Lambda}{m} 
+ c^{(2), {\rm fin}}_{\gamma \gamma (00)} , 
%---------------
\end{eqnarray}
\end{subequations}
%---------------
where $\beta_0 = \frac{11}{3} C_A - \frac{4}{3} T_F n_f$, $n_f$ is the number
of light quark flavors, $\alpha_s(\mu_R)$ is the strong coupling in the
$\overline{\rm MS}$ scheme at scale $\mu_R$ with $n_f$ flavors, and 
$c^{(2), {\rm fin}}_{\gamma \gamma (00)}$ is known numerically
as~\cite{Sang:2015uxg} 
%---------------
\begin{subequations}
\label{eq:sdcs_c0gamgam_fin}
\begin{eqnarray}
%---------------
c^{(2), {\rm fin}}_{\gamma \gamma (00)} |_{n_f =3} 
&=& -2.40611 + 0.07997 i, \\
c^{(2), {\rm fin}}_{\gamma \gamma (00)} |_{n_f =4} 
&=& -2.39666 + 0.11534 i. 
%---------------
\end{eqnarray}
\end{subequations}
%---------------
The numerical results in eq.~(\ref{eq:sdcs_c0gamgam_fin}) include the
contributions from the heavy quark loop. 

The NRQCD factorization formula for the two-photon decay rate of $\chi_{Q2}$
($Q=c$ or $b$) is~\cite{Bodwin:1994jh} 
%---------------
\begin{equation}
%---------------
\Gamma(\chi_{Q 2} \to \gamma \gamma) =
\frac{16 \pi e_Q^4 \alpha^2}{5 m_{\chi_{Q2}}}
\left( | c_{\gamma \gamma (22)} |^2 + | c_{\gamma \gamma (20)} |^2 \right)
\frac{\langle \chi_{Q2} | {\cal O}(^3P_2) | \chi_{Q2} \rangle}{m^3} + O(v^2),
%---------------
\end{equation}
%---------------
where $m_{\chi_{Q2}}$ is the mass of the $\chi_{Q 2}$. 
A factor of $m_{\chi_{Q0}}$ in the denominator 
comes from the two-body phase space. The short-distance
coefficients $c_{\gamma \gamma (22)}$ and $c_{\gamma \gamma (20)}$ correspond
to the two helicity amplitudes, which add at the squared amplitude level. 
They are given by~\cite{Barbieri:1980yp,
Barbieri:1981xz, Bodwin:1994jh, Sang:2015uxg}
%---------------
\begin{subequations}
\begin{eqnarray}
%---------------
c_{\gamma \gamma (22)}
&=& 1+ \alpha_s (\mu_R) c_{\gamma \gamma (22)} ^{(1)}
+ \alpha_s^2(\mu_R) c_{\gamma \gamma (22)} ^{(2)} + O(\alpha_s^3),
\\
c_{\gamma \gamma (20)}
&=& \alpha_s (\mu_R) c_{\gamma \gamma (20)} ^{(1)}
+ O(\alpha_s^2),
%---------------
\end{eqnarray}
\end{subequations}
%---------------
and
%---------------
\begin{subequations}
\begin{eqnarray}
%---------------
c_{\gamma \gamma (22)} ^{(1)}
&=& - \frac{2 C_F}{\pi}, 
\quad 
c_{\gamma \gamma (20)} ^{(1)}
= 
- \frac{C_F}{\sqrt{6} \pi} \left( \frac{3 \pi^2}{8} - 6 \log 2 + 1 \right), 
\\
c_{\gamma \gamma (22)} ^{(2)}
&=&
- \frac{2 C_F}{\pi^2} \frac{\beta_0}{4} 
\log \frac{\mu_R^2}{m^2}
- \gamma_{^3P_2}^{(2)} \log \frac{\Lambda}{m}
+ c^{(2), {\rm fin}}_{\gamma \gamma (22)} .
%---------------
\end{eqnarray}
\end{subequations}
%---------------
The finite part $c^{(2), {\rm fin}}_{\gamma \gamma (22)}$ is known numerically
as~\cite{Sang:2015uxg} 
%---------------
\begin{subequations}
\label{eq:sdcs_c2gamgam_fin}
\begin{eqnarray}
%---------------
c^{(2), {\rm fin}}_{\gamma \gamma (22)} |_{n_f =3}
&=& -3.11689 - 0.11467 i, \\
c^{(2), {\rm fin}}_{\gamma \gamma (22)} |_{n_f =4}
&=& -3.37915 - 0.83616 i,
%---------------
\end{eqnarray}
\end{subequations}
%---------------
which include the contributions from the heavy quark loop. 
While the order-$\alpha_s^2$ correction to $c_{\gamma \gamma (20)}$ has also
been computed in ref.~\cite{Sang:2015uxg}, this contributes to the decay rate
from order $\alpha_s^3$, and hence, is neglected in this work.  We also note
that to order-$\alpha_s^2$ accuracy, the numerical size of $c_{\gamma \gamma
(20)}$ is tiny compared to $c_{\gamma \gamma (22)}$~\cite{Sang:2015uxg}. 

The production cross sections $\sigma(e^+ e^- \to \chi_{QJ} + \gamma)$ for
$J=0$, 1, and 2 at the collision energy $\sqrt{s}$ are given at leading order
in $v$ by~\cite{Chung:2008km} 
%---------------
\begin{eqnarray}
%---------------
\sigma(e^+ e^- \to \chi_{QJ} + \gamma) 
&=& {\sigma}_{QJ}^{(0)} (s, m) 
\left[ 1 + \alpha_s (\mu_R) \hat{\sigma}_{QJ}^{(1)} (r) + 
\alpha_s^2(\mu_R) \hat{\sigma}_{QJ}^{(2)} (r) + O(\alpha_s^3) \right]
\nonumber \\ && \times 
\langle \chi_{QJ} | {\cal O}(^3P_J) | \chi_{QJ} \rangle + O(v^2),
%---------------
\end{eqnarray}
%---------------
where the tree-level short-distance coefficients are given
by~\cite{Chung:2008km} 
%---------------
\begin{subequations}
\label{eq:sdcs_cs_lo}
\begin{eqnarray}
%---------------
{\sigma}_{Q0}^{(0)} (s, m)
&=& \frac{(4 \pi)^3 \alpha^3 e_Q^4 (1-3 r)^2}
{18\pi m^3 s^2 (1-r)},
\\
{\sigma}_{Q1}^{(0)} (s, m)
&=& \frac{(4 \pi)^3 \alpha^3 e_Q^4 (1+r)}
{3\pi m^3 s^2 (1-r)},
\\
{\sigma}_{Q2}^{(0)} (s, m)
&=& \frac{(4 \pi)^3 \alpha^3 e_Q^4 (1+3r+6 r^2)}
{9\pi m^3 s^2 (1-r)},
%---------------
\end{eqnarray}
\end{subequations}
%---------------
where $r = 4 m^2/s$. 
The order-$\alpha_s$ corrections are known analytically as~\cite{Sang:2009jc}
%---------------
\begin{subequations}
\label{eq:sdcs_cs_nlo}
\begin{eqnarray}
%---------------
\hat{\sigma}_{Q0}^{(1)} (r) &=& \frac{C_0^0 (r)}{\pi}, 
\\
\hat{\sigma}_{Q1}^{(1)} (r) &=& \frac{C_1^0 (r) + r C_1^1 (r)}{\pi(1+r)} 
\\
\hat{\sigma}_{Q2}^{(1)} (r) &=& 
\frac{C_2^0 (r) + 3 r C_2^1 (r) + 6 r^2 C_2^2 (r)}{\pi (1+3 r+6 r^2)}, 
%---------------
\end{eqnarray}
\end{subequations}
%---------------
where explicit expressions for $C_i^j(r)$ are given in ref.~\cite{Sang:2009jc}. 
The order-$\alpha_s^2$ corrections are given by 
%---------------
\begin{equation}
%---------------
\hat{\sigma}_{QJ}^{(2)} (r) 
= \frac{\beta_0}{4 \pi} \hat{\sigma}_{QJ}^{(1)}(r) \log \frac{\mu_R^2}{4 m^2} 
-2 \gamma_{^3P_J}^{(2)} \log \frac{\Lambda}{m} 
+ 
\hat{\sigma}_{QJ}^{(2), {\rm fin}} (r), 
%---------------
\end{equation}
%---------------
where the $\hat{\sigma}_{QJ}^{(2), {\rm fin}} (r)$ have been 
computed numerically 
in ref.~\cite{Sang:2020fql} for $Q=c$ at $\sqrt{s} = 10.58$~GeV and $n_f = 3$ 
for two different choices of charm quark masses $m = 1.4$~GeV 
and $m = 1.68$~GeV: 
%---------------
\begin{subequations}
\label{eq:sdcs_cs_nnlo}
\begin{eqnarray}
%---------------
\hat{\sigma}_{c0}^{(2), {\rm fin}} (r) |_{m=1.4{\rm ~GeV}} &=& -0.799,
\\
\hat{\sigma}_{c1}^{(2), {\rm fin}} (r) |_{m=1.4{\rm ~GeV}} &=& -2.358,
\\
\hat{\sigma}_{c2}^{(2), {\rm fin}} (r) |_{m=1.4{\rm ~GeV}} &=& -4.285,
\\
\hat{\sigma}_{c0}^{(2), {\rm fin}} (r) |_{m=1.68{\rm ~GeV}} &=& -0.857, 
\\
\hat{\sigma}_{c1}^{(2), {\rm fin}} (r) |_{m=1.68{\rm ~GeV}} &=& -2.487,
\\
\hat{\sigma}_{c2}^{(2), {\rm fin}} (r) |_{m=1.68{\rm ~GeV}} &=& -5.047,
%---------------
\end{eqnarray}
\end{subequations}
%---------------
where the contribution from the heavy quark loop is included. 
We compute numerical values of $\hat{\sigma}_{cJ}^{(2), {\rm fin}} (r)$ 
for other values of
$m$ by assuming that $\hat{\sigma}_{cJ}^{(2), {\rm fin}} (r)$ 
changes linearly under small
variations of $m$. Unfortunately, the numerical results given in 
ref.~\cite{Sang:2020fql} does not generalize to 
$\hat{\sigma}_{QJ}^{(2), {\rm fin}} (r)$ at different values of $\sqrt{s}$ 
or for values of $m$ that deviate too much from the charm quark mass. 

The decay constants $f_{\chi_{Q0}}$ and $f_{\chi_{Q1}}$ are given in NRQCD
by~\cite{Kniehl:2006qw}
%---------------
\begin{subequations}
\label{eq:decayconsts_nrqcd}
\begin{eqnarray}
%---------------
f_{\chi_{Q0}}^{\rm OS} &=& \frac{2}{\sqrt{m_{\chi_{Q0}}}} 
\left( 1 + \alpha_s(\mu_R) c_s^{(1)} + \alpha_s^2(\mu_R) c_s^{(2)} 
+ O(\alpha_s^3) \right) 
\frac{\langle \chi_{Q0} | {\cal O} (^3P_0) | \chi_{Q0} \rangle}{m}, 
\\
f_{\chi_{Q2}} &=& \frac{\sqrt{6}}{\sqrt{m_{\chi_{Q1}}}} 
\left( 1 + \alpha_s(\mu_R) c_a^{(1)} + \alpha_s^2(\mu_R) c_a^{(2)} 
+ O(\alpha_s^3) \right) 
\frac{\langle \chi_{Q1} | {\cal O} (^3P_1) | \chi_{Q1} \rangle}{m}, 
%---------------
\end{eqnarray}
\end{subequations}
%---------------
where the denominator factors $\sqrt{m_{\chi_{QJ}}}$ come from the fact that
the decay constants are defined in QCD with relativistically normalized states,
while the NRQCD LDMEs are defined with nonrelativistic normalization. 
The superscript OS on $f_{\chi_{Q0}}$ implies that the scalar decay constant
has been renormalized in the on-shell scheme. 
The short-distance coefficients at order $\alpha_s$ and $\alpha_s^2$ are given
by~\cite{Kniehl:2006qw}
%---------------
\begin{subequations}
\label{eq:decayconsts_sdcs}
\begin{eqnarray}
%---------------
c_s^{(1)} &=& \frac{C_F}{2 \pi}, 
\\ 
c_s^{(2)} &=& \frac{C_F}{2 \pi^2} \frac{\beta_0}{4} \log \frac{\mu_R^2}{m^2}
+ \frac{C_F^2}{\pi^2} \left( \frac{5}{16} - \frac{37}{8} \zeta_2 
+ 3 \zeta_2 \log 2 - \frac{11}{4} \zeta_3 - 2 \zeta_2 \log
\frac{\Lambda^2}{m^2} \right) 
\nonumber \\ && 
+ \frac{C_F C_A}{\pi^2} \left( \frac{49}{144} + \frac{1}{8} \zeta_2 - 3 \zeta_2
\log 2- \frac{5}{4} \zeta_3 - \frac{1}{2} \zeta_2 \log \frac{\Lambda^2}{m^2}
\right) 
\nonumber \\ && 
+ \frac{C_F T_F}{\pi^2} \left( -\frac{5}{36} n_f + \frac{121}{36} -2 \zeta_2
+ X_{\rm sing}^{(s)} \right), 
\\
c_a^{(1)} &=& - \frac{C_F}{\pi},
\\
c_a^{(2)} &=& - \frac{C_F}{\pi^2} \frac{\beta_0}{4} \log \frac{\mu_R^2}{m^2}
+ \frac{C_F^2}{\pi^2} \left( \frac{23}{24} - \frac{27}{4} \zeta_2 +
\frac{19}{4} \zeta_2 \log 2 - \frac{27}{16} \zeta_3 - \frac{5}{4} \zeta_2 
\log \frac{\Lambda^2}{m^2} \right)
\nonumber \\ &&
+ \frac{C_F C_A}{\pi^2} \left( - \frac{101}{72} + \frac{35}{24} \zeta_2 -
\frac{7}{2} \zeta_2 \log 2 - \frac{9}{8} \zeta_3 - \frac{1}{2} \zeta_2 
\log \frac{\Lambda^2}{m^2} \right)
\nonumber \\ &&
+ \frac{C_F T_F}{\pi^2} \left( \frac{7}{18} n_f + \frac{20}{9} - \frac{4}{3}
\zeta_2 + X_{\rm sing}^{(a)} \right),
%---------------
\end{eqnarray}
\end{subequations}
%---------------
where $\zeta_s = \sum_{n=1}^\infty \frac{1}{n^s}$,
and 
%---------------
\begin{subequations}
\label{eq:decayconsts_sdcs_sing}
\begin{eqnarray}
%---------------
X_{\rm sing}^{(s)} &=& 
\frac{2}{3} - \frac{29}{12} \zeta_2 + 4 \zeta_2 \log 2-\log 2+ \frac{\pi i}{2},
\\
X_{\rm sing}^{(a)} &=& 
- \frac{23}{12} \zeta_2 + 4 \zeta_2 \log 2 -2 \log 2 + \frac{2}{3} \log^2 2 + i
\pi \left( 1 - \frac{2}{3} \log 2 \right). 
%---------------
\end{eqnarray}
\end{subequations}
%---------------
We note that the results for $c_a^{(1)}$ and $c_a^{(2)}$ in 
eqs.~(\ref{eq:decayconsts_sdcs}) are computed in 
na\"ive dimensional regularization, where $\gamma_5$ commutes with every
$\gamma$ matrix, except in $X^{(a)}_{\rm sing}$, where the t'Hooft-Veltman
scheme was used~\cite{Kniehl:2006qw}.

%==============================================================================
\section{Potentials in perturbative QCD} 
\label{appendix:potentials}
%==============================================================================

In this appendix, we list the short-distance behaviors of the potentials, which
are obtained from perturbative QCD. 
In perturbative QCD, the static potential is given 
at leading order in $\alpha_s$ by 
%---------------
\begin{equation}
\label{eq:staticpotential_pert}
%---------------
V^{(0)} (r) \big|_{\rm pert}
= - \frac{\alpha_s (\mu) C_F}{r} + O(\alpha_s^2). 
%---------------
\end{equation}
%---------------
where $\alpha_s = \alpha_s(\mu)$ is the $\overline{\rm MS}$-renormalized QCD
coupling constant at scale $\mu$. Corrections of relative order 
$\alpha_s$ and $\alpha_s^2$ have been computed 
in~\cite{Fischler:1977yf, Schroder:1998vy}, 
and corrections of relative order $\alpha_s^3$ are given in 
refs.~\cite{Brambilla:1999qa, Kniehl:1999ud, Smirnov:2008pn, Anzai:2009tm,
Smirnov:2009fh, Pineda:2011dg}. In this paper, we include corrections to the
static potential up to relative order $\alpha_s^2$.  

The forms of the $1/m$ and $1/m^2$ potentials generally depend on the 
matching scheme in which the potentials are determined. 
In on-shell matching, where we match on-shell $S$-matrix elements in
NRQCD and pNRQCD in momentum space, we obtain~\cite{Gupta:1982kp, 
Pantaleone:1987qh, Titard:1993nn, Manohar:2000hj, 
Kniehl:2001ju, Kniehl:2002br,Beneke:1999qg,Beneke:2013jia}
%---------------
\begin{subequations}
\label{eq:potentials_pert}
\begin{eqnarray}
%---------------
V^{(1)} (r) \big|_{\rm pert}^{\rm OS}
&=& \frac{\alpha_s^2 C_F (\tfrac{1}{2} C_F - C_A)}{2 r^2}
 + O(\alpha_s^3), 
\\
V_r^{(2)} (r) \big|_{\rm pert}^{\rm OS}
&=& 0 + O(\alpha_s^2), \\
\label{eq:velpotential_pert}
V_{p^2}^{(2)} (r) \big|_{\rm pert}^{\rm OS}
&=& - \frac{\alpha_s C_F}{r} + O(\alpha_s^2), \\
V_{L^2}^{(2)} (r) \big|_{\rm pert}^{\rm OS}
&=& 0 + O(\alpha_s^2), \\
\label{eq:spinpotential_pert}
V_{S^2}^{(2)} (r) \big|_{\rm pert}^{\rm OS}
&=& \frac{4 \pi \alpha_s C_F}{3} \delta^{(3)} (\bm{r}) 
+ O(\alpha_s^2), 
\\
V_{S_{12}}^{(2)} (r) \big|_{\rm pert}^{\rm OS}
&=& \frac{\alpha_s C_F}{4 r^3} 
+ O(\alpha_s^2), 
\\
V_{so}^{(2)} (r) \big|_{\rm pert}^{\rm OS}
&=& \frac{3 \alpha_s C_F}{2 r^3} 
+ O(\alpha_s^2). 
%---------------
\end{eqnarray}
\end{subequations}
%---------------
We use the superscript OS to denote the on-shell matching scheme. 

In Wilson-loop matching, the potentials are given in terms of the rectangular
Wilson loop $W_{r \times T}$ with spatial size $r$ and time extension $T$, with
insertions of gluon fields~\cite{Brambilla:2000gk, Pineda:2000sz}. 
The short-distance behavior of the potentials in Wilson loop matching can be
obtained by computing the nonperturbative definitions in perturbative 
QCD~\cite{Peset:2015vvi}. 
We list the results at leading nonvanishing orders in $\alpha_s$: 
%---------------
\begin{subequations}
\label{eq:potentials_wilson_short}
\begin{eqnarray}
%---------------
V^{(1)} (r) \big|_{\rm pert}^{\rm WL} &=& 
- \frac{\alpha_s^2 C_F C_A}{2 r^2} + O(\alpha_s^3), 
\\
V^{(2)}_{r} (r) \big|_{\rm pert}^{\rm WL} &=& 
\pi \alpha_s C_F \delta^{(3)} (\bm{r}) + O(\alpha_s^2), 
\\
V_{p^2}^{(2)} (r) \big|_{\rm pert}^{\rm WL} &=& 
- \frac{\alpha_s C_F}{r} + O(\alpha_s^2), \\
V_{L^2}^{(2)} (r) \big|_{\rm pert}^{\rm WL}
&=& \frac{\alpha_s C_F}{2 r^3} + O(\alpha_s^2), \\
V_{S^2}^{(2)} (r) \big|_{\rm pert}^{\rm WL} &=& 
\frac{4 \pi \alpha_s C_F}{3} \delta^{(3)} (\bm{r}) 
+ O(\alpha_s^2), 
\\
V_{S_{12}}^{(2)} (r) \big|_{\rm pert}^{\rm WL}
&=& \frac{\alpha_s C_F}{4 r^3} 
+ O(\alpha_s^2), 
\\
V_{so}^{(2)} (r) \big|_{\rm pert}^{\rm WL}
&=& \frac{3 \alpha_s C_F}{2 r^3} 
+ O(\alpha_s^2). 
%---------------
\end{eqnarray}
\end{subequations}
%---------------
The superscript WL denotes that the potential is obtained in Wilson loop
matching.

The potentials from on-shell matching in eq.~(\ref{eq:potentials_pert}) and
the potentials from Wilson loop matching in
eq.~(\ref{eq:potentials_wilson_short}) are related by unitary transformations.
The effect of the unitary transformations on $P$-wave wavefunctions are 
described in section~\ref{sec:unitary_transformations}. 

The momentum-space potentials in the on-shell matching scheme can be 
computed in $d$ dimensions, which are suitable for calculations in DR. 
For calculations in this work, 
we only need the color singlet projection of the potential. 
If we include the $1/m$ and $1/m^2$ potentials at leading nonvanishing orders
in $\alpha_s$, we obtain~\cite{Beneke:1999qg, Beneke:2013jia} 
%---------------
\begin{eqnarray}
\label{eq:pot_pert_DR}
%---------------
V(\bm{p}',\bm{p}) &=& - \frac{4 \pi \alpha_s C_F}{\bm{q}^2} 
+ \delta \tilde{V}_C(\bm{q}^2) 
+ 
\frac{\pi^2 \alpha_s^2 C_F }{m |\bm{q}|} 
\left( \frac{\Lambda^2}{\bm{q}^2}
\right)^\epsilon \left( \frac{C_F}{2} (1-2 \epsilon) - C_A (1-\epsilon) \right)
c_\epsilon 
\nonumber \\ && 
- \frac{2 \pi \alpha_s C_F (\bm{p}^2+\bm{p}'^2)}{m^2 \bm{q}^2} 
+ \frac{2 \pi \alpha_s C_F}{m^2} 
+ \frac{\pi \alpha_s C_F}{4 m^2 \bm{q}^2} 
[\sigma_i, \sigma_j] q_j \otimes [\sigma_i, \sigma_k] q_k
\nonumber \\ && 
- \frac{3 \pi \alpha_s C_F}{2 m^2 \bm{q}^2} 
\big( [\sigma_i, \sigma_j] q_i p_j \otimes 1 - 1 \otimes [\sigma_i, \sigma_j]
q_i p_j \big)
- (2 \pi)^{d-1} \delta^{(d-1)} (\bm{q}) \frac{\bm{p}^4}{4 m^3}, 
%---------------
\end{eqnarray}
%---------------
where $\bm{q} = \bm{p}'-\bm{p}$ and 
%---------------
\begin{equation}
\label{eq:cep}
%---------------
c_\epsilon = \frac{e^{\gamma_{\rm E} \epsilon} \Gamma(\frac{1}{2}-\epsilon)^2
\Gamma(\frac{1}{2}+\epsilon)}{\pi^{3/2} \Gamma(1-2 \epsilon)}. 
%---------------
\end{equation}
%---------------
The $\Lambda$ in eq.~(\ref{eq:pot_pert_DR}) comes from associating a factor of 
$\left( \Lambda^2 \frac{e^{\gamma_{\rm E}}}{4 \pi} \right)^\epsilon$ with each
loop integral. 
The last term in eq.~(\ref{eq:pot_pert_DR}) comes from the relativistic
correction to the kinetic energy. 
The term $\delta \tilde{V}_C(\bm{q}^2)$ encodes the loop corrections to the
static potential, whose explicit expressions are not needed in this paper. 
We identify the spin-dependent terms proportional to 
$[\sigma_i, \sigma_j] q_j \otimes [\sigma_i, \sigma_k] q_k$ and 
$[\sigma_i, \sigma_j] q_i p_j \otimes 1 - 1 \otimes [\sigma_i, \sigma_j]
q_i p_j $ as the hyperfine and spin-orbit terms, respectively. 
It can be shown that eq.~(\ref{eq:pot_pert_DR}) reproduces the position-space
expressions in eqs.~(\ref{eq:potentials_pert}) in the on-shell scheme (the
$\bm{p}$ and $\bm{p}'$-independent term and the isotropic part of the 
hyperfine term combine
to give the $\bm{S}^2$ term in the position-space expression). 

Now we obtain the $d$-dimensional expression for $\delta \tilde {\cal V}
(\bm{p}',\bm{p})$ from eq.~(\ref{eq:pot_pert_DR}) by reducing the $\bm{p}^2$ and
$\bm{p}'^2$ by using the $d$-dimensional Lippmann-Schwinger equation and the
Schr\"odinger equation. We obtain 
%---------------
\begin{eqnarray}
\label{eq:potvar_pert_DR}
%---------------
\delta \tilde {\cal V} (\bm{p}',\bm{p}) &=& 
\frac{\pi^2 \alpha_s^2 C_F }{m |\bm{q}|}
\left( \frac{\Lambda^2}{\bm{q}^2}
\right)^\epsilon \left( \frac{C_F}{2} (1-2 \epsilon) - C_A (1-\epsilon) \right)
c_\epsilon
\nonumber \\ &&
+ \frac{1}{m} \int_{\bm{k}} \frac{4 \pi \alpha_s C_F}{\bm{k}^2} 
\tilde V_{\rm LO} (\bm{k}-\bm{q})
- \frac{1}{4 m} \int_{\bm{k}} \tilde V_{\rm LO} (\bm{k}) 
\tilde V_{\rm LO} (\bm{k}-\bm{q})
\nonumber \\ &&
+ \frac{2 \pi \alpha_s C_F}{m^2}
+ \frac{\pi \alpha_s C_F}{4 m^2 \bm{q}^2}
[\sigma_i, \sigma_j] q_j \otimes [\sigma_i, \sigma_k] q_k
\nonumber \\ &&
- \frac{3 \pi \alpha_s C_F}{2 m^2 \bm{q}^2}
\big( [\sigma_i, \sigma_j] q_i p_j \otimes 1 - 1 \otimes [\sigma_i, \sigma_j]
q_i p_j \big), 
%---------------
\end{eqnarray}
%---------------
where $\tilde V_{\rm LO} (\bm{q}) = - 4 \pi \alpha_s C_F/\bm{q}^2$ in
perturbative QCD.

%==============================================================================
\section{Matrix elements of the spin-spin potential}
\label{appendix:s12}
%==============================================================================

In this section we discuss the angular matrix elements of 
the spin-spin potential on $P$-wave states. 
While the angular matrix elements of $S_{12}$ are generally known, 
we focus on its action on $P$-wave
LO wavefunctions $\Psi_n^{\rm LO} (\bm{r})$ in the Cartesian basis, which is
also useful in analytical calculations of the scheme conversion coefficient. 
In the Cartesian basis, the dependencies of $P$-wave 
wavefunctions on the angles of $\bm{r}$ and $Q \bar Q$ spin can be written as 
$1_{2 \times 2} \hat{\bm{r}}$, 
$\hat{\bm{r}} \cdot \bm{\sigma}$, 
$\hat{\bm{r}}^{[i} {\sigma}^{j]}$, 
and 
$\hat{\bm{r}}^{(i} {\sigma}^{j)}$ 
for the $^1P_1$, $^3P_0$, $^3P_1$, and $^3P_2$ states, respectively. 
Then, the application of $S_{12}$ on a $^1P_1$ state can be computed as 
%-------------
\begin{equation}
%-------------
S_{12} \hat{\bm{r}} 
= - (3 \hat{\bm{r}} \cdot \bm{\sigma} \hat{\bm{r}} \cdot \bm{\sigma}
- \bm{\sigma} \cdot \bm{\sigma}) \hat{\bm{r}} = 0. 
%-------------
\end{equation}
%-------------
For $^3P_0$ and $^3P_1$ states, we have 
%-------------
\begin{equation}
%-------------
S_{12} \hat{\bm{r}} \cdot \bm{\sigma}
= -3 \hat{\bm{r}} \cdot \bm{\sigma} (\hat{\bm{r}} \cdot \bm{\sigma} )
\hat{\bm{r}} \cdot \bm{\sigma}
+ {\sigma}^i (\hat{\bm{r}} \cdot \bm{\sigma}) {\sigma}^i
= - 4 \hat{\bm{r}} \cdot \bm{\sigma}, 
%-------------
\end{equation}
%-------------
%-------------
\begin{equation}
%-------------
S_{12} \hat{\bm{r}}^{[i} {\sigma}^{j]}
= - 3 \hat{\bm{r}} \cdot \bm{\sigma} ( \hat{\bm{r}}^{[i} {\sigma}^{j]} )
\hat{\bm{r}} \cdot \bm{\sigma}
+ {\sigma}^k ( \hat{\bm{r}}^{[i} {\sigma}^{j]} ) {\sigma}^k
= 2 \hat{\bm{r}}^{[i} {\sigma}^{j]}. 
%-------------
\end{equation}
%-------------
Finally, for a $^3P_2$ state, we have 
%-------------
\begin{eqnarray}
%-------------
S_{12} \hat{\bm{r}}^{(i} {\sigma}^{j)}
&=& - 3 \hat{\bm{r}} \cdot \bm{\sigma} ( \hat{\bm{r}}^{(i} {\sigma}^{j)} )
\hat{\bm{r}} \cdot \bm{\sigma}
+ {\sigma}^k ( \hat{\bm{r}}^{(i} {\sigma}^{j)} ) {\sigma}^k
\nonumber \\
&=& 
- \frac{2}{5} \hat{\bm{r}}^{(i} \bm{\sigma}^{j)} 
+ \frac{6}{5} f^{ij} (\hat{\bm{r}}, \bm{\sigma}), 
%-------------
\end{eqnarray}
%-------------
where 
%---------------
\begin{equation}
\label{eq:Fwavetensor}
%---------------
f^{ij} (\hat{\bm{r}}, \bm{\sigma})
= (\delta^{ij} - 5 \hat{\bm{r}}^i \hat{\bm{r}}^j )
\bm{\sigma} \cdot \hat{\bm{r}}
+ \hat{\bm{r}}^i \sigma^j
+ \hat{\bm{r}}^j \sigma^i
%---------------
\end{equation}
%---------------
is a $F$-wave contribution of total spin 2. This contribution is orthogonal to
the $P$-wave wavefunctions, that is, 
$\delta^{ij} f^{ij} (\hat{\bm{r}}, \bm{\sigma})
= \hat{\bm{r}}^{[i} {\sigma}^{j]} f^{ij} (\hat{\bm{r}}, \bm{\sigma})
= \hat{\bm{r}}^{(i} \bm{\sigma}^{j)} f^{ij} (\hat{\bm{r}}, \bm{\sigma})
= 0$. 
While $S_{12}$ induces transitions between states with
different orbital angular momentum, $S_{12}$ is diagonal 
within the $P$-wave block, with diagonal elements 
$0, 4, -2$, and $2/5$ for angular momentum states with quantum numbers 
$^1P_1$, $^3P_0$, $^3P_1$, and $^3P_2$, respectively. 
As we have argued in section~\ref{sec:wavefunctions}, 
the $F$-wave contribution does not affect the
LDMEs to first order in the QMPT, and so the off-diagonal matrix elements 
of $S_{12}$ can be neglected 
in our calculation of the wavefunctions at the origin.

%==============================================================================
\section{Logarithmically divergent tensor integrals}
\label{appendix:tensorint}
%==============================================================================

In this appendix, we compute the logarithmically divergent two-loop 
tensor integrals in eqs.~(\ref{eq:tenint_DR}) and (\ref{eq:tenint_r}). 
We compute the two-loop integrals as nested one-loop integrals, 
by first integrating over $\bm{p}$, and then integrating over $\bm{p}'$. 
We note that the integration over $\bm{p}$ can be done in the same way 
in DR and in finite-$r$ regularization, as long as we work in $d-1$ spatial
dimensions and expand in powers of $\epsilon$ for the finite-$r$
regularized integral. 
The integrals can be simplified by using the fact that we only need the
difference between DR and finite-$r$ regularization, so that any UV-finite
contribution that appear commonly in both regularizations can be neglected. 
Since the tensor integrals are at most logarithmically divergent, we can make
any changes in the integrands as long as the large $\bm{p}$ and $\bm{p}'$
behaviors are kept unaltered. 
We note that in the integral over $\bm{p}$, 
%---------------
\begin{equation}
%---------------
\frac{1}{E-\bm{p}^2/m}
= 
- \frac{m}{\bm{p}^2}
- \frac{m}{ \bm{p}^4/(m E) - \bm{p}^2}. 
%---------------
\end{equation}
%---------------
Since the second term does not produce UV divergences, it can be neglected in
our calculation of the integrals. This removes the scale $E$ from the
integration over $\bm{p}$, and leaves only the scale $\bm{p}'^2$, which
simplifies the calculation. 
We do not make similar modifications for the $\bm{p}'$ integral, because if we
set $E =0$ everywhere, the tensor integrals develop logarithmic IR divergences, 
which must
be regulated in the same way in both DR and finite-$r$ regularization. 
While this is in principle possible, for example by regulating the IR
divergences in DR, this requires computing the finite-$r$ regularized integral
in $d-1$ spatial dimensions, which is more difficult than keeping $E$ nonzero
in the $\bm{p}'$ integral. 

In $J_{3b}$ and $J_{3c}$, the integration over
$\bm{p}$ contains power IR divergences. We regulate the IR divergences
dimensionally in both DR and finite-$r$ regularized integrals. 
Unlike logarithmic divergences, power divergences are subtracted
automatically by expanding in powers of $\epsilon$, and so, 
we can compute the finite-$r$ regularized integral over $\bm{p}'$ in $3$
spatial dimensions.

\subsection[$J_{3a}$]{\boldmath $J_{3a}$} 

We compute the integral over $\bm{p}$ using Feynman parametrization. 
%---------------
\begin{eqnarray}
%---------------
\int_{\bm{p}}
\frac{1}{(\bm{p}-\bm{p}')^2}
\frac{\bm{p}^j \bm{p}^k \bm{p}^l}{(\bm{p}^2)^3}
&=&
3 \int_0^1 dx \, (1-x)^2 \int_{\bm{p}}
\frac{
\bm{p}^j \bm{p}^k \bm{p}^l}{[\bm{p}^2 - 2 x \bm{p}\cdot \bm{p}' + x \bm{p}'^2]^4}
\nonumber \\ &=&
3 \int_0^1 dx \, (1-x)^2 \int_{\bm{p}}
\frac{
\frac{x}{d-1}
( \bm{p}'^j \delta^{kl} +\bm{p}'^k \delta^{jl}
+ \bm{p}'^l \delta^{jk}) \bm{p}^2
+ x^3 \bm{p}'^j \bm{p}'^k \bm{p}'^l
}{[\bm{p}^2 + x (1-x) \bm{p}'^2 ]^4}
\nonumber \\
&=&
\Lambda^{2 \epsilon} e^{\epsilon \gamma_{\rm E}}
\frac{2^{2 \epsilon-6} \Gamma(\frac{1}{2}-\epsilon) 
\Gamma(\frac{3}{2}+\epsilon) 
}
{\pi \Gamma(1-\epsilon)} 
\nonumber \\ && \hspace{5ex}  \times 
\bigg[ 
(3+2 \epsilon) 
\frac{\bm{p}'^j \bm{p}'^k \bm{p}'^l }{( \bm{p}'^2)^{5/2+\epsilon}}
+ 
\frac{ \bm{p}'^j \delta^{kl} + \bm{p}'^k \delta^{jl} + \bm{p}'^l \delta^{jk}}
{(\bm{p}'^2 )^{3/2+\epsilon}}
\bigg].
%---------------
\end{eqnarray}
%---------------
The integrals over $\bm{p}'$ in DR are evaluated as 
%---------------
\begin{eqnarray}
\label{eq:pprimerank3_DR}
%---------------
&& \hspace{-8ex} 
\int_{\bm{p}'} \frac{\bm{p}'^i \bm{p}'^j \bm{p}'^k \bm{p}'^l}{
(\bm{p}'^2-m E) ( \bm{p}'^2 )^{5/2+\epsilon}}
=
\frac{\Gamma(7/2+\epsilon)}{\Gamma(5/2+\epsilon)}
\int_0^1 dz
\int_{\bm{p}'} \frac{z^{3/2+\epsilon} \bm{p}'^i \bm{p}'^j \bm{p}'^k \bm{p}'^l}{
[ \bm{p}'^2 - (1-z) m E ]^{7/2+\epsilon}}
\nonumber\\
&=&
(\delta^{ij} \delta^{kl}
+ \delta^{ik} \delta^{jl}
+ \delta^{il} \delta^{jk})
\frac{\Lambda^{2 \epsilon} e^{\epsilon \gamma_{\rm E}}}
{4(4 \pi)^{3/2}}
\frac{\Gamma(2 \epsilon) \Gamma(1-2 \epsilon)}{\Gamma(7/2-\epsilon) 
(- m E)^{-2 \epsilon}}, 
%---------------
\end{eqnarray}
%---------------
and 
%---------------
\begin{equation}
\label{eq:pprimerank1_DR}
%---------------
\int_{\bm{p}'} \frac{\bm{p}'^i \bm{p}'^j}{
(\bm{p}'^2-m E) ( \bm{p}'^2 )^{3/2+\epsilon}}
=
\delta^{ij}
\frac{\Lambda^{2 \epsilon} e^{\epsilon \gamma_{\rm E}}}{2(4 \pi)^{3/2}}
\frac{\Gamma(2 \epsilon) \Gamma(1-2 \epsilon)}{\Gamma(5/2-\epsilon)
(- m E)^{-2 \epsilon}}, 
%---------------
\end{equation}
%---------------
so that 
%---------------
\begin{eqnarray}
%---------------
\int_{\bm{p}'} \int_{\bm{p}} \frac{\bm{p}'^i \bm{p}^j \bm{p}^k \bm{p}^l}
{(\bm{p}'^2-m E)(\bm{p}-\bm{p}')^2(\bm{p}^2)^3}
&=&
\frac{
\delta^{ij} \delta^{kl} + \delta^{ik} \delta^{jl} + \delta^{il} \delta^{jk}
}{1920 \pi^2 } 
\nonumber \\ && \times 
\left[ \frac{1}{\epsilon_{\rm UV}} 
+2 \log \left( - \frac{\Lambda^2}{m E} \right) + \frac{76}{15} 
+O(\epsilon) \right]. 
%---------------
\end{eqnarray}
%---------------
On the other hand, the finite-$r$ regulated integrals over $\bm{p}'$ 
are computed as 
%---------------
\begin{eqnarray}
\label{eq:pprimerank3_r}
%---------------
\left( -i \frac{\partial}{\partial r} \right) 
\int_{\bm{p}'} \frac{e^{i \bm{p}' \cdot \bm{r}} 
\bm{p}'^j \bm{p}'^k \bm{p}'^l
}{(\bm{p}'^2-m E) ( \bm{p}'^2 )^{5/2}}
\bigg|_{|\bm{r}|=r_0}
&=& 
\left( -i \frac{\partial}{\partial r} \right) 
\frac{\Gamma(7/2)}{\Gamma(5/2)} 
\int_0^1 dz \, z^{3/2} 
\int_{\bm{p}'} \frac{e^{i \bm{p}' \cdot \bm{r}} 
\bm{p}'^j \bm{p}'^k \bm{p}'^l
}{[ \bm{p}'^2-(1-z) m E]^{7/2}}
\nonumber \\
&=&
- \frac{1}{30 \pi^2}
\bigg\{
\hat{\bm{r}}^j \hat{\bm{r}}^k \hat{\bm{r}}^l
+ (\hat{\bm{r}}^j \delta^{kl}
+ \hat{\bm{r}}^k \delta^{lj}
+ \hat{\bm{r}}^l \delta^{jk} )
\nonumber \\ &&
\times \left[- \frac{8}{15} + \gamma_{\rm E} 
+ \frac{1}{2} \log (-r_0^2 m E)
\right]
\bigg\},
%---------------
\end{eqnarray}
%---------------
and 
%---------------
\begin{equation}
\label{eq:pprimerank1_r}
%---------------
\left( -i \frac{\partial}{\partial r} \right)
\int_{\bm{p}'} \frac{e^{i \bm{p}' \cdot \bm{r}} \bm{p}'^j 
}{(\bm{p}'^2-m E) ( \bm{p}'^2 )^{3/2}}
\bigg|_{|\bm{r}|=r_0}
=
-\frac{1}{6 \pi^2} 
\hat{\bm{r}}^j 
\left[ -\frac{1}{3} + \gamma_{\rm E} 
+\frac{1}{2} \log (- r_0^2 m E) \right] ,
%---------------
\end{equation}
%---------------
which give 
%---------------
\begin{eqnarray}
%---------------
\int_{\bm{p}'} \int_{\bm{p}} \frac{
e^{i \bm{p}' \cdot \bm{r}}
\hat{\bm{r}} \cdot \bm{p}' \bm{p}^j \bm{p}^k \bm{p}^l}
{(\bm{p}'^2-m E)(\bm{p}-\bm{p}')^2(\bm{p}^2)^3}
\bigg|_{|\bm{r}|=r_0}
&=&
- \frac{ ( \hat{\bm{r}}^j \delta^{kl} + \hat{\bm{r}}^k \delta^{jl}
+ \hat{\bm{r}}^l \delta^{jk} ) 
}{1920 \pi^2} 
\nonumber \\ && \times 
\left[
- \frac{49}{30} 
+ 4  \gamma_{\rm E} 
+ 2 \log (- r_0^2 m E) 
\right]
\nonumber \\ && - 
\frac{\hat{\bm{r}}^i \hat{\bm{r}}^j \hat{\bm{r}}^k}{1280 \pi^2} + O(r_0).
%---------------
\end{eqnarray}
%---------------
From these we obtain 
%---------------
\begin{eqnarray}
%---------------
\frac{1}{m^2} \hat{\bm{r}}^i \left( J_{3a}^{ijkl} |_{(r_0)}
-J_{3a}^{ijkl} |_{\rm DR} \right)
&=&
- \frac{( \hat{\bm{r}}^j \delta^{kl} + \hat{\bm{r}}^k \delta^{jl}
+ \hat{\bm{r}}^l \delta^{jk} )}{480 \pi^2}
\left[ \frac{1}{4 \epsilon_{\rm UV}} + \frac{103}{120}
+ \log (\Lambda r_0 e^{\gamma_{\rm E}})
\right]
\nonumber \\ &&
- \frac{1}{1280 \pi^2}
 \hat{\bm{r}}^j \hat{\bm{r}}^k \hat{\bm{r}}^l.
%---------------
\end{eqnarray}
%---------------

\subsection[$J_{3b}$]{\boldmath $J_{3b}$} 

We first integrate over $\bm{p}$. This integral contains a power IR divergence, 
which we regulate in DR. 
%---------------
\begin{eqnarray}
%---------------
&& \hspace{-5ex} 
\int_{\bm{p}} \frac{ (\bm{p}'-\bm{p})^j (\bm{p}'-\bm{p})^k \bm{p}^l
}{[(\bm{p}'-\bm{p})^2] (\bm{p}^2)^3}
=
\frac{\Gamma(4)}{\Gamma(3)} 
\int_0^1 dx 
\int_{\bm{p}} \frac{(1-x)^2 
 (\bm{p}'-\bm{p})^j (\bm{p}'-\bm{p})^k \bm{p}^l
}{[ \bm{p}^2 -2 x \bm{p} \cdot \bm{p}' +x \bm{p}'^2 ]^4} 
\nonumber \\
&=&
\frac{\Lambda^{2 \epsilon}e^{\epsilon \gamma_{\rm E}} }{(4 \pi)^{3/2}}
\frac{\Gamma(5/2+\epsilon)}{\Gamma(3)}
\int_0^1 dx
\frac{
x (1-x)^4 \bm{p}'^j \bm{p}'^k \bm{p}'^l
}{[ x (1-x) \bm{p}'^2 ]^{5/2+\epsilon}}
\nonumber \\ &&
+ \frac{\Lambda^{2 \epsilon}e^{\epsilon \gamma_{\rm E}} }{2 (4 \pi)^{3/2}}
\frac{\Gamma(3/2+\epsilon)}{\Gamma(3)}
\int_0^1 dx (1-x)^2
\frac{
x \delta^{jk} \bm{p}'^l
-(1-x) \bm{p}'^k \delta^{jl}
- (1-x)\bm{p}'^j \delta^{kl}
}{[ x (1-x) \bm{p}'^2 ]^{3/2+\epsilon}}
\nonumber \\
&=&
\Lambda^{2 \epsilon}e^{\epsilon \gamma_{\rm E}} 
\frac{2^{2 \epsilon-7}  \Gamma(1/2+\epsilon)
\Gamma(1/2-\epsilon)}{\pi \Gamma(1-\epsilon)}
\nonumber \\ && \times 
\bigg[ 
(4 \epsilon^2-9)
\frac{
\bm{p}'^j \bm{p}'^k \bm{p}'^l
}{[ \bm{p}'^2 ]^{5/2+\epsilon}}
+ \frac{(1+2 \epsilon) \delta^{jk} \bm{p}'^l
+ (3-2 \epsilon)
[ \bm{p}'^k \delta^{jl} + \bm{p}'^j \delta^{kl}]
}{[ \bm{p}'^2 ]^{3/2+\epsilon}}
\bigg]. 
%---------------
\end{eqnarray}
%---------------
Since this expression does not contain poles when expanded in powers of 
$\epsilon$, there are no logarithmic IR divergences, and we can set
$\epsilon=0$ in calculations in finite-$r$ regularization, which subtracts the
power IR divergence. 
By using eqs.~(\ref{eq:pprimerank3_DR}) and (\ref{eq:pprimerank1_DR}), 
we obtain in DR 
%---------------
\begin{eqnarray}
\label{eq:J3bint_DR} 
%---------------
&& \hspace{-8ex} 
\int_{\bm{p}'} \int_{\bm{p}} 
\frac{\bm{p}'^i (\bm{p}'-\bm{p})^j (\bm{p}'-\bm{p})^k \bm{p}^l
}{(\bm{p}^2-mE) (\bm{p}'-\bm{p})^2 (\bm{p}^2)^3}
\nonumber \\ 
&=& 
\frac{3 \delta^{ij} \delta^{kl} + 3 \delta^{ik} \delta^{jl}
-2 \delta^{il} \delta^{jk}}{7680 \pi^2 \epsilon_{\rm UV}} 
+ \frac{(\delta^{ij} \delta^{kl} + \delta^{ik} \delta^{jl})
}{6400\pi^2} 
\left[ 1 + 5 \log \left( - \frac{\Lambda^2}{m E} \right) \right]
\nonumber \\ && 
- \frac{\delta^{il} \delta^{jk}}{28800 \pi^2} 
\left[ 8 + 15 \log \left( - \frac{\Lambda^2}{m E} \right) \right]
+ O(\epsilon), 
%---------------
\end{eqnarray}
%---------------
where the $1/\epsilon_{\rm UV}$ pole comes from the UV divergence of the
$\bm{p}'$ integral. 
Similarly, by using eqs.~(\ref{eq:pprimerank3_r}) and
(\ref{eq:pprimerank1_r}) we obtain 
in finite-$r$ regularization 
%---------------
\begin{eqnarray}
\label{eq:J3bint_finiter} 
%---------------
&& \hspace{-8ex} 
\int_{\bm{p}'} \int_{\bm{p}}
\frac{e^{i \bm{p}' \cdot \bm{r}} 
\hat{\bm{r}} \cdot \bm{p}' (\bm{p}'-\bm{p})^j (\bm{p}'-\bm{p})^k \bm{p}^l
}{(\bm{p}^2-mE) (\bm{p}'-\bm{p})^2 (\bm{p}^2)^3}
\bigg|_{|\bm{r}|=r_0}
\nonumber \\ 
&=&
\frac{\hat{\bm{r}}^j \delta^{kl} + \hat{\bm{r}}^k \delta^{jl}
}{6400\pi^2}
\left[ \frac{1}{3} 
- 10 \gamma_{\rm E}
- 5 \log \left( -r_0^2 m E \right) \right]
\nonumber \\ &&
- \frac{\hat{\bm{r}}^{l} \delta^{jk}}{28800 \pi^2}
\left[ 
\frac{47}{2} 
- 30 \gamma_{\rm E}
- 15 \log \left( -r_0^2 m E \right) \right]
+ \frac{3}{1280 \pi^2} 
\hat{\bm{r}}^j \hat{\bm{r}}^k \hat{\bm{r}}^l
+ O(r_0).
%---------------
\end{eqnarray}
%---------------
By subtracting eq.~(\ref{eq:J3bint_DR}) from eq.~(\ref{eq:J3bint_finiter}) 
we obtain 
%---------------
\begin{eqnarray}
%---------------
\frac{1}{m^2} 
\hat{\bm{r}}^i \left( J_{3b}^{ijkl}|_{(r_0)} - J_{3b}^{ijkl}|_{\rm DR} \right)
&=&
- \frac{3 \hat{\bm{r}}^j \delta^{kl} + 3 \hat{\bm{r}}^k \delta^{jl}
-2 \hat{\bm{r}}^l \delta^{jk}}{7680 \pi^2 \epsilon_{\rm UV}}
\nonumber \\ &&
+ \frac{\hat{\bm{r}}^j \delta^{kl} + \hat{\bm{r}}^k \delta^{jl}
}{6400\pi^2}
\left[ -\frac{2}{3}
- 10 \log \left( \Lambda r_0 e^{\gamma_{\rm E}} \right) \right]
\nonumber \\ &&
- \frac{\hat{\bm{r}}^{l} \delta^{jk}}{28800 \pi^2}
\left[
\frac{31}{2}
- 30 \log \left( \Lambda r_0 e^{ \gamma_{\rm E}} \right) \right]
\nonumber \\ &&
+ \frac{3}{1280 \pi^2}
\hat{\bm{r}}^j \hat{\bm{r}}^k \hat{\bm{r}}^l
+ O(\epsilon,r_0).
%---------------
\end{eqnarray}
%---------------

\subsection[$J_{3c}$]{\boldmath $J_{3c}$} 

We again first integrate over $\bm{p}$, 
and we regulate the power IR divergence in DR.
%---------------
\begin{eqnarray}
%---------------
&& \hspace{-5ex} 
\int_{\bm{p}} \frac{ (\bm{p}'-\bm{p})^j \bm{p}^k \bm{p}^l
}{[(\bm{p}'-\bm{p})^2] (\bm{p}^2)^3}
= 
\frac{\Gamma(4)}{\Gamma(3)} 
\int_0^1 dx 
\int_{\bm{p}} \frac{(1-x)^2 
 (\bm{p}'-\bm{p})^j \bm{p}^k \bm{p}^l
}{[ \bm{p}^2 -2 x \bm{p} \cdot \bm{p}' +x \bm{p}'^2 ]^4} 
\nonumber \\
&=&
- \frac{\Lambda^{2 \epsilon}e^{\epsilon \gamma_{\rm E}} }{(4 \pi)^{3/2}}
\frac{\Gamma(5/2+\epsilon)}{\Gamma(3)}
\int_0^1 dx
\frac{
x (1-x)^4 \bm{p}'^j \bm{p}'^k \bm{p}'^l
}{[ x (1-x) \bm{p}'^2 ]^{5/2+\epsilon}}
\nonumber \\ &&
- \frac{\Lambda^{2 \epsilon}e^{\epsilon \gamma_{\rm E}} }{2 (4 \pi)^{3/2}}
\frac{\Gamma(3/2+\epsilon)}{\Gamma(3)}
\int_0^1 dx (1-x)^2
\frac{
x \delta^{jk} \bm{p}'^l
+x \bm{p}'^k \delta^{jl}
- (1-x)\bm{p}'^j \delta^{kl}
}{[ x (1-x) \bm{p}'^2 ]^{3/2+\epsilon}}
\nonumber \\
&=&
- \Lambda^{2 \epsilon}e^{\epsilon \gamma_{\rm E}} 
\frac{2^{2 \epsilon-7}  \Gamma(1/2+\epsilon)
\Gamma(1/2-\epsilon)}{\pi \Gamma(1-\epsilon)}
\nonumber \\ && \times 
\bigg[ 
(4 \epsilon^2-9)
\frac{
\bm{p}'^j \bm{p}'^k \bm{p}'^l
}{[ \bm{p}'^2 ]^{5/2+\epsilon}}
+ \frac{(1+2 \epsilon) (\delta^{jk} \bm{p}'^l + \bm{p}'^k \delta^{jl})
+ (3-2 \epsilon) \bm{p}'^j \delta^{kl}
}{[ \bm{p}'^2 ]^{3/2+\epsilon}}
\bigg]. 
%---------------
\end{eqnarray}
%---------------
Again, this expression does not contain poles in $\epsilon$, so that we can
set $\epsilon=0$ in the finite-$r$ regularized integral, which subtracts the
power IR divergence. 
By using eqs.~(\ref{eq:pprimerank3_DR}) and (\ref{eq:pprimerank1_DR}),
we obtain in DR
%---------------
\begin{eqnarray}
\label{eq:J3cresult_DR}
%---------------
&& \hspace{-8ex}
\int_{\bm{p}'} \int_{\bm{p}}
\frac{\bm{p}'^i (\bm{p}'-\bm{p})^j \bm{p}^k \bm{p}^l
}{(\bm{p}^2-mE) (\bm{p}'-\bm{p})^2 (\bm{p}^2)^3}
\nonumber \\
&=&
- \frac{3 \delta^{ij} \delta^{kl} - 2 \delta^{ik} \delta^{jl}
-2 \delta^{il} \delta^{jk}}{7680 \pi^2 \epsilon_{\rm UV}}
+ \frac{\delta^{ik} \delta^{jl} + \delta^{il} \delta^{jk}
}{28800\pi^2}
\left[ 8 + 15 \log \left( - \frac{\Lambda^2}{m E} \right) \right]
\nonumber \\ &&
- \frac{\delta^{ij} \delta^{kl}}{6400 \pi^2}
\left[ 1 + 5 \log \left( - \frac{\Lambda^2}{m E} \right) \right]
+ O(\epsilon).
%---------------
\end{eqnarray}
%---------------
Similarly, by using eqs.~(\ref{eq:pprimerank3_r}) and
(\ref{eq:pprimerank1_r}) we obtain
in finite-$r$ regularization
%---------------
\begin{eqnarray}
\label{eq:J3cresult_finiter}
%---------------
&& \hspace{-8ex}
\int_{\bm{p}'} \int_{\bm{p}}
\frac{e^{i \bm{p}' \cdot \bm{r}}
\hat{\bm{r}} \cdot \bm{p}' (\bm{p}'-\bm{p})^j \bm{p}^k \bm{p}^l
}{(\bm{p}^2-mE) (\bm{p}'-\bm{p})^2 (\bm{p}^2)^3}
\bigg|_{|\bm{r}|=r_0}
\nonumber \\
&=&
\frac{\hat{\bm{r}}^k \delta^{jl} + \hat{\bm{r}}^l \delta^{jk}
}{28800\pi^2}
\left[ \frac{47}{2} - 30 \gamma_{\rm E} 
- 15 \log \left( -r_0^2 m E\right) \right]
\nonumber \\ &&
- \frac{\hat{\bm{r}}^{j} \delta^{kl}}{6400 \pi^2}
\left[
\frac{1}{3} -10 \gamma_{\rm E}  
- 5 \log \left( -r_0^2 m E \right) \right]
- \frac{3}{1280 \pi^2}
\hat{\bm{r}}^j \hat{\bm{r}}^k \hat{\bm{r}}^l
+ O(r_0).
%---------------
\end{eqnarray}
%---------------
By subtracting eq.~(\ref{eq:J3cresult_DR}) from 
eq.~(\ref{eq:J3cresult_finiter}) we obtain
%---------------
\begin{eqnarray}
%---------------
\frac{1}{m^2}
\hat{\bm{r}}^i \left( J_{3c}^{ijkl}|_{(r_0)} - J_{3c}^{ijkl}|_{\rm DR} \right)
&=&
\frac{3 \hat{\bm{r}}^j \delta^{kl} - 2 \hat{\bm{r}}^k \delta^{jl}
-2 \hat{\bm{r}}^l \delta^{jk}}{7680 \pi^2 \epsilon}
\nonumber \\ && 
+ \frac{\hat{\bm{r}}^k \delta^{jl} + \hat{\bm{r}}^l \delta^{jk}
}{28800\pi^2}
\left[ \frac{31}{2}
- 30 \log \left( \Lambda r_0 e^{\gamma_{\rm E}} \right) \right]
\nonumber \\ &&
- \frac{\hat{\bm{r}}^{j} \delta^{kl}}{6400 \pi^2}
\left[
- \frac{2}{3}  
- 10 \log \left( \Lambda r_0 e^{\gamma_{\rm E}} \right) \right]
\nonumber \\ &&
- \frac{3}{1280 \pi^2}
\hat{\bm{r}}^j \hat{\bm{r}}^k \hat{\bm{r}}^l. 
%---------------
\end{eqnarray}
%---------------

%==============================================================================
\section{\boldmath Convergence of $1/m$ corrections to wavefunctions}
\label{appendix:1mallorder}
%==============================================================================

In this appendix, we test the validity of the Rayleigh-Schr\"odinger
perturbation theory at first order in computation of the corrections from the
$1/m$ potential to $S$- and $P$-wave wavefunctions. 
The radial equation that includes the $1/m$ potential reads 
%---------------
\begin{equation}
\label{eq:radial_1m}
%---------------
\left[ - \frac{1}{m} \left( \frac{\partial^2}{\partial r^2}
+ \frac{2}{r} \frac{\partial}{\partial r} \right) + V_{\rm LO} (r)
+ \frac{V^{(1)} (r)}{m} 
+ \frac{L (L+1)}{m r^2} \right] R(r) 
= E \, R(r), 
%---------------
\end{equation}
%---------------
where $L = 0$ and 1 for $S$ and $P$-wave states, respectively. 
To first order in the Rayleigh-Schr\"odinger perturbation theory, 
$R(r)$ is computed from 
%---------------
\begin{equation}
\label{eq:radial_1m_qmpt}
%---------------
R(r)|_{\textrm{first order}} = R_{\rm LO} (r) - 
\int_0^\infty dr' \, r'^2 \, \hat{G}_n^L (r,r') \frac{V^{(1)} (r')}{m} 
R_{\rm LO} (r'),
%---------------
\end{equation}
%---------------
where $\hat{G}_n^L (r,r')$ is the contribution to the reduced Green's function 
from orbital angular momentum $L$ defined through the relation 
%---------------
\begin{equation}
%---------------
\hat{G}_n (\bm{r},\bm{r}')  
= 
\sum_{L=0}^\infty \sum_{M=-L}^{+L} 
\hat{G}_n^L (r,r') Y_L^{M} (\hat{\bm{r}}) Y_L^{M*} (\hat{\bm{r}}'), 
%---------------
\end{equation}
%---------------
and $R_{\rm LO}(r)$ is the bound-state solution of the LO radial equation 
%---------------
\begin{equation}
\label{eq:radial_lo}
%---------------
\left[ - \frac{1}{m} \left( \frac{\partial^2}{\partial r^2}
+ \frac{2}{r} \frac{\partial}{\partial r} \right) + V_{\rm LO} (r)
+ \frac{L (L+1)}{m r^2} \right] R_{\rm LO}(r)
= E_n^{\rm LO} \, R_{\rm LO} (r).
%---------------
\end{equation}
%---------------
The validity of eq.~(\ref{eq:radial_1m_qmpt}) can be tested by comparing it 
with the solution of eq.~(\ref{eq:radial_1m}). 
We find the two lowest $S$-wave and $P$-wave bound-state solutions of 
eq.~(\ref{eq:radial_1m}) 
by using the modified Crank-Nicolson method in ref.~\cite{Kang:2006jd}. 
We use $m=1.316$~GeV and $\alpha_s = \alpha_s(\mu_R = 2.5{\rm ~GeV})$, which
are used in our numerical calculations of charmonium wavefunctions. 
We use the expression for $V^{(1)} (r)$ given in the Wilson-loop 
matching scheme in eq.~(\ref{eq:1mpotential_WL_longdistancematch}). 
For the computation of eq.~(\ref{eq:radial_1m_qmpt}), we use the method
developed in section~\ref{sec:redgreen_numerical} to compute the radial
integral.

%%%%%%%%%%%%%%%%%%%%%%%%%%%%%%%%%%%%%%%%%%%%%%%%%%%%%%%%%%%%%%%%%%%%%%%%%%%%%%%
\begin{figure}[tbp]
\centering
\includegraphics[width=.45\textwidth]{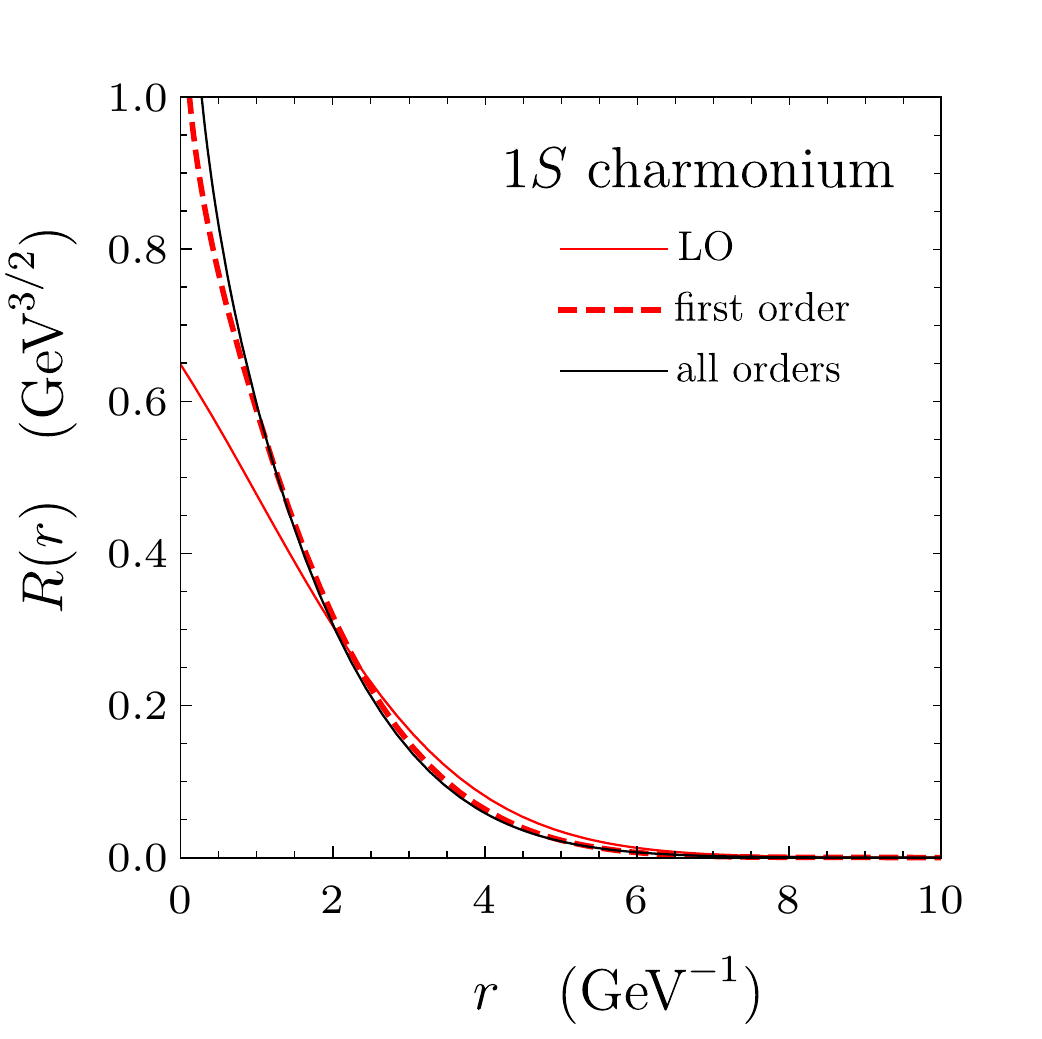}
\includegraphics[width=.465\textwidth]{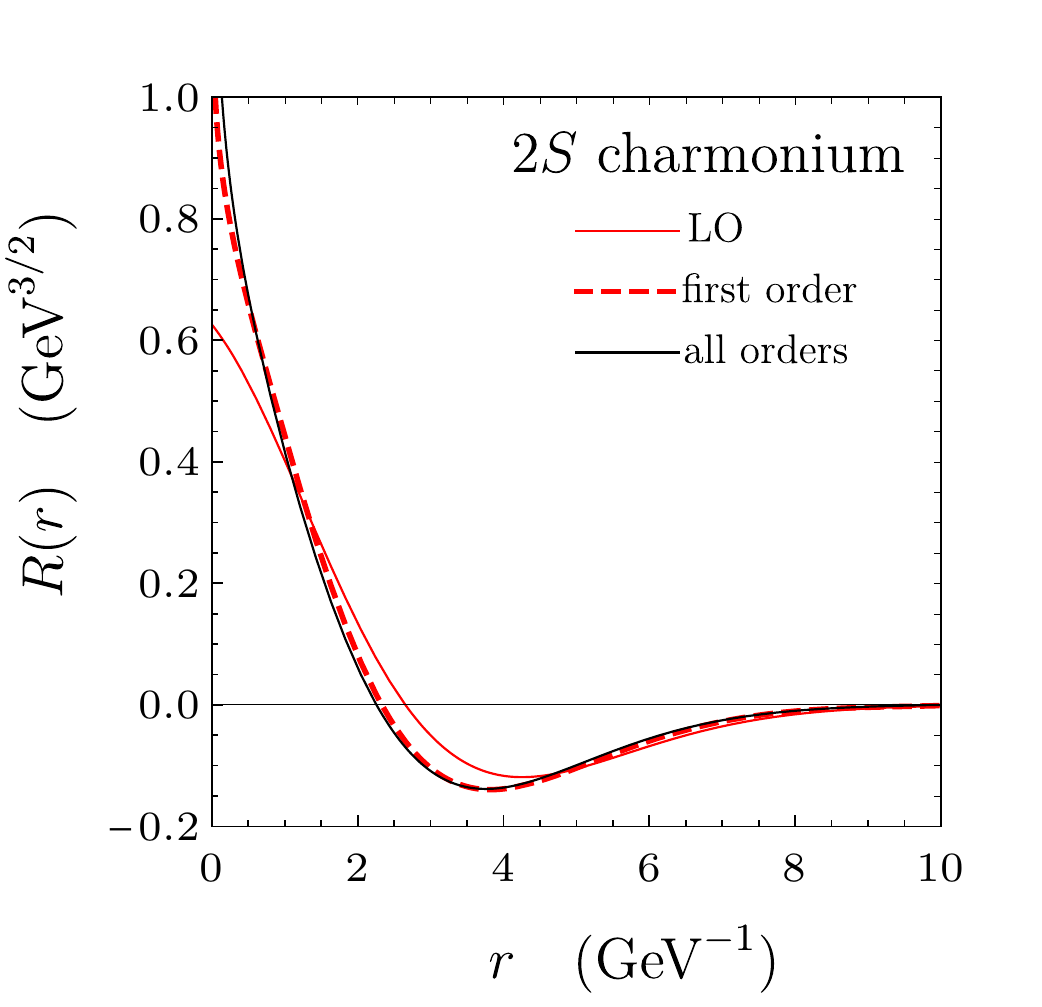}
\caption{\label{fig:charmswave1mconv}
Radial wavefunctions for $1S$ and $2S$ states including the effect of the $1/m$
potential to first order in Rayleigh-Schr\"odinger perturbation theory (red
dashed lines) compared to the all-orders calculation (black solid lines). 
The LO radial wavefunctions are shown as red solid lines for comparison. 
}
\end{figure}
%%%%%%%%%%%%%%%%%%%%%%%%%%%%%%%%%%%%%%%%%%%%%%%%%%%%%%%%%%%%%%%%%%%%%%%%%%%%%%%

We show the radial wavefunctions for the $1S$ and $2S$ states in
fig.~\ref{fig:charmswave1mconv}. For both $1S$ and $2S$ states, the radial
wavefunctions computed to first order in the Rayleigh-Schr\"odinger perturbation
theory agree well with the all-orders calculation, showing that the corrections
at first order reproduce the bulk of the all-orders correction from the 
$1/m$ potential. The small deviations at $r \lesssim 1/m$ 
arise from the fact that 
the all-orders calculation includes the logarithmic divergences of the form 
$(\alpha_s^2 \log r)^n$ coming from the $1/m$ potential 
to higher orders in $\alpha_s$, while the first-order calculation includes
the short-distance divergences only at leading order (order $\alpha_s^2$). 
As these are
short-distance effects that must be subtracted order by order in
$\alpha_s$ through renormalization, 
the agreements in the region $r > 1/m$ is sufficient to confirm the validity of
the Rayleigh-Schr\"odinger perturbation theory at first order. 

%%%%%%%%%%%%%%%%%%%%%%%%%%%%%%%%%%%%%%%%%%%%%%%%%%%%%%%%%%%%%%%%%%%%%%%%%%%%%%%
\begin{figure}[tbp]
\centering
\includegraphics[width=.45\textwidth]{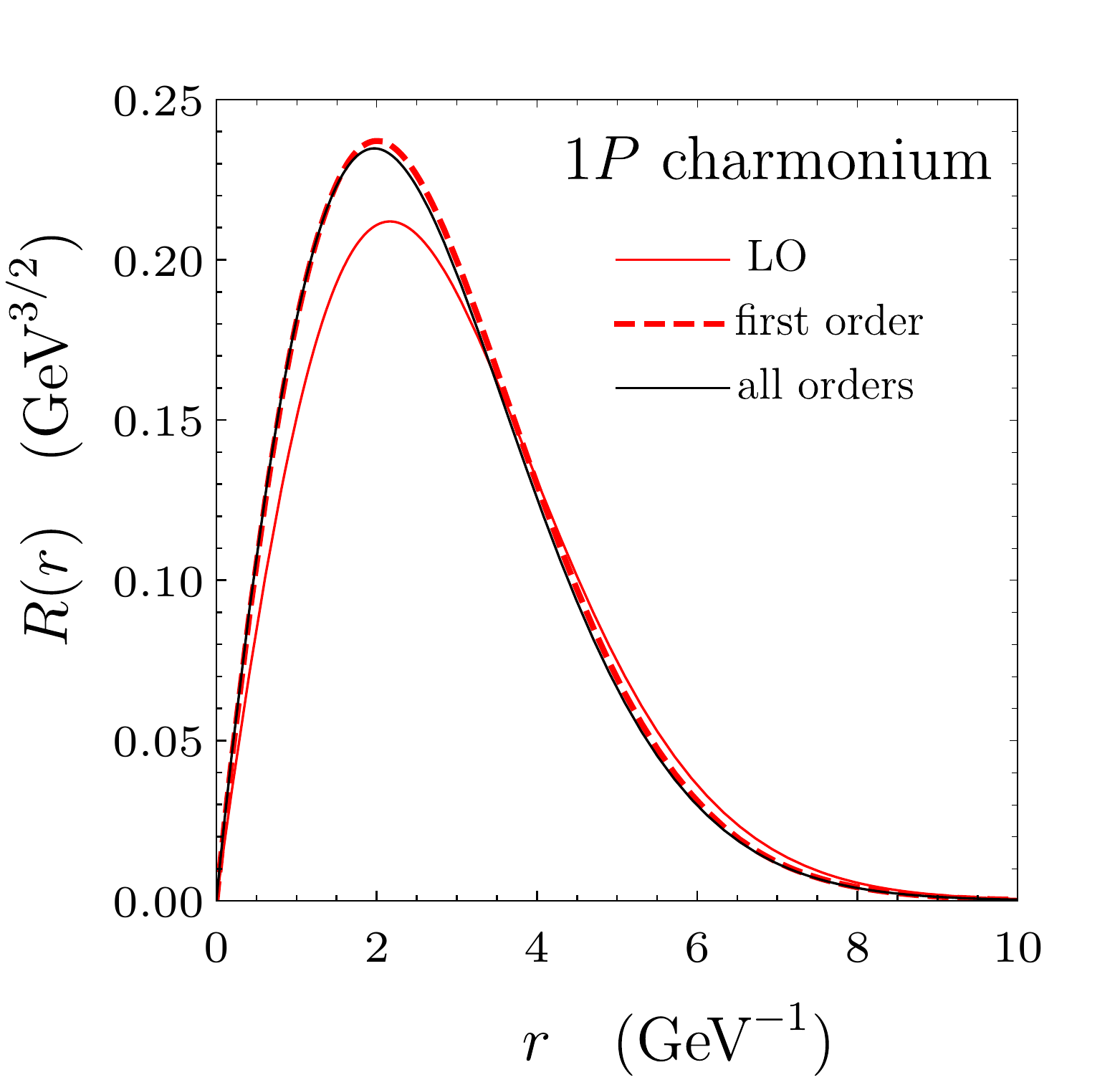}
\includegraphics[width=.465\textwidth]{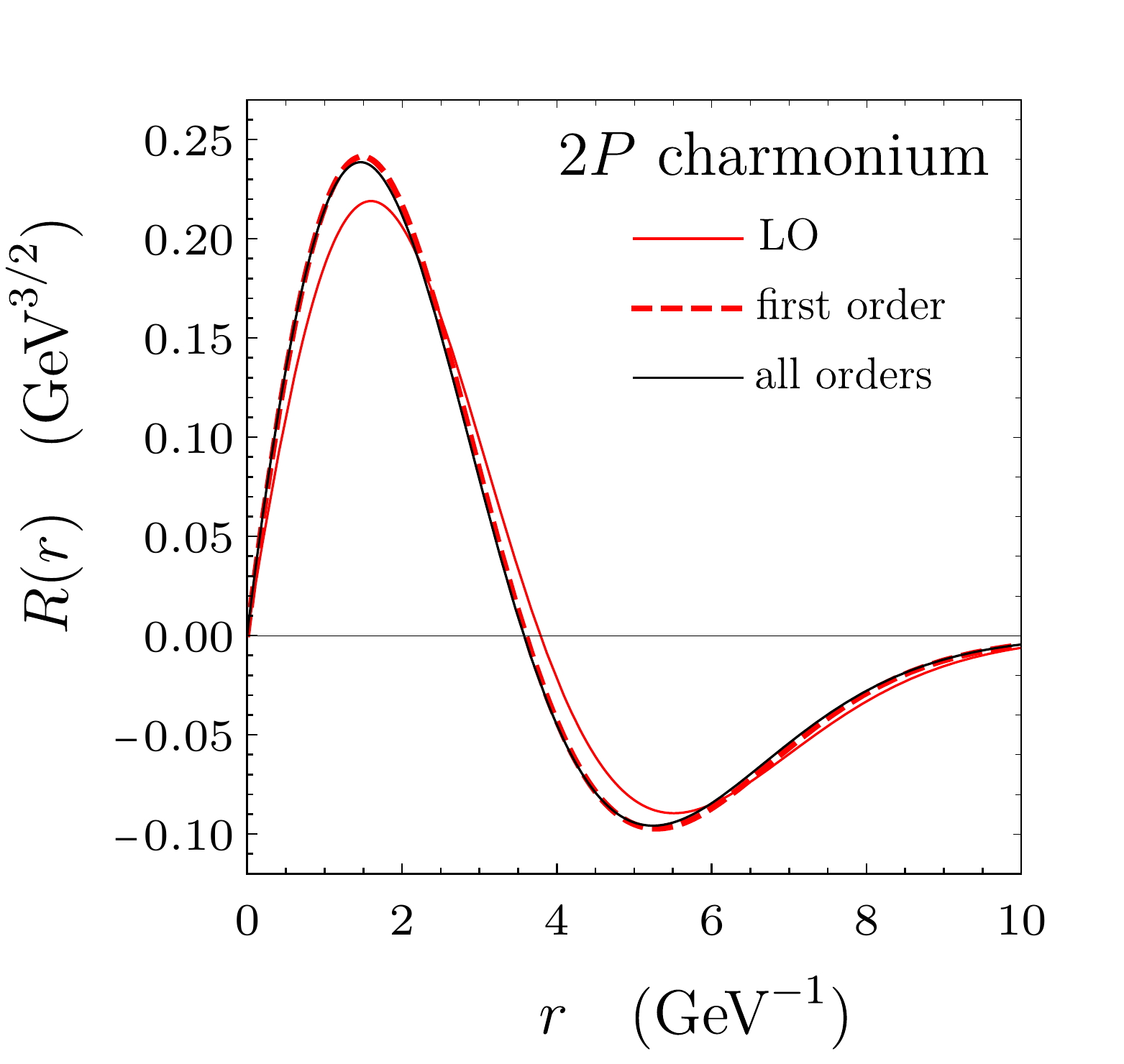}
\caption{\label{fig:charmpwave1mconv}
Radial wavefunctions for $1P$ and $2P$ states including the effect of the $1/m$
potential to first order in Rayleigh-Schr\"odinger perturbation theory (red
dashed lines) compared to the all-orders calculation (black solid lines).
The LO radial wavefunctions are shown as red solid lines for comparison.
}
\end{figure}
%%%%%%%%%%%%%%%%%%%%%%%%%%%%%%%%%%%%%%%%%%%%%%%%%%%%%%%%%%%%%%%%%%%%%%%%%%%%%%%

Similarly to the $S$-wave case,  
we show the radial wavefunctions for the $1P$ and $2P$ states in 
fig.~\ref{fig:charmpwave1mconv}. 
Again, for both $1P$ and $2P$ states, the radial
wavefunctions computed to first order in the Rayleigh-Schr\"odinger perturbation
theory agree well with the all-orders calculation, showing that also for the
$P$-wave states, the corrections
at first order reproduce the bulk of the all-orders corrections from the
$1/m$ potential.

By repeating the same analysis for $m=4.743$~GeV and 
$\alpha_s = \alpha_s(\mu_R = 5{\rm ~GeV})$, 
we find that the agreements between
first-order calculations and all-orders calculations are 
even better for heavier quark mass. 
Therefore, we conclude that computing the effect of the $1/m$ potential by
using Rayleigh-Schr\"odinger perturbation theory order by order in the $1/m$
expansion is well justified.

%==============================================================================
\section{\boldmath $P$-wave wavefunctions in perturbative QCD}
\label{appendix:pertQCD}
%==============================================================================

If we work strictly in perturbative QCD, the LO potential is just the Coulomb
potential $V_{\rm LO}(r) = - \alpha_s C_F/r$, and in this case, the solutions of
the LO Schr\"odinger equation, as well as the Green's function, 
can be found analytically~\cite{doi:10.1063/1.1703733}. 
For the $L=1$ case, the $P$-wave Green's function is given by 
%---------------
\begin{eqnarray}
\label{eq:PwaveGreen_analytic}
%---------------
G^{L=1} (r',r;E) &=& 
\frac{2 \alpha_s C_F m^3 E}{3} (1-\lambda) \Gamma(-\lambda) 
\, r_< \, r_> 
\exp\left( -  \frac{1}{2 \lambda} \alpha_s C_F m (r_<+r_>) \right) 
\nonumber \\ && \times 
{}_1F_1 (2 - \lambda; 4 ; \alpha_s C_F m r_</\lambda) 
\, U (2 - \lambda; 4 ; \alpha_s C_F m r_>/\lambda),
%---------------
\end{eqnarray}
%---------------
where $r_< = \min(r,r')$, $r_> = \max(r,r')$,
$\lambda = \alpha_s C_F/\sqrt{-4 E/m}$, and 
%---------------
\begin{subequations}
\begin{eqnarray}
%---------------
{}_1F_1 (a;b;z) &=& \sum_{k=0}^\infty \frac{(a)_k}{(b)_k} \frac{z^k}{k!}, \\
U(a;b;z) &=& \frac{1}{\Gamma(a)} \int_0^\infty dt \, e^{-z t} t^{a-1} 
(1+t)^{b-a-1}. 
%---------------
\end{eqnarray}
\end{subequations}
%---------------
This result can be obtained from eq.~(\ref{eq:greenf_numerical}) by using the 
expressions for $u_<(r)$ and $u_>(r)$ that are given in terms of the 
two linearly independent Whittaker functions. 
The $P$-wave bound states can be identified by the 
singularities at $\lambda -1 = 1$, 2, 3, $\ldots$. 
The reduced Green's function can be obtained from
eq.~(\ref{eq:PwaveGreen_analytic}), for example 
by computing the subtraction term in eq.~(\ref{eq:reducedgreen_relation1})
from the residue of eq.~(\ref{eq:PwaveGreen_analytic}) at positive integer
values of $\lambda-1$, which also provides the corresponding LO radial
wavefunction. 
Once the expressions for the reduced Green's functions are obtained, it is 
in principle possible to compute the perturbative QCD corrections to the 
$P$-wave wavefunctions
at the origin, especially the $\overline{\rm MS}$-renormalized non-Coulombic
corrections $\delta_\Psi^{\rm NC}$, 
by evaluating the position-space integrals in 
eq.~(\ref{eq:delpsi_expressions}). 
While an analytical calculation of $\delta_\Psi^{\rm NC}$ is outside the scope
of this paper, 
it is a simple task to compute numerically the position-space integrals and
obtain values for $\delta_\Psi^{\rm NC}$ in perturbative QCD. We tabulate
values of $\delta_\Psi^{\rm NC}$ at the $\overline{\rm MS}$ scale $\Lambda=m$
for various values of the strong coupling $\alpha_s$ in
table~\ref{tab:pert_delNC}. 
Note that in perturbative QCD, $\delta_\Psi^{\rm NC}$ is independent of $m$
when $\Lambda=m$. 
These results can be useful in two-loop calculations of decay and production
rates of $P$-wave bound states in perturbative QCD and in 
weakly coupled pNRQCD. 

%%%%%%%%%%%%%%%%%%%%%%%%%%%%%%%%%%%%%%%%%%%%%%%%%%%%%%%%%%%%%%%%%%%%%%%%%%%%%%%
\begin{table}[tbp]
\centering
\begin{tabular}{|c|c|c|c|c|c|}
\hline
$\alpha_s$ & State & 
$\delta_\Psi^{\rm NC}|_{^3P_0}$ & $\delta_\Psi^{\rm NC}|_{^3P_1}$ &
$\delta_\Psi^{\rm NC}|_{^3P_2}$ & $\delta_\Psi^{\rm NC}|_{^1P_1}$ \\
\hline
\multirow{3}{*}{0.20} & 
$1P$ & 0.318 & 0.258 & 0.201 & 0.236 \\ \cline{2-6} & 
$2P$ & 0.296 & 0.243 & 0.190 & 0.222 \\ \cline{2-6} & 
$3P$ & 0.286 & 0.236 & 0.185 & 0.216 \\ \hline
\multirow{3}{*}{0.22} & 
$1P$ & 0.376 & 0.306 & 0.238 & 0.279 \\ \cline{2-6} & 
$2P$ & 0.350 & 0.287 & 0.225 & 0.263 \\ \cline{2-6} & 
$3P$ & 0.338 & 0.279 & 0.219 & 0.255 \\ \hline
\multirow{3}{*}{0.24} & 
$1P$ & 0.438 & 0.357 & 0.278 & 0.326 \\ \cline{2-6} & 
$2P$ & 0.407 & 0.335 & 0.262 & 0.306 \\ \cline{2-6} & 
$3P$ & 0.392 & 0.324 & 0.255 & 0.297 \\ \hline
\multirow{3}{*}{0.26} & 
$1P$ & 0.504 & 0.412 & 0.320 & 0.376 \\ \cline{2-6} & 
$2P$ & 0.468 & 0.385 & 0.302 & 0.353 \\ \cline{2-6} & 
$3P$ & 0.450 & 0.373 & 0.293 & 0.341 \\ \hline
\multirow{3}{*}{0.28} & 
$1P$ & 0.574 & 0.469 & 0.365 & 0.428 \\ \cline{2-6} & 
$2P$ & 0.532 & 0.439 & 0.344 & 0.402 \\ \cline{2-6} & 
$3P$ & 0.511 & 0.424 & 0.334 & 0.388 \\ \hline
\multirow{3}{*}{0.30} & 
$1P$ & 0.648 & 0.530 & 0.413 & 0.484 \\ \cline{2-6} & 
$2P$ & 0.599 & 0.495 & 0.389 & 0.453 \\ \cline{2-6} & 
$3P$ & 0.575 & 0.478 & 0.376 & 0.438 \\ \hline
\multirow{3}{*}{0.32} & 
$1P$ & 0.725 & 0.594 & 0.463 & 0.542 \\ \cline{2-6} & 
$2P$ & 0.669 & 0.554 & 0.435 & 0.507 \\ \cline{2-6} & 
$3P$ & 0.642 & 0.534 & 0.421 & 0.490 \\ \hline
\multirow{3}{*}{0.34} & 
$1P$ & 0.805 & 0.660 & 0.515 & 0.603 \\ \cline{2-6} & 
$2P$ & 0.742 & 0.615 & 0.484 & 0.564 \\ \cline{2-6} & 
$3P$ & 0.711 & 0.593 & 0.468 & 0.544 \\ \hline
\multirow{3}{*}{0.36} & 
$1P$ & 0.889 & 0.730 & 0.570 & 0.667 \\ \cline{2-6} & 
$2P$ & 0.818 & 0.679 & 0.535 & 0.623 \\ \cline{2-6} & 
$3P$ & 0.784 & 0.654 & 0.517 & 0.600 \\ \hline
\end{tabular}
\caption{\label{tab:pert_delNC}
Perturbative QCD results for 
non-Coulombic corrections to the $P$-wave wavefunctions at the origin 
$\delta_\Psi^{\rm NC}$ 
in the $\overline{\rm MS}$ scheme at scale $\Lambda=m$ for $1P$, $2P$, and $3P$
states. 
}
\end{table}
%%%%%%%%%%%%%%%%%%%%%%%%%%%%%%%%%%%%%%%%%%%%%%%%%%%%%%%%%%%%%%%%%%%%%%%%%%%%%%%

%\begin{thebibliography}{99}
\bibliography{pwave_drpnrqcd_v3.bib}
\bibliographystyle{JHEP}

%\end{thebibliography}
\end{document}